\documentclass{article}
\usepackage{arxiv}
\usepackage[utf8]{inputenc} 
\usepackage[T1]{fontenc}    
\usepackage{url}            
\usepackage{booktabs}       
\usepackage{amsfonts}       
\usepackage{nicefrac}       
\usepackage{microtype}      
\usepackage{graphicx}
\usepackage{placeins}
\usepackage{amsmath}
\usepackage{amssymb}
\newcommand{\emb}{\mathcal E}
\newcommand{\R}{{\mathbb R}}
\DeclareMathOperator*{\argmin}{argmin}

\title{Evaluating Node Embeddings of Complex Networks}

\author{
Arash Dehghan-Kooshkghazi\thanks{Department of Mathematics, Toronto Metropolitan University, Toronto, ON, Canada; e-mail: \texttt{arash.dehghan@ryerson.ca}}
\And
Bogumi\l{} Kami\'nski\thanks{Decision Analysis and Support Unit, SGH Warsaw School of Economics, Warsaw, Poland; e-mail: \texttt{bogumil.kaminski@sgh.waw.pl}}
\And
\L{}ukasz Krai\'nski\thanks{Decision Analysis and Support Unit, SGH Warsaw School of Economics, Warsaw, Poland; e-mail: \texttt{lkrain@sgh.waw.pl}}
\And
Pawe\l{}~Pra\l{}at\thanks{Department of Mathematics, Toronto Metropolitan University, Toronto, ON, Canada; e-mail: \texttt{pralat@ryerson.ca}}
\And
Fran\c{c}ois Th\'eberge\thanks{Tutte Institute for Mathematics and Computing, Ottawa, ON, Canada; email: \texttt{theberge@ieee.org}}
}

\date{\today}

\begin{document}

\maketitle

\begin{abstract}
Graph embedding is a transformation of nodes of a graph into a set of vectors. A~good embedding should capture the graph topology, node-to-node relationship, and other relevant information about the graph, its subgraphs, and nodes. If these objectives are achieved, an embedding is a meaningful, understandable, compressed representations of a network that can be used for other machine learning tools such as node classification, community detection, or link prediction. 

In this paper, we do a series of extensive experiments with selected graph embedding algorithms, both on real-world networks as well as artificially generated ones. Based on those experiments we formulate the following general conclusions. First, we confirm the main problem of node embeddings that is rather well-known to practitioners but less documented in the literature. There exist many algorithms available to choose from which use different techniques and have various parameters that may be tuned, the dimension being one of them. One needs to ensure that embeddings describe the properties of the underlying graphs well but, as our experiments confirm, it highly depends on properties of the network at hand and the given application in mind. As a result, selecting the best embedding is a challenging task and very often requires domain experts. Since investigating embeddings in a supervised manner is computationally expensive, there is a need for an unsupervised tool that is able to select a handful of promising embeddings for future (supervised) investigation. A general framework, introduced recently in the literature and easily available on GitHub repository, provides one of the very first tools for an unsupervised graph embedding comparison by assigning the ``divergence score'' to embeddings with a goal of distinguishing good from bad ones. We show that the divergence score strongly correlates with the quality of embeddings by investigating three main applications of node embeddings: node classification, community detection, and link prediction.
\end{abstract}

\section{Introduction}

Networks (often called graphs in mathematical literature, especially in combinatorics) are commonly used representations that are able to capture relational information within data such as friendships in social media, hyperlinks between web pages, or common interests between users. In some cases, the data is naturally represented as a network---consider, for example, a power grid network or a network of airline flight connections between airports. However, increasingly there is a need to capture contextual relations beyond the obvious ones, for example, one might want to model predecessor words in the text. Graphs allow us to specify a structure and a context of the problem and that structure should be included in the representation used in dedicated ML algorithms.

The goal of many machine learning applications is to make predictions or discover new patterns using graph-structured data as feature information. For example, one might want to better understand the role of a particular researcher within a collaboration network, similarity between users interacting on Amazon or Yelp, classify proteins in a biological interaction network, or recommend new users or products to users of some social media. As a result, the study of networks has emerged in diverse disciplines as a means of analyzing these complex relational data. Capturing aspects of a complex system as a graph can bring physical insights and predictive power~\cite{Newman_book,Barabasi2016,kaminski2021mining}. 

Network Geometry is a rapidly developing approach in Network Science~\cite{Bianconi2015} which further abstracts the system by modelling the nodes of the network as points in a geometric space. There are many successful examples of this approach that include latent space models~\cite{Hoff2002}, and connections between geometry and network clustering and community structure~\cite{Krioukov2016,Zuev2015}. Very often, these geometric representations naturally correspond to physical space, such as when modelling wireless networks or when networks are embedded in some geographic space~\cite{Gastner2006,Expert2011}. See~\cite{Janssen2010} for more details about applying spatial graphs to model complex networks.

Regardless of whether there is any natural interpretation or not, in order to extract useful structural information from graphs, one might want to try to embed them in a geometric space by assigning coordinates to each node such that nearby nodes are more likely to share an edge than those far from each other. In particular, in the case of link prediction, a good embedding should have the property that most of the network’s edges can be predicted from the coordinates of the nodes. On the other hand, in the case of node classification, one might want to include information about the global position of a node in the graph or the structure of the node's local graph neighbourhood. Other applications might require different properties to be preserved. As a result, there are many embedding algorithms (based on techniques from linear algebra, random walks, or deep learning) and the list constantly grows. Moreover, many of these algorithms have various parameters that can be carefully tuned to generate embeddings in some multidimensional spaces, possibly in different dimensions. Hence, unfortunately, in the absence of a general-purpose representation for graphs, graph embedding very often requires domain experts to craft features or to use specialized feature selection algorithms.

\medskip

Though graph embeddings continue to gain importance and popularity, there exist other tools and techniques to handle relational data directly. However, it is often more efficient to create an embedding that can be seen as a form of feature engineering in which each node is mapped to a feature vector. This approach has a number of advantages. For example, if additional node features are available, then they can simply be merged with the embedded node representations to form a richer representation of the data. Moreover, there are many scalable machine learning tools available to handle datasets represented as feature vectors. The many applications of graph embedding include link prediction (finding missing edges, e.g.\ identifying which researchers are likely to write a paper together, or which book should be recommended to a given reader), node classification (e.g.\ predicting the unknown label for some nodes given known labels for other nodes), clustering (e.g.\ finding groups of users with common interest). Embedding in low dimension is also useful for visualization (in 2 or 3 dimensions) and data compression. Several applications of graph embeddings to real world problems are described in~\cite{survey}, including computer vision, recommender systems, and natural language processing (NLP), to name a few. Another comprehensive review of important applications of embeddings can be found in~\cite{hamilton2018}, including those in visualization, community detection, node classification, and link prediction tasks. While most of the applications are inherently transductive, inductive approaches have also been proposed recently. For example, in GraphSAGE~\cite{hamilton2017} node feature data can be used to generate embeddings for previously unseen nodes. The idea behind embeddings can also be linked to propositionalization techniques developed in the Inductive Logic Programming (ILP) field, see e.g.~\cite{lavrac2020}. The major difference between these two approaches is that propositionalization typically encodes data in symbolic space while embeddings work in numeric space. The consequence is that embeddings, typically at the cost of lower interpretability, have lower sparsity of the representation. As a result, they are more space efficient and so can be easily used as a source features for modern machine learning algorithms such as deep neural networks. Thus, arguably, embeddings are considered to be more promising in terms of predictive performance, efficiency and scalability~\cite{lavrac2020}. Having said that, as discussed earlier, one may easily enrich the embeddings by including some domain-specific, high-level relations constructed using ILP which cannot be easily discovered by other means. This, in turn, might improve the quality of algorithms that use embeddings.

\medskip

Let us now highlight the main contribution of this paper. Embeddings prove to be useful in various scenarios but, in order for algorithms designed for a specific application to perform well, it is crucial to feed them with an appropriate, high-quality, embedding. Indeed, the well-known concept in data science / machine learning---“garbage in, garbage out” (GIGO)---says that flawed or nonsense input data produces nonsense output. Selecting the best embedding can be done in a supervised way by careful examination of all potential candidates. Unfortunately, there are more than 100 embedding algorithms having their own parameters that can be tuned, the dimension being only one of them. Moreover, most of these algorithms are randomized and stability is rather low---the quality of resulting embeddings varies even when the set of parameters is fixed. Unfortunately, this creates a huge computational challenge.
Here are the main questions we try to answer in this research project. 
\emph{Is there a way to evaluate a large family of embeddings in an unsupervised way?}
If this is possible, even if the evaluation is not perfect, then it would be a useful tool that would allow the analyst to pre-select a few embeddings for future, supervised, and more careful investigation. But, it raises the followup question:
\emph{Is it safe to trust this pre-selection process? Are we sure that no good quality embedding is lost?}
For many other unsupervised machine learning tasks such as clustering, dimensionality reduction, or graph community detection, unsupervised quality measures have been proposed in the literature (e.g.\ silhouette measure for evaluation of results of cluster analysis~\cite{sil}). We argue that such approach may also be useful for evaluation of graph embeddings.

In order to answer these questions, we perform a detailed study of a number of popular graph embedding algorithms: \textbf{node2vec}, \textbf{VERSE}, \textbf{LINE}, \textbf{Deep Walk}, \textbf{HOPE}, and \textbf{SDNE} (see Section~\ref{sec:algo}). We evaluate embeddings using a general framework that assigns the “divergence score” to each embedding which, in an unsupervised learning fashion, distinguishes good from bad embeddings (see Section~\ref{sec:framework}). According to our knowledge, this framework is the very first attempt of unsupervised evaluation of graph embeddings but, since they gain importance quickly, it is expected to see more approaches in the near future. We start with experiments on real-world networks (see Section~\ref{sec:real_world}) but in order to understand how basic statistics affect the ``divergence score'' of embeddings we also perform a series of tests on synthetic graphs generated by the \textbf{ABCD} model, similar to the well-known and widely used \textbf{LFR} (see Section~\ref{sec:ABCD}). In particular, the summary of our experiments and conclusions for practitioners can be found in Sub-section~\ref{sec:summary}. In short, if one needs to pick one embedding algorithm before running the experiments, then \textbf{node2vec} is the best choice as it performed best in our tests. Having said that, there is no single winner in all tests.

Finally, let us mention that evaluating embedding algorithms is a subjective task. Our experiments are based on the ``divergence score'' that proved to be a useful tool in a number of applied projects we were personally involved with but clearly we are biased. Since it is the first measure introduced in the literature, it is important to answer the second question raised earlier: \emph{Can we trust the framework?} In order to convince readers without prior experience with the framework, we finish the paper with a few experiments to show that there is a strong correlation between the ``divergence score'' and the quality of the selected machine learning tools that use embeddings as an input: classification, community detection, and link prediction (see Section~\ref{sec:justification}). We hope that after reading the paper the conclusion will be apparent: it is best to generate a relatively large family of embeddings (possibly using various algorithms and suitably tuned parameters) and then use the benchmarking framework to pre-select a few candidates. Then, one should make an informed decision, taking into account a trade-off between the quality of the embedding and its dimension that affects the speed and memory requirement. Using the framework is especially recommended in unsupervised learning contexts, for example, anomaly or community detection. 

The results presented in this paper are closely related to our earlier work presented in \cite{Embedding_Complex_Networks,Embedding_Complex_Networks2} where we introduced the ``divergence score''. Our earlier papers focused on motivations behind the design of the proposed measure, its mathematical derivation, and performance considerations of its computation. In this paper, we extend the previously published results in two ways. First, we apply the ``divergence score'' to draw empirical conclusions about properties of embeddings generated by selected popular algorithms. We check how stable the results are by varying both hyperparameters of embedding algorithms as well as the structure and size of the embedded graphs. Second, we compare the recommendations obtained based on the ``divergence score'' against their performance in a number of classical graph mining applications (such as node classification, community detection, and link prediction) to confirm that, indeed, it is a useful tool allowing to rank embeddings.


\section{Node Embedding Algorithms}\label{sec:algo}

There are over 100 algorithms proposed in the literature for node embeddings which are based on various approaches such as random walks, linear algebra and deep learning~\cite{GoyalFerrara2018}. 
Moreover, many of these algorithms have various parameters that can be carefully tuned to generate embeddings in some multidimensional spaces, possibly in different dimensions. 
For our experiments, we selected 6 popular algorithms that span different families.
All but one of them (\textbf{VERSE}\footnote{\texttt{https://github.com/xgfs/verse/}}) are taken from the \textbf{OpenNE} framework\footnote{\texttt{https://github.com/thunlp/OpenNE}}.


The first two algorithms, \textbf{Deep Walk}~\cite{DeepWalk} and \textbf{node2vec}~\cite{node2vec}, are based on random walks performed on the graph. This approach was successfully used in \textbf{Natural Language Processing} (\textbf{NLP}); for example the \textbf{Word2Vec} algorithm~\cite{word2vec} is based on the assumption that ``words are known by the company they keep''. For a given word, embedding is achieved by looking at words appearing close to each other as defined by context windows (groups of consecutive words). For graphs, the nodes play the role of words and ``sentences'' are constructed via random walks. The exact procedure how one performs such random walks differs for the two algorithms we selected.

In the \textbf{Deep Walk} algorithm, the family of walks is sampled by performing random walks on graph $G$, typically between 32 and 64 per node, and for some fixed length. The walks are then used as sentences. For each node $v_i$, the algorithm tries to find an embedding $e_i$ of $v_i$ that maximizes the approximated likelihood of observing the nodes in its context windows obtained from generated walks, assuming independence of observations. 
We set all parameters to their default values, namely, 
number of walks: 10,
walk length: 80,
workers: 8,
window size: 10.

In \textbf{node2vec}, biased random walks are defined via two main parameters.
The {\it return parameter} ($p$) controls the likelihood of immediately revisiting a node in the random walk. Setting it to a high value ensures that we are less likely to sample an already-visited node in the following two steps.
The {\it in-out parameter} ($q$) allows the search to differentiate between inward and outward nodes so we can smoothly interpolate between breadth-first-search (BFS) and depth-first search (DFS) exploration. 
We set all parameters to their default values, namely, 
number of walks: 10,
walk length: 80,
workers: 8,
window size: 10,
p: 1,
q: 1.


There are several deep learning methods successfully used for embedding nodes in a graph. One of those, \textbf{Structural Deep Network Embedding} (\textbf{SDNE})~\cite{SDNE}, is an \textbf{autoencoder}, a type of artificial neural network that is a commonly used deep learning model for representing complex objects such as images. The goal is to represent objects of interest in lower dimension in such a way that the original object can be reconstructed as best as possible from its low dimensional vector representation. Autoencoders are trained to minimize reconstruction errors (such as squared errors), often referred to as the loss function. \textbf{SDNE} aims at preserving both the first and the second order proximity: first order proximity is derived directly from weights of the edges while the second order indicates similarity between nodes' neighbourhoods.
We changed the number of neurons at each encoder layer from its default value of 1000 to 128, as it was consistently producing very poor results. The remaining parameters were set to their default values, namely, 
alpha: 1e-6,
beta: 5,
Nu1 (l1-loss of weights in autoencoder): 1e-5,
Nu2 (l2-loss of weights in autoencoder): 1e-4,
batch size: 200,
learning rate: 0.01.


Several embedding algorithms are based on linear algebra. 
The \textbf{High Order Proximity} preserved \textbf{Embedding} algorithm (\textbf{HOPE})~\cite{HOPE} is aimed at embedding nodes in directed graphs, but can also be used for undirected graphs. 
For every node $v_i$, we define two embeddings, $e_{s,i}$ and $e_{t,i}$, the source and, respectively, the target embedding. Let $\mathbf{E}_s$ and $\mathbf{E}_t$ be the corresponding matrices of the source and the target embeddings. The loss function for \textbf{HOPE}, for a given proximity matrix $\mathbf{S}$, is defined as follows:
$$
\Phi(\mathbf{E}_s, \mathbf{E}_t) = ||\mathbf{S}- \mathbf{E}_s^T\mathbf{E}_t||_F,
$$
where $||\cdot||_F$ is the \textbf{Frobenius norm} that is a natural and straightforward extension of the Euclidean norm to matrices.
There are several choices for the proximity measure matrix $\mathbf{S}$ such as Katz similarity, common neighbours, or Adamic-Adar. We used the default proximity measure, common neighbours.


The next algorithm, \textbf{Large-scale Information Network Embedding (LINE)}~\cite{LINE}, is an efficient method for node embedding which explicitly defines two functions to encode the first and the second order proximity.
In order to capture the first order proximity, the joint probability distribution is defined for a pair of nodes based on their embeddings.
The method is similar for the second order proximity. In this case, each node $v_i$ is assigned with a source and a target embedding vectors, $e_{s,i}$ and $e_{t,i}$, and the conditional probability distribution is considered for a target of a random edge sampled from the set of edges having one endpoint in $v_i$.
We set all parameters to their default values, namely, 
batch size: 1000,
epoch: 10,
negative ratio: 5,
order: 3,
label file: no labels used,
CLF ratio: 0.5,
auto save: true.


Finally, \textbf{VERtex Similarity Embeddings (VERSE)}~\cite{verse} is a simple, versatile, and memory-efficient method that derives graph embeddings explicitly calibrated to preserve the distributions of a selected node-to-node similarity measure. It is a general framework that learns any similarity measures among nodes via training a simple, yet expressive, single-layer neural network. This includes popular similarity measures such as personalized PageRank, SimRank, and adjacency similarity. We used the default proximity measure, personalized PageRank. We also set all remaining parameters to their default values, namely, 
alpha: 0.85,
learning rate: 0.0025,
threads: 4,
nsamples:~3.

\section{An Unsupervised Framework for Comparing Graph Embeddings}\label{sec:framework}

Evaluating graph embedding algorithms is a challenging task. This subjective process depends on a specific application of the embedding at hand, and typically requires ad-hoc experiments and tests performed by the domain experts. However, in the recent papers~\cite{Embedding_Complex_Networks,Embedding_Complex_Networks2}, the ``divergence score'' was proposed that can be assigned to outcomes of the embedding algorithms to help distinguish good ones from bad ones. This general framework provides a tool for an unsupervised graph embedding comparison and is available at the GitHub repository\footnote{\texttt{https://github.com/KrainskiL/CGE.jl}}. To the best of our knowledge, it is the first tool of this nature.

In order to justify why we use the framework in our experiments and to build an intuition, let us try to answer the following related question: \emph{What do we expect from a good embedding?} One natural and desired property is to require that based on a good embedding one should be able to predict most of the network's edges from the coordinates of the nodes in the embedded space. One typically expects that if two nodes are far away from each other, then the chance they are adjacent in the graph is smaller compared to another pair of nodes that are close to each other. But, of course, in any real-world network there are some sporadic long edges and some nodes that are close to each other are not adjacent. Due to this fact, in the framework we use an embedding algorithm is not considered good when it only pays attention to local properties such as the existence of particular edges (microscopic point of view) but rather the expectation is that it is able to capture some global properties such as the number of edges induced by some relatively large subsets of nodes (macroscopic point of view). So, how one may evaluate if the global structure is consistent with our expectations and intuition without considering individual pairs of nodes?

The approach that is proposed works as follows. First, some dense parts of the graph need to be identified by a good graph clustering algorithm. By default, the framework uses the \textbf{Ensemble Clustering algorithm for Graphs} (\textbf{ECG}\footnote{\texttt{https://github.com/ftheberge/graph-partition-and-measures}}) which is based on the classical \textbf{Louvain} algorithm and the concept of consensus clustering~\cite{Poulin2019}. This algorithm is known to have good stability but the choice of graph clustering algorithm is flexible and it was empirically verified that it does not affect the outcome of the process as long as the set of nodes is partitioned into clusters such that there are substantially more edges captured within clusters than between them. The clusters that are found provide the desired macroscopic point of view of the graph. Note that for this task only information about the graph $G$ is used; in particular, the embedding is not used at all. We then consider the graph from a different point of view. Using the \textbf{Geometric Chung-Lu} (\textbf{GCL}) model, based on the degree distribution of the graph and the embedding, we compute the expected number of edges within each cluster found earlier, as well as between them. The embedding is scored by computing a divergence score between these expected number of edges and the actual number of edges present in the graph. To measure dissimilarity between the two corresponding probability distributions, the well-known and widely used Jensen--Shannon divergence measure was used. It can be viewed as a smoothed version of the Kullback-Leibler divergence. Let us also mention that GCL is defined for any dimension and so the framework is able to compare embeddings in different dimensions based exclusively on their predictive power, ignoring their complexities. See Appendix~\ref{sec:divergence} and \cite{Embedding_Complex_Networks,Embedding_Complex_Networks2} for more details.

\medskip

In order to see the framework ``in action'', we perform a small experiment with the well-known College Football real-world network with known community structures. This graph represents the schedule of United States football games between Division IA colleges during the regular season in Fall 2000~\cite{Football}. The data consists of 115 teams (nodes) and 613 games (edges). The teams are divided into conferences containing 8--12 teams each. In general, games are more frequent between members of the same conference than between members of different conferences, with teams playing an average of about seven intra-conference games and four inter-conference games in the 2000 season. There are a few exceptions to this rule, as detailed in~\cite{Lu2018}: one of the conferences is really a group of independent teams, one conference is really broken into two groups, and 3 other teams play mainly against teams from other conferences. We refer to those as {\it outlying} nodes, which we represent with a distinctive triangular shape.
 
\begin{figure}[ht]
\begin{center}
\includegraphics[width=8cm]{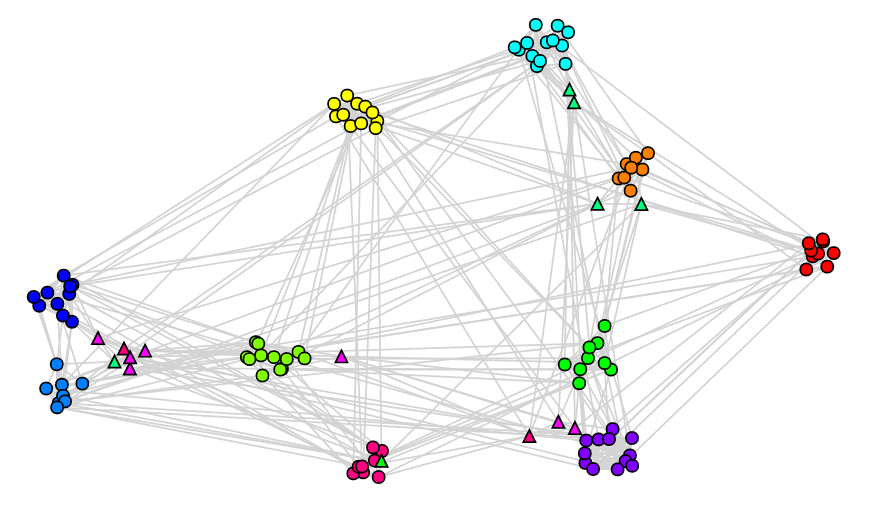}
\includegraphics[width=8cm]{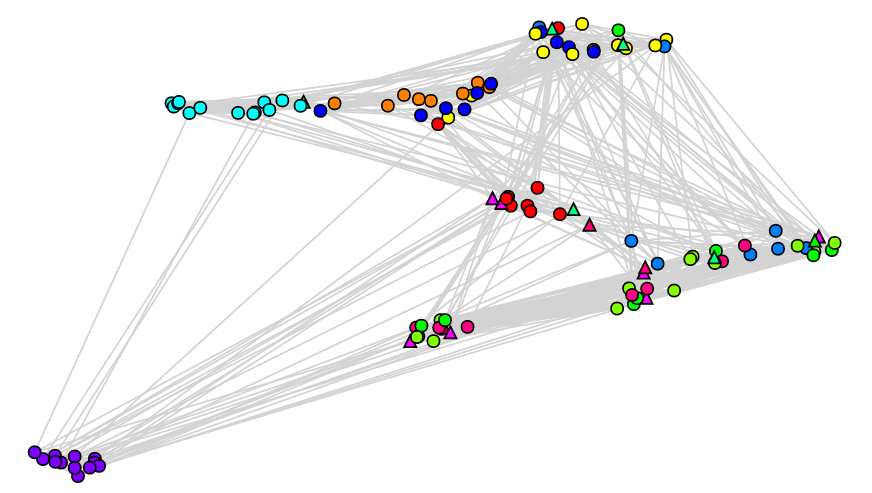}
\end{center}
\caption{The \textbf{College Football Graph}: we show the best (left) and the worst (right) scoring embedding.}
\label{fig:foot_best}
\end{figure}

In order to illustrate the application of the framework, we ran various embedding algorithms in different dimensions and sets of parameters on the Football dataset. In Figure~\ref{fig:foot_best}, we show the best and worst scoring embeddings based on the divergence score. The colours of nodes correspond to the conferences, and the triangular shaped nodes correspond to outlying nodes as observed earlier. The communities are very clear in the left plot while in the right plot, only a few communities are clearly grouped together.

To visualize the embeddings in high dimensions we needed to perform dimension reduction that seeks to produce a low dimensional representation of high dimensional data that preserves relevant structure. We used the \textbf{Uniform Manifold Approximation and Projection} (\textbf{UMAP}\footnote{\texttt{https://github.com/lmcinnes/umap}})~\cite{UMAP}, a novel manifold learning technique for dimension reduction. \textbf{UMAP} is constructed from a theoretical framework based in Riemannian geometry and algebraic topology. It provides a practical scalable algorithm that applies to real world datasets.

Finally, let us make a comment that not all embeddings proposed in the literature try to capture an information about edges. Some algorithms indeed try to preserve edges whereas others care about some different structural properties; for example, they might try to map together nodes with similar functions or role within the network. Having said that, many important applications a data scientist needs to deal with in everyday work require preserving (global) edge densities and the framework favours embeddings that do a good job from that perspective. We come back to this discussion in Section~\ref{sec:justification} and justify using the framework more.

\section{Experiments on Real-World Graphs}\label{sec:real_world} 

Let us start with evaluating the performance of the selected graph embedding algorithms run on a few real-world networks. For our experiments, we selected four networks. Before we briefly describe these datasets, let us present a few of their statistics.

\begin{table}[!htb]
\center
\begin{tabular}{|c|c|c|c|c|}
\hline
\textbf{Property}      & \textbf{Mouse Brain} & \textbf{Airports} & \textbf{Email-EU} & \textbf{Github} \\ \hline
Nodes                  & 1029                 & 464               & 986               & 37700           \\ \hline
Edges                  & 1700                 & 7595              & 16017             & 288996          \\ \hline
Density                & 0.00321              & 0.07071           & 0.03298           & 0.0041          \\ \hline
Maximum degree         & 153                  & 175               & 342               & 9458            \\ \hline
Minimum degree         & 1                    & 1                 & 1                 & 1               \\ \hline
Average degree         & 3.304                    & 32.737            & 32.489            & 15.331          \\ \hline
Assortativity          & -0.215               & -0.055            & -0.025            & -0.075          \\ \hline
Number of triangles    & 0                    & 100358            & 104395            & 523782          \\ \hline
Global clustering coefficient & 0                    & 0.476             & 0.266             & 0.012           \\ \hline
Maximum $k$-core         & 5                    & 50                & 34                & 34              \\ \hline
Number of components        & 20                    & 2                & 1                & 1              \\ \hline
Diameter         & 12                    & 7                & 7                & 11              \\ \hline
Average path length      & 4.913                    & 2.455                & 2.588                & 3.246              \\ \hline
\end{tabular}
\caption{Some statistics of the four networks we experimented with.}
\end{table}

\subsubsection*{Mouse Brain Graph\protect\footnote{\texttt{http://networkrepository.com/bn-mouse-kasthuri-graph-v4.php}}}

This graph represents the mouse brain. Nodes represent regions of the brain and edges represent neuronal fiber tracts that connect one node to another. One interesting feature of this graph (which, in fact, was the main reason to select this graph for our experiments) is that it contains no triangles, something that is very rare in general. Indeed, many social networks (and to a lesser degree other networks) exhibit relatively large clustering coefficient which can be described as the overall probability for the network to have adjacent nodes interconnected, thus revealing the existence of tightly connected communities (or clusters, subgroups, cliques). This network has the clustering coefficient equal to zero and so it will allow us to understand the effect of this graph parameter on the quality of the embedding algorithms.


\subsubsection*{Airports Graph\protect\footnote{\texttt{https://github.com/ftheberge/GraphMiningNotebooks/tree/master/Datasets/Airports}}}

This graph contains information about flights between airports based on a record of more than 3.5 million US Domestic Flights from 1990 to 2009. It has been taken from OpenFlights website which have a huge database of different travelling mediums across the globe. The nodes are represented by the 3-letter airport codes; the latitude and the longitude as well as the state and the city are also available. The edges are directed with weights representing the total volume of passengers between the two airports. 


\subsubsection*{GitHub Graph\protect\footnote{\texttt{https://github.com/benedekrozemberczki/MUSAE}}}

This graph is a large social network of GitHub developers which was collected from the public API in June 2019. Nodes correspond to developers who have starred at least 10 repositories and edges represent mutual follower relationships between them.  The node features are extracted based on the location, repositories starred, employer and e-mail address. In particular, the set of nodes was partitioned into web developers and machine learning developers, feature derived from the job title of each user. As a result, this network is suitable for experiments on binary node classification---one might want to predict whether the GitHub user is a web or a machine learning developer. We ignore this partition in our experiments and work with the entire graph.


\subsubsection*{Email-EU Graph\protect\footnote{\texttt{https://snap.stanford.edu/data/email-Eu-core-temporal.html}}}

The network was generated using email data from a large European research institution. Emails are anonymized and there is an edge between $u$ and $v$ if person $u$ sent person $v$ at least one email. The dataset does not contain incoming messages from or outgoing messages to the rest of the world.  More importantly, it contains ``ground-truth'' community memberships of the nodes indicating which of 42 departments at the research institute individuals belong to. As a result, this dataset is suitable for experiments aiming to detect communities but we ignore this external knowledge in our experiments. 


\subsection{Specification of the Experiments and Results} 

Let us present the general approach that we use to test each of the four networks we selected to experiment with. For a given embedding algorithm $A$ and a given dimension $d \in \{4, 8, 16, 32, 64, 128\}$, we independently run the algorithm 30 times. This is done to not only measure how good the algorithms are but also their stability. Recall that 4 out of 6 algorithms we test are randomized. In order to do that, we compute the average divergence score $a_{A,d}$ and the standard deviation $s_{A,d}$. 

\begin{figure}[!htb]
\includegraphics[width=0.49\textwidth]{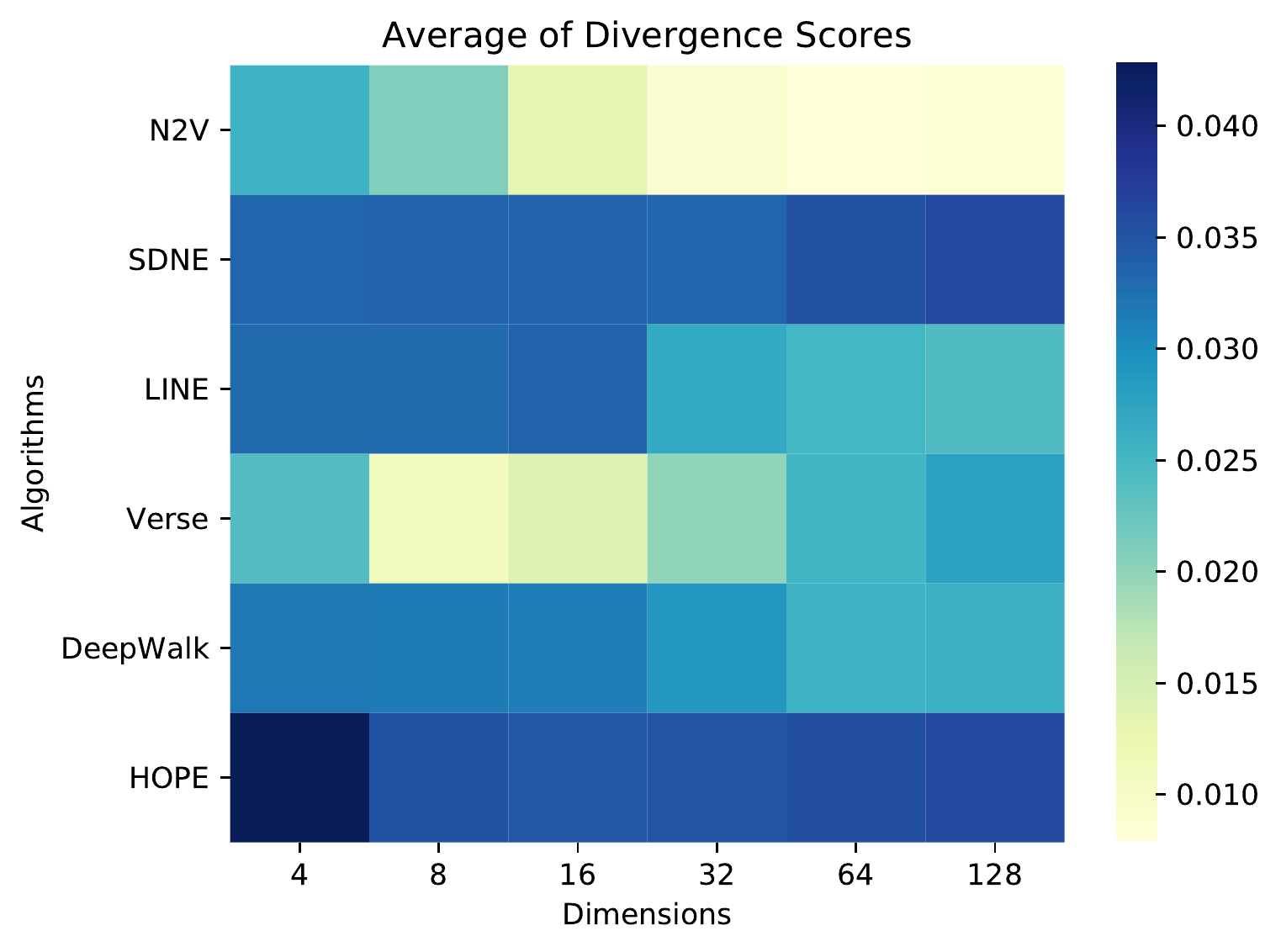}
\includegraphics[width=0.52\textwidth]{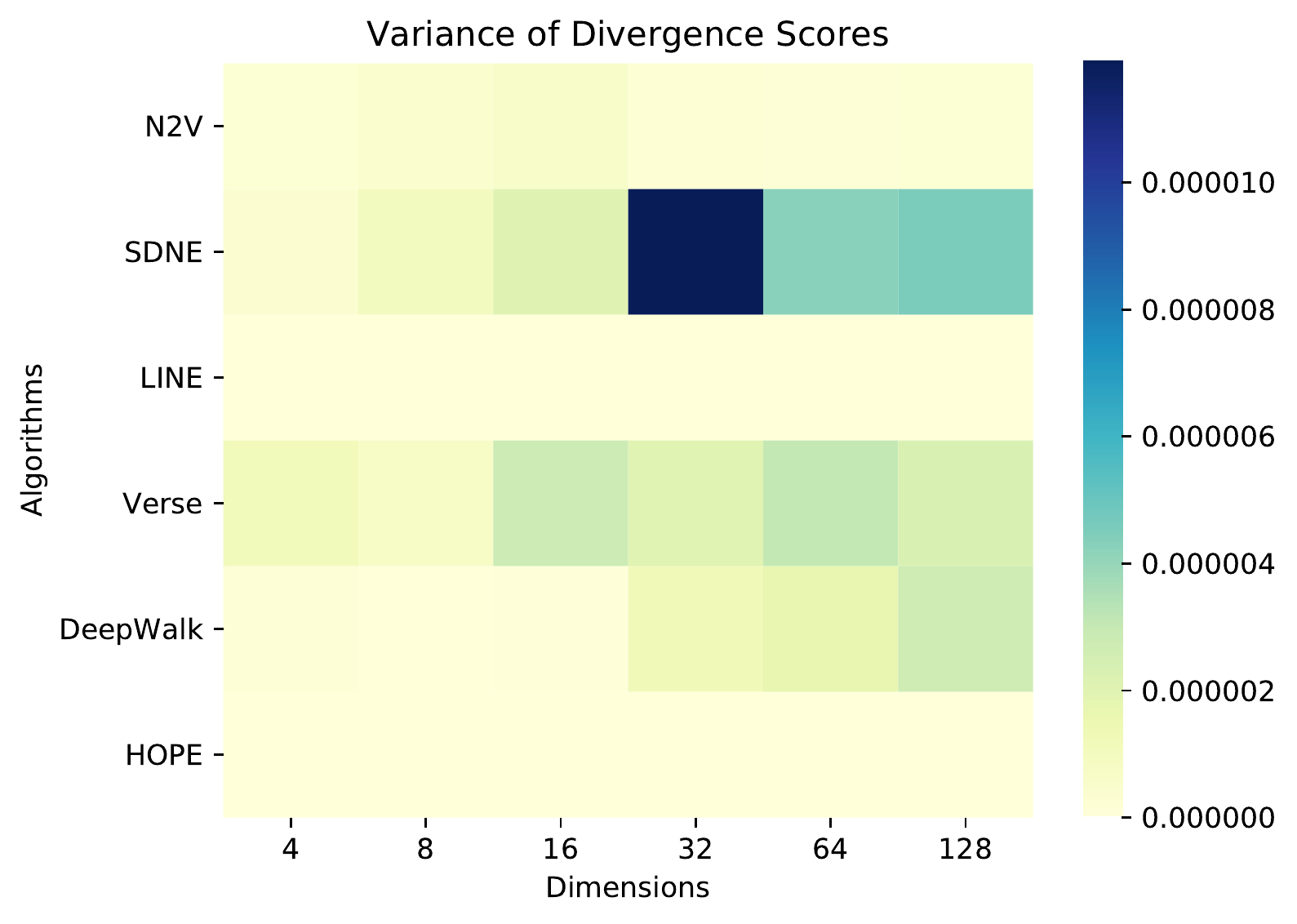}
\caption{\textbf{Mouse Brain Graph}: Average and Standard Deviation of the Divergence Score (Heat Map)}
\label{fig:Mouse}
\end{figure}

\begin{figure}[!htb]
\includegraphics[width=0.505\textwidth]{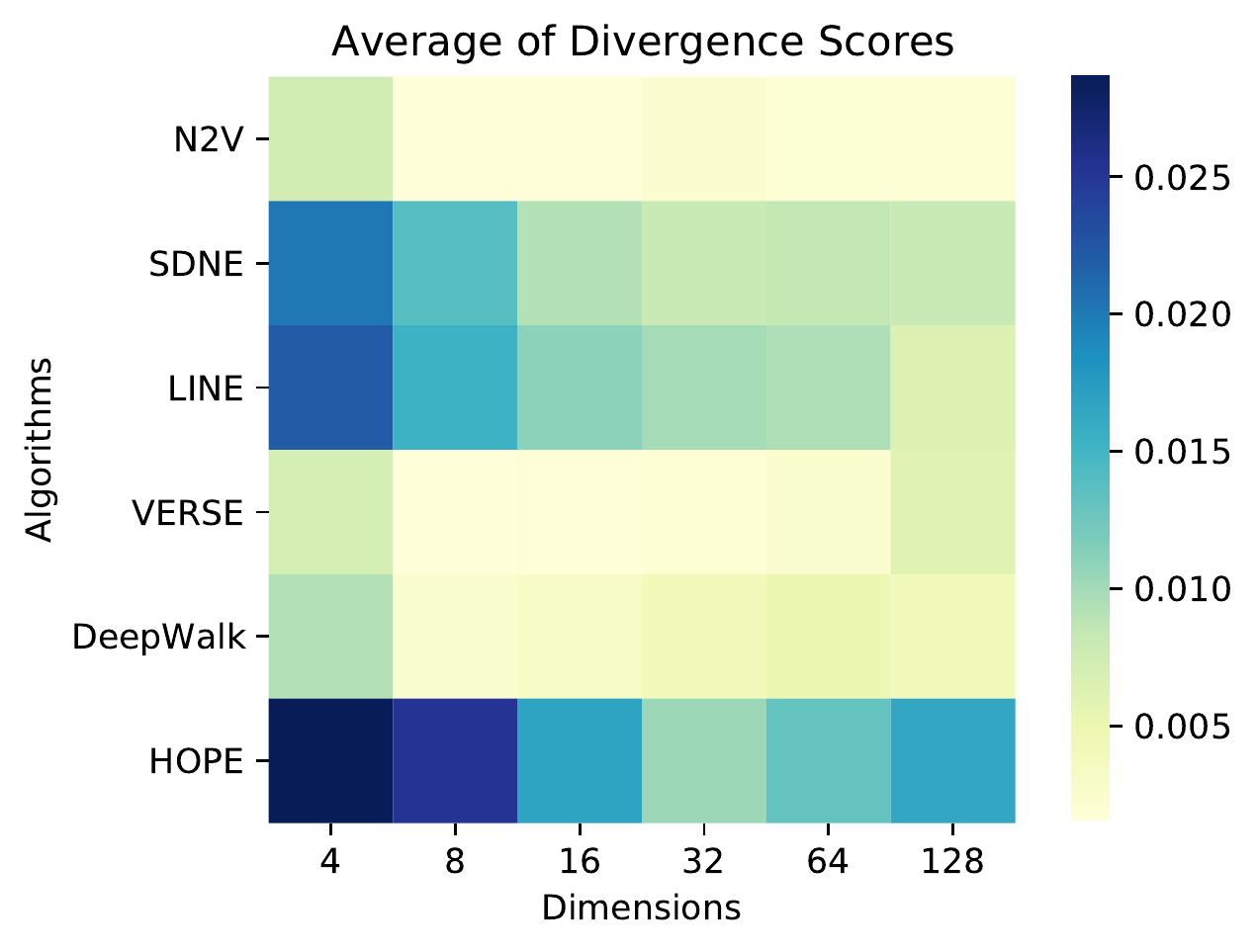}
\includegraphics[width=0.485\textwidth]{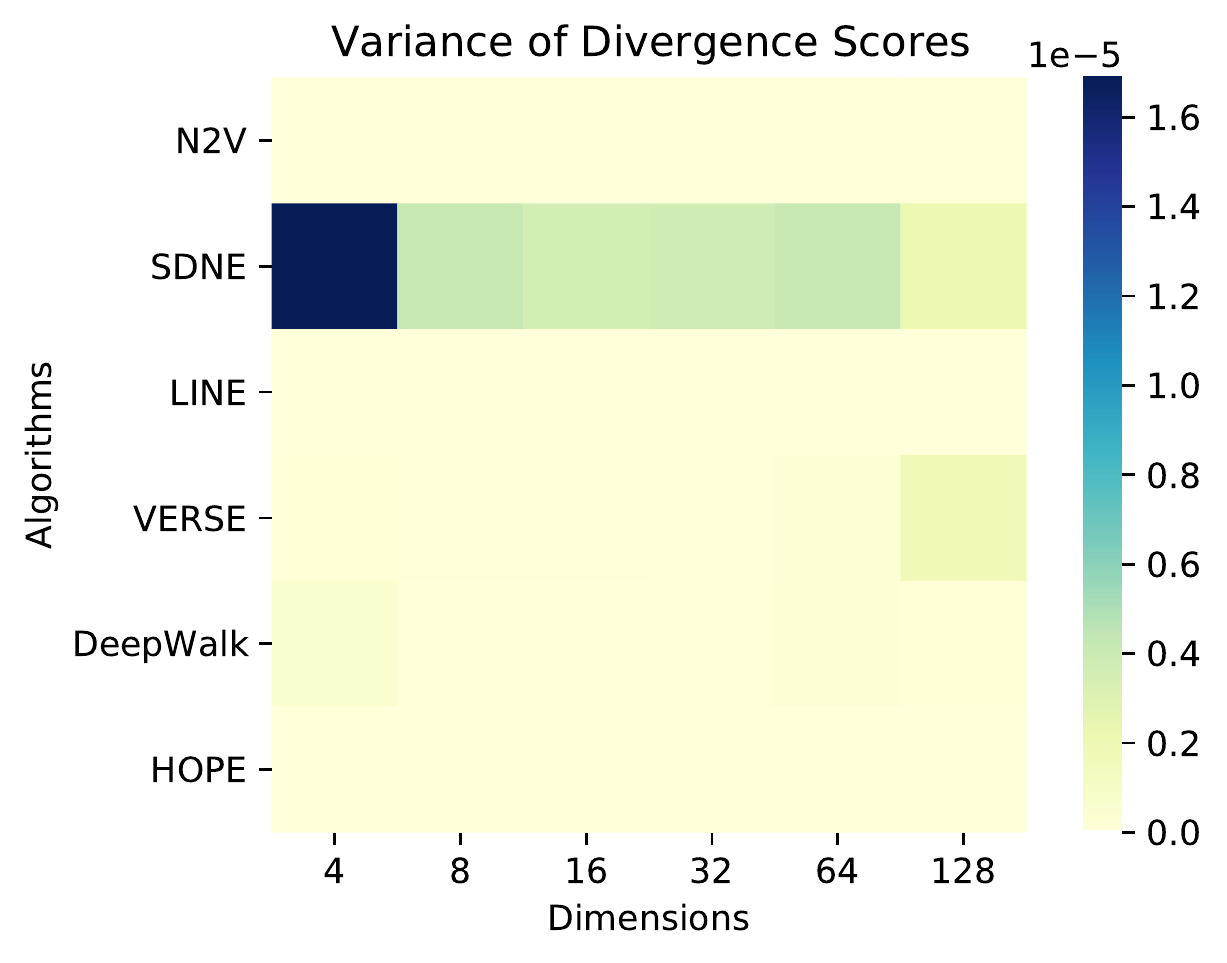}
\caption{\textbf{Email-EU Graph}: Average and Standard Deviation of the Divergence Score (Heat Map)}
\label{fig:email}
\end{figure}

The results are presented in Figures~\ref{fig:Mouse}--\ref{fig:email} in the form of a heat-map: for each algorithm $A$ ($y$ axis) and each dimension $d$ ($x$ axis), the corresponding square is presented in light colour if the divergence score $a_{A,d}$ is small (that is, the embedding scores well according to the benchmark framework), and dark colours are used if the divergence score is large (that is, the embedding does not score well). The same approach is used to visualize the behaviour of the standard deviation $s_{A,d}$.


\begin{figure}[!htb]
\includegraphics[width=0.505\textwidth]{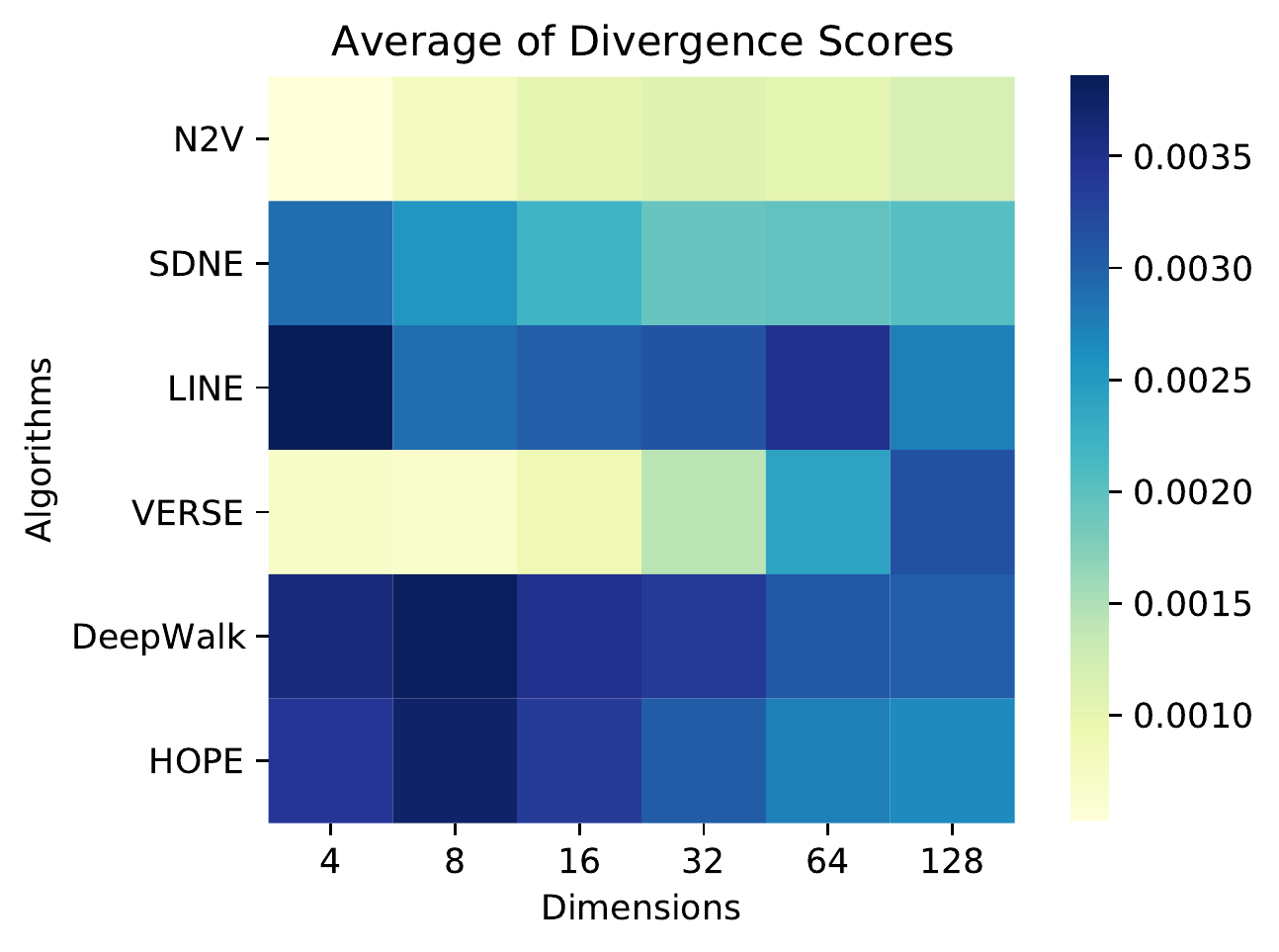}
\includegraphics[width=0.485\textwidth]{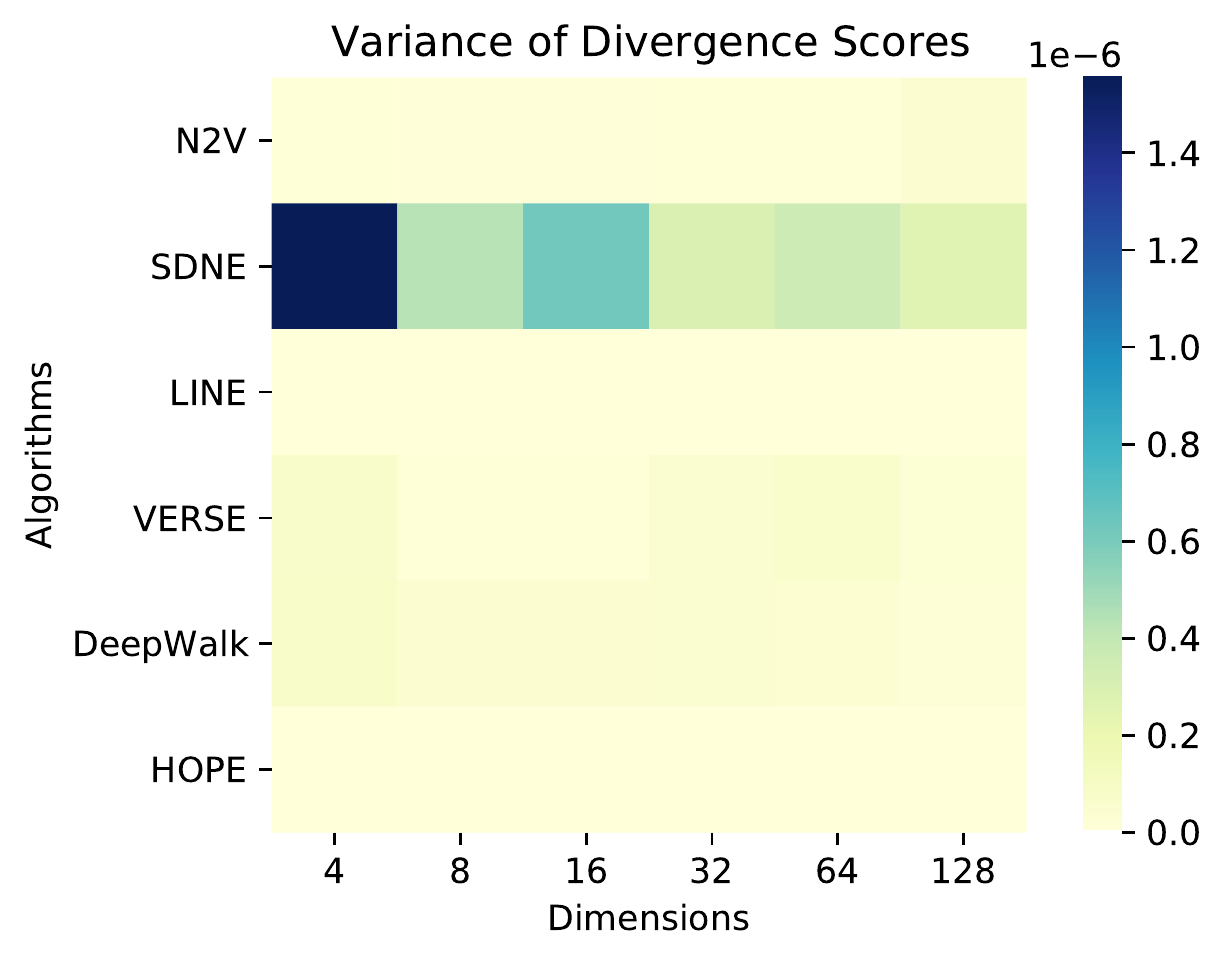}
\caption{\textbf{Airports Graph}: Average and Standard Deviation of the Divergence Score (Heat Map)}
\end{figure}

\begin{figure}[!htb]
\includegraphics[width=0.52\textwidth]{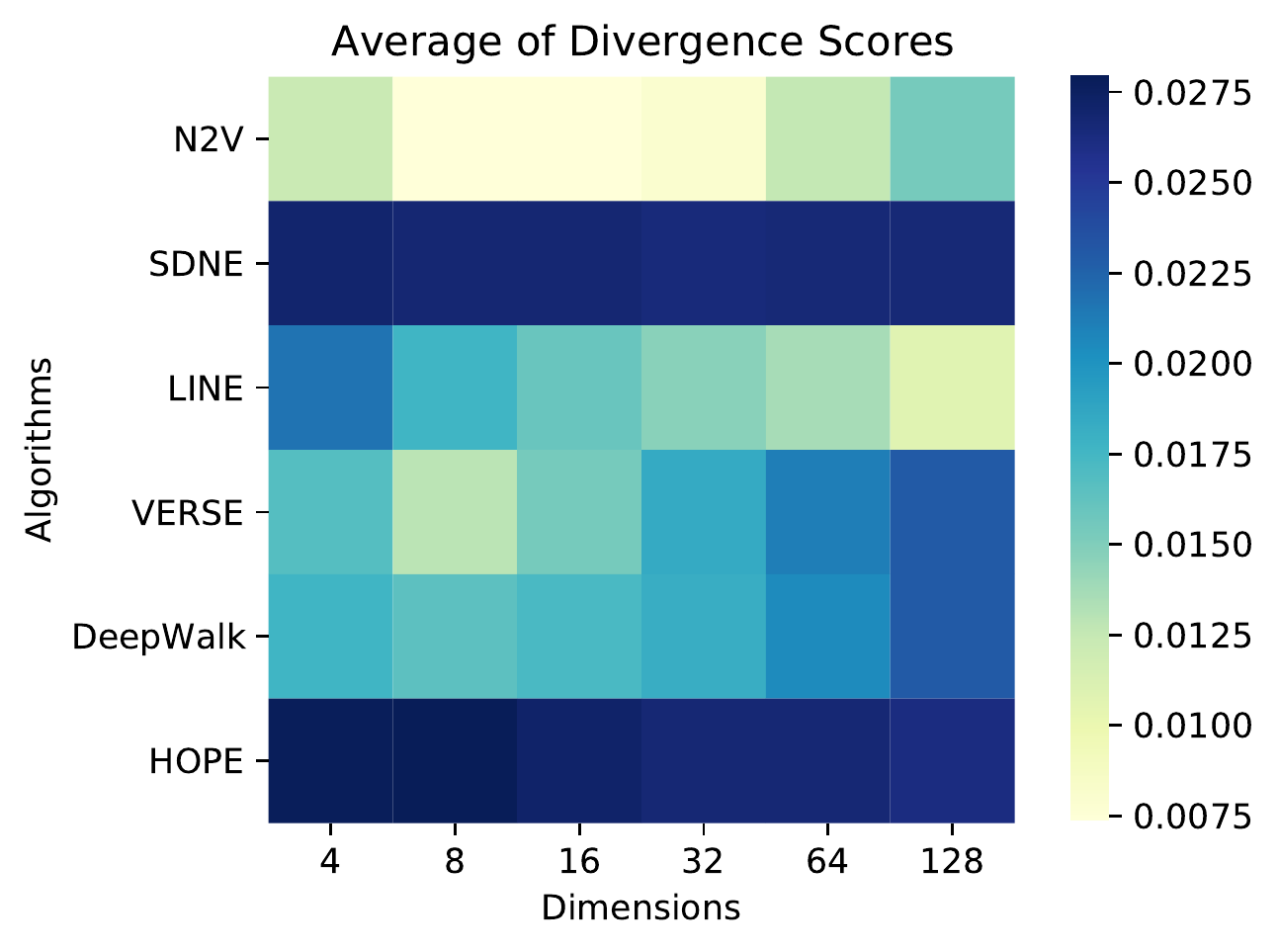}
\includegraphics[width=0.48\textwidth]{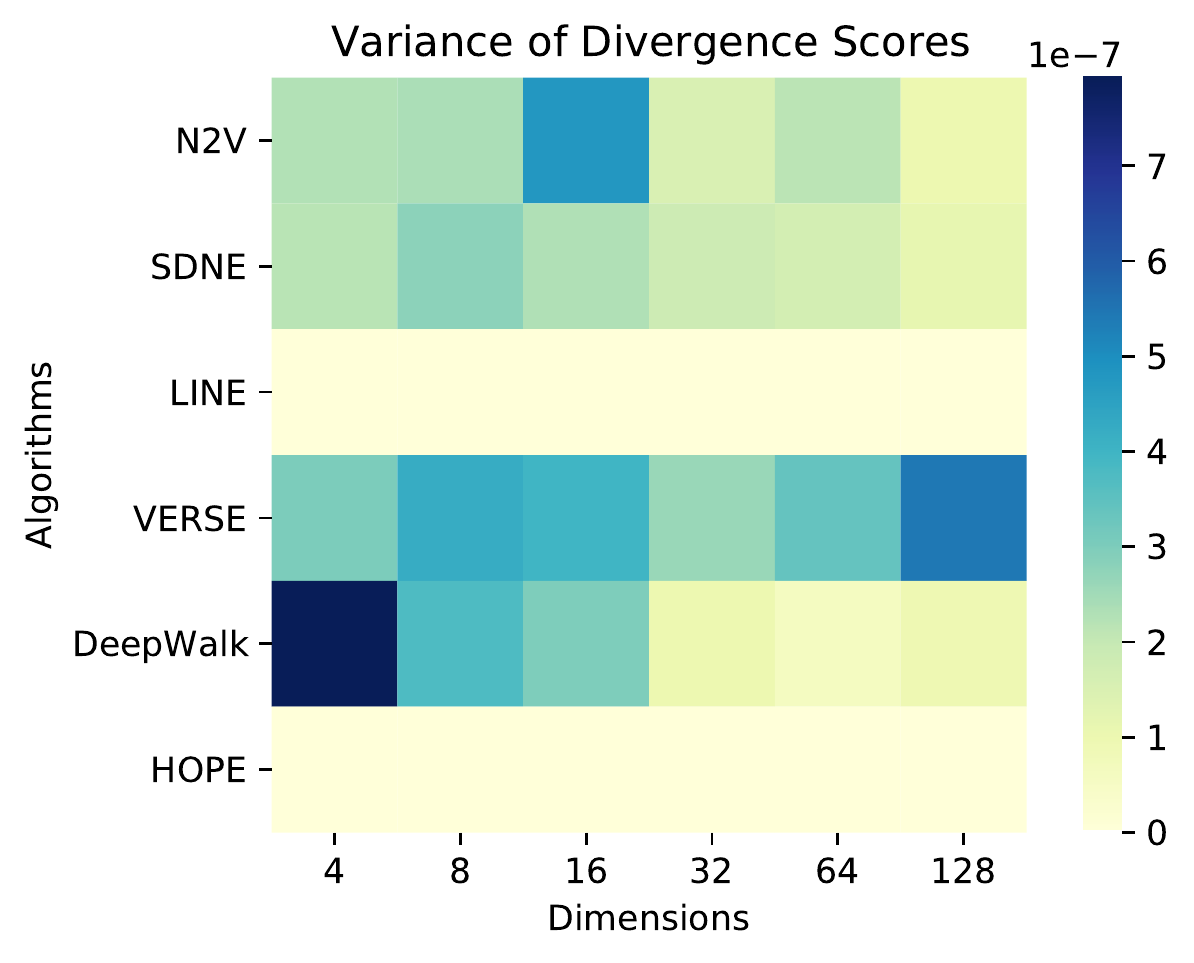}
\caption{\textbf{GitHub Graph}: Average and Standard Deviation of the Divergence Score (Heat Map)}
\end{figure}

The results are generally improving when the dimension increases, with the exception of the \textbf{VERSE} algorithm that, surprisingly, seems to perform better in lower dimensions. Dimensions 4 and 8 are too small to capture the ``shape'' of the networks that is manifested via relatively large divergence scores obtained for such embeddings. In all four networks, \textbf{node2vec} consistently generated embeddings that scored the best with \textbf{VERSE} (in lower dimensions) taking the second place. \textbf{HOPE} end \textbf{SDNE} did not perform very well which might suggest that such algorithms do not aim to preserve global densities (property that the benchmarking framework tries to evaluate) but some other aspects of the network. The score of the \textbf{DeepWalk} algorithm varies a lot from network to network; it is very bad for the Airport Graph but performs well on the Email-EU Graph. With regards to stability, \textbf{SDNE} seems to be the least stable of the algorithms we tested and \textbf{node2vec} wins one more time---it seems to be not only consistently good but also quite stable.

\section{Experiments on Synthetic Graphs} \label{sec:ABCD}

As commonly done in the literature as well as in the applied world, we analyze how the selected embedding algorithms perform on artificially constructed networks with communities (see, for example,~\cite{Lu2018}). By doing this we may flexibly change the characteristics of the network (such as its size, the number of clusters, the degree distribution, etc.) and assess the impact of these changes on the results. A popular example of such network generator is the \textbf{LFR} benchmark proposed by Lancichinetti, Fortunato, and Radicchi~\cite{LFR} that produces synthetic graphs resembling real world graphs. For our experiments, we use an alternative random graph model, namely, the \textbf{A}rtificial \textbf{B}enchmark for \textbf{C}ommunity \textbf{D}etection (\textbf{ABCD}\footnote{\texttt{https://github.com/bkamins/ABCDGraphGenerator.jl}}) graph~\cite{ABCD}. In both benchmarks, the size of each community is drawn from a power-law distribution, as is the degree of each node. As a result, both benchmarks produce graphs with similar properties. The main reason for using \textbf{ABCD} instead of \textbf{LFR} is that the mixing parameter $\mu$, the main parameter of the \textbf{LFR} model guiding the strength of the communities, has a non-obvious interpretation and so can lead to unnaturally-defined networks. Another reason is that \textbf{ABCD} is faster than \textbf{LFR} and can be easily parallelized~\cite{ABCDe} (\textbf{ABCDe}\footnote{\texttt{https://github.com/tolcz/ABCDeGraphGenerator.jl}}). Moreover, due to its simplicity, it is possible to analyze it theoretically~\cite{ABCD-theory}. Such results, despite the fact that often asymptotic in nature, may shed some light on the behaviour of embedding algorithms on real-world networks. See Appendix~\ref{sec:ABCDapp} for more details.

\subsection{Parameters of the Model}

The \textbf{ABCD} model has a number of parameters that can be independently tuned and so it is suitable for testing which properties of real networks affect the quality of embedding algorithms. In order to do that, we fix all parameters but one and then investigate how sensitive the algorithms are with respect to the selected parameter. Of course, it might be the case that the quality of a given algorithm depends on some specific combination of parameters but such more subtle correlations are more challenging to detect and so it is left for a future research. Here are the parameters we want to investigate as well as their default values.

\begin{itemize}
\item \textbf{Size of the network}:  
$n$ is the number of nodes in the graph. The default value is $n=10{,}000$.
\item \textbf{Degree distribution}: 
$\gamma$ is the (negative) exponent of the power-law degree distribution. The default value is $\gamma=2.5$. The default degree sequence is generated \emph{in advance} and used for \emph{all} experiments that use the default settings of $n=10{,}000$, and $\Delta \approx n^{1/(\gamma-1)}$ (the maximum degree). It is generated with $\gamma=2.5$, $\delta=5$ (the minimum degree) and $\Delta=464 \approx n^{1/(\gamma-1)} = n^{2/3}$. Of course, if $n$ or $\Delta$ changes, then the degree sequence has to be re-generated but it is then used for all experiments with that choice of parameters $n$ and $\Delta$. 
\item \textbf{Maximum degree}: 
$\Delta$ is the maximum degree in the graph. The default value is $\Delta \approx n^{1/(\gamma-1)}$ which corresponds to the so-called \textbf{natural cut-off}. This specific value ensures that the expected number of nodes of degree at least $\Delta$ is close to 1.
\item \textbf{Level of noise}: 
$\xi$ is the mixing parameter that controls the fraction of edges between communities.  Essentially, this parameter may be viewed as the amount of noise in the graph. In one extreme case, if $\xi=0$, then all the edges are within communities. On the other hand, if $\xi=1$, then communities are not present in the graph and edges are simply wired randomly, regardless of the assignment of nodes into communities. The default value for our experiments is $\xi=0.2$.
\item \textbf{Community sizes}: 
$\beta$ is the (negative) exponent of the distribution of community sizes. In order to test other parameters in a rather easy set-up, instead of generating the sequence of community sizes randomly, by default we simply consider 5 large communities with the following distribution: 30\%, 25\%, 20\%, 15\%, 10\%. More importantly, this choice reduces the contribution to the variance from the \textbf{ABCD} model and, as a result, the experiments concentrate on the stability of the embedding algorithms used instead of the random graph model. We discuss it more in Subsection~\ref{sec:beta}.
\end{itemize}

There are a few other parameters of the \textbf{ABCD} model that we do not investigate as they should not substantially affect the behaviour of the embedding algorithms. As mentioned above, the minimum degree is set to be $\delta = 5$, the minimum and the maximum community sizes are set to be  50 and 1000, respectively. The configuration model was used to generate underlying graphs with the global variant of the model. For more details about the \textbf{ABCD} model we direct the reader to~\cite{ABCD}.

The synthetic networks used in our experiments are relatively small ($n=10{,}000$). The reason for this is that we test a large grid of parameters and for a given set of parameters, a large number of experiments need to be performed due to two sources of randomness, the first one associated with random graphs \textbf{ABCD} and the other associated with randomized embedding algorithms (see below for more details). As a result, the experiments took approximately 20{,}000 vCPU-hours and larger graphs would be impossible to investigate with such precision (see the last section for more details). In order to make sure generated graphs are large enough to capture the behaviour of both synthetic networks and associated embeddings well, we performed a few selected experiments on larger networks and the outcome was consistent with earlier findings. We direct the interested reader to the GitHub repository with all results (see the last section). Let us also stress the fact that the bottleneck is with generating thousands of embeddings, not with evaluating them by the framework which is scalable~\cite{Embedding_Complex_Networks2}.


\bigskip

\begin{figure}[!htb]
\includegraphics[width=0.48\textwidth]{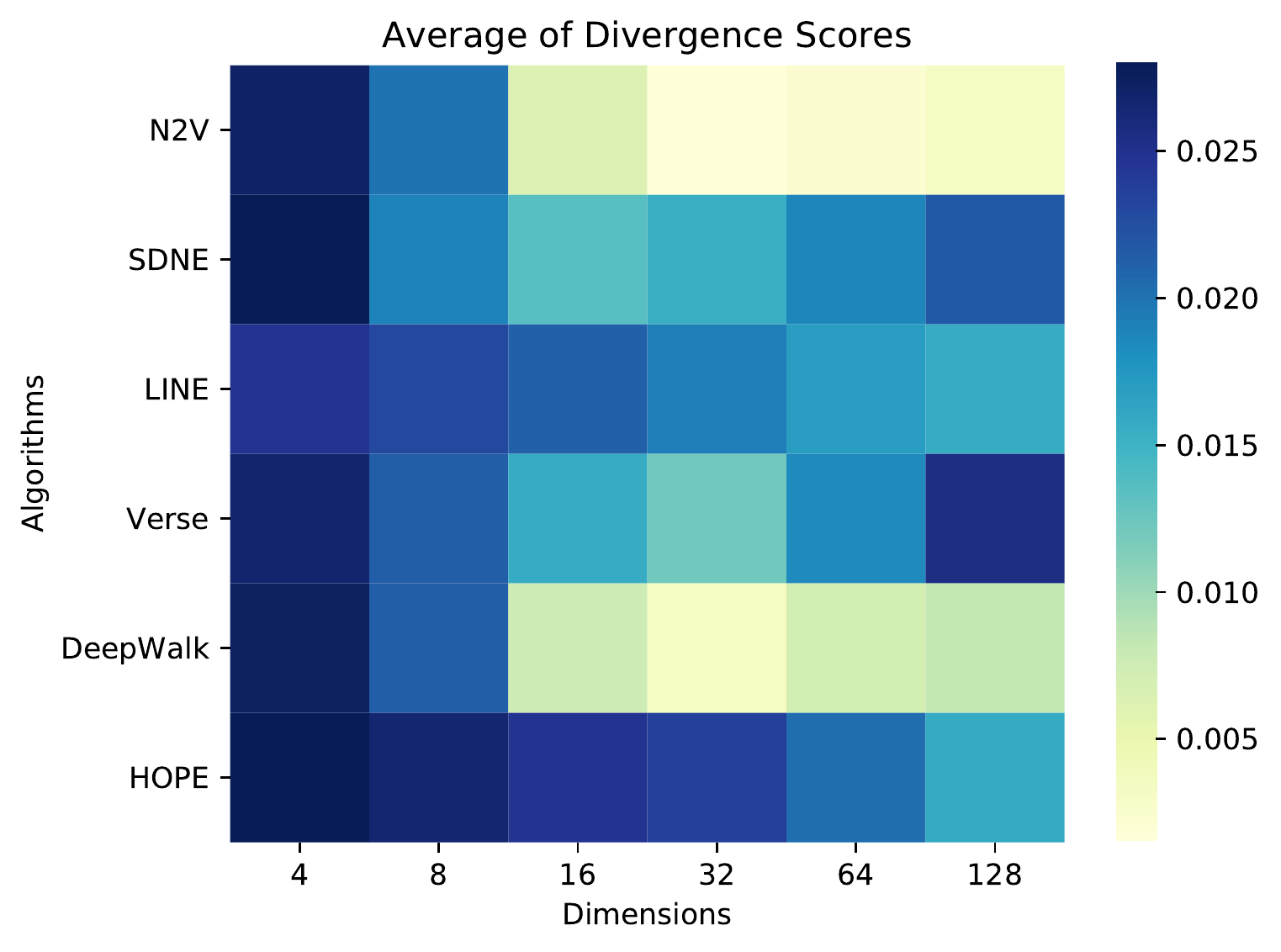}
\includegraphics[width=0.52\textwidth]{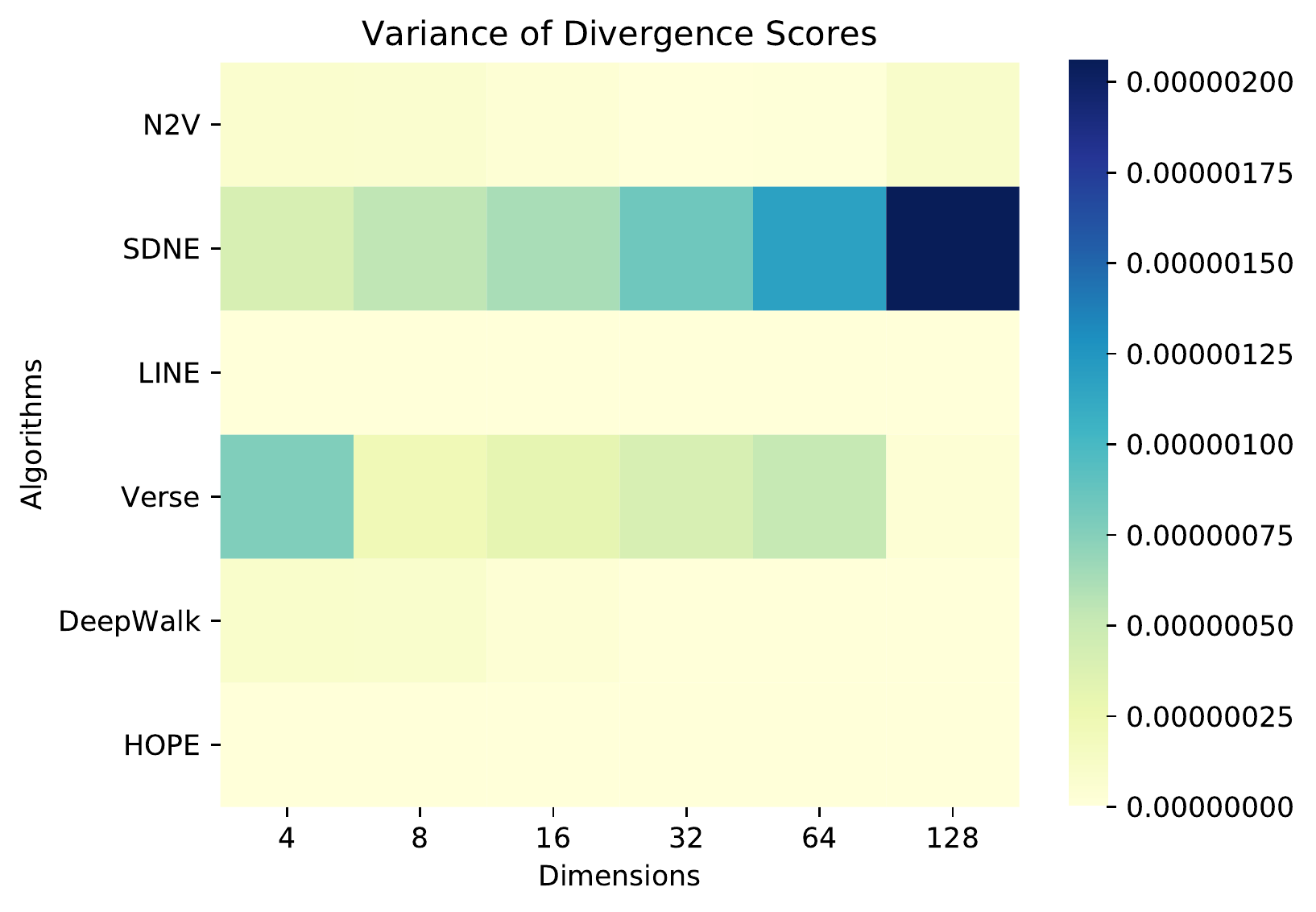}
\caption{\textbf{ABCD}: Average and Standard Deviation of the Divergence Score (Heat Map)}
\label{fig:ABCD}
\end{figure}

Before we move on to the more detailed experiments investigating each parameter of the model independently, let us repeat the experiment we did for the four selected real-world networks in the previous section on a single instance of the \textbf{ABCD} graph. This graph was generated with all parameters of the model set to their default values except the one controlling the community sizes that was fixed to $\beta = 1.5$. The results of this experiment are presented in Figure~\ref{fig:ABCD}. 


The conclusion is similar to what we observed in experiments with real graphs in the previous section. However, embeddings in lower dimensions (4 and 8) seem to be even more challenging than before and \textbf{VERSE} is not performing as good as before with \textbf{DeepWalk} taking its second place.

\subsection{Specification of the Experiments} 

In this section, we present a general approach that is used to test the five parameters mentioned above. Some specific modifications, if necessary, are explained below. In order to test how a given parameter of the \textbf{ABCD} model affects the divergence score, we pick $\ell$ values of this parameter to test (typically, $\ell=10$) and assign default values to the remaining parameters. 

Note that there are two sources of randomness involved in the process, one coming from the graph generation process and the second one coming from the embedding algorithm (indeed, 4 out of 6 algorithms we selected for testing are randomized algorithms). In order to investigate which source of randomness affects the divergence score more, in each experiment we generate $10\ell$ \textbf{ABCD} graphs. For a given value of the tested parameter $x$, we generated a family of 10 graphs $\mathcal{F}_{x}$, independently sampled by the model but with the same set of parameters.

For a given algorithm $A$, a given dimension $d \in \{32,64,128\}$, and a given parameter $x$, we independently run the algorithm on the 10 graphs from $\mathcal{F}_{x}$, 10 times for each graph. We compute the average divergence score $a_{A,d}(x)$ and the standard deviation $s_{A,d}(x)$ (over $100$ experiments; 10 graphs, 10 embeddings per graph). In order to see how the quality of embeddings change for a given graph, for each $G \in \mathcal{F}_{x}$, we additionally compute the average score $a_{A,d}(x, G)$, and the standard deviation $s_{A,d}(x, G)$ (over $10$ experiments; 10 embeddings of $G$). 

For each experiment, the variance is decomposed into two components related to the two sources of randomness based on the classical \textbf{ANOVA} method (a statistical test generalizing the \textbf{$t$-test}). The total sum of squared residuals $SS_T$ can be viewed as the sum of graph-specific sum of squares $SS_G$ and embedding-specific sum of squares $SS_E$.  In the tables below, we report $SS_T$ as well as the ratio $r_E$ based on the two decomposition elements, namely, $r_E = \frac{SS_E}{SS_T}.$ Clearly, $0 \le r_E \le 1$. Values of $r_E$ close to zero indicate that the noise coming from the graph generation is significantly larger than the noise related to the embeddings. On the other hand, values close to $0.5$ imply that both sources of randomness contribute roughly equally to the variance. Let us also note that two of our embedding algorithms (namely, \textbf{LINE} and \textbf{HOPE}) are deterministic and so $e_E=0$ for these algorithms. 

Finally, note that $SS_T$ can be alternatively decomposed into $\ell$ pieces corresponding to the $\ell$ values of the parameter $x$ tested. A natural question is then to see if all pieces equally contribute to $SS_T$ or maybe only a few of them contribute in a non-negligible way. To answer this questions we computed the average correlation between the corresponding ratio $r_E$ and parameter $x$. We got very small correlations, namely, $0.0009$ ($n$), $-0.0065$ ($\gamma$), $-0.0017$ ($\Delta$), $0.0245$ ($\beta$), $-0.0040$ ($\xi$).

\medskip

The results of our experiments are presented in the following form.
\begin{itemize}
\item [Plot 1.] For each embedding algorithm $A$, we plot 3 functions $a_{A,d}(x)$ for the selected dimensions $d \in \{32,64,128\}$ as a function of the tested parameter $x$. We additionally display confidence bands: $a_{A,d}(x) \pm  s_{A,d}(x)$.
\item [Plot 2.] For each dimension $d \in \{32,64,128\}$, we plot 6 functions $a_{A,d}(x)$ for all embedding algorithms $A$ as a function of the tested parameter $x$. We additionally display confidence bands: $a_{A,d}(x) \pm  s_{A,d}(x)$. 
\item [Plot 3.] For each embedding algorithm $A$, we plot average $s_{A,d}(x, G)$ (over 10 graphs) for the 3 selected dimensions $d \in \{32,64,128\}$ as a function of the tested parameter $x$.
\item [Plot 4.] For each dimension $d \in \{32,64,128\}$, we plot average $s_{A,d}(x, G)$ (over 10 graphs) for all 6 embedding algorithms $A$ as a function of the tested parameter $x$.
\item [Plot 5.] For each embedding algorithm $A$, we generate one plot as follows. For all values of the tested parameter $x$, all dimensions $d$, and all graphs $G \in \mathcal{F}_{x}$ (300 points), plot $(a_{A,d}(x, G), s_{A,d}(x, G))$. This way we test whether graphs with large divergence scores produce more variable results.
\end{itemize}

\medskip

Now, we are ready to discuss results of the experiments. For convenience and easier comparison, all plots are shown together on Figures~\ref{fig:n}--\ref{fig:beta} in the Appendix.

\subsection{Size of the Network ($n$)}

In this experiment, we study how sensitive the evaluated embedding algorithms are with respect to $n$, the size of the network. We consider $\ell = 10$ different values of the corresponding parameter: $n \in \{1000, 2000, \ldots, 10000\}$. The remaining parameters are set to their default values. (Note that, in particular, the exponent of the power-law degree distribution is set to $\gamma=2.5$ and the maximum degree is set to $\Delta = n^{1/(\gamma-1)} = n^{2/3}$. However, since the degree distribution is a function of $n$, it has to be re-generated for each tested value of $n$.) The plots are presented on Figure~\ref{fig:n}, the decomposition of the variance can be found in Table~\ref{tab:n1} and the speed of the embedding algorithms is reported in Table~\ref{tab:n2}.

\begin{table}[htbp!]
\begin{center}
\caption{Size of the Network ($n$) --- Decomposition of the Variance: $r_E$ / $SS_T$ (in $10^{-5}$)}
\label{tab:n1}
\begin{tabular}{l|c|c|c|c|c|c|}
\cline{2-7}
& \multicolumn{6}{c|}{\textbf{Algorithm}}                                                           \\ \hline
\multicolumn{1}{|l|}{\textbf{Dim}} & \textbf{node2vec} & \textbf{DeepWalk} & \textbf{LINE} & \textbf{HOPE} & \textbf{SDNE} & \textbf{VERSE} \\ \hline
\multicolumn{1}{|l|}{\textbf{32}}  & 0.24/0.08          & 0.35/3.34    &0/127.03 & 0/80.72         & 0.36/122.95            & 0.27/0.86            \\ \hline
\multicolumn{1}{|l|}{\textbf{64}}  & 0.25/0.07          & 0.30/1.53    &0/89.77 &   0/85.83        & 0.36/126.1            & 0.36/2.44           \\ \hline
\multicolumn{1}{|l|}{\textbf{128}} & 0.50/0.13          & 0.24/0.89    &0/75.89 &  0/86.47         & 0.28/135.37            & 0.34/7.94            \\ \hline
\end{tabular}
\end{center}
\end{table}

\begin{table}[htbp!]
\begin{center}
\caption{Size of the Network ($n$) --- the Average Running Time (in Seconds)}
\label{tab:n2}
\begin{tabular}{l|c|c|c|c|c|c|}
\cline{2-7}
& \multicolumn{6}{c|}{\textbf{Algorithm}}                                                           \\ \hline
\multicolumn{1}{|l|}{\textbf{Dim}} & \textbf{node2vec} & \textbf{DeepWalk} & \textbf{LINE} & \textbf{HOPE} & \textbf{SDNE} & \textbf{VERSE} \\ \hline
\multicolumn{1}{|l|}{\textbf{32}}  & 81          & 86              & 89            & 15            & 9            & 53            \\ \hline
\multicolumn{1}{|l|}{\textbf{64}}  & 84          & 97              & 91           & 16            & 10            & 67            \\ \hline
\multicolumn{1}{|l|}{\textbf{128}} & 92          & 126              & 94           & 18           & 10            & 98            \\ \hline
\end{tabular}
\end{center}
\end{table}

The divergence score is rather stable as $n$ increases, that is, it does not drastically change with the size of the network (Plots 1 and 2 on Figure~\ref{fig:n}). This is, of course, a desired property which indicates that the embeddings capture the ``big picture'' of the network, focusing on its structure and topology. \textbf{node2vec} and \textbf{VERSE} (in low dimension) perform best whereas \textbf{SDNE} appears to be the worst one. Let us observe that \textbf{DeepWalk} and \textbf{LINE} (\textbf{HOPE} to some degree too) improve their quality after increasing the dimension whereas large dimension (128), somewhat surprisingly, hurts the remaining algorithms. Stability of the embedding algorithms is also indicated by relatively small standard deviations with the exception of \textbf{SDNE} which performs not so well from that perspective (Plots 3 and 4). Finally, \textbf{DeepWalk}, \textbf{node2vec}, and \textbf{VERSE} exhibit a positive correlation between the average divergence score and the corresponding standard deviation (Plot 5). Again, somewhat surprisingly, \textbf{SDNE} does not behave as expected and shows no correlation between the average divergence score and its standard deviation. 

The decomposition of the variance shows that the contribution from the two sources of randomness is comparable. In terms of speed, \textbf{SDNE} and \textbf{HOPE} are the fastest, followed by \textbf{VERSE}. There is also a slight increase of the running time with respect to the dimension; with the most visible difference for \textbf{VERSE} and \textbf{DeepWalk}. 

\subsection{Degree Distribution ($\gamma$)} 

In this experiment, we investigate the behaviour of embedding algorithms for different degree distributions. We consider $\ell = 10$ different values of the corresponding parameter: $\gamma \in \{2.1, 2.2, \ldots, 3.0\}$. The remaining parameters are set to their default values. (However, despite the fact that the maximum degree is set to $\Delta \approx n^{1/(\gamma-1)}$, it is a function of $\gamma$ which changes in this experiment; in particular, $\Delta$ decreases when $\gamma$ increases.) The plots are presented on Figure~\ref{fig:gamma}, the decomposition of the variance can be found in Table~\ref{tab:gamma1} and the speed of the embedding algorithms is reported in Table~\ref{tab:gamma2}.

\begin{table}[htbp!]
\begin{center}
\caption{Degree Distribution ($\gamma$) --- Decomposition of the Variance: $r_E$ / $SS_T$ (in $10^{-5}$)}
\label{tab:gamma1}
\begin{tabular}{l|c|c|c|c|c|c|}
\cline{2-7}
& \multicolumn{6}{c|}{\textbf{Algorithm}}                                                           \\ \hline
\multicolumn{1}{|l|}{\textbf{Dim}} & \textbf{node2vec} & \textbf{DeepWalk} & \textbf{LINE} & \textbf{HOPE} & \textbf{SDNE} & \textbf{VERSE} \\ \hline
\multicolumn{1}{|l|}{\textbf{32}}  & 0.45/0.05          & 0.49/1.42   &0/54.44 & 0/12.97         & 0.63/38.66           & 0.44/0.03            \\ \hline
\multicolumn{1}{|l|}{\textbf{64}}  & 0.27/0.02          & 0.37/0.59    &0/42.64 &   0/19.04        & 0.61/38.54            & 0.82/0.32          \\ \hline
\multicolumn{1}{|l|}{\textbf{128}} & 0.31/0.02          & 0.14/0.25    &0/26.31 &  0/24.75         & 0.51/44.09           & 0.75/3.00            \\ \hline
\end{tabular}
\end{center}
\end{table}

\begin{table}[htbp!]
\begin{center}
\caption{Degree Distribution ($\gamma$) --- the Average Running Time (in Seconds)}
\label{tab:gamma2}
\begin{tabular}{l|c|c|c|c|c|c|}
\cline{2-7}
& \multicolumn{6}{c|}{\textbf{Algorithm}}                                                           \\ \hline
\multicolumn{1}{|l|}{\textbf{Dim}} & \textbf{node2vec} & \textbf{DeepWalk} & \textbf{LINE} & \textbf{HOPE} & \textbf{SDNE} & \textbf{VERSE} \\ \hline
\multicolumn{1}{|l|}{\textbf{32}}  & 175          & 206               & 90            & 58            & 15            & 102           \\ \hline
\multicolumn{1}{|l|}{\textbf{64}}  & 180          & 236               & 92           & 63            & 16            & 126            \\ \hline
\multicolumn{1}{|l|}{\textbf{128}} & 202          & 316               & 97           & 70           & 16            & 195            \\ \hline
\end{tabular}
\end{center}
\end{table}

Let us first note that the global density of the graph as well as the maximum degree $\Delta$ decrease as $\gamma$ increases. Hence, it seems natural to expect that the divergence score should be getting worse (that is, increasing) for large values of $\gamma$, as sparser random graphs are less ``predictable'' and so more challenging to embed (for example, there might be some unusually sparse or dense regions that occur by pure randomness). However, this behaviour is present only for \textbf{SDNE} and \textbf{HOPE}; in particular, the quality of \textbf{DeepWalk} visibly improves for large values of $\gamma$ (Plots 1 and 2 in Figure~\ref{fig:gamma}). This might mean that \textbf{DeepWalk} has a problem with embedding nodes of large degree. As before, \textbf{node2vec} and \textbf{VERSE} (in low dimension) perform best whereas \textbf{SDNE} appears to be the worst one. We also consistently see the peculiar property that some algorithms (such as \textbf{VERSE} and \textbf{SDNE}) perform worse in higher dimension. \textbf{SDNE} continues to be unstable with large values of the standard deviation (Plots 3 and 4) and with no correlation between the average value and the standard deviation (Plot 5). 

The decomposition of the variance remains comparable. \textbf{SDNE} continues to be the fastest, \textbf{HOPE} slows down in comparison to the earlier experiment with the graph sizes, and \textbf{LINE} speeds up; however, the order of the algorithms remains the same. The dimension continues to slow down the algorithms but no visible change can be detected; from the complexity point of view, the results are consistent with the ones we discussed in the previous section.

\subsection{Maximum Degree ($\Delta$)} 


In this experiment we investigate the maximum degree $\Delta = n^x$ by considering $\ell = 10$ different values of parameter $x \in \{0.25, 0.30, \ldots, 0.70\}$. The remaining parameters are set to their default values.
The plots are presented on Figure~\ref{fig:delta}, the decomposition of the variance can be found in Table~\ref{tab:delta1} and the speed of the embedding algorithms is reported in Table~\ref{tab:delta2}.


\begin{table}[htbp!]
\begin{center}
\caption{Maximum Degree ($\Delta$) --- Decomposition of the Variance: $r_E$ / $SS_T$ (in $10^{-5}$)}
\label{tab:delta1}
\begin{tabular}{l|c|c|c|c|c|c|}
\cline{2-7}
& \multicolumn{6}{c|}{\textbf{Algorithm}}                                                           \\ \hline
\multicolumn{1}{|l|}{\textbf{Dim}} & \textbf{node2vec} & \textbf{DeepWalk} & \textbf{LINE} & \textbf{HOPE} & \textbf{SDNE} & \textbf{VERSE} \\ \hline
\multicolumn{1}{|l|}{\textbf{32}}  & 0.37/0.06          & 0.45/1.19   &0/79.35 & 0/36.05         & 0.40/57.55           & 0.34/0.05            \\ \hline
\multicolumn{1}{|l|}{\textbf{64}}  & 0.36/0.04          & 0.26/0.72    &0/45.52 &   0/39.32        & 0.38/60.20            & 0.56/0.22          \\ \hline
\multicolumn{1}{|l|}{\textbf{128}} & 0.31/0.04          & 0.08/0.49    &0/45.61 &  0/38.3         & 0.35/61.74           & 0.75/6.75            \\ \hline
\end{tabular}
\end{center}
\end{table}

\begin{table}[htbp!]
\begin{center}
\caption{Maximum Degree ($\Delta$) --- the Average Running Time (in Seconds)}
\label{tab:delta2}
\begin{tabular}{l|c|c|c|c|c|c|}
\cline{2-7}
& \multicolumn{6}{c|}{\textbf{Algorithm}}                                                           \\ \hline
\multicolumn{1}{|l|}{\textbf{Dim}} & \textbf{node2vec} & \textbf{DeepWalk} & \textbf{LINE} & \textbf{HOPE} & \textbf{SDNE} & \textbf{VERSE} \\ \hline
\multicolumn{1}{|l|}{\textbf{32}}  & 176          & 209               & 90            & 58            & 14            & 97           \\ \hline
\multicolumn{1}{|l|}{\textbf{64}}  & 181          & 236               & 92           & 62            & 15            & 125            \\ \hline
\multicolumn{1}{|l|}{\textbf{128}} & 199          & 312               & 98           & 71           & 15            & 207            \\ \hline
\end{tabular}
\end{center}
\end{table}

Let us first note that the global density of the graph slightly increases as $\Delta$ increases. Hence, from that perspective, we experience a similar behaviour as with the decreasing of $\gamma$ from our previous experiment. Hence, it seems natural to expect that the divergence score should behave similarly to that of the previous experiment, with the degree distribution that is modelled by the parameter $\gamma$ (that is, increasing functions should be now decreasing and vice versa). This behaviour is certainly present for \textbf{DeepWalk}; as observed before, the quality of \textbf{DeepWalk} visibly drops (especially in low dimension) when large degree nodes are present suggesting that they create a problem for this particular embedding algorithm. Other algorithms do not show similar duality between the two experiments. It implies (perhaps not surprisingly) that the qualities of these algorithms cannot be simply deduced based on the density of the graph or the maximum degree. It seems that the quality depends in some non-trivial way on the degree distribution (Plots 1 and 2 in Figure~\ref{fig:delta}). Global comparison of the algorithms (ranking, decomposition of the variance, and speed) remains the same as in the previous subsection.

\subsection{Level of Noise ($\xi$)} 


In this experiment we investigate the level of noise by considering $\ell = 10$ different values of the corresponding parameter: $\xi \in \{0.1, 0.2, \ldots, 1.0\}$. The remaining parameters are set to their default values.
The plots are presented on Figure~\ref{fig:xi}, the decomposition of the variance can be found in Table~\ref{tab:xi1} and the speed of the embedding algorithms is reported in Table~\ref{tab:xi2}.


\begin{table}[htbp!]
\begin{center}
\caption{Level of Noise ($\xi$) --- Decomposition of the Variance: $r_E$ / $SS_T$ (in $10^{-5}$)}
\label{tab:xi1}
\begin{tabular}{l|c|c|c|c|c|c|}
\cline{2-7}
& \multicolumn{6}{c|}{\textbf{Algorithm}}                                                           \\ \hline
\multicolumn{1}{|l|}{\textbf{Dim}} & \textbf{node2vec} & \textbf{DeepWalk} & \textbf{LINE} & \textbf{HOPE} & \textbf{SDNE} & \textbf{VERSE} \\ \hline
\multicolumn{1}{|l|}{\textbf{32}}  & 0.56/2.61          & 0.38/4.33   &0/60.81 & 0/48.72         & 0.25/56.86           & 0.47/5.24            \\ \hline
\multicolumn{1}{|l|}{\textbf{64}}  & 0.76/3.90          & 0.50/5.55    &0/38.59 &   0/42.16        & 0.20/62.67            & 0.48/3.23          \\ \hline
\multicolumn{1}{|l|}{\textbf{128}} & 0.65/6.12          & 0.59/6.84    &0/31.43 &  0/35.12         & 0.22/59.83           & 0.41/4.57            \\ \hline
\end{tabular}
\end{center}
\end{table}

\begin{table}[htbp!]
\begin{center}
\caption{Level of Noise ($\xi$) --- the Average Running Time (in Seconds)}
\label{tab:xi2}
\begin{tabular}{l|c|c|c|c|c|c|}
\cline{2-7}
                                   & \multicolumn{6}{c|}{\textbf{Algorithm}}                                                           \\ \hline
\multicolumn{1}{|l|}{\textbf{Dim}} & \textbf{node2vec} & \textbf{DeepWalk} & \textbf{LINE} & \textbf{HOPE} & \textbf{SDNE} & \textbf{VERSE} \\ \hline
\multicolumn{1}{|l|}{\textbf{32}}  & 175          & 211               & 88            & 59            & 15            & 97            \\ \hline
\multicolumn{1}{|l|}{\textbf{64}}  & 178          & 241               & 93           & 63            & 16            & 135            \\ \hline
\multicolumn{1}{|l|}{\textbf{128}} & 204          & 303               & 99           & 68           & 16            & 191            \\ \hline
\end{tabular}
\end{center}
\end{table}

The level of noise present in the graph is an important aspect of the embedding algorithms which is confirmed by this experiment. For low level of noise (modelled by small value of parameter $\xi$), communities are easy to identify and to extract from the graph and so it should be relatively easy to embed the graph preserving the community structure. As a result, one would expect all algorithms to score well and have small divergence score for such values of $\xi$. On the other hand, for values of $\xi$ close to one, the graph is very close to the random graph with a given degree distribution and no communities. For such graphs, no matter how nodes are embedded in space, the densities between ``communities'' and within them are going to be very close to the corresponding expected values in the null-model. Hence, such graphs should score well again but the interpretation is different: all algorithms perform predictably bad, given this impossible task of preserving community structure. On the other hand, graphs with values of $\xi$ between zero and one are challenging to properly embed and so one would expect that the divergence score generates ``inverted-$v$ shape'' as a function of $\xi$. \textbf{node2vec} and \textbf{VERSE} exhibit such shape and, again, these two algorithms win again (\textbf{VERSE} gets worse for $\xi \in \{0.6, 0.7, 0.8\}$). Oddly, \textbf{SDNE}, \textbf{HOPE} and \textbf{LINE} perform badly for very low level of noise (Plots 1 and 2 in Figure~\ref{fig:xi}). This could possibly be explained by too local nature of such algorithms. Local algorithms, in the presence of low level of noise, embed nodes based on the knowledge coming exclusively from their corresponding communities. As a result, communities are embedded almost independently and so such algorithms do a poor job of separating the communities. The behaviour of \textbf{DeepWalk} is even more challenging to explain.

It is worth pointing out that one striking difference between Plots 2 in Figure~\ref{fig:xi} and the corresponding plots for the other parameters tested is that there is no visible difference between the three dimensions that we evaluated. This indicates that the dimension affects the divergence score by a great deal. The dimension still affects the performance of the algorithms but its influence is much weaker than the parameter $\xi$ tested in this experiment. Let us also note that there is a strong correlation between the average score and the standard deviation (Plot 5) but \textbf{SDNE} continues to be the most unstable (Plots 3 and 4).

The decomposition of the variance remains comparable to the earlier experiments. \textbf{SDNE} continues to be the fastest and \textbf{DeepWalk} was the slowest. The dimension continues to slow down the algorithms but no visible change can be detected. Let us also mention that during this experiment we were forced to switch to more powerful machines as \textbf{HOPE} with dimension 128 ran out of memory which indicates that large level of noise is not only challenging from the quality of the embeddings point of view but also from the computational one. 

\subsection{Community Sizes ($\beta$)}\label{sec:beta} 


In this experiment we investigate the distribution of community sizes by considering $\ell = 10$ different values of the corresponding parameter: $\beta \in \{1.1, 1.2, \ldots, 2.0\}$. The remaining parameters are set to their default values.
The plots are presented on Figure~\ref{fig:beta}, the decomposition of the variance can be found in Table~\ref{tab:beta1} and the speed of the embedding algorithms is reported in Table~\ref{tab:beta2}.

\begin{table}[htbp!]
\center
\caption{Distribution of Communities as a Function of Parameter $\beta$}
\label{tab:beta0}
\begin{tabular}{l|c|c|c|c|c|c|c|c|c|c|}
\cline{2-11}
& \multicolumn{10}{c|}{\textbf{$\beta$}}        \\ \hline
\multicolumn{1}{|l|}{\textbf{Statistics}}          & \textbf{1.1} & \textbf{1.2} & \textbf{1.3} & \textbf{1.4} & \textbf{1.5} & \textbf{1.6} & \textbf{1.7} & \textbf{1.8} & \textbf{1.9} & \textbf{2.0} \\ \hline
\multicolumn{1}{|l|}{\textbf{No. communities}}     & 34           & 38           & 39           & 42           & 48           & 44           & 56           & 60           & 65           & 64           \\ \hline
\multicolumn{1}{|l|}{\textbf{Min community size}}  & 50           & 50           & 50           & 50           & 50           & 50           & 50           & 50           & 50           & 50           \\ \hline
\multicolumn{1}{|l|}{\textbf{Max community size}}  & 983          & 977          & 979          & 968          & 977          & 997          & 978          & 939          & 972          & 988          \\ \hline
\multicolumn{1}{|l|}{\textbf{Mean community size}} & 292.4        & 259.1        & 252.5        & 233.1        & 204.9        & 227.3        & 176.4        & 165.6        & 152.2        & 154.6        \\ \hline
\end{tabular}
\end{table}

\begin{table}[htbp!]
\begin{center}
\caption{Community Sizes ($\beta$) --- Decomposition of the Variance: $r_E$ / $SS_T$ (in $10^{-5}$)}
\label{tab:beta1}
\begin{tabular}{l|c|c|c|c|c|c|}
\cline{2-7}
& \multicolumn{6}{c|}{\textbf{Algorithm}}                                                           \\ \hline
\multicolumn{1}{|l|}{\textbf{Dim}} & \textbf{node2vec} & \textbf{DeepWalk} & \textbf{LINE} & \textbf{HOPE} & \textbf{SDNE} & \textbf{VERSE} \\ \hline
\multicolumn{1}{|l|}{\textbf{32}}  & 0.01/18.82          & 0.02/29.45   &0/1397.83 & 0/1262.72         & 0.04/1427.12           & 0.02/1046.42            \\ \hline
\multicolumn{1}{|l|}{\textbf{64}}  & 0.14/12.37          & 0.01/125.11    &0/1330.70 &   0/1280.63        & 0.04/1412.63            & 0.02/1392.63          \\ \hline
\multicolumn{1}{|l|}{\textbf{128}} & 0.06/131.39          & 0.002/184.07    &0/1235.29 &  0/1301.64         & 0.04/1460.59           & 0.01/1257.91            \\ \hline
\end{tabular}
\end{center}
\end{table}

\begin{table}[htbp!]
\begin{center}
\caption{Community Sizes ($\beta$) --- the Average Running Time (in Seconds)}
\label{tab:beta2}
\begin{tabular}{l|c|c|c|c|c|c|}
\cline{2-7}
& \multicolumn{6}{c|}{\textbf{Algorithm}}                                                           \\ \hline
\multicolumn{1}{|l|}{\textbf{Dim}} & \textbf{node2vec} & \textbf{DeepWalk} & \textbf{LINE} & \textbf{HOPE} & \textbf{SDNE} & \textbf{VERSE} \\ \hline
\multicolumn{1}{|l|}{\textbf{32}}  & 169          & 217               & 87            & 60            & 14            & 99            \\ \hline
\multicolumn{1}{|l|}{\textbf{64}}  & 182          & 239               & 91           & 62            & 15            & 134            \\ \hline
\multicolumn{1}{|l|}{\textbf{128}} & 206          & 297               & 102           & 68           & 16            & 188            \\ \hline
\end{tabular}
\end{center}
\end{table}

Note that the number of communities increases (and so the average community size decreases) when $\beta$ increases---see Table~\ref{tab:beta0}. Since a graph with large number of small communities is difficult to embed, one would expect that the divergence score should get worse for large values of $\beta$. All algorithms we tested confirm this intuition. \textbf{DeepWalk}, \textbf{node2vec}, and \textbf{VERSE} perform better in smaller dimension which suggests that it should be the choice for graphs with a large number of communities. The divergence score for the remaining three algorithms is not affected by the choice of the dimension. \textbf{VERSE}, which used to perform well before, gets worse, \textbf{DeepWalk} improves, but \textbf{node2vec} is still winning (Plots 1 and 2 in Figure~\ref{fig:beta}). \textbf{SDNE} continues to be unstable but still is the fastest with \textbf{LINE} and \textbf{HOPE} being roughly 5 times slower; \textbf{DeepWalk} is the slowest. As expected, the dimension continues to slow the algorithms down. 

The most drastic difference is for the decomposition of the variance. In all earlier experiments, we used to report large values of the parameter $r_E$. This time, $r_E$ is by an order of magnitude smaller. As mentioned earlier when we introduced this ratio, this indicates that the main contribution to the variance comes from the randomness of the graph generation process, not from the randomized embedding algorithms. This is expected as the distribution of the community sizes is the most fragile parameter of the \textbf{ABCD} model (in fact, any model, including \textbf{LFR}). Indeed, with non-negligible probability the number of components might be substantially different for two graphs with the same set of parameters but independently generated. That was the main reason why we fixed the number of communities to 5 communities instead of generating one sequence of communities with, say, $\beta = 1.5$ and keeping it for all experiments. This way we balanced the two sources of randomness, from the graph generation process and from embedding algorithms themselves. 

\subsection{Summary of Experimental Results}\label{sec:summary}


In this section, we tested the 5 most important parameters of the \textbf{ABCD} model with the hope to better understand the influence of basic statistics of real-world networks on the quality of embedding algorithms measured by the divergence score. In general, the distribution of community sizes, controlled by the parameter $\beta$ in the \textbf{ABCD} model, has the single largest impact on the results. Hence, we decided to fix 5 community sizes throughout, except when testing $\beta$ itself; in that experiment, there are many more communities and we observe that the divergence score increases with $\beta$. The impact is not as clear and strong when the size of the network ($n$) or the degree distribution ($\gamma$ and $\Delta$) changes.

\textbf{node2vec} was constantly winning in all the experiments we performed. \textbf{VERSE} was good but not in the presence of many small communities or large level of noise, and only in low dimension. These observations are also observed in the experiments with real graphs we presented earlier. \textbf{SDNE} seems to be the most unpredictable algorithm but it may be the case because it aims to capture different aspects of the graph such as the role of particular nodes within network instead of preserving densities between communities. 

The conclusion is that \textbf{node2vec} algorithm seems to be a good first choice to try in general but, depending on the graph that needs to be embedded, other algorithms give similarly good results. Moreover, \textbf{node2vec} has a number of parameters itself which potentially might produce even better outcomes. We investigate this aspect in the next set of experiments. In general, there is no clear winner (specific algorithm with a specific set of parameters) that works best and so there is a need to be able to guide our choice of the algorithm and its parameters by a trustworthy and unsupervised benchmark framework.

Execution time of all algorithms is sensitive with respect to the dimension. However, the increase of the computing time is significantly larger for \textbf{node2vec}, \textbf{DeepWalk} and \textbf{VERSE} than for \textbf{LINE}, \textbf{HOPE} and \textbf{SDNE}. \textbf{node2vec} and \textbf{DeepWalk} are the most computationally intensive, however, they support parallel execution and their run-time can be controlled by available CPU cores. \textbf{SDNE} is constantly the fastest from all tested algorithms; roughly 20 times faster than \textbf{DeepWalk}.

\subsection{node2vec} 

In these final experiments on the synthetic graphs, we focus on \textbf{node2vec}, the embedding algorithm that seems to perform best for both real-world networks we experimented with as well as synthetic graphs generated by the \textbf{ABCD} model. The goal is to investigate how the behaviour of \textbf{node2vec} changes for various parameters of this algorithm. 

\begin{figure}[!htb]
\begin{center}
\includegraphics[width=0.4\textwidth]{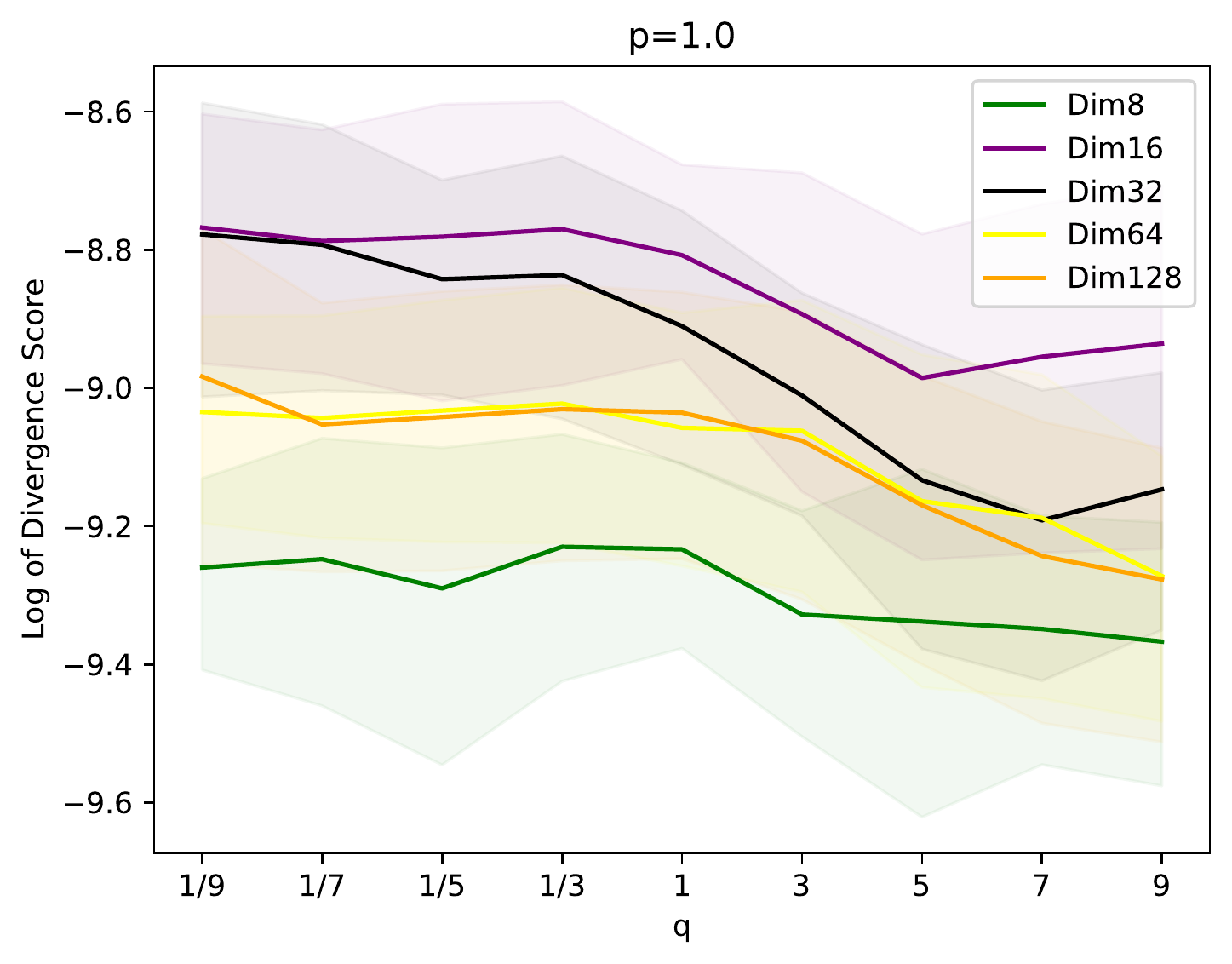}
\includegraphics[width=0.4\textwidth]{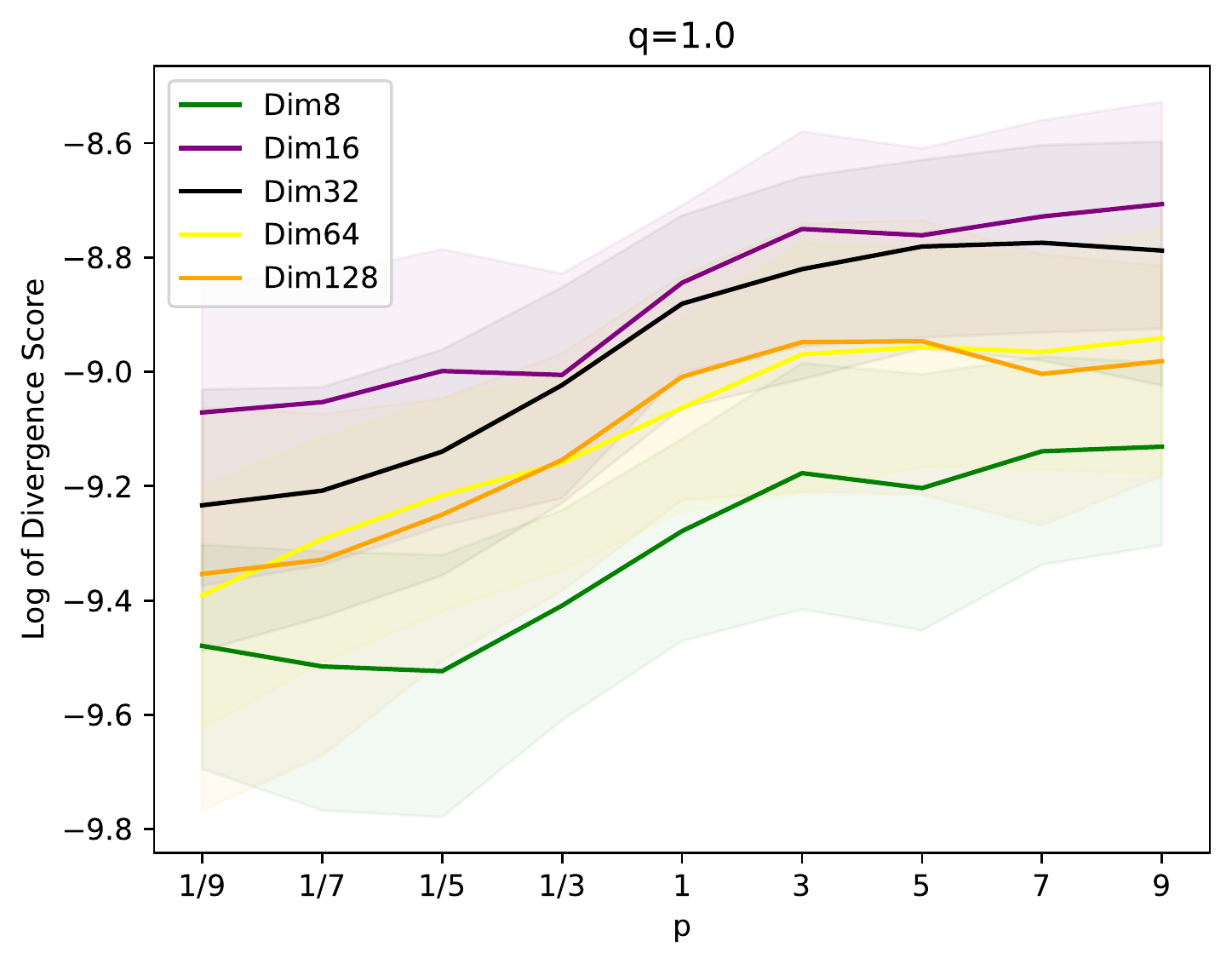}
\caption{\textbf{node2vec}: the influence of the return parameter $p$ and the in-out parameter $q$ (plots 6 and 7).}
\label{fig:n2v_pq}
\end{center}
\end{figure}

Recall that the {\it return parameter} $p$ controls the likelihood of 2-hop redundancy in the corresponding random walk. Large values of this parameter decrease the probability that already-visited node is sampled in the following two steps. On the other hand, small values of the {\it in-out parameter} $q$ encourages \textbf{DFS}-like exploration whereas large values can be used to emulate \textbf{BFS}-like exploration. 
In our first experiment, we independently generated $5$ \textbf{ABCD} graphs using the default set of parameters, family $\mathcal{F}$. For a given dimension $d \in \{4, 8, 16, 32,64,128\}$, a given set of parameters $(p,q)$ of \textbf{node2vec}, we independently run the algorithm on 5 graphs from $\mathcal{F}$, 5 times for each graph. We computed the average divergence score $a_{node2vec,d}(p,q)$ and the standard deviation $s_{node2vec,d}(p,q)$ (over $25$ experiments; 5 graphs, 5 embeddings per graph). 
The results are presented in Figure~\ref{fig:n2v_pq} in the following form.
\begin{itemize}
\item [Plot 6.] We plot 5 functions $a_{node2vec,d}(p,q)$ for the selected dimensions $d \in \{8, 16, 32, 64, 128\}$ with the return parameter fixed to $p=1$ as a function of the parameter $q \in \{1, 3, 5, 7, 9, 1/3, 1/5, 1/7, 1/9\}$. We additionally display confidence bands: $a_{node2vec,d}(p,q) \pm  s_{node2vec,d}(p,q)$. 
\item [Plot 7.] We plot 5 functions $a_{node2vec,d}(p,q)$ for the selected dimensions $d \in \{8, 16, 32, 64, 128\}$ with the in-out parameter fixed to $q=1$ as a function of the parameter $p \in \{1, 3, 5, 7, 9, 1/3, 1/5, 1/7, 1/9\}$. We additionally display confidence bands: $a_{node2vec,d}(p,q) \pm  s_{node2vec,d}(p,q)$.
\end{itemize}

\begin{figure}[!htb]
\begin{center}
\includegraphics[width=0.4\textwidth]{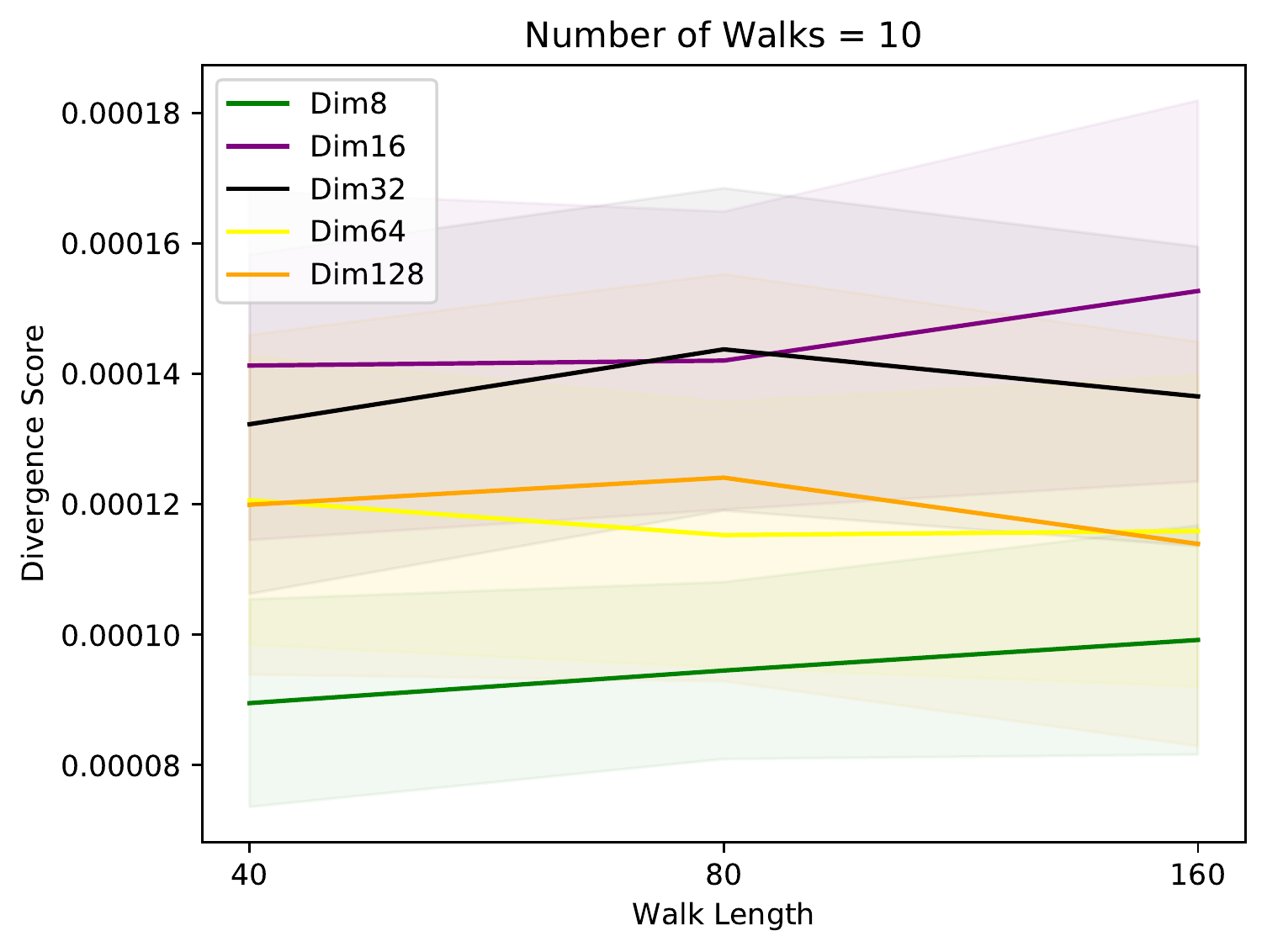}
\includegraphics[width=0.4\textwidth]{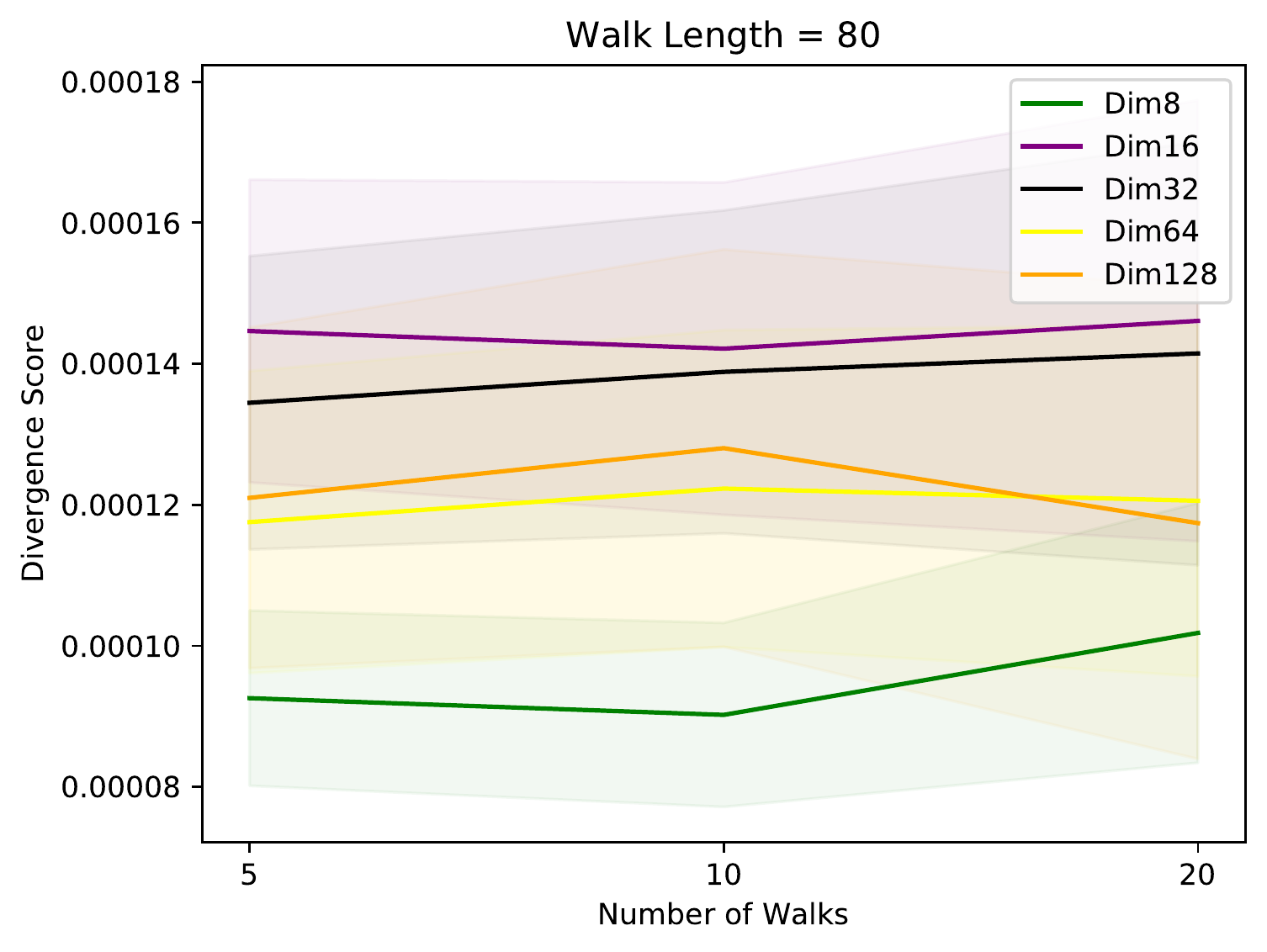}
\caption{\textbf{node2vec}: the influence of the number of walks $k$ and the walk length $w$ (plots 8 and 9).}
\label{fig:n2v_walks}
\end{center}
\end{figure}

In the second experiment, we investigate the influence of the second set of parameters: the \emph{number of walks} which has the default value of $k=10$, and the \emph{walk length} that is set to $w=80$ by default. 
The results are presented in Figure~\ref{fig:n2v_walks} in the following form.

\begin{itemize}
\item [Plot 8.] We plot 5 functions $a_{node2vec,d}(k,w)$ for the selected dimensions $d \in \{8, 16, 32, 64, 128\}$ with the number of walks fixed to $k=10$ as a function of the walk length $w \in \{40, 80, 160\}$. We additionally display confidence bands: $a_{node2vec,d}(k,w) \pm  s_{node2vec,d}(k,w)$.
\item [Plot 9.] We plot 5 functions $a_{node2vec,d}(k,w)$ for the selected dimensions $d \in \{8, 16, 32, 64, 128\}$ with the walk length fixed to $w=80$ as a function of the number of walks $k \in \{5, 10, 20\}$. We additionally display confidence bands: $a_{node2vec,d}(k,w) \pm  s_{node2vec,d}(k,w)$.
\end{itemize}

The results of the experiments are consistently very good. We see a slight improvement of the divergence score when the ratio between $q$ and $p$ increases. On the other hand, the number of walks and the walk length seem to be even more stable without a visible improvements. One surprising thing to notice is that dimension 8 gave the best result, better than dimension 128. The remaining dimensions behave as expected: the divergence score improves as the dimension increases. 

Finally, let us stress that these results are performed only on the \textbf{ABCD} model with a given set of default parameters defined at the beginning of this section (in particular, in the presence of a relatively low level of noise) but the fact that the result are stable and very good is another reason to use \textbf{node2vec} as a default embedding algorithm.

\section{$\ldots$But Can One Trust the Framework?}\label{sec:justification}

In all experiments we performed in this paper we measured the quality of embedding algorithms by computing the divergence score returned by the benchmarking framework. The divergence score measures to which degree the following natural and desired properties are satisfied. Embeddings that score well extract enough information from the graph that allows one to reconstruct the number of edges between communities as well as within them. In particular, pairs of nodes that are close in the embedded space often tend to be adjacent and vice versa---there are some sporadic long edges but they are not so common. (See Section~\ref{sec:framework} for a longer discussion.)

However, despite the fact that the definition of the divergence score seems to be natural, the following important questions arise: \emph{Should one trust the divergence store in making decisions whether or not to use a given outcome of the embedding algorithm?} What if the framework favours embeddings that perform poorly when feed as an input for some machine learning algorithm? In order to answer this question, we highlight a few of the most common applications of graph embedding algorithms and show that the performance of the corresponding tools highly depend on the divergence score. Of course, the list is not intended to be complete and there are many other important potential applications one might want to explore. In order to measure the quality of embeddings with respect to the three selected applications, we use some ad-hoc but rather standard supervised methods. 

\subsection{Nodes Classification}

Node classification is an example of a semi-supervised learning algorithm where labels are only available for a small fraction of nodes and the goal is to label the remaining set of nodes based on this small initial seed set. This is a situation often observed in, for example, social networks in which labels might indicate a user’s interests, beliefs, or demographic characteristics. There could be many reasons for labels not to be available for a large fraction of nodes, for example, a user’s demographic information might not be available to protect their privacy. Our task is then to infer missing labels based on the small set of labeled nodes and the structure of the graph.

Since embedding algorithms can be viewed as the process of extracting features of the nodes from the structure of the graph, one may reduce the problem to a classical machine learning predictive modelling classification problem for the set of vectors. There are many algorithms, such as logistic regression, $k$-nearest neighbours, decision trees, XGBoost, support vector machine, etc., for any potential scenario that one might be interested in, including binary, multi-class, and multi-label classifications.

For our experiment, we used the synthetic \textbf{ABCD} graph with $n=10{,}000$ and all parameters set to their default values (in particular, the configuration model with the global variant was used) except parameter $\beta$ that was set to $\beta=1.5$. The reason for deviating from the default 5 communities is to introduce more challenging scenario for the classifier with a larger number of communities (42 communities were generated by the random model). Indeed, the community of each node of this graph is its ground-truth community provided by the \textbf{ABCD} model and the goal of the classifier is to predict them. 

The set of nodes was randomly partitioned into a training set (with 75\% of the nodes) and a test set (with the remaining 25\%). For each embedding algorithm $A$ and each dimension $d \in \{4, 8, 16, 32, 64, 128\}$, we used the labels from the training set and their embeddings to train the \textbf{Gradient Boosted Trees} (\textbf{XGBoost}) model. This model was used as it is integrated with a number of packages making it easy to use, and the model itself has some additional nice features which distinguish it from other gradient boosting algorithms. Having said that, let us stress that our goal in this experiment is not to classify nodes as best as possible but rather to detect a possible correlation between the quality of a classifier and the divergence score of embeddings used. Similar conclusions can be derived by using other models which might or might not give better results. Following the same argument, \textbf{XGBoost} was used with default hyper-parameters values and no tuning was done as part of the experiment. We applied the model on the embeddings of the nodes from the test set to predict the labels of the corresponding nodes. Based on the ground-truth, we then computed the overall accuracy (the fraction of predictions our model got right). 

\begin{figure}[!htb]
\begin{center}
\includegraphics[width=0.49\textwidth]{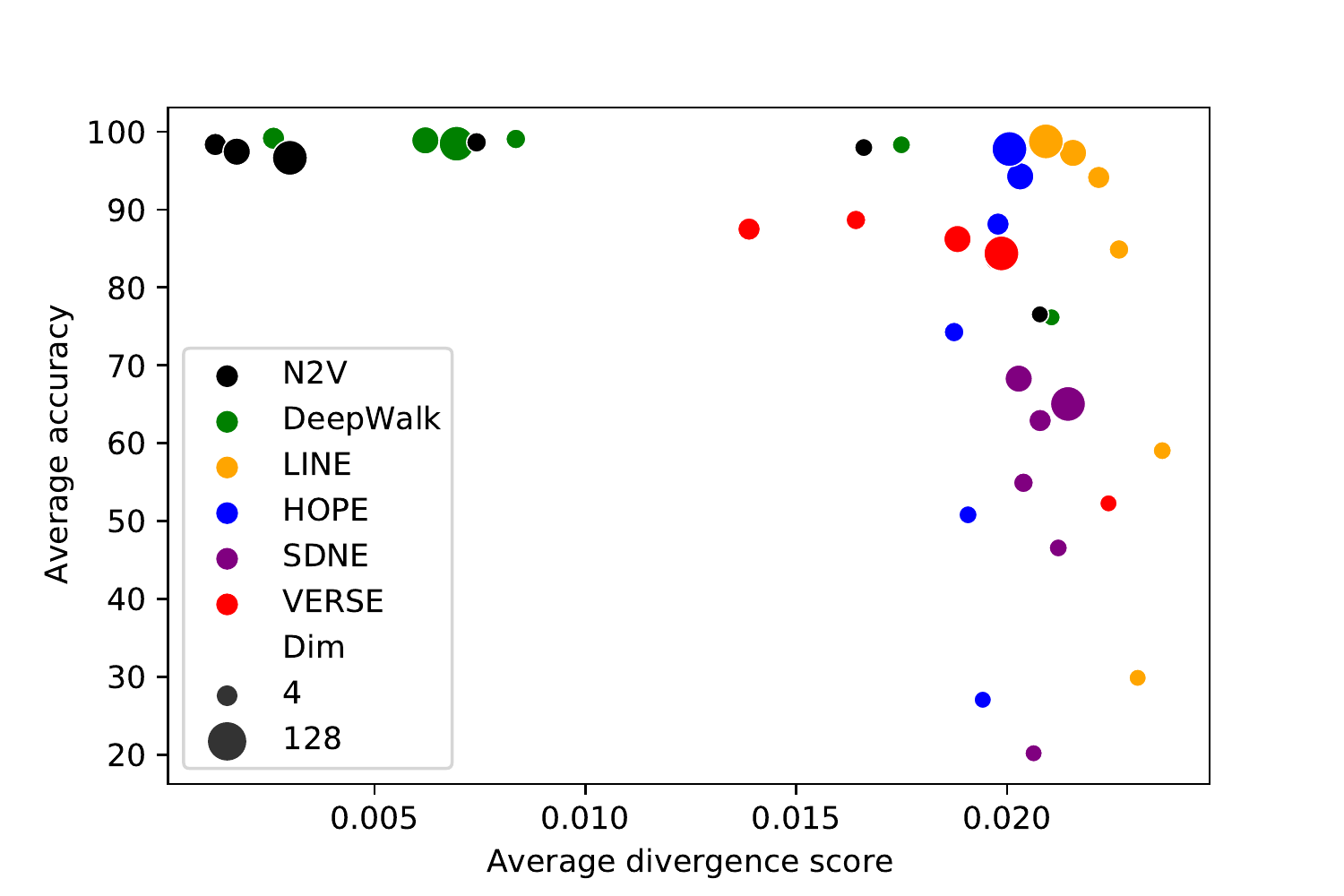}
\includegraphics[width=0.49\textwidth]{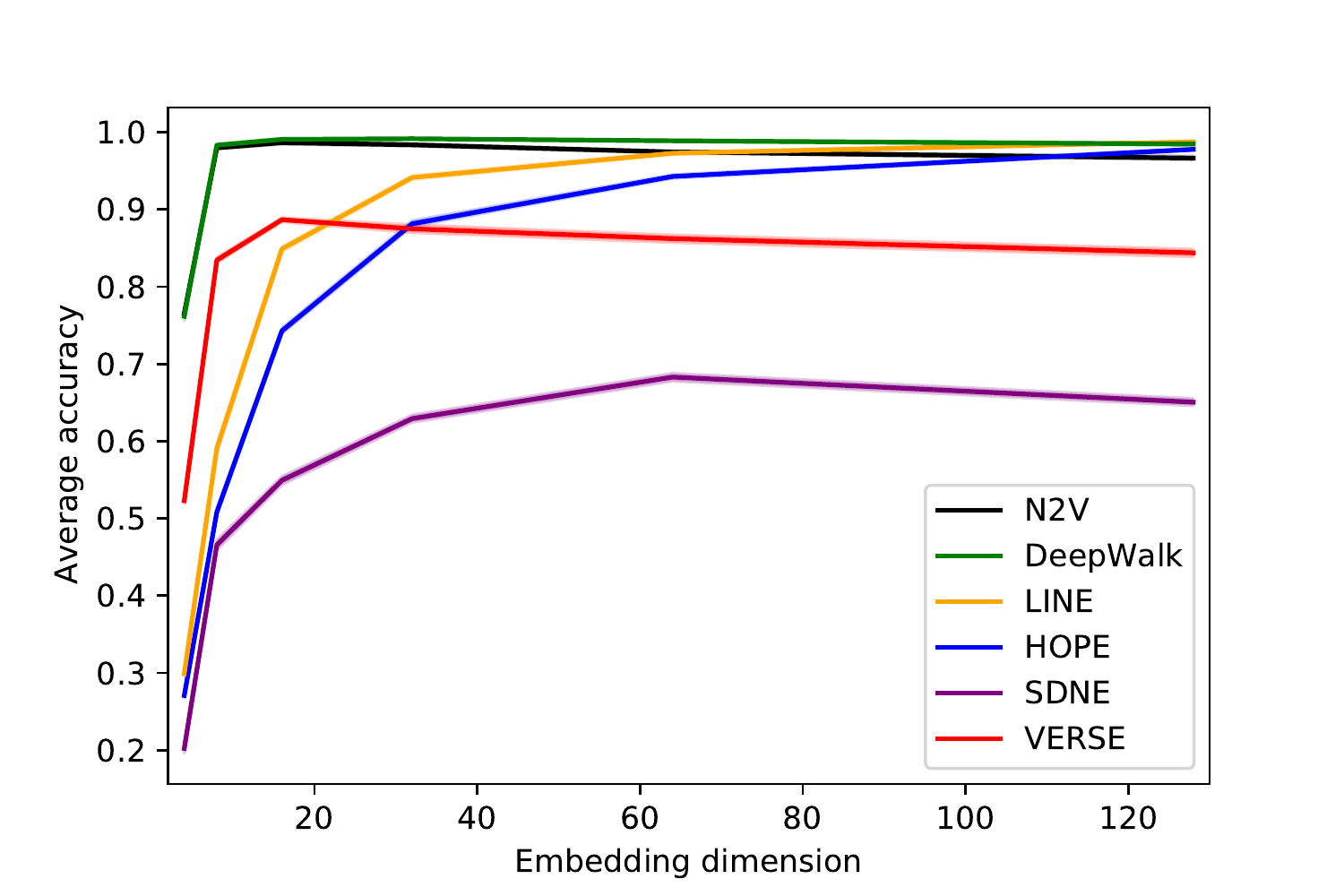}
\caption{\textbf{Nodes classification}: relation between the accuracy and the divergence score (left) and the dimension of the embedding (right).}\label{exp7}
\end{center}
\end{figure}

For a given algorithm $A$ and a given dimension $d$, we repeated the above procedure 10 times, independently and randomly splitting the set of nodes into training and test sets. The average accuracy depends on the choice of $A$ and $d$. More importantly, there seems to be a strong correlation between the accuracy and the quality of the embedding based on the framework, the divergence score---see Figure~\ref{exp7} (left). Indeed, the correlation coefficient is equal to $-0.52$, which shows that there is a significant correlation between the two metrics but the relation seems to be non-linear. In particular, embeddings with low divergence score achieve high accuracy whereas embeddings with high divergence exhibit varying accuracy. In other words, large divergence score does not imply that the performance of node classification algorithm is going to be poor but low divergence score seems to guarantee the success. We also noticed that, in general, the dimension cannot be too small (4 or 8) but the average accuracy quickly stabilizes and there is no need to use embeddings in very large dimensions---see Figure~\ref{exp7} (right). 

\subsection{Community Detection}

There are various techniques and algorithms for detecting communities in networks. Node embeddings provide an alternative tool for clustering related nodes, or they may be used to tune and to improve the graph tools with providing additional, complementary information. Indeed, since each node is associated with a real-valued vector embedded in $d$-dimensional space, one may alternatively ignore the initial graph and apply some generic clustering algorithm to the set of associated vectors. Clustering points seems to be a much easier task and is a well-studied area of research with many scalable algorithms, such as $k$-means or DBSCAN, that are easily available for use. 

\begin{figure}[!htb]
\begin{center}
\includegraphics[width=0.47\textwidth]{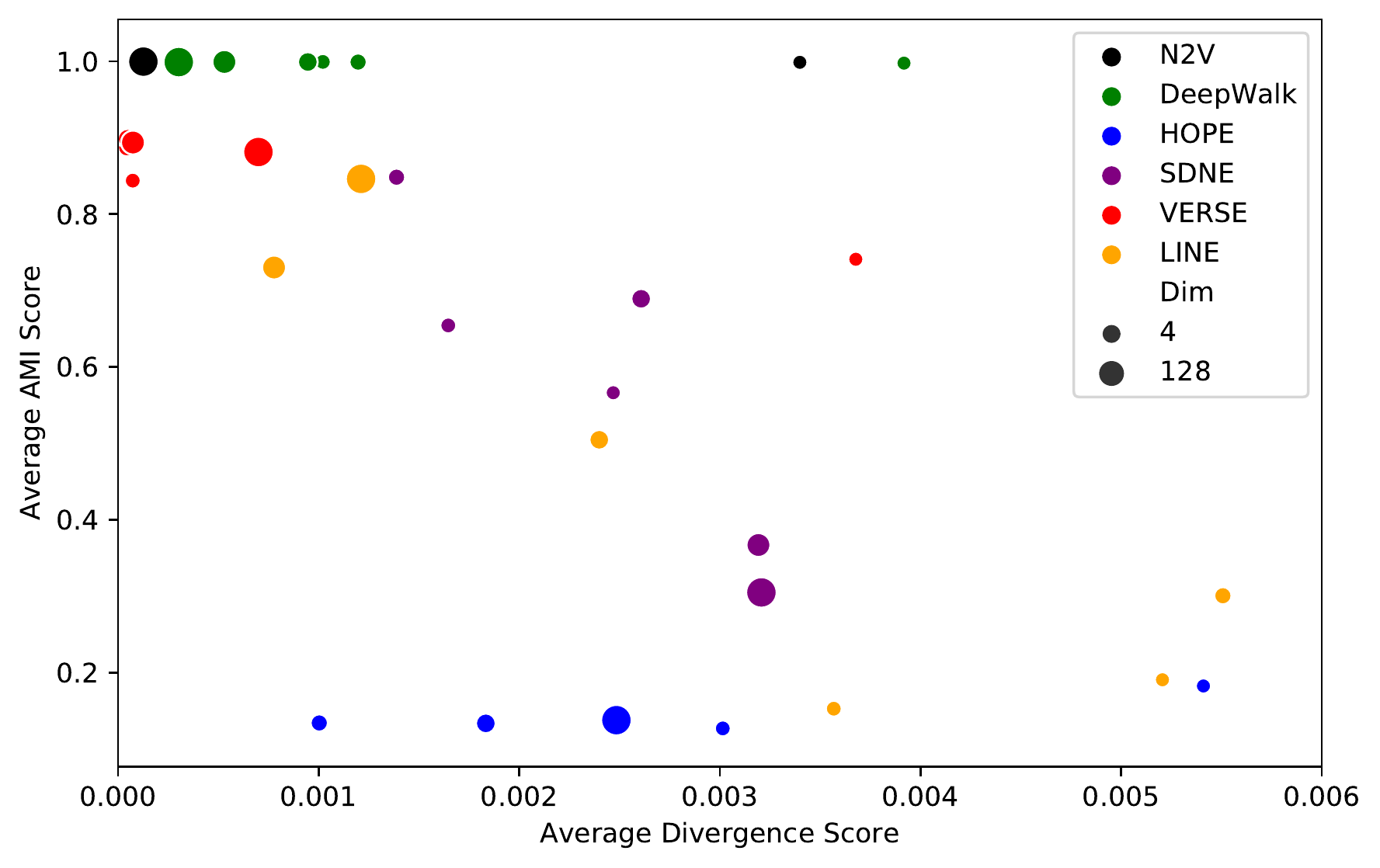}
\includegraphics[width=0.52\textwidth]{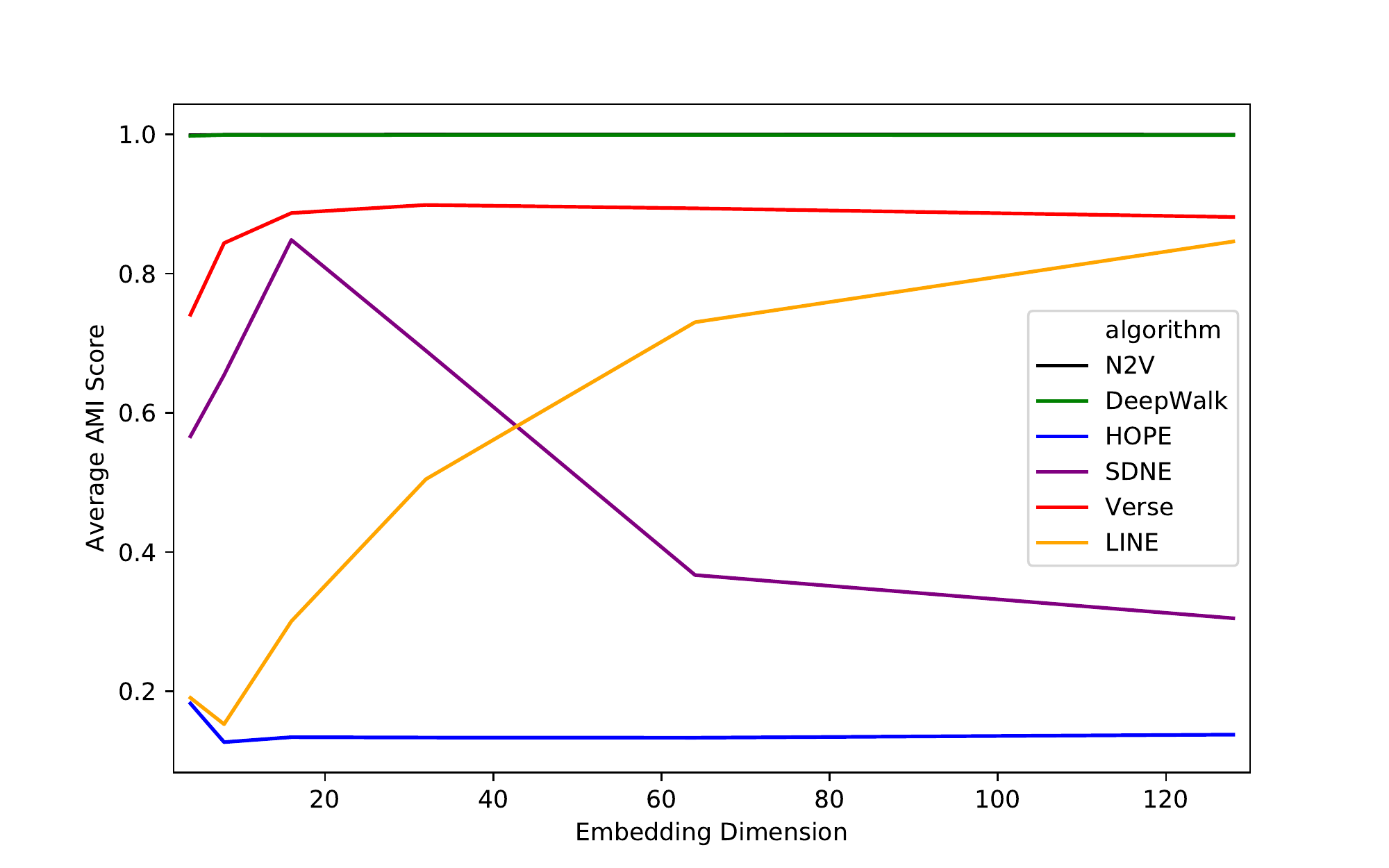}
\caption{\textbf{Community Detection}: relation between the AMI score and the divergence score (left) and the dimension of the embedding (right).} \label{fig:community_detection}
\end{center}
\end{figure}

For our experiment, we used the synthetic \textbf{ABCD} graph with $n=10{,}000$ and all parameters set to their default values. For a given algorithm $A$ and a given dimension $d \in \{4, 8, 16, 32, 64, 128\}$, we independently run 10 times \textbf{$k$-means} algorithm with the correct number of clusters (namely, $k=5$). As we discussed in the previous experiment, our goal is to investigate if there is a correlation between the divergence score and the quality of the embedding and we do not aim to detect communities as best as possible. Hence, we use the ``vanilla'' $k$-means algorithm instead of some more advanced tools such as, for example, DB-scan. In order to compare the results with the ground-truth community structure generated by the \textbf{ABCD} model, we compute the average \textbf{Adjusted Mutual Information} (\textbf{AMI}) score, a widely used measure based on information theory. As before, there is a correlation between the average AMI score and the average divergence score---see Figure~\ref{fig:community_detection} (left). The correlation between the two measures is equal to $-0.64$. There seems to be one outlier with a very low AMI score, \textbf{HOPE}; the correlation without this algorithm is $-0.73$. The quality of \textbf{LINE} increases with the dimension whereas \textbf{SDNE} does the opposite---see Figure~\ref{fig:community_detection} (right). The remaining four algorithms are invariant from that perspective. 


\subsection{Link Prediction}

Node embeddings can be successfully used to predict missing links or to predict links that are likely to be formed in the future. Indeed, networks are often constructed from the observed interactions between nodes, which may be incomplete or inaccurate. In particular, the situation of missing links is typical in the analysis of biological networks in which verifying the existence of links between nodes requires experiments that are expensive and might not be accurate. Moreover, a task that is closely related to link prediction is the main ingredient of recommendation systems. The goal might be to predict missing friendship links in social networks or to recommend new friends. Another task might be to predict new links between users and possible products that they may like.

Once nodes are embedded in $d$-dimensional space, one may use the distance between the corresponding vectors to make the prediction. Nodes that are close to each other in the embedded space but are not adjacent might get connected in the near future as they seem to be similar to each other. On the other hand, since networks that we typically mine are dynamic, one might be interested in predicting which links will become inactive; for example, which users on Instagram a given user might want to unfollow in the near future. A natural guess would be to pick nodes that are far in the embedded space as it indicates that the nodes are dissimilar. 

As before, for our experiment we used the synthetic \textbf{ABCD} graph $G$ with $n=10{,}000$ and all parameters set to their default values. This time, we randomly select 10\% of the edges of $G$, set $E$, and remove them, thus forming a new graph $G' = G \setminus E$. Then we take another random sample of non-adjacent pairs of nodes in $G$, set $E'$. Both classes have the same number of pairs of nodes so that the test set created in such a way is balanced. Our goal is to train a model that uses the embedding of graph $G'$ to detect which pairs of nodes in $E \cup E'$ are adjacent in $G$.

For a given algorithm $A$ and a given dimension $d \in \{4, 8, 16, 32, 64, 128\}$, we find the embedding of graph $G'$. A natural strategy would be to consider all pairs of adjacent nodes in $G'$ (the positive class) as well as a random subset of pairs of non-adjacent nodes (the negative class) and use some standard tools such as the logistic regression model on such training set. Such tools combine embeddings of two nodes into a feature vector to be used for prediction. The output is an estimation of the probability for the positive class in the training data set. However, in the spirit of the two earlier experiments, we keep it simple and compute the $L_2$-distance between pairs of the selected edges and non-edges from $E \cup E'$. 
Based on this, we compute the class membership score using the following simple formula: for each $uv \in E \cup E'$,
$$ 
p (uv) =1-\frac{d(u,v)}{d_{\max}}
$$ 
where $d(uv)$ is the $L_2$-distance between nodes $u$ and $v$, and $d_{\max} = \max_{uv \in E \cup E'} d(uv)$ is maximum distance in the entire set. In particular, as desired, nodes that are close to each other are predicted to be adjacent with high probability and nodes that are far away from each other are most likely non-adjacent. In order to measure the quality of this simple model, we use the \textbf{Area Under the ROC Curve} (\textbf{AUC}) that provides a measure of separability as it tells us how capable the model is of distinguishing between the two classes. Indeed, \textbf{AUC} is bounded from above by 1 and can be interpreted as probability that a random positive observation has a higher predicted probability than a random negative observation. 

\begin{figure}[!htb]
\begin{center}
\includegraphics[width=0.49\textwidth]{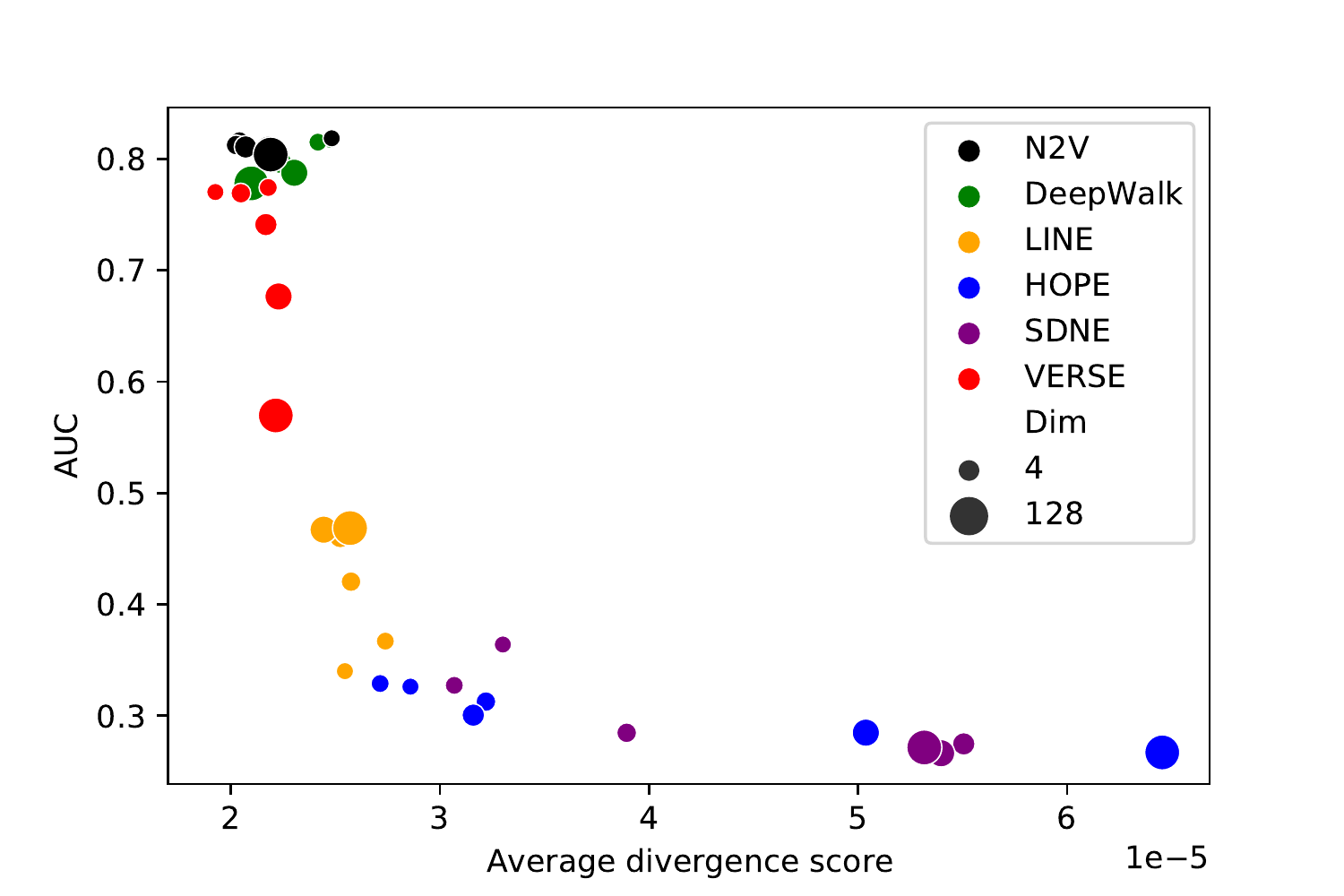}
\includegraphics[width=0.49\textwidth]{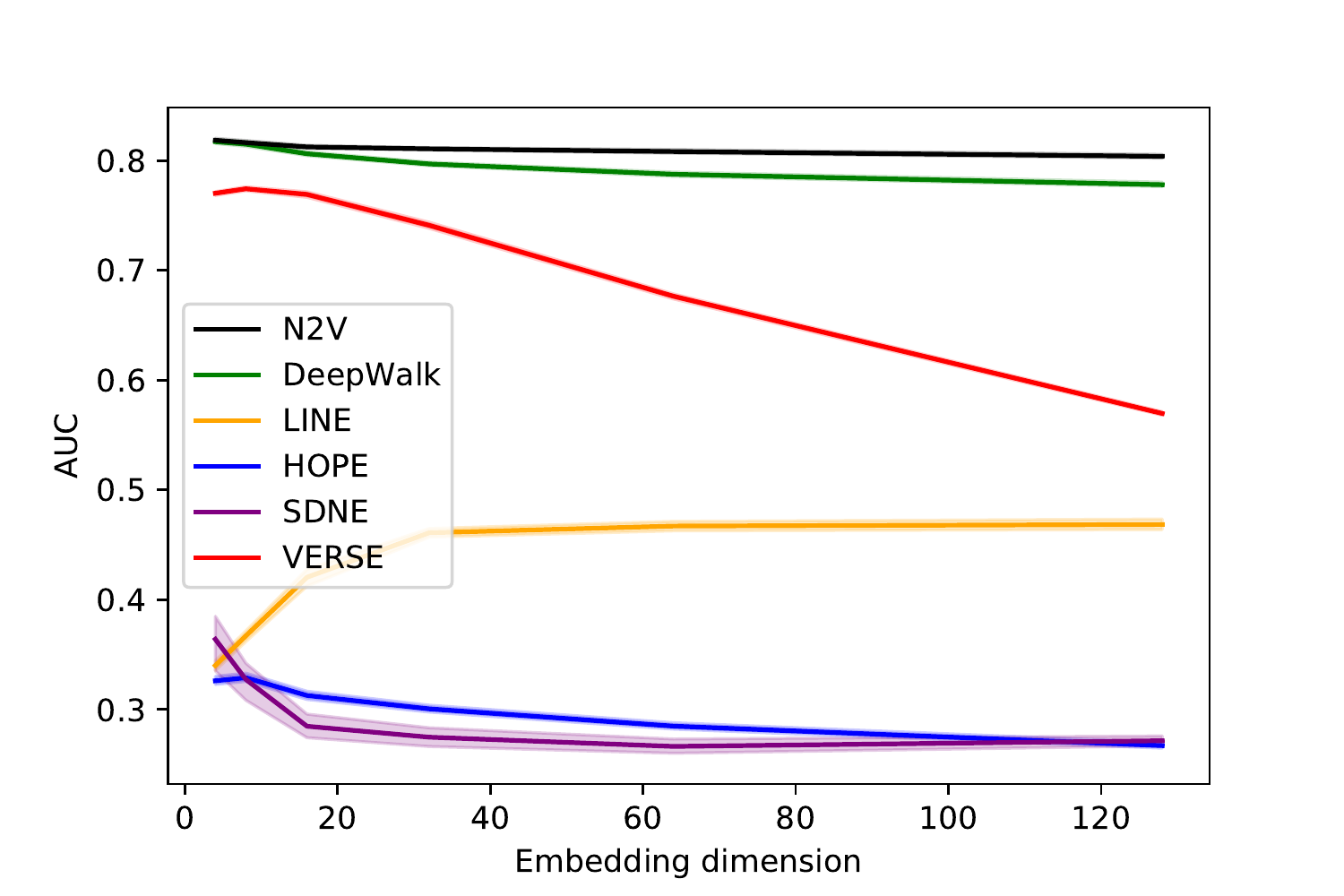}
\caption{\textbf{Link Prediction}: relation between the AUC score and the divergence score.} \label{fig:link_prediction}
\end{center}
\end{figure}

We repeat the above procedure 10 times for each algorithm $A$ and dimension $d$, each time independently selecting 10\% of edges of $G$ to form graph $G'$. We compute the average \textbf{AUC} score as well as the average divergence score. As in the previous experiments, there is a strong correlation between the \textbf{AUC} score and the divergence score: the correlation coefficient is equal to $-0.72$. Indeed, Figure~\ref{fig:link_prediction} (left) shows that the relation is not linear but there is a clear trend, as expected and as desired. Surprisingly, some of the embedding algorithms achieve lower \textbf{AUC} scores in higher dimensions but the difference is significant only for \textbf{VERSE}---see Figure~\ref{fig:link_prediction} (right). However, note that our simple classifier rely exclusively on distances while embeddings in higher dimensions may capture other features that might be used by more sophisticated classifiers. In any case, these results confirm that more useful information is kept by embeddings that score well using the benchmarking framework. 

\begin{figure}[!htb]
\begin{center}
\includegraphics[width=0.49\textwidth]{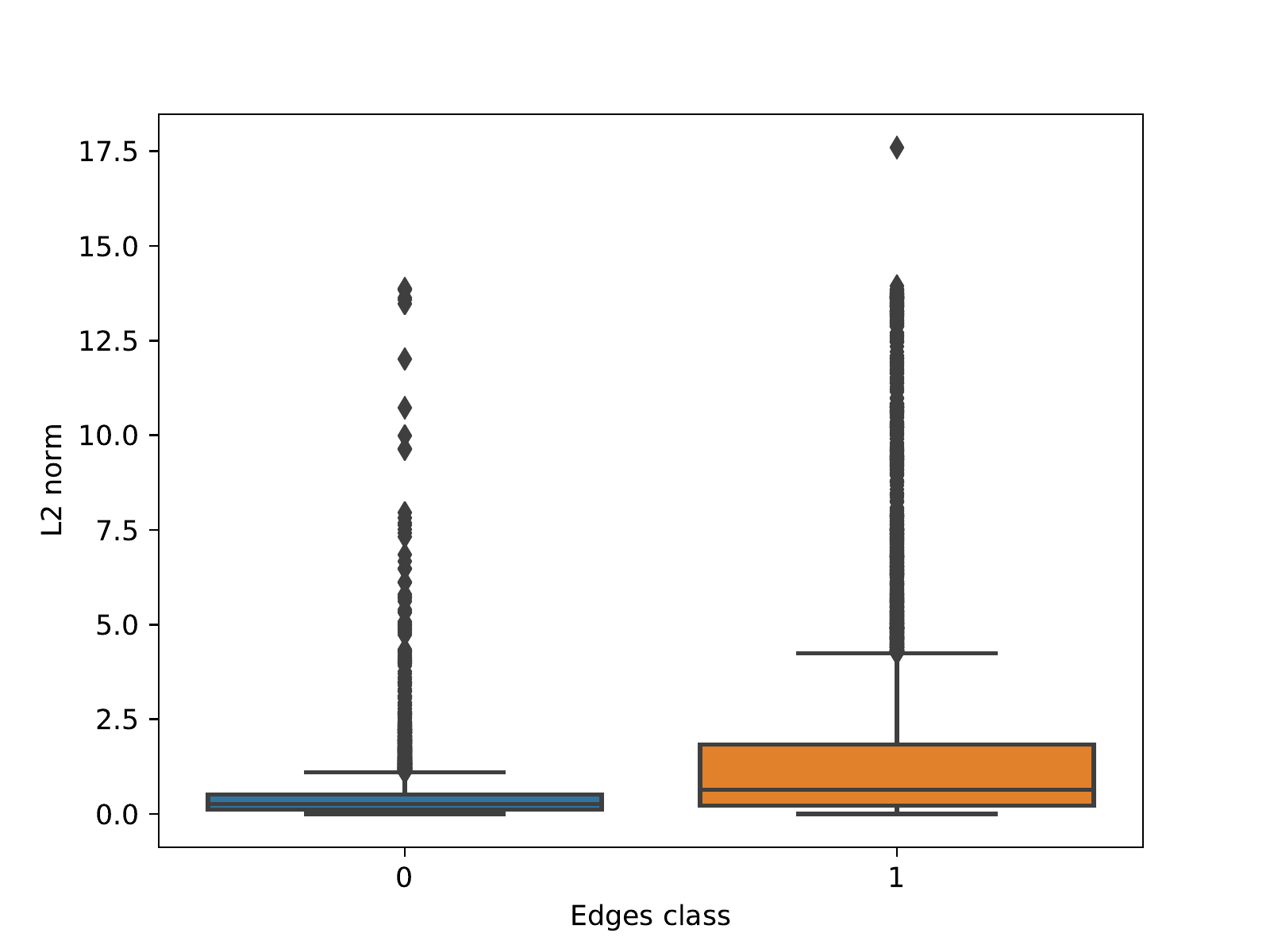}
\includegraphics[width=0.49\textwidth]{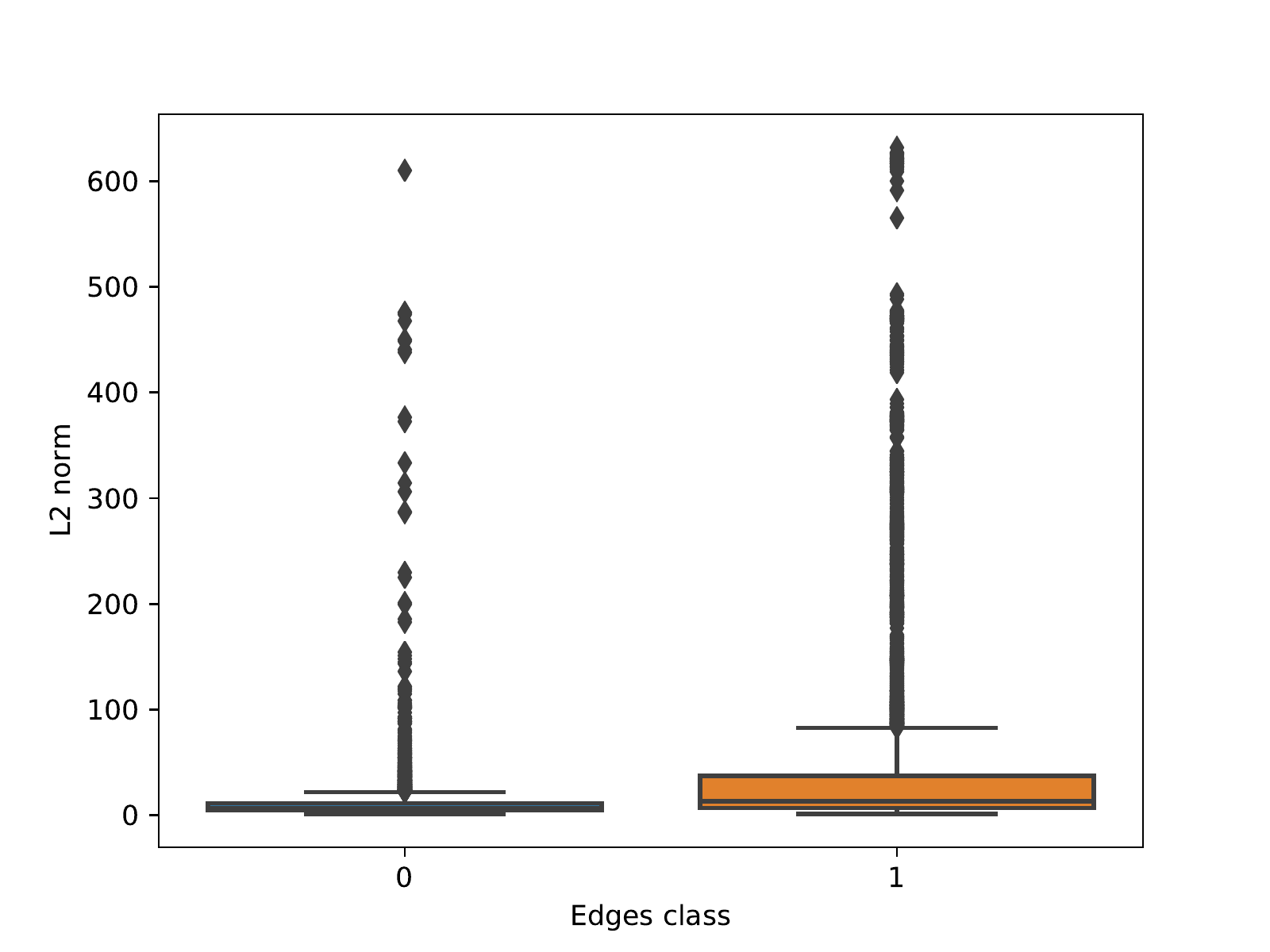}
\caption{Two of the worst embeddings according to the divergence score---\textbf{HOPE}, $d=4$ and \textbf{SDNE}, $d=128$.}\label{fig:worst}
\end{center}
\end{figure}

\begin{figure}[!htb]
\begin{center}
\includegraphics[width=0.49\textwidth]{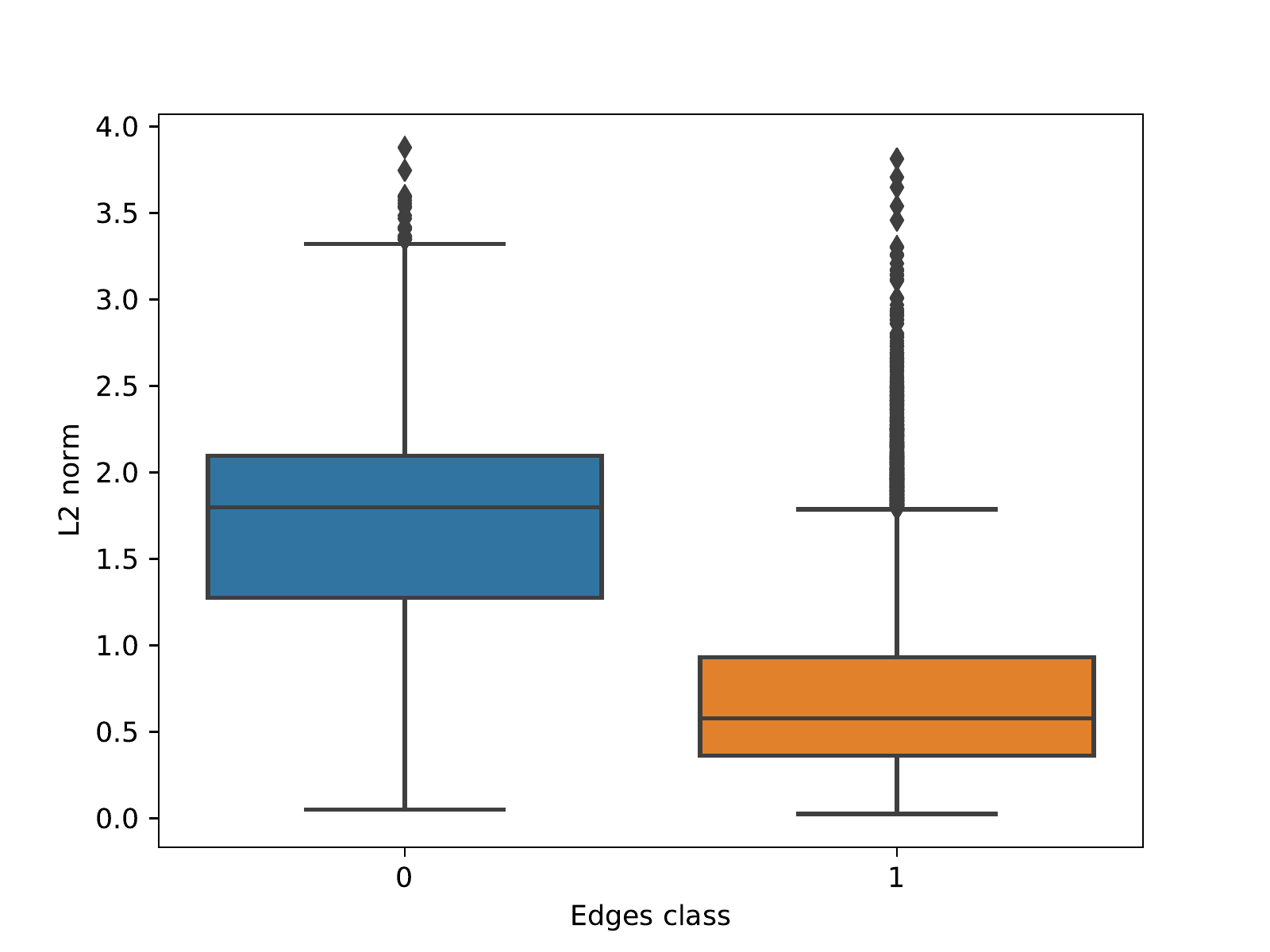}
\includegraphics[width=0.49\textwidth]{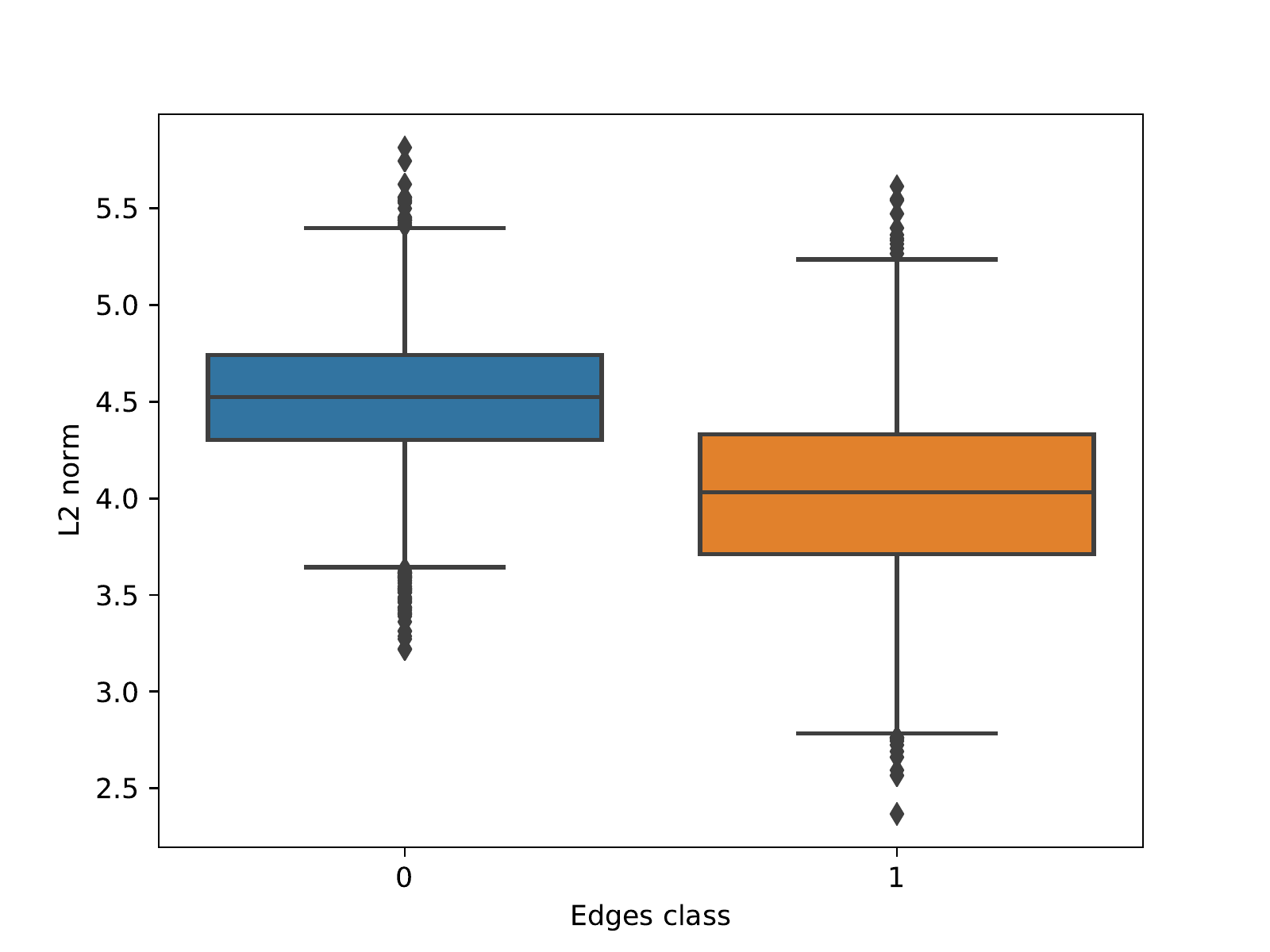}
\caption{Two of the best embeddings according to the divergence score---\textbf{node2vec}, $d=4$ and \textbf{node2vec}, $d=128$.}\label{fig:best}
\end{center}
\end{figure}

To further support this conclusion, we investigated two embeddings that scored well by the framework (\textbf{node2vec}, $d=4$ and \textbf{node2vec}, $d=128$) and two that were ranked as the worst ones (\textbf{HOPE}, $d=4$ and \textbf{SDNE}, $d=128$)---see Table~\ref{tab:performance}. For each of them, we plot a distribution of lengths independently for $E$ and $E'$---see Figures~\ref{fig:worst} and~\ref{fig:best}. Good embeddings keep adjacent nodes close to each other whereas bad embeddings actually do the opposite.

\begin{table}[!htb]
\center
\caption{Performance metrics for distance-based link prediction.}
\begin{tabular}{|l|l|l|l|}
\hline
\textbf{Embedding} & \textbf{AUC} & \textbf{Accuracy} & \textbf{Divergence} \\ \hline
node2vec $d=4$           & 0.82         & 0.79              & 2.46 x $10^{-5}$               \\ \hline
node2vec $d=128$        & 0.81         & 0.74              & 1.92 x $10^{-5}$              \\ \hline
HOPE $d=4$        & 0.31         & 0.37              & 3.55 x $10^{-5}$              \\ \hline
SDNE $d=128$      & 0.29         & 0.35              & 4.56 x $10^{-5}$              \\ \hline
\end{tabular}
\label{tab:performance}
\end{table}

\section{Summary: $\ldots$Yes! One Can Trust the Framework!}

Node embedding is an important tool to extract useful information from graphs. There are many excellent algorithms proposed in the literature but the quality of their outcomes depends on the structure of the network that one aims to process. As a group of researchers and practitioners that often use embedding algorithms, in this project we aimed to investigate various algorithms (using different techniques to build them) to be able to make a better and more informed choices which ones to use. The conclusion we converged to is to use \textbf{node2vec} as a default choice---this algorithm constantly works good for both real world networks as well as synthetically generated ones. Having said that, the other algorithms were often at least comparable if not slightly better, but the competitors change from experiment to experiment. Moreover, each algorithm (including \textbf{node2vec}) has a number of parameters one can tune, the dimension being only one of them.

In light of this unclear best choice of the algorithm and its parameters, we recommend to use the benchmarking framework to make that decision in an unsupervised way, without manually inspecting the quality of the generated embeddings. In order to support this recommendation, we performed a number of experiments in which we apply some classical tools to important machine learning tasks and measured if there is a correlation between the divergence score returned by the benchmarking framework and the quality of the tool that uses embeddings to guide the process. A strong correlation between the two measures supports the recommended approach.

Having said that, there are some natural followup questions that can be asked and experiments to be performed. Let us mention about two of them. We measured how the divergence score depends on some simple statistics of the graph such as the level of noise or degree distribution. In order to achieve it, we experimented with the \textbf{ABCD} graph in which we fixed all but one parameters and vary the one that is not fixed. To get a more detailed understanding of the effect on the divergence score, it would be interesting to see how a combination of two parameters affect the quality of the embeddings. Second experiment that we would like to suggest is related to the applications of node embeddings. In this paper we tested three natural applications (nodes classification, community detection, and link prediction). Another important application is to detecting anomalies. It would be interesting to see how the quality of various anomaly detection algorithms depends on the divergence score. 

\section{Acknowledgement}

Experiments were conducted using \textbf{SOSCIP}\footnote{\texttt{https://www.soscip.org/}} Cloud infrastructure.  Launched in 2012, the \textbf{SOSCIP} consortium is a collaboration between Ontario’s research-intensive post-secondary institutions and small- and medium-sized enterprises (SMEs) across the province. Working together with the partners, \textbf{SOSCIP} is driving the uptake of AI and data science solutions and enabling the development of a knowledge-based and innovative economy in Ontario by supporting technical skill development and delivering high-quality outcomes. \textbf{SOSCIP} supports industrial-academic collaborative research projects through partnership-building services and access to leading-edge advanced computing platforms, fuelling innovation across every sector of Ontario’s economy.

For our experiments, we used Compute G4-x8 (8 vCPUs,	32 GB RAM) machines and Ubuntu 18.04 operating system. Computation used for experimentation and calibration of the scripts took approximately 20{,}000 vCPU-hours. The scripts and results can be found on GitHub repository\footnote{\texttt{https://github.com/arash-dehghan/EmbeddingComplexNetworks}}.

\section{Appendix}

\subsection{Definition of the Divergence Score}\label{sec:divergence}

In this section we provide a mathematical definition of the divergence score. It is a shortened version of definition given in~\cite{Embedding_Complex_Networks}.

Given a graph $G=(V,E)$, its degree distribution $\textbf{w}$ on $V$, and an embedding $\emb : V \to \R^k$ of its vertices in $k$-dimensional space, we perform the five steps detailed below
to obtain $\Delta_\emb(G)$, a \emph{divergence score} for the embedding.

\medskip \noindent \textbf{Step 1:} Run some clustering algorithm such as the Ensemble Clustering for Graphs (ECG) on $G$ to obtain a partition $\textbf{C}$ of the vertex set $V$ into $\ell$ communities $C_1, \ldots, C_\ell$. 

\medskip \noindent \textbf{Step 2:} For each $i \in [\ell]$, let $c_{i}$ be the proportion of edges of $G$ with both endpoints in $C_i$. Similarly, for each $1 \le i < j \le \ell$, let $c_{i,j}$ be the proportion of edges of $G$ with one endpoint in $C_i$ and the other one in $C_j$. Let
\begin{equation}
\bar{\textbf{c}} = (c_{1,2},\ldots, c_{1,\ell}, c_{2,3}, \ldots, c_{2,\ell}, \ldots, c_{\ell-1,\ell} )
\quad \text{and }
\hat{\textbf{c}} = (c_1, \ldots, c_\ell)
\label{eq:c}
\end{equation}
be two vectors with a total of $\binom{\ell}{2} + \ell = \binom{\ell+1}{2}$ entries which together sum to one. These {\it graph vectors} characterize the partition $\textbf{C}$ from the perspective of the graph $G$.

\medskip \noindent \textbf{Step 3:} For a given parameter $\alpha \in \R_+$ and the same vertex partition $\textbf{C}$, we consider $\mathcal{G}(\textbf{w}, \emb, \alpha)$, the \textbf{Geometric Chung-Lu model}. For each $1 \le i < j \le \ell$, we compute $b_{i,j}$, the expected proportion of edges of $\mathcal{G}(\textbf{w}, \emb, \alpha)$ with one endpoint in $C_i$ and the other one in $C_j$. Similarly, for each $i \in [\ell]$, let $b_i$ be the expected proportion of edges within $C_i$. That gives us another two vectors
\begin{equation}
\bar{\textbf{b}}_\emb(\alpha) = (b_{1,2},\ldots, b_{1,\ell}, b_{2,3}, \ldots, b_{2,\ell}, \ldots, b_{\ell-1,\ell} )  \qquad \text{ and } \qquad
\hat{\textbf{b}}_\emb(\alpha) = (b_{1},\ldots,  b_{\ell} )
\label{eq:b}
\end{equation}
with a total of $\binom{\ell+1}{2}$ entries which together sum to one. These \emph{model vectors} characterize the partition $\textbf{C}$ from the perspective of the embedding $\emb$.

\medskip \noindent \textbf{Step 4:} Compute the Jensen–Shannon divergence between the two pairs of vectors, that is, between $\bar{\textbf{c}}$ and $\bar{\textbf{b}}_\emb(\alpha)$, and between $\hat{\textbf{c}}$ and $\hat{\textbf{b}}_\emb(\alpha)$, in order to measure how well the model $\mathcal{G}(\textbf{w}, \emb, \alpha)$ fits the graph $G$. Let $\Delta_\alpha$ be a weighted average of the two distances.

\medskip \noindent \textbf{Step 5:} Select $\hat{\alpha} = \argmin_{\alpha} \Delta_\alpha$, and define the \emph{divergence score} for embedding $\emb$ on $G$ as: $\Delta_\emb(G) = \Delta_{\hat{\alpha}}$.

\subsection{ABCD Random Graph Model with Community Structure}\label{sec:ABCDapp}

In this section, we briefly discuss the {\bf ABCD} models. It is a short description taken from~\cite{ABCDe}; details can be found in~\cite{ABCD} or in~\cite{ABCD-theory}.
As in {\bf LFR}, for a given number of nodes $n$, we start by generating a power law distribution both for the degrees and community sizes. Those are governed by the power law exponent parameters $(\gamma,\beta)$. We also provide additional information to the model, again as it is done in {\bf LFR}, namely, the average and the maximum degree, and the range for the community sizes. The user may alternatively provide a specific degree distribution and/or community sizes.

For each community, we generate a random {\it community} subgraph on the nodes from a given community using either the \textbf{configuration model} which preserves the exact degree distribution, or the \textbf{Chung-Lu model} which preserves the expected degree distribution.
On top of it, we independently generate a {\it background} random graph on all the nodes. Everything is tuned properly so that the degree distribution of the union of all graphs follows the desired degree distribution (only in expectation in the case of the Chung-Lu variant).
The mixing parameter $\xi$ guides the proportion of edges which are generated via the background graph. In particular, in the two extreme cases, when $\xi=1$ the graph has no community structure while if $\xi=0$, then we get disjoint communities.
In order to generate simple graphs, we may have to do some re-sampling or edge re-wiring, which are described in~\cite{ABCD}.

During this process, larger communities will additionally get some more internal edges due to the background graph. As argued in~\cite{ABCD}, this ``global'' variant of the model is more natural and so we recommend it. However, in order to provide a variant where the expected proportion of internal edges is exactly the same for every community (as it is done in {\bf LFR}), we also provide a ``local'' variant of {\bf ABCD} in which the mixing parameter $\xi$ is automatically adjusted for every community.

\begin{figure}
\centering
\includegraphics[scale=0.5]{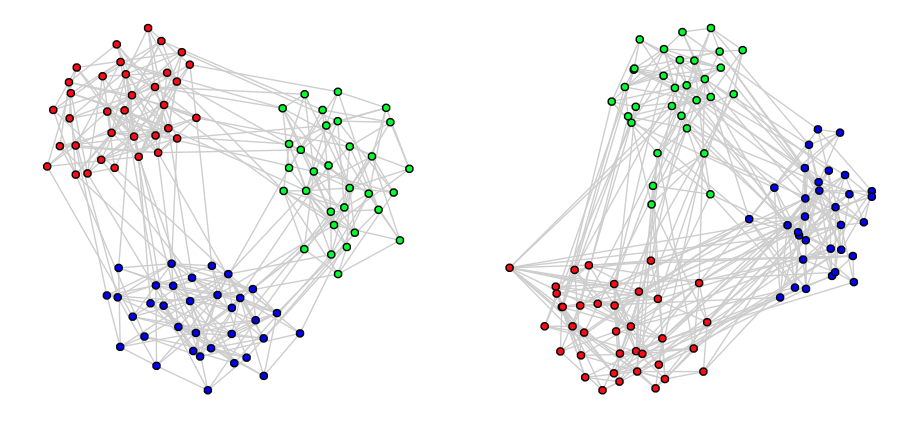}
\caption{Two examples of \textbf{ABCD} graphs with low level of noise ($\xi=0.2$, left) and higher level of noise ($\xi=0.4$, right).}
\label{fig:examples}
\end{figure}

Two examples of \textbf{ABCD} graphs on $n=100$ nodes are presented in Figure~\ref{fig:examples}.
Degree distribution was generated with power law exponent $\gamma=2.5$ with minimum and maximum values 5 and 15, respectively. Community sizes were generated with power law exponent $\beta = 1.5$ with minimum and maximum values 30 and 50, respectively; communities are shown in different colours. The global variant and the configuration model was used to generate the graphs. The left plot has the mixing parameter set $\xi=0.2$ while the ``noisier'' graph on the right plot has the parameter fixed to $\xi=0.4$.

\subsection{Visualizations of the Experiments}

\begin{figure}[htbp!]
\begin{center}
\includegraphics[width=0.3\textwidth]{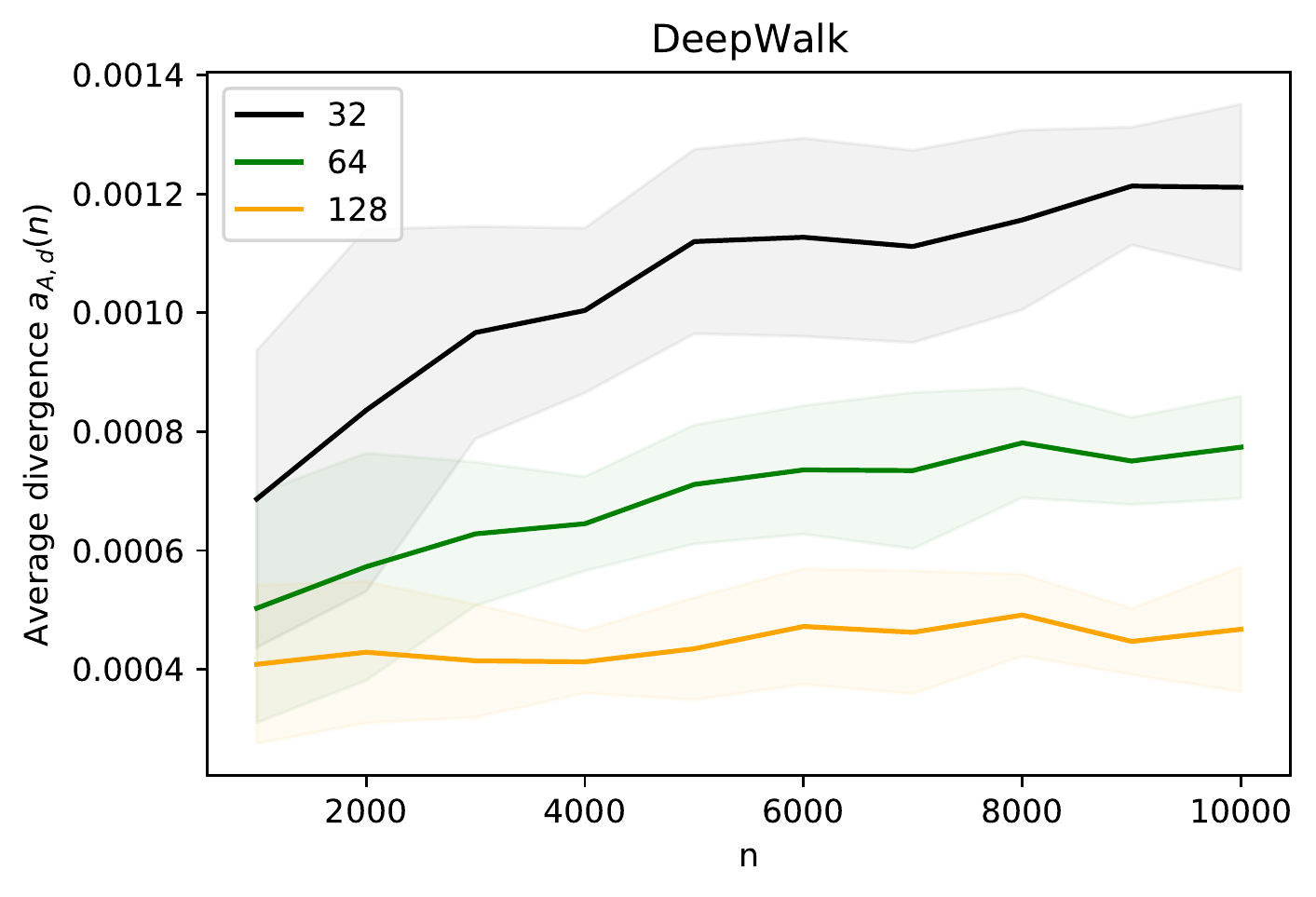}
\includegraphics[width=0.3\textwidth]{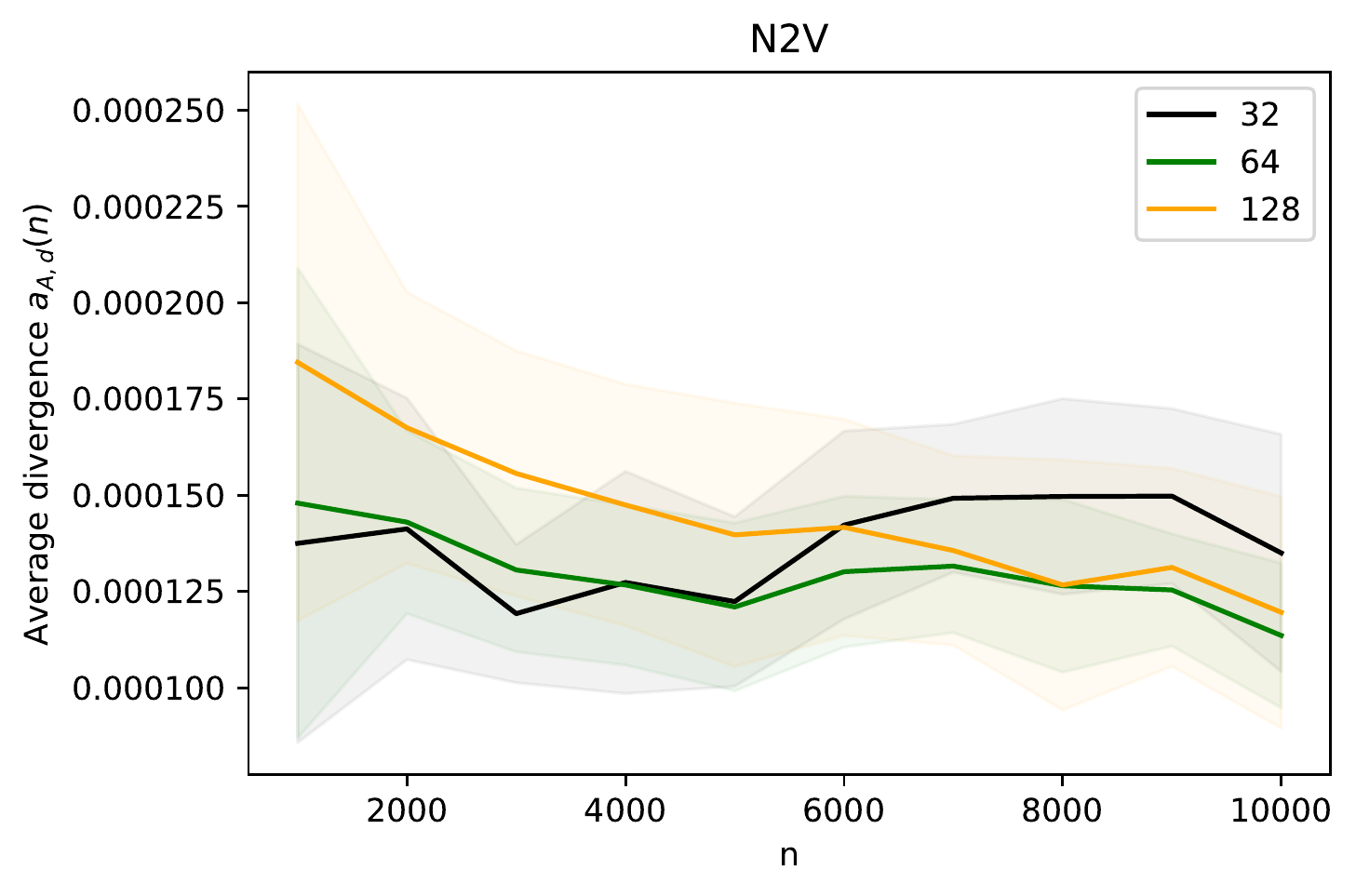}
\includegraphics[width=0.3\textwidth]{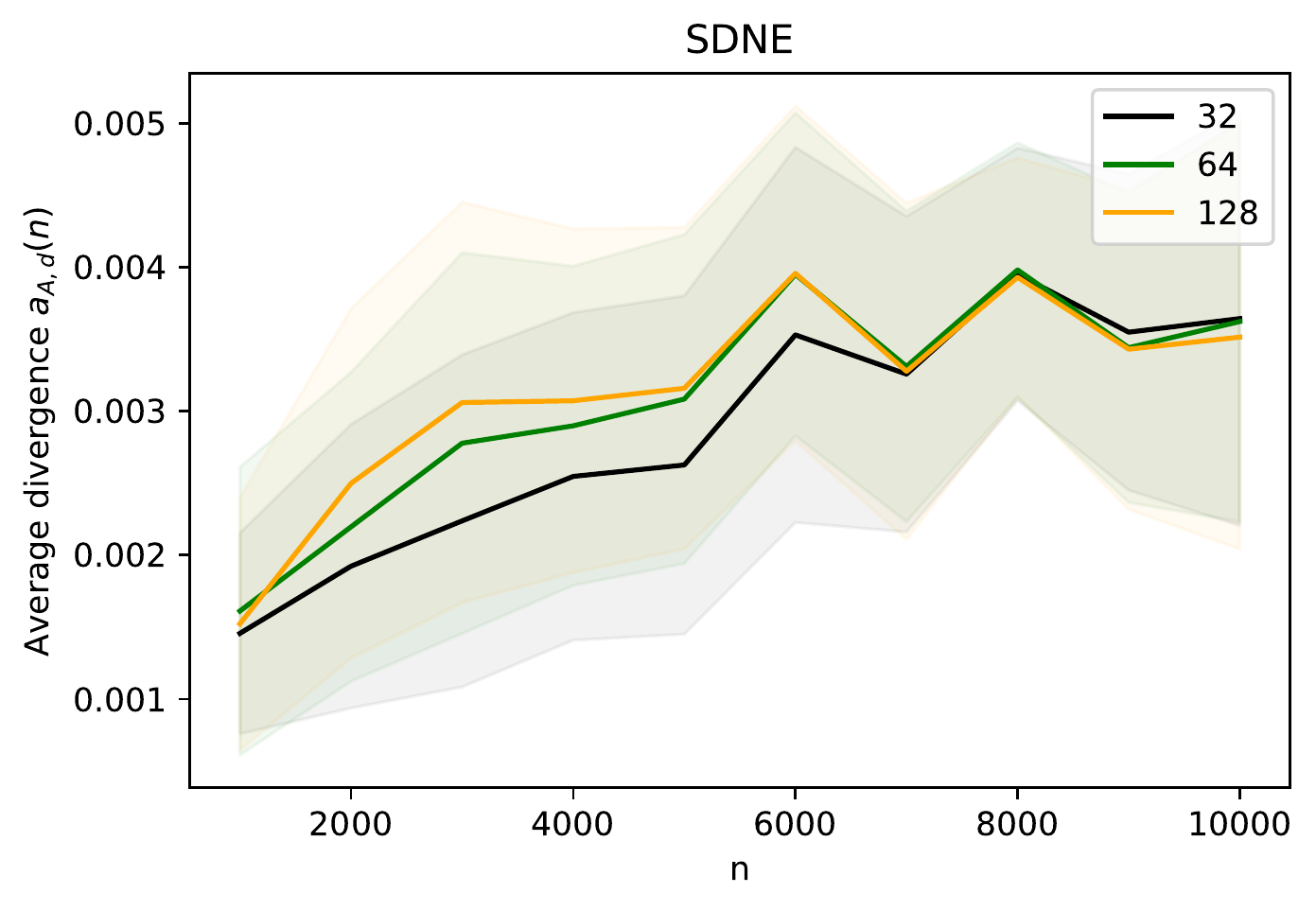} \\
\includegraphics[width=0.3\textwidth]{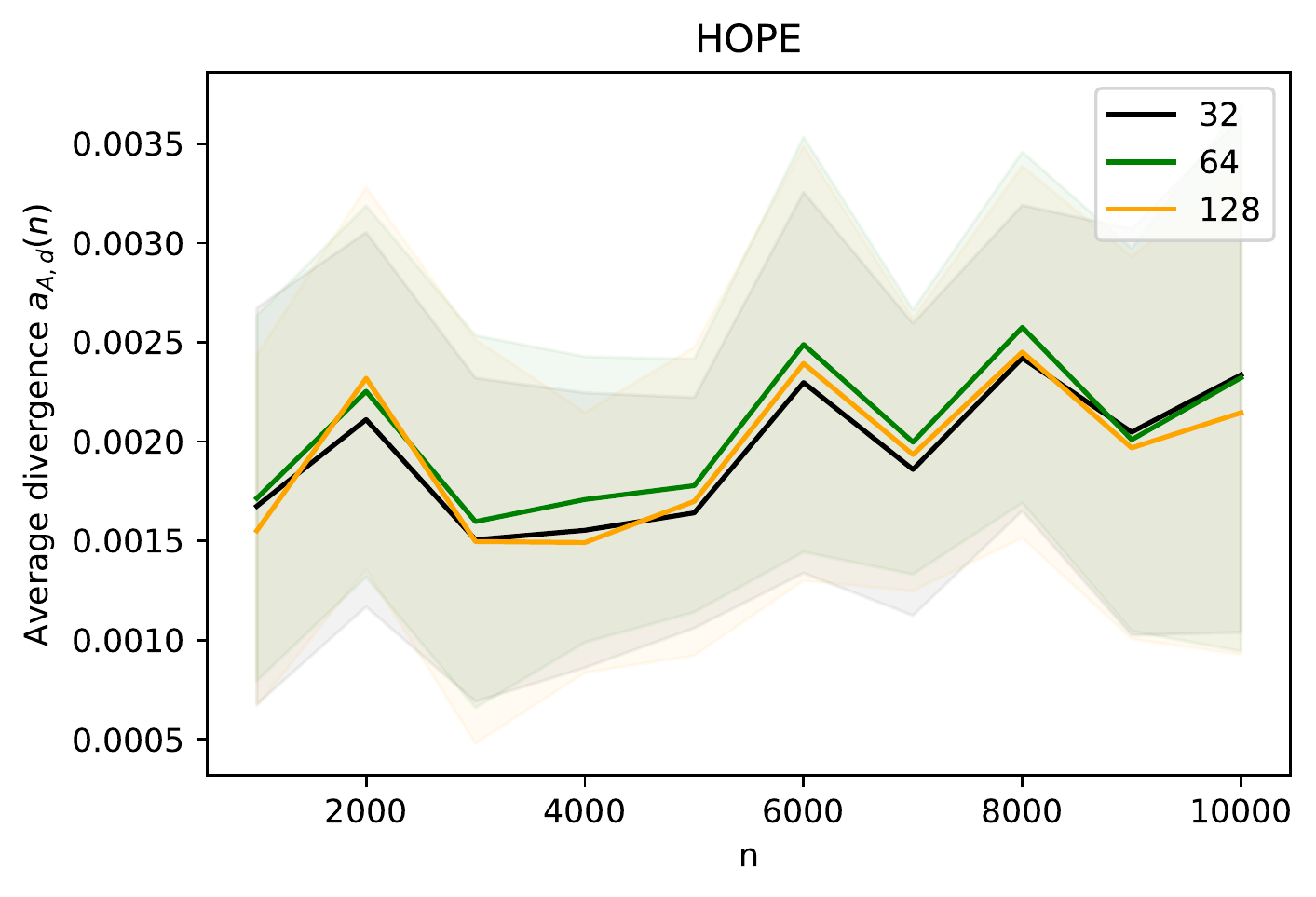}
\includegraphics[width=0.3\textwidth]{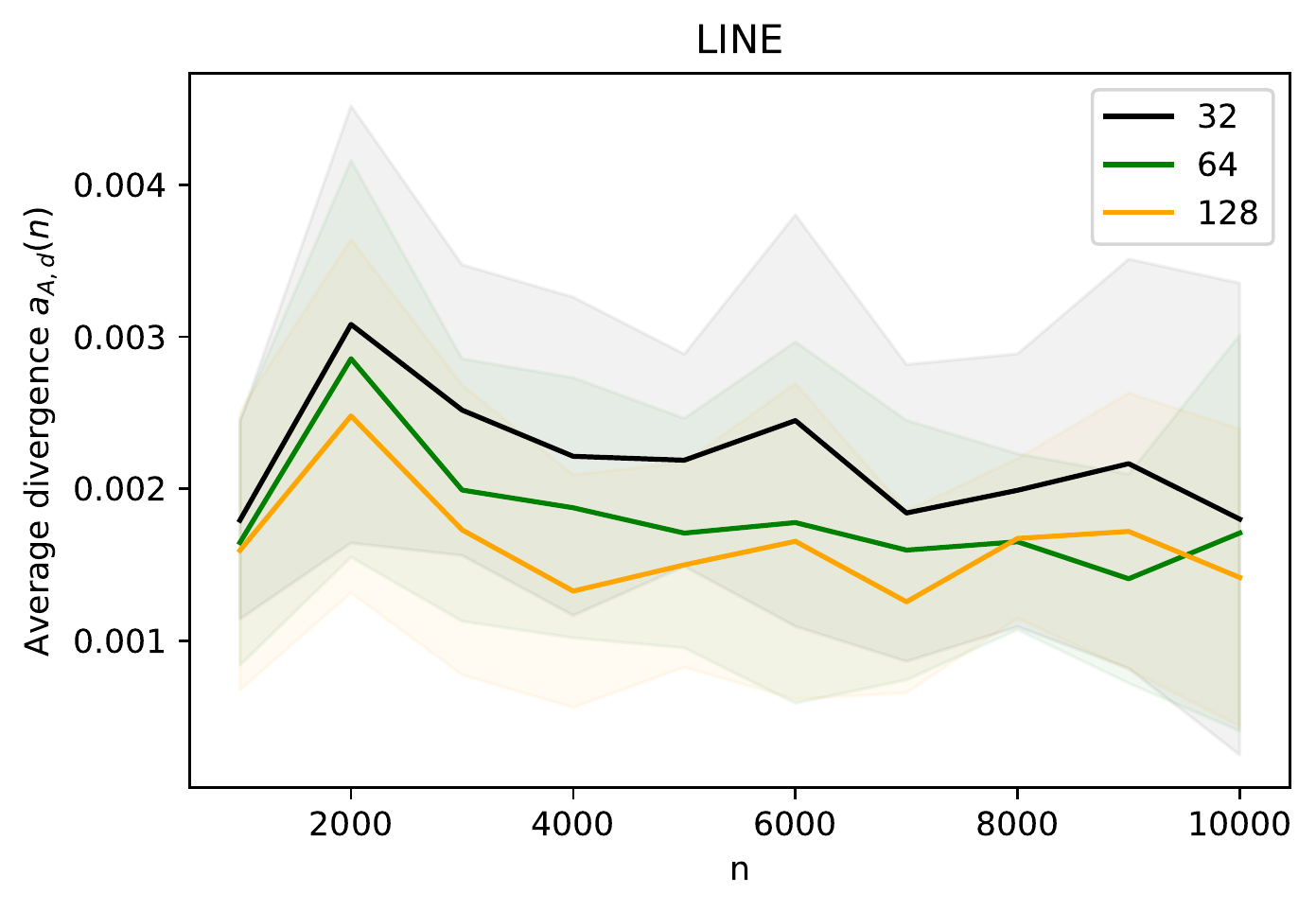}
\includegraphics[width=0.3\textwidth]{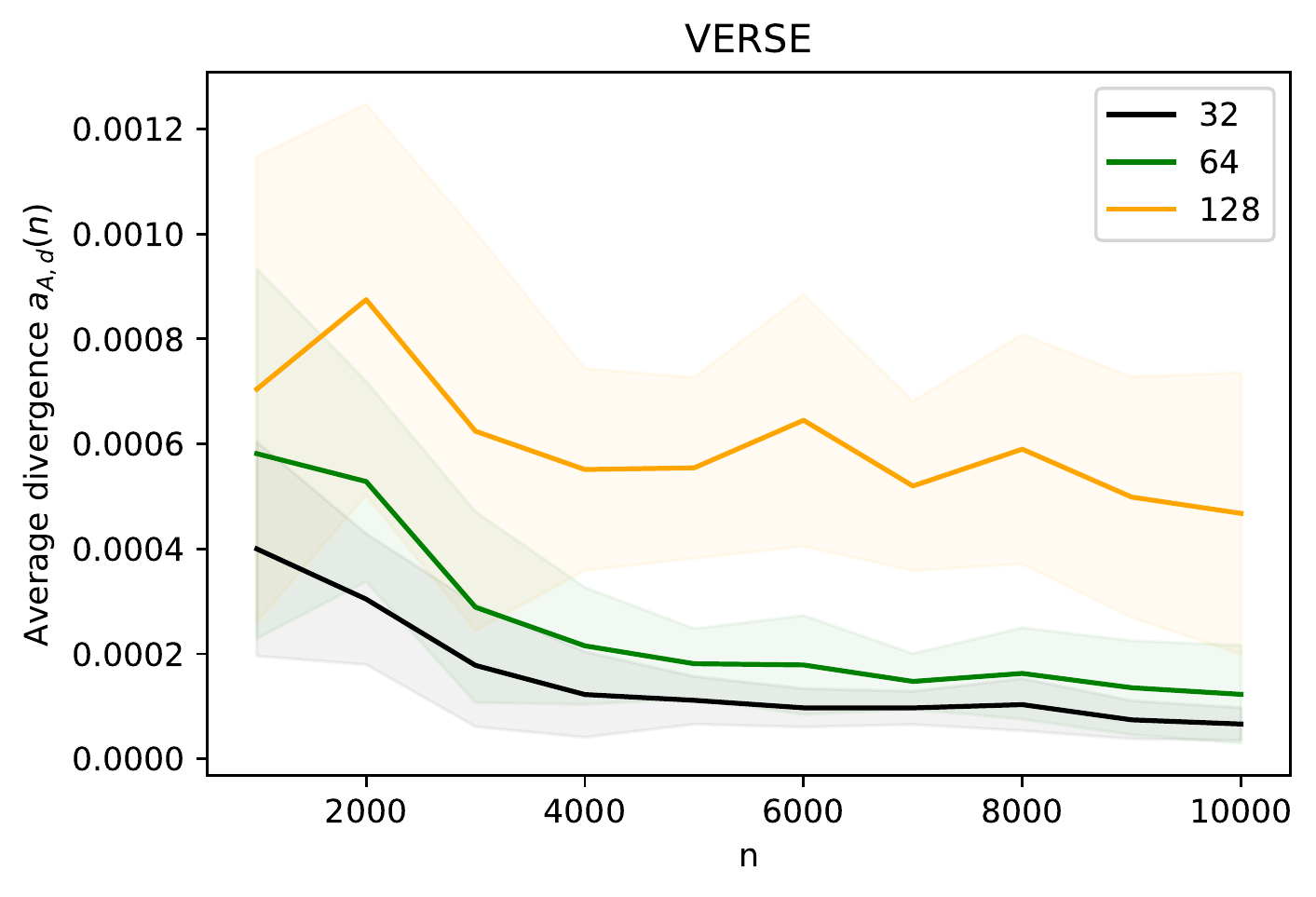} \\
\includegraphics[width=0.3\textwidth]{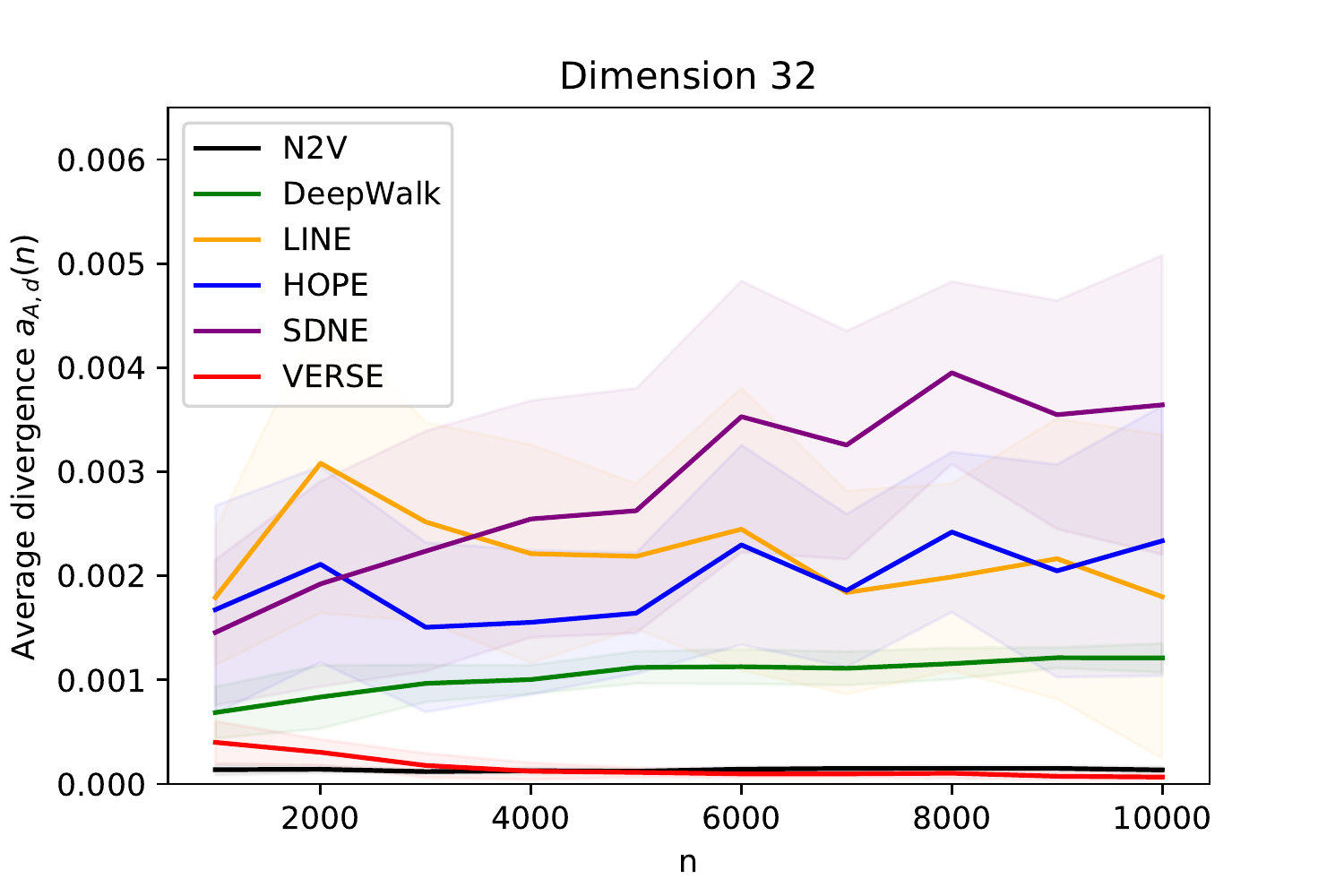}
\includegraphics[width=0.3\textwidth]{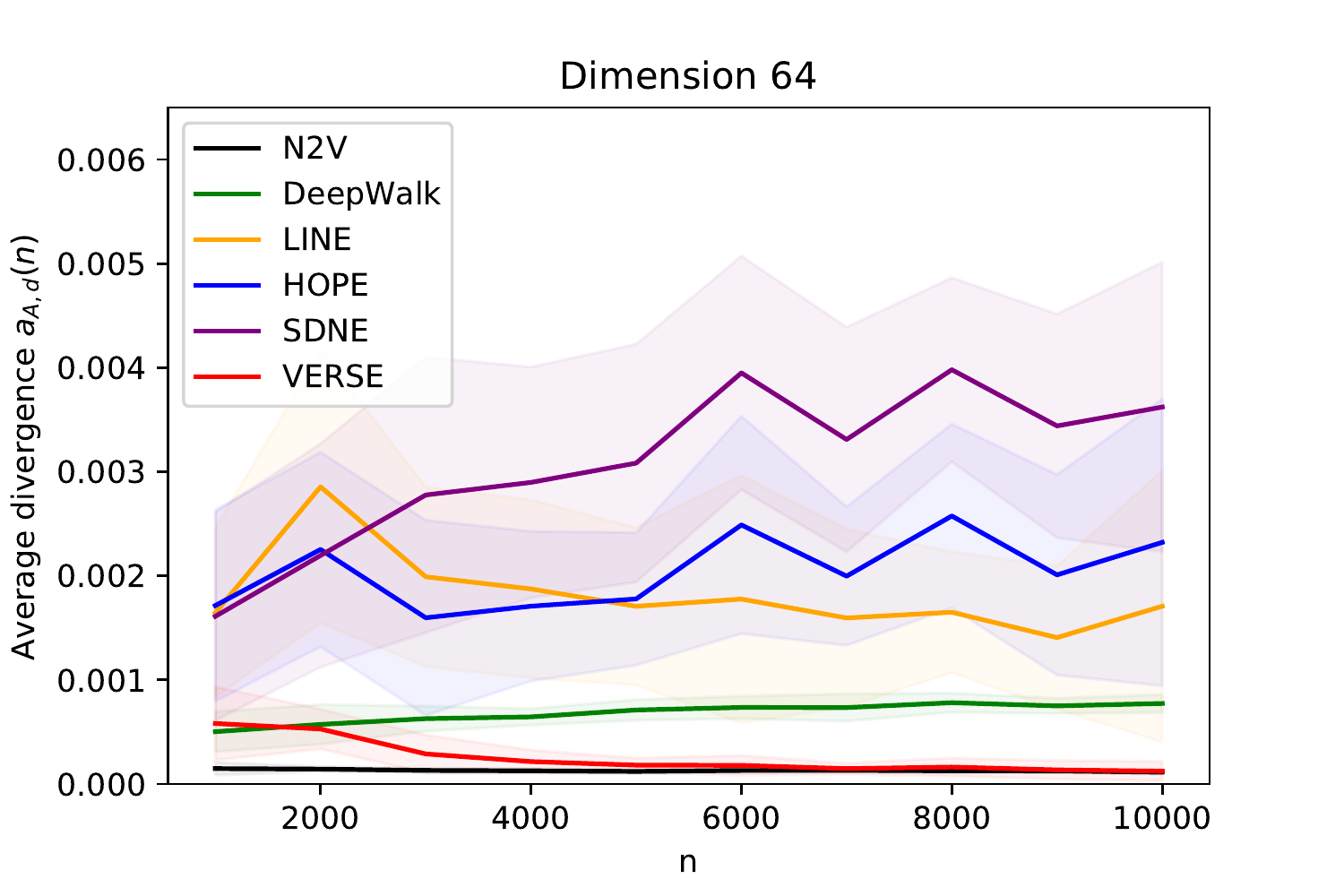}
\includegraphics[width=0.3\textwidth]{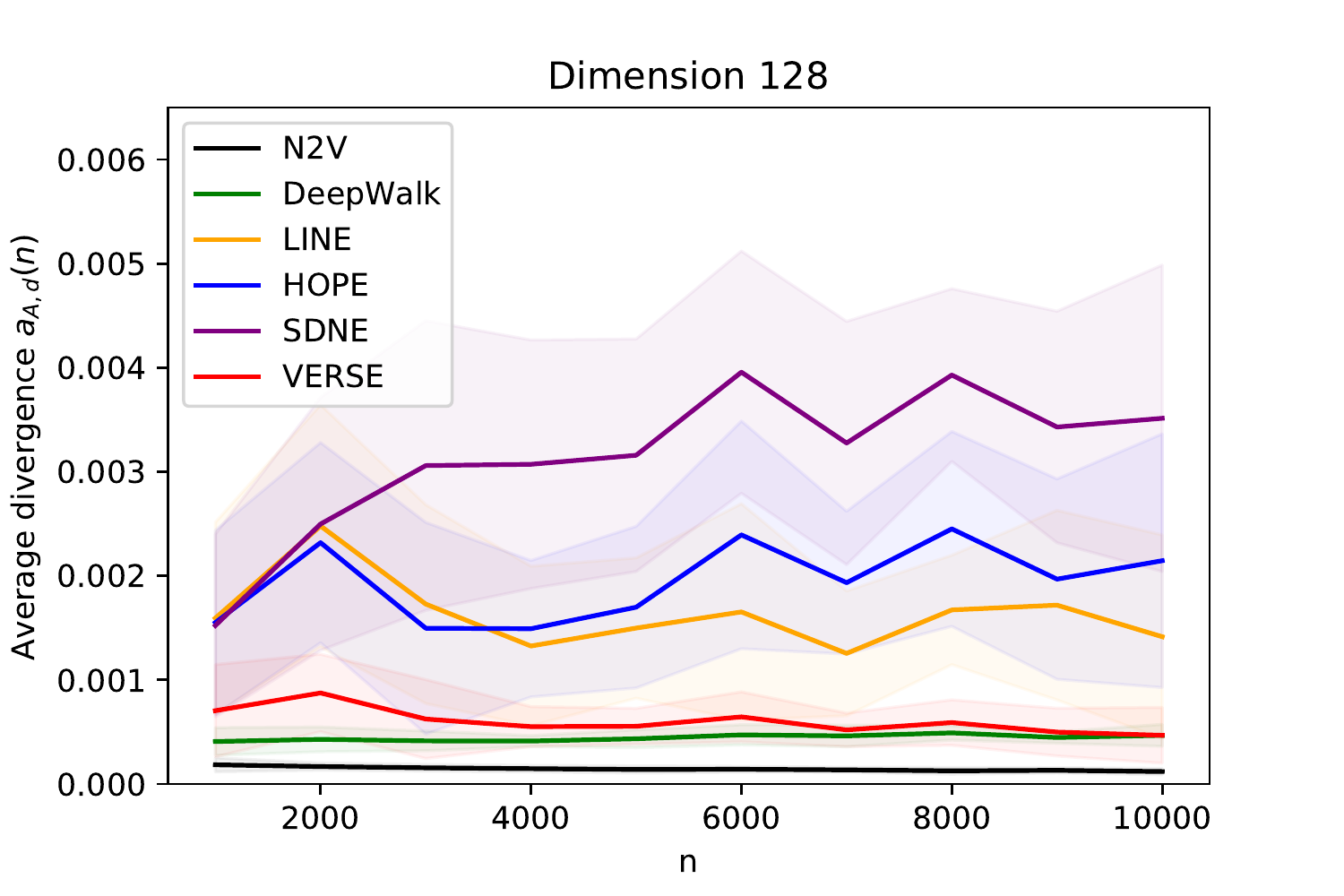} \\
Plots 1 and  2: $a_{A,d}(n) \pm  s_{A,d}(n)$ \\

\includegraphics[width=0.2\textwidth]{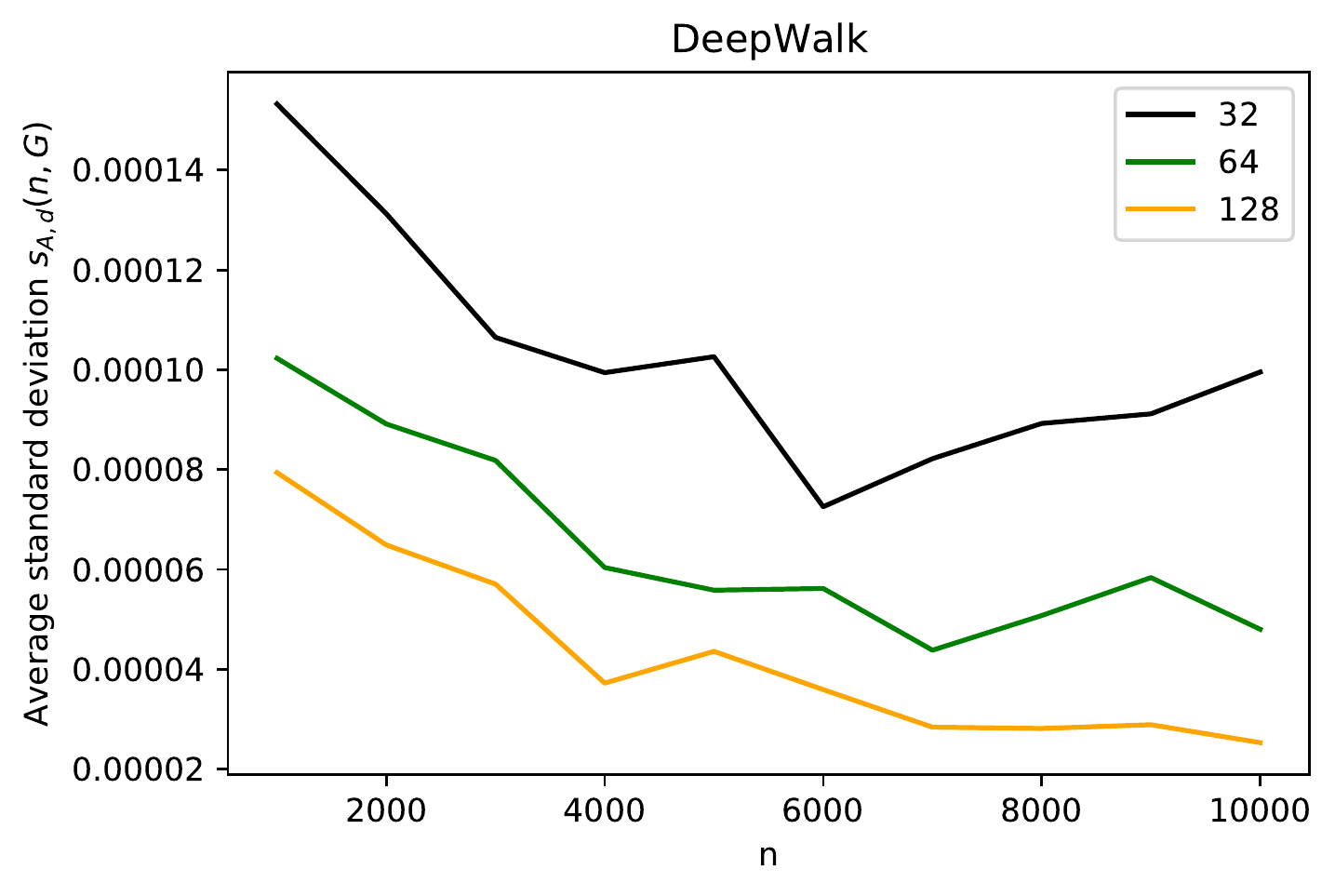}
\includegraphics[width=0.2\textwidth]{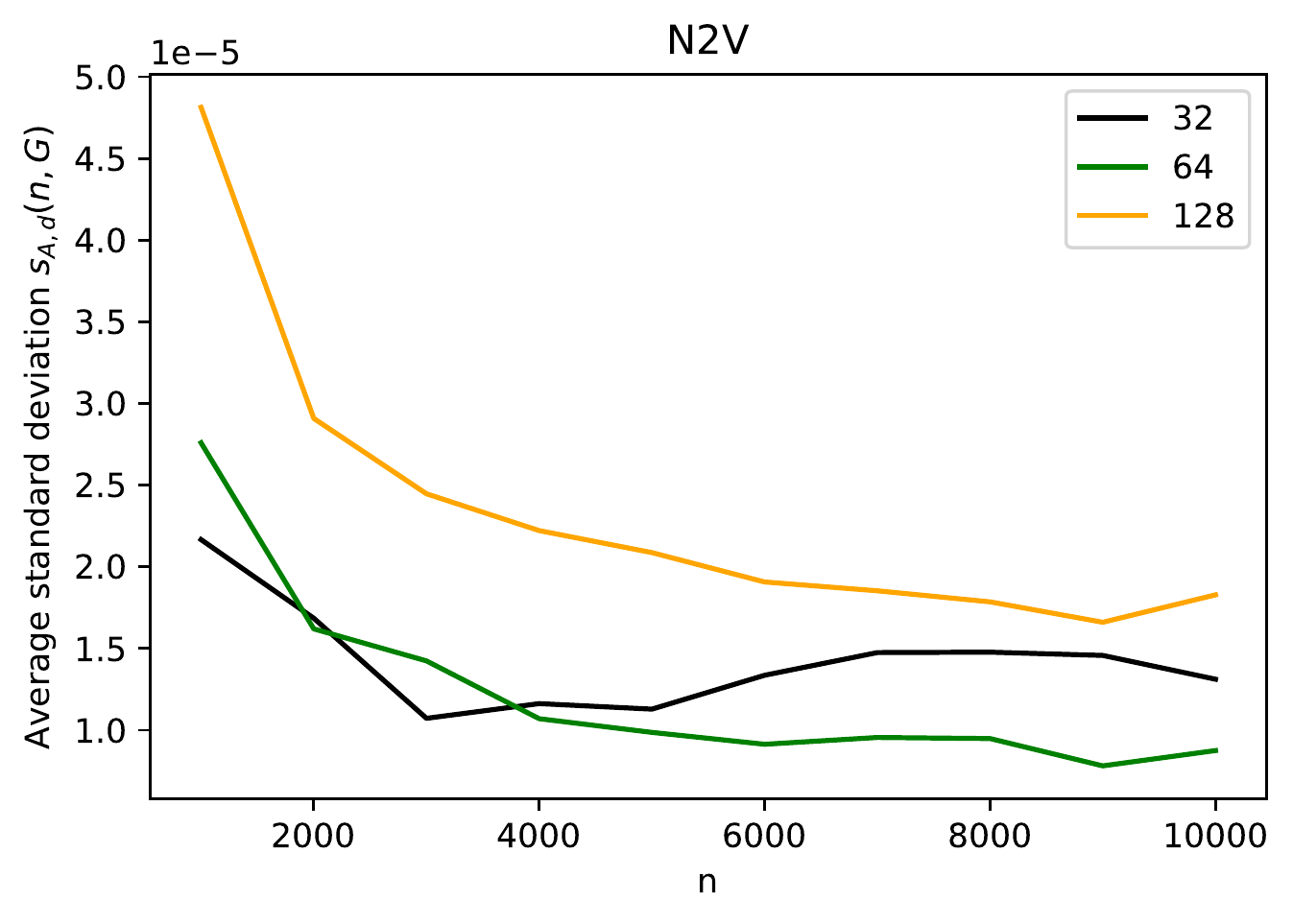}
\includegraphics[width=0.2\textwidth]{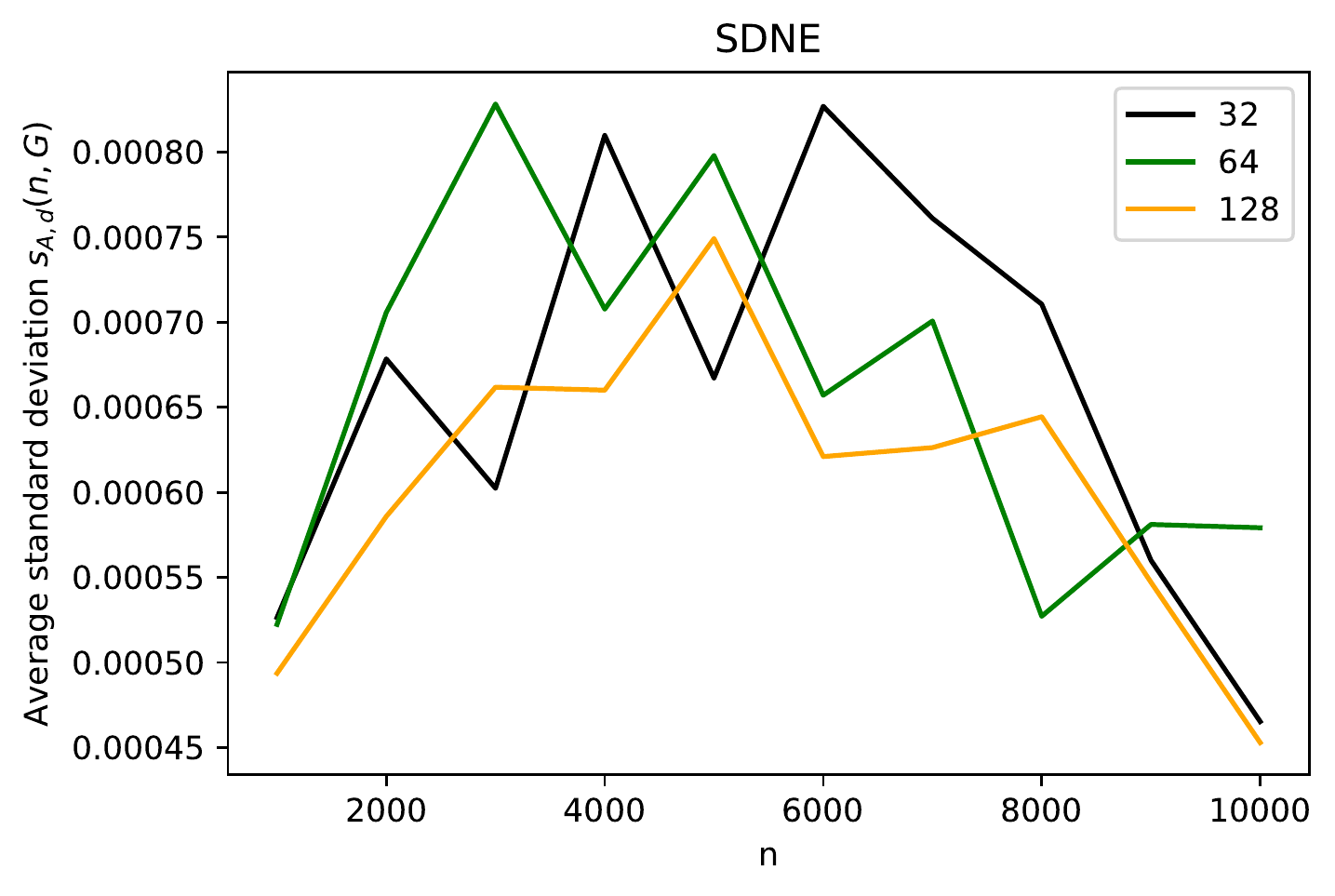}
\includegraphics[width=0.2\textwidth]{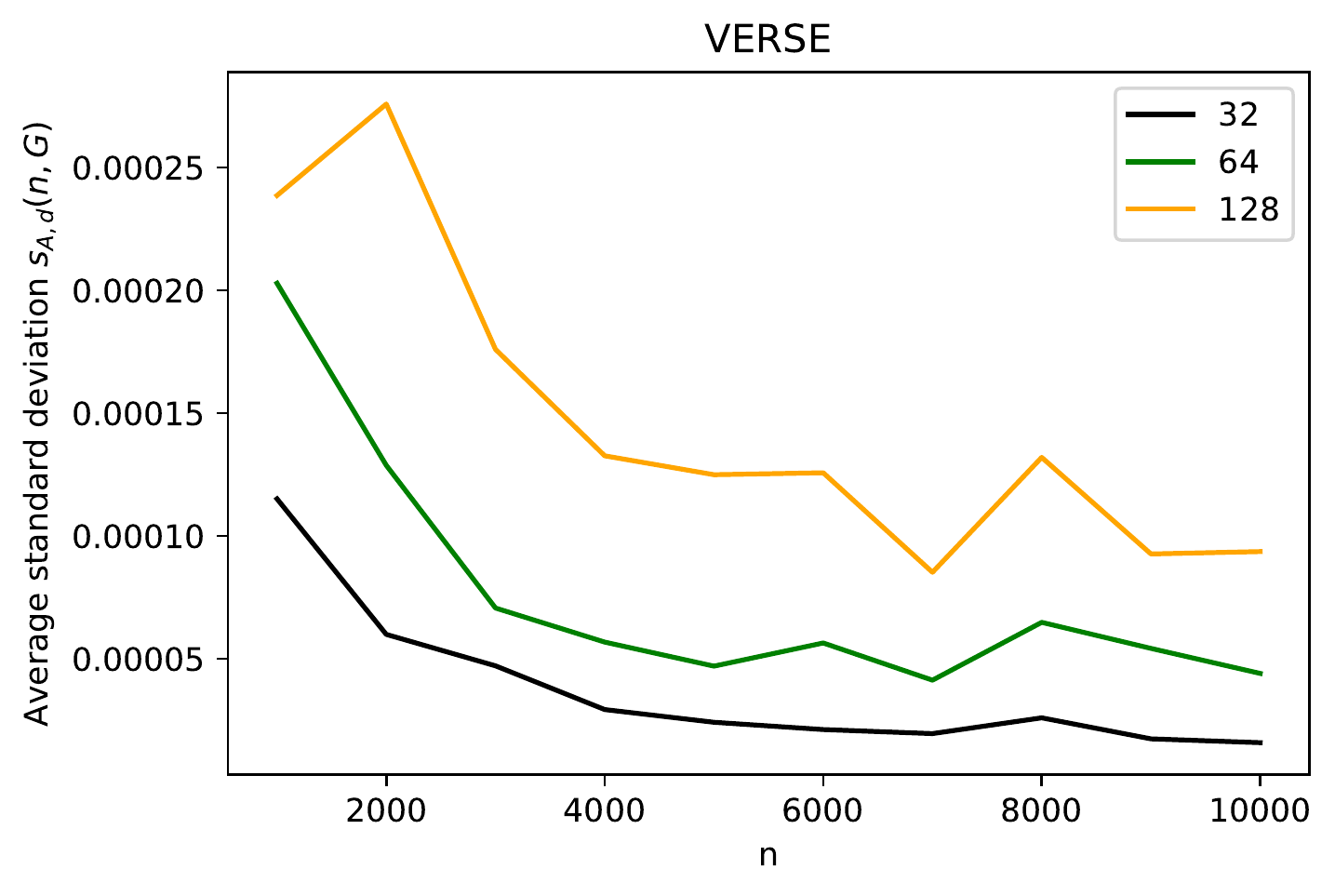} \\
\includegraphics[width=0.3\textwidth]{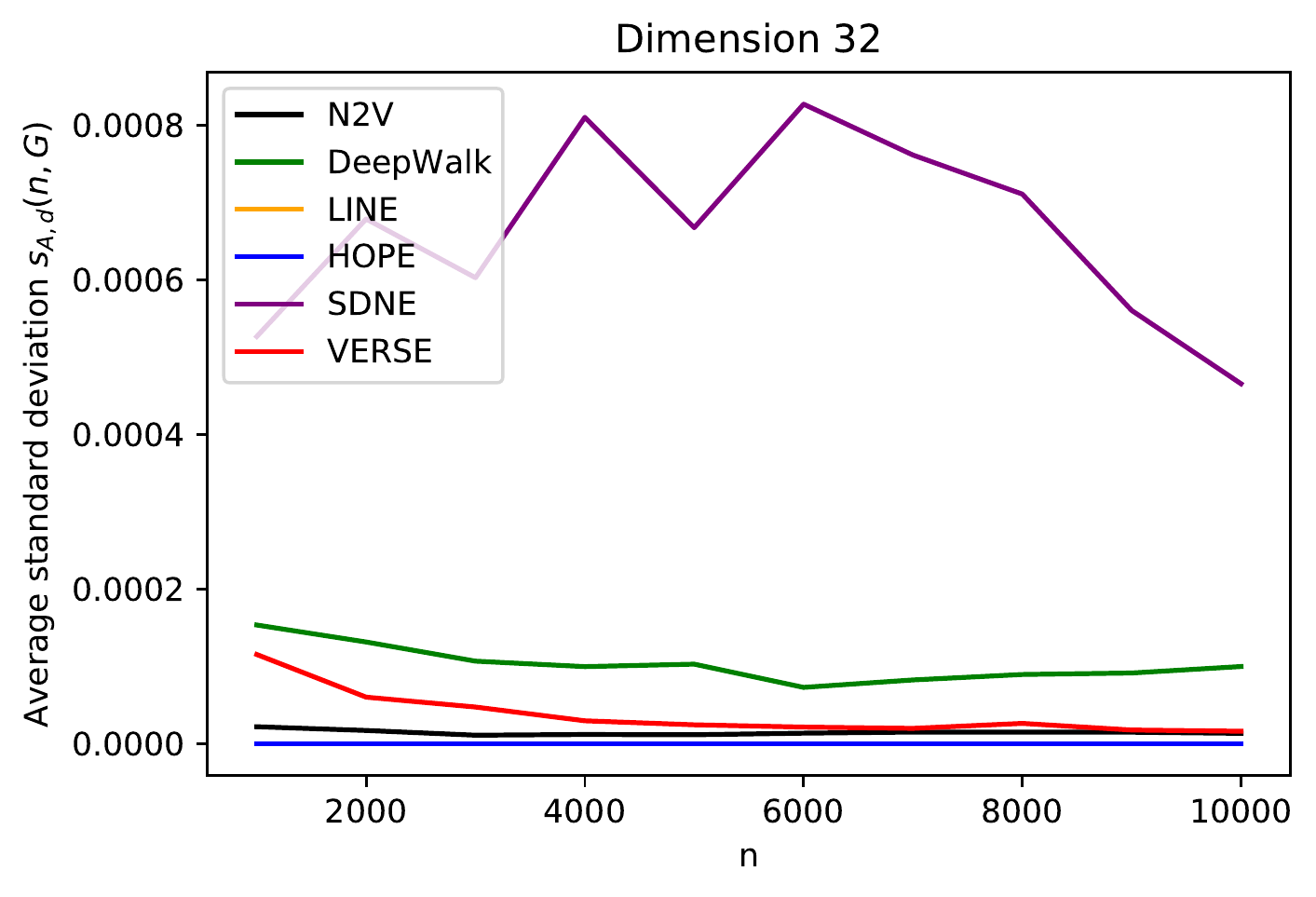}
\includegraphics[width=0.3\textwidth]{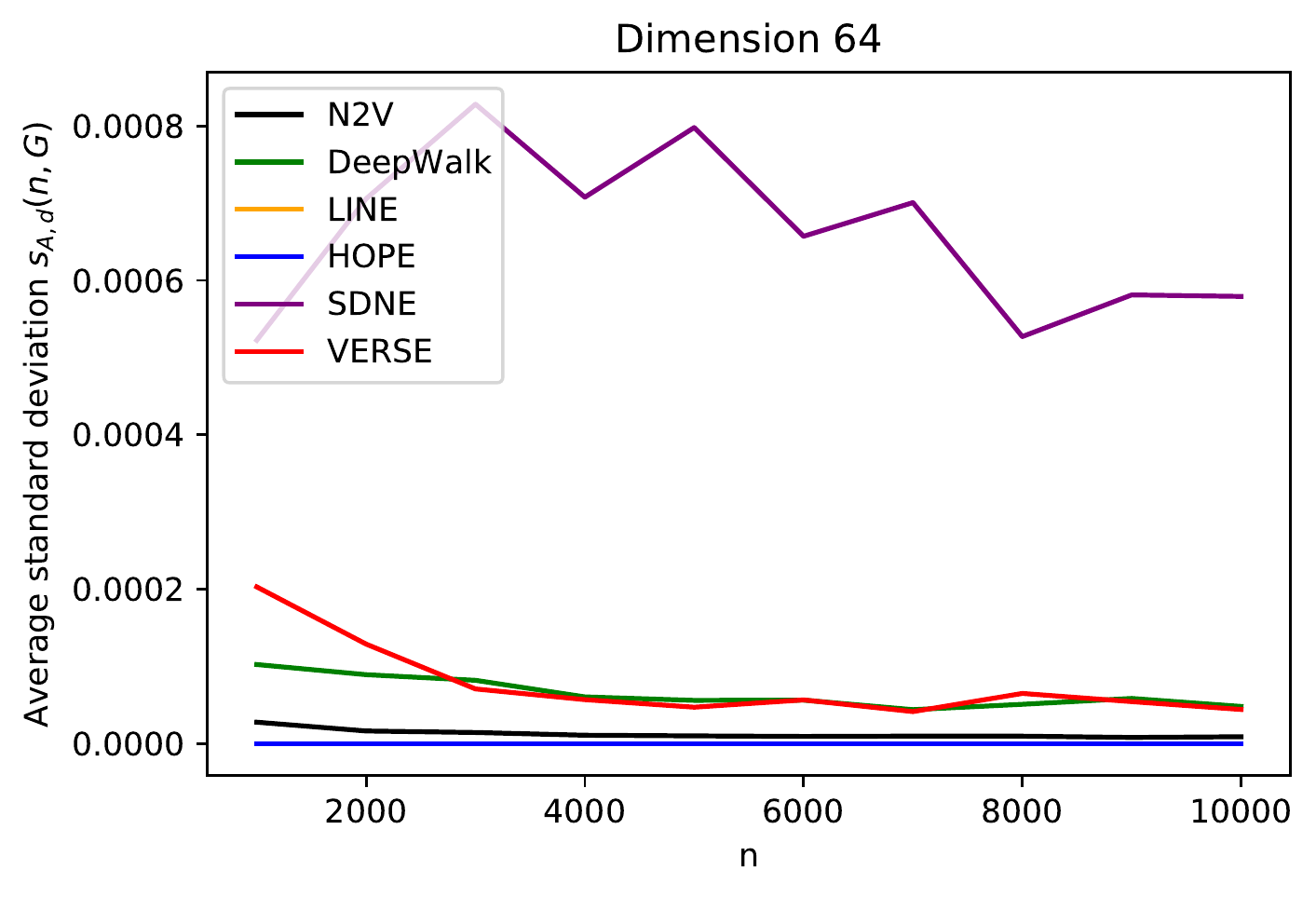}
\includegraphics[width=0.3\textwidth]{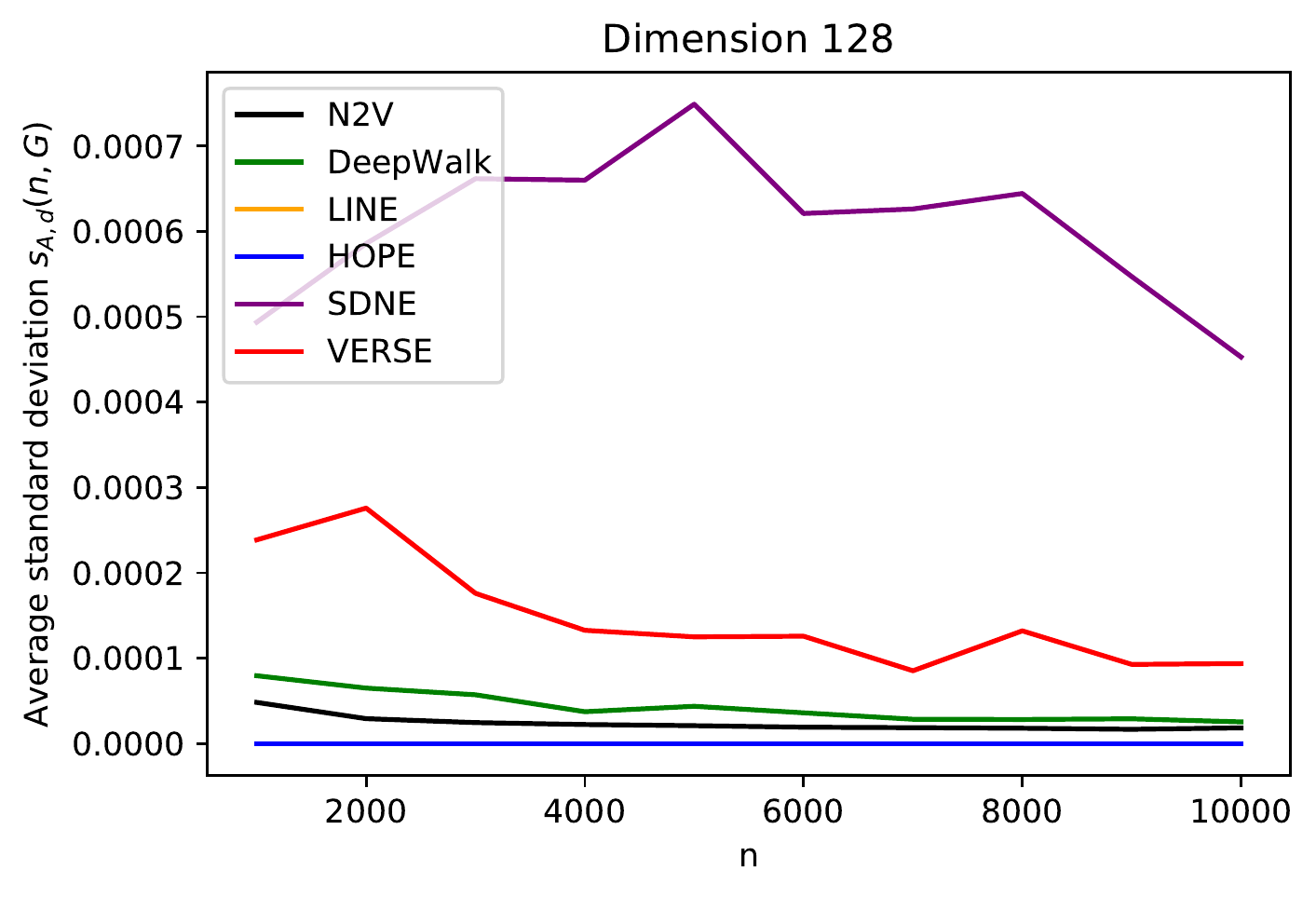} \\
Plots 3 and 4: average $s_{A,d}(n, G)$ (over 10 graphs)  \\

\includegraphics[width=0.2\textwidth]{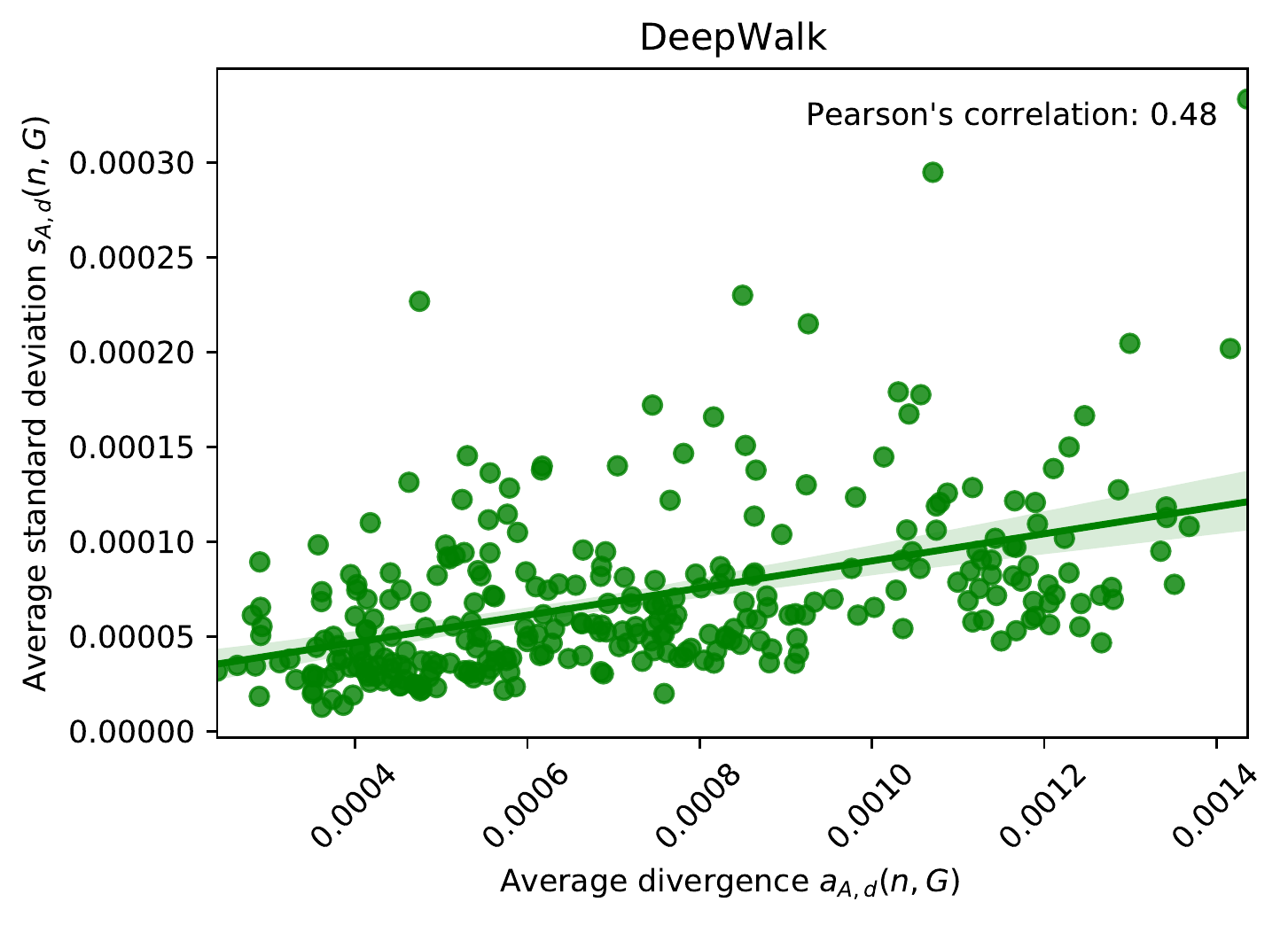}
\includegraphics[width=0.2\textwidth]{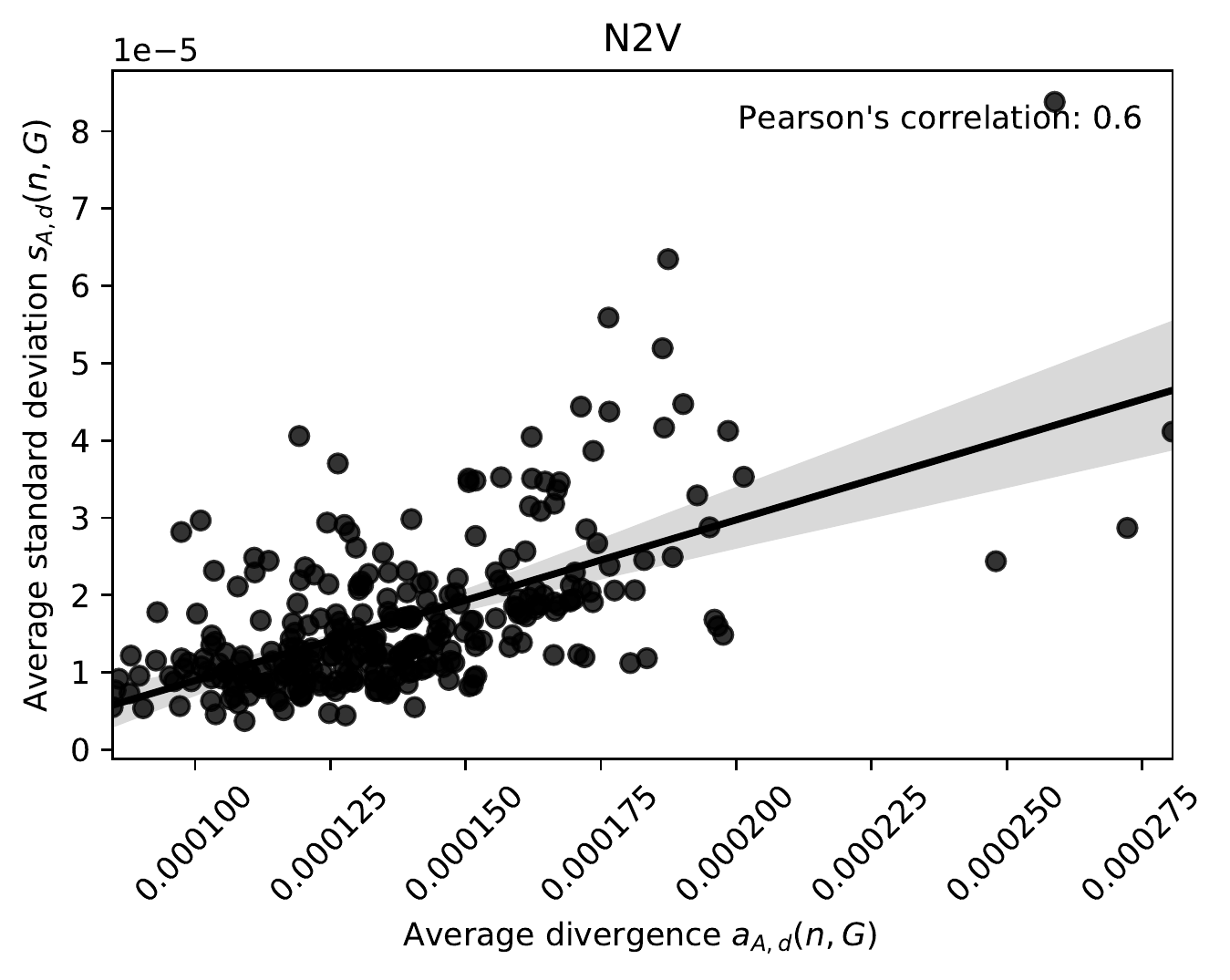}
\includegraphics[width=0.2\textwidth]{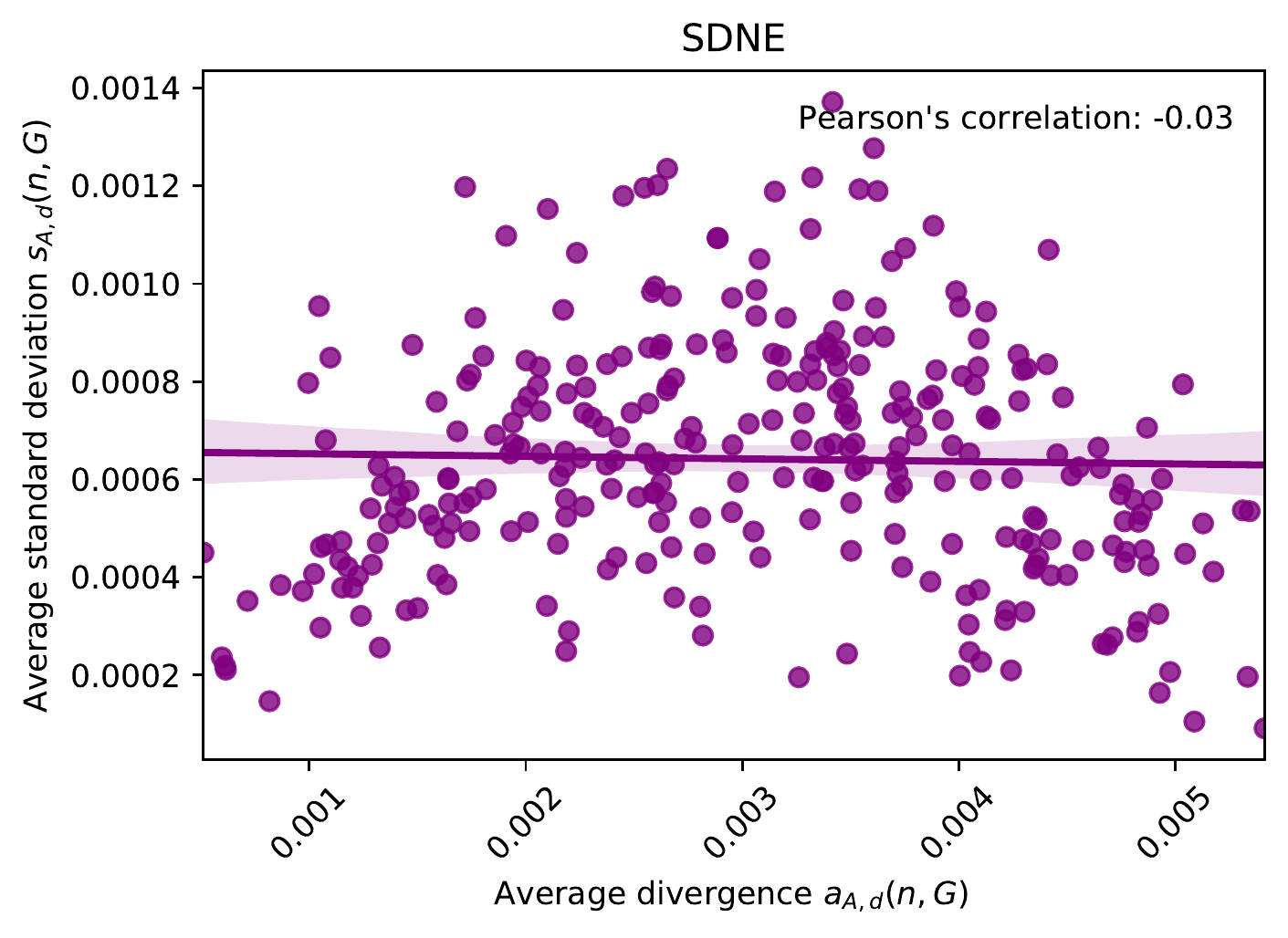}
\includegraphics[width=0.2\textwidth]{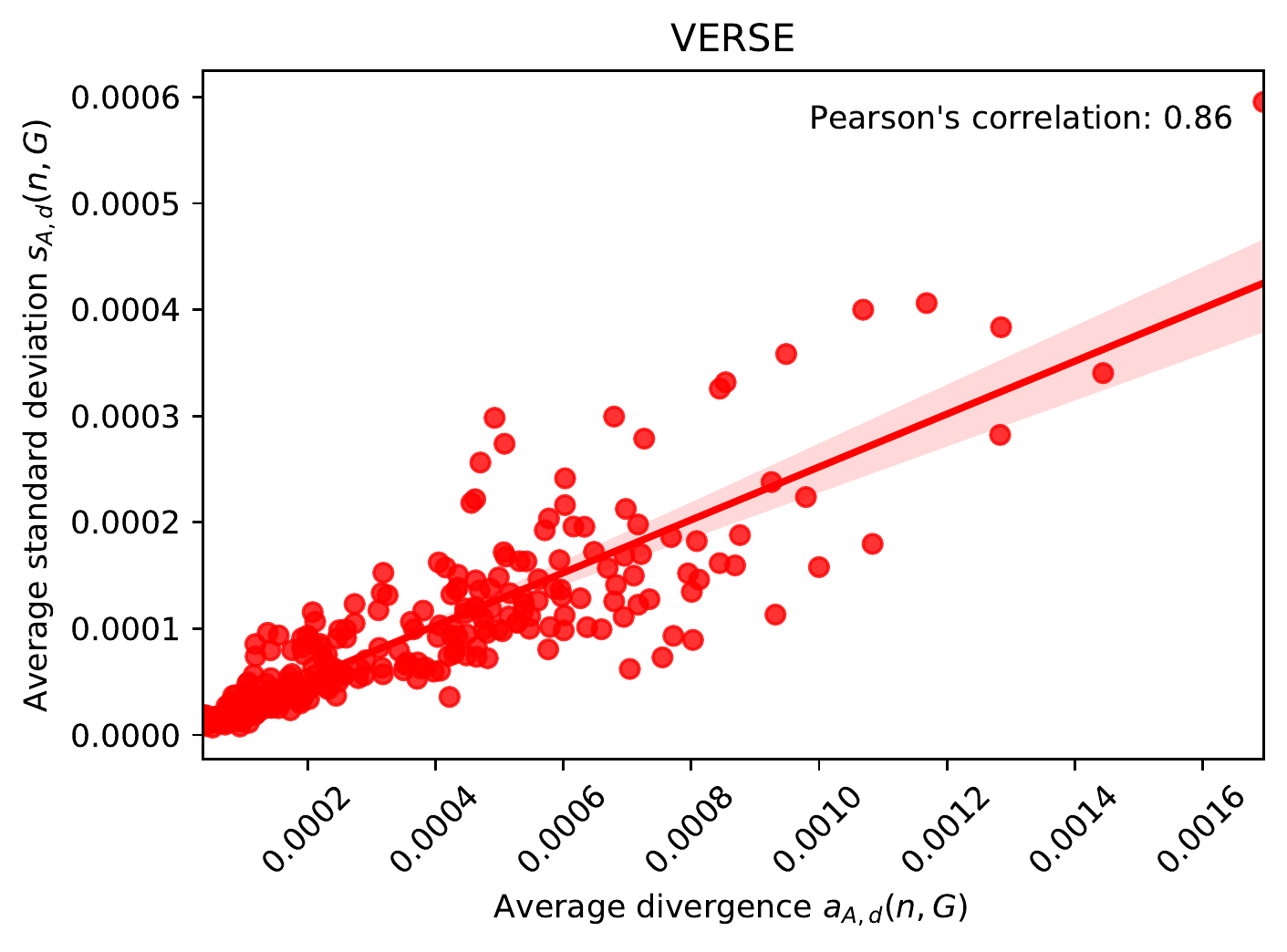} \\
Plot 5: correlation between $a_{A,d}(n, G)$ and $s_{A,d}(n, G)$ 
\caption{Size of the Network ($n$)}\label{fig:n}
\end{center}
\end{figure}

\begin{figure}[htbp!]
\begin{center}
\includegraphics[width=0.3\textwidth]{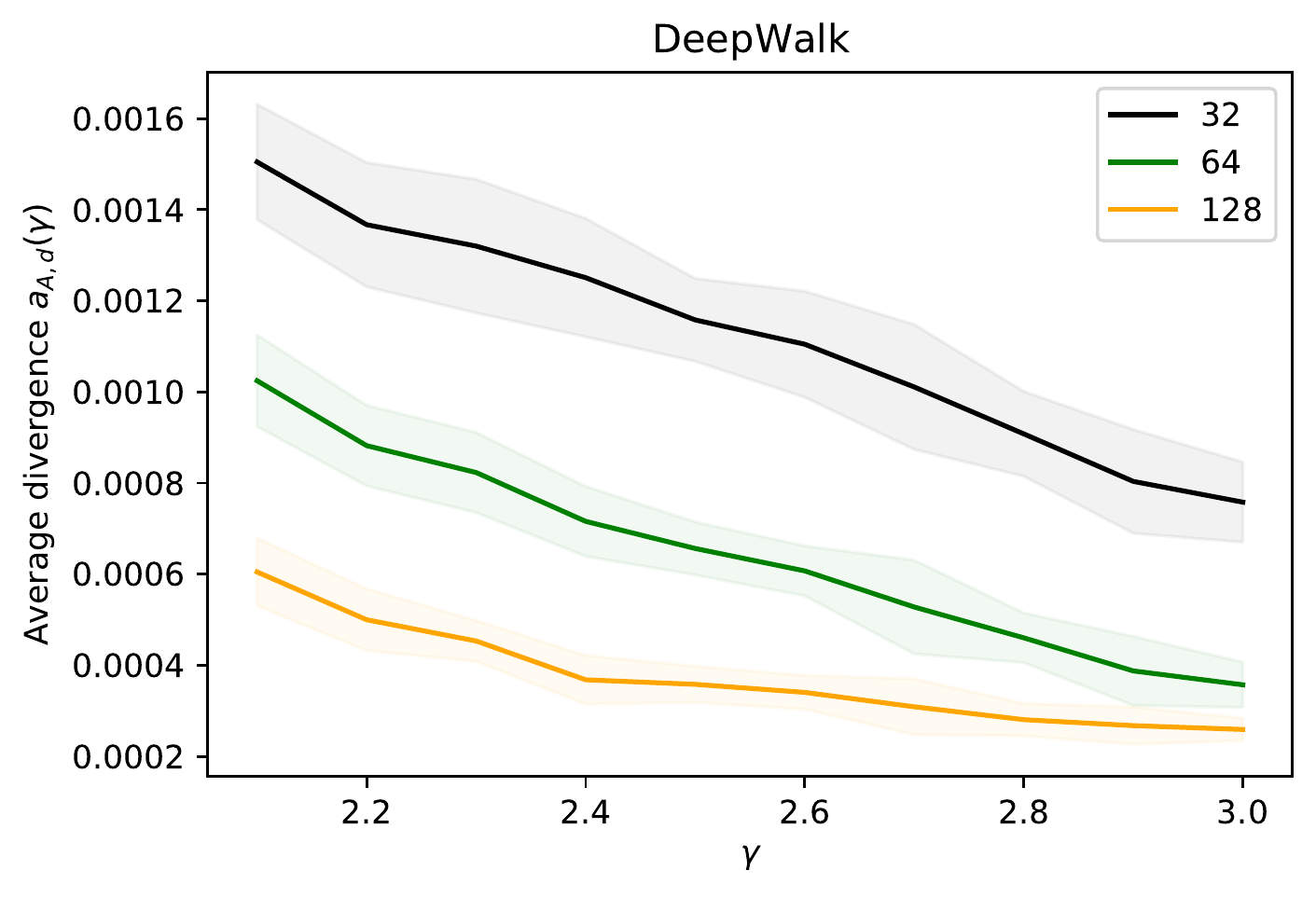}
\includegraphics[width=0.3\textwidth]{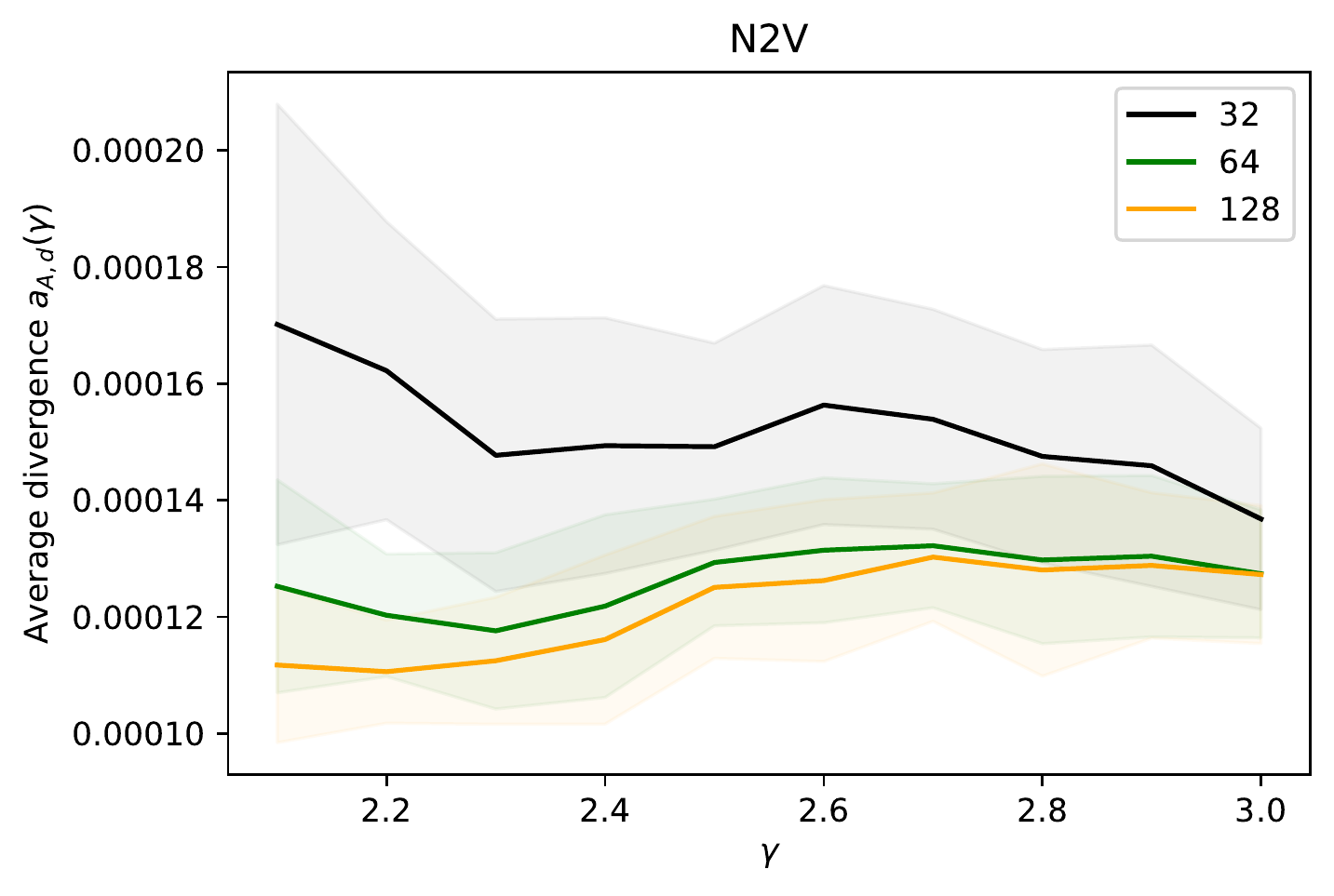}
\includegraphics[width=0.3\textwidth]{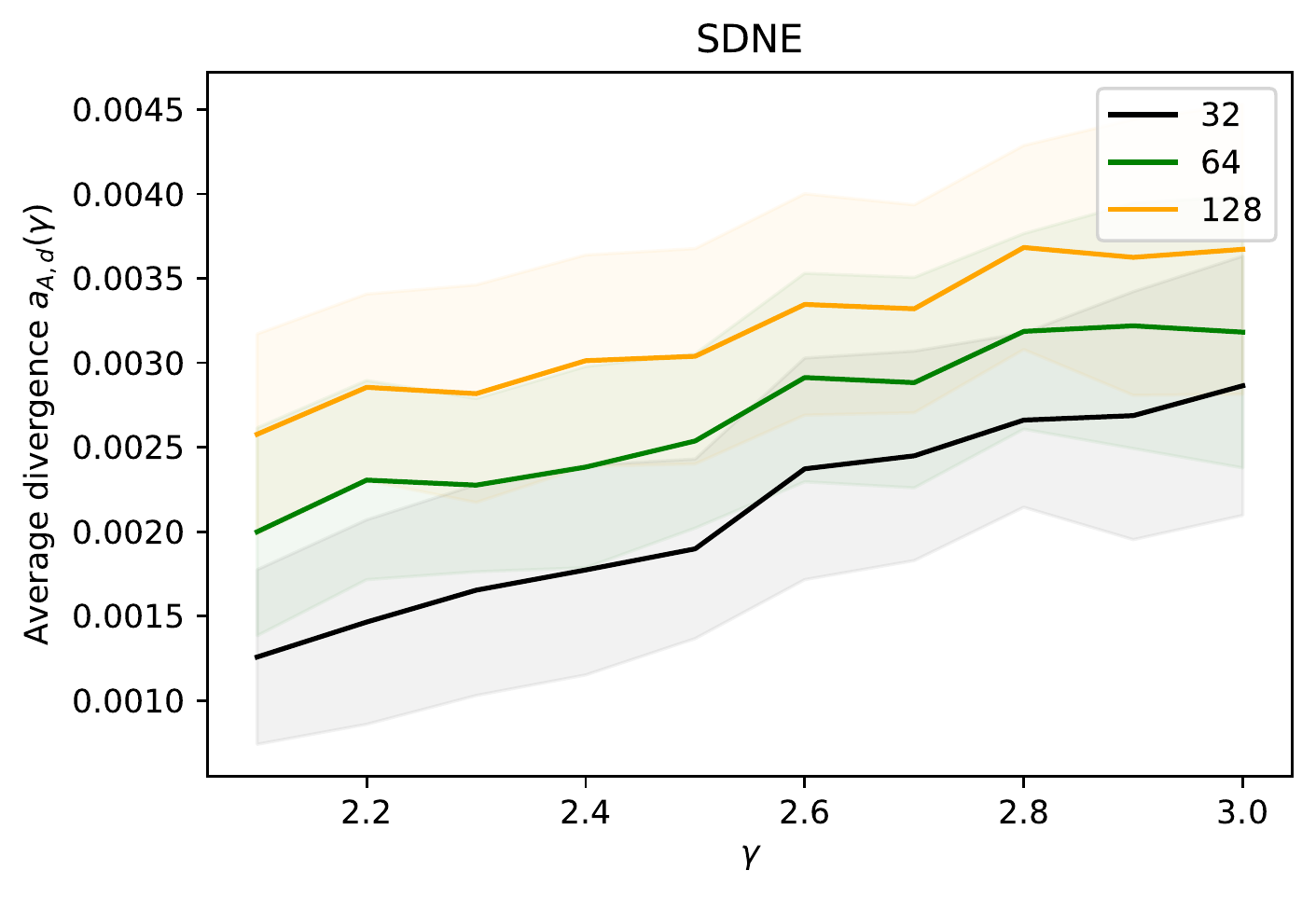} \\
\includegraphics[width=0.3\textwidth]{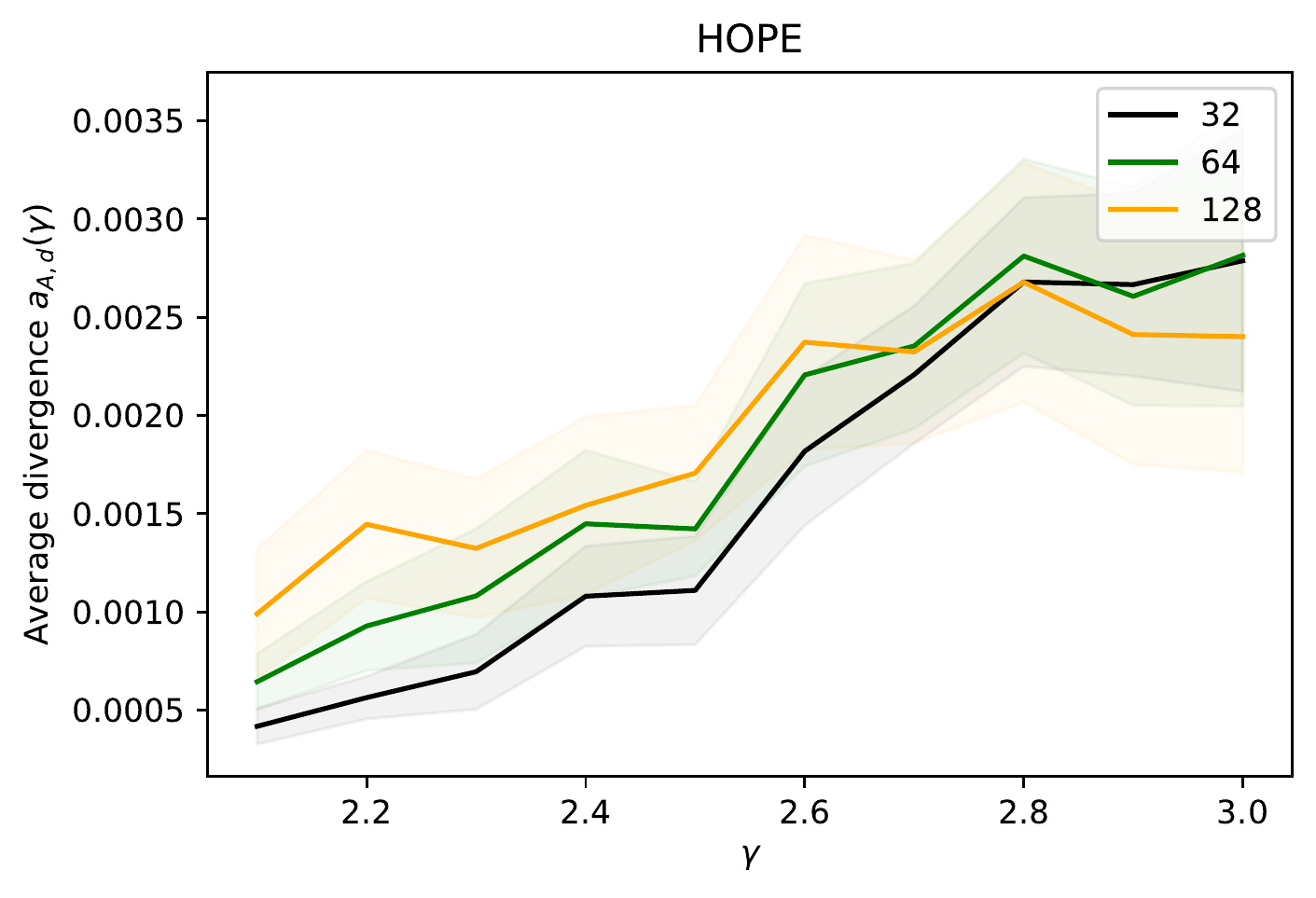}
\includegraphics[width=0.3\textwidth]{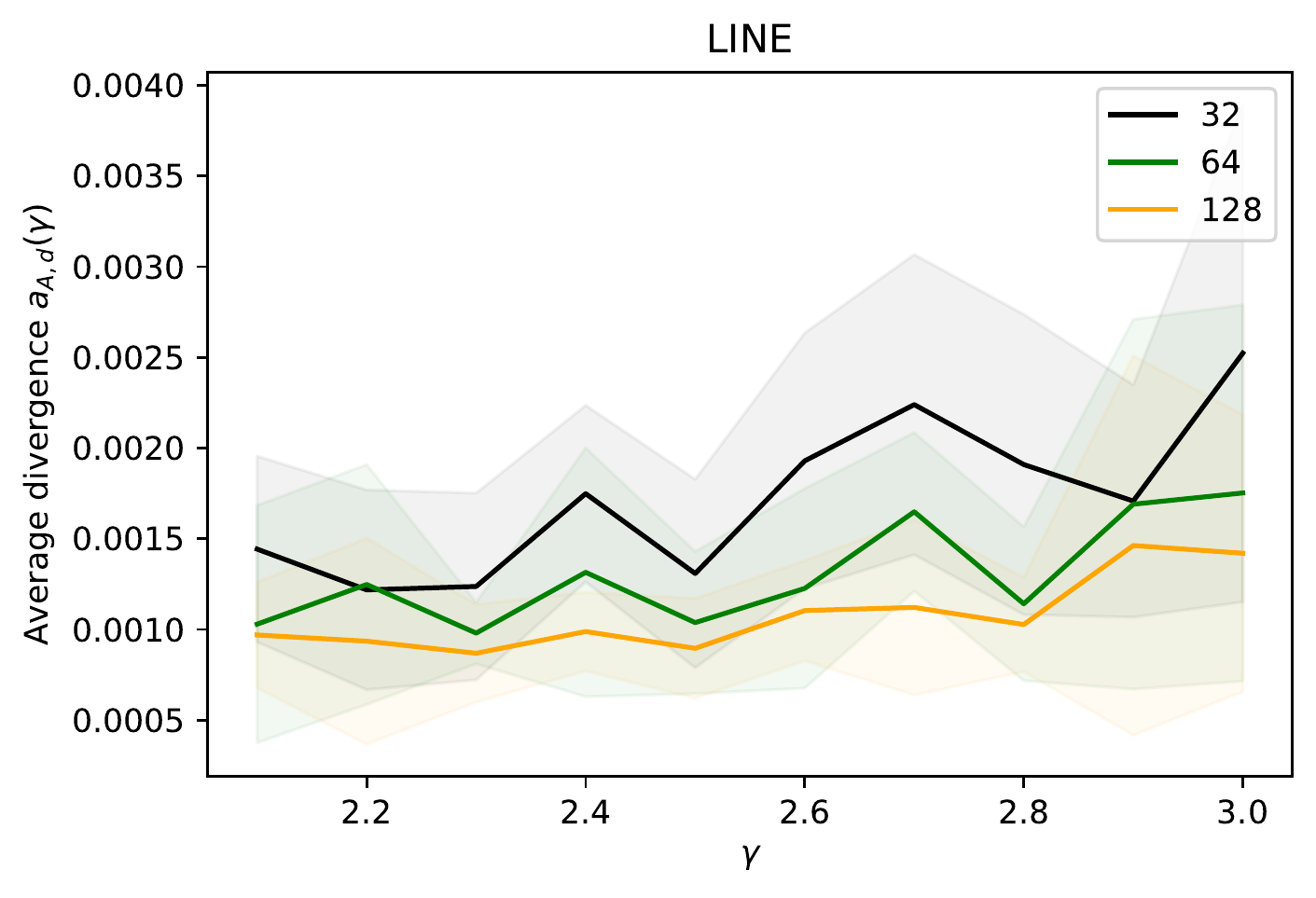}
\includegraphics[width=0.3\textwidth]{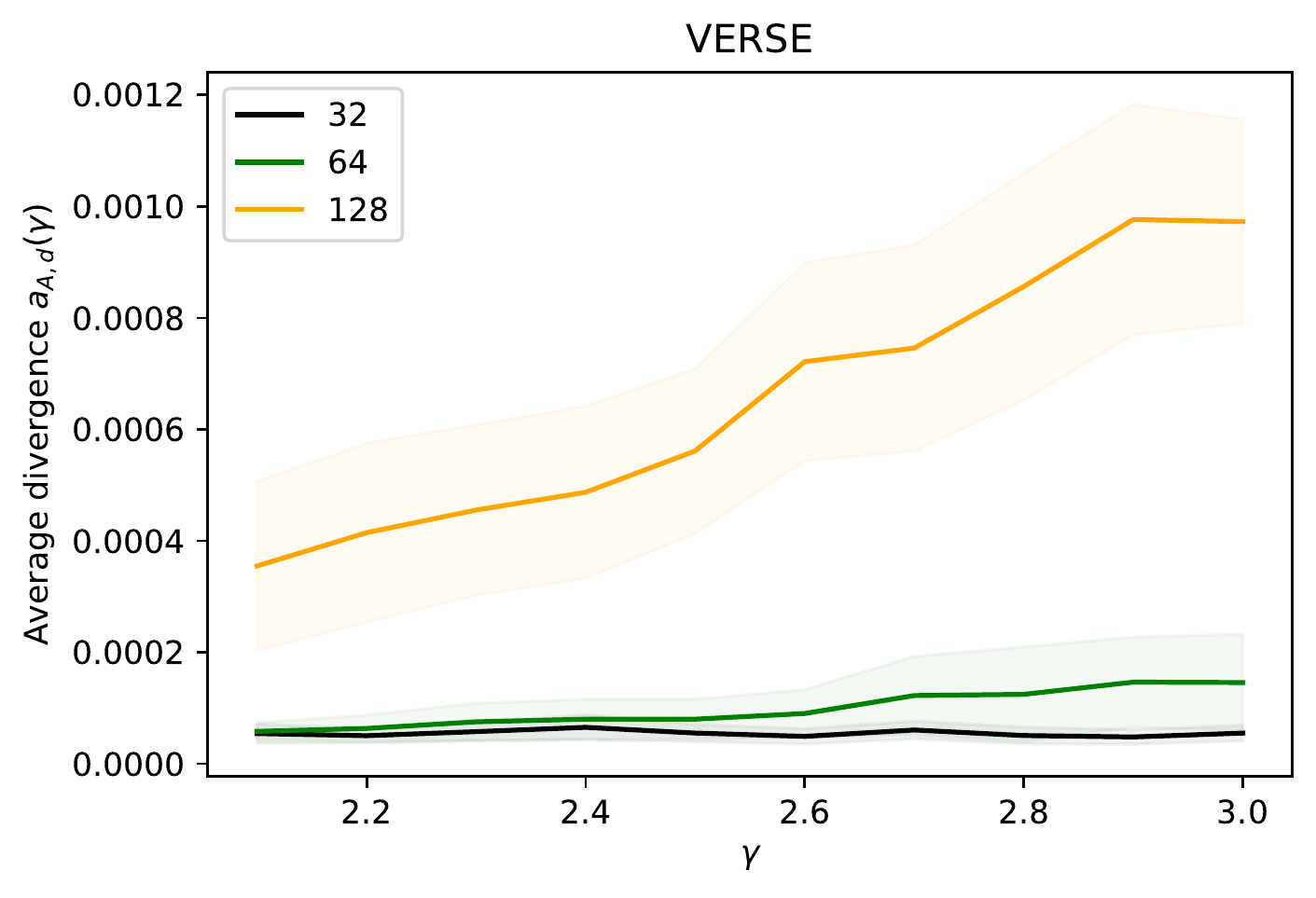} \\
\includegraphics[width=0.3\textwidth]{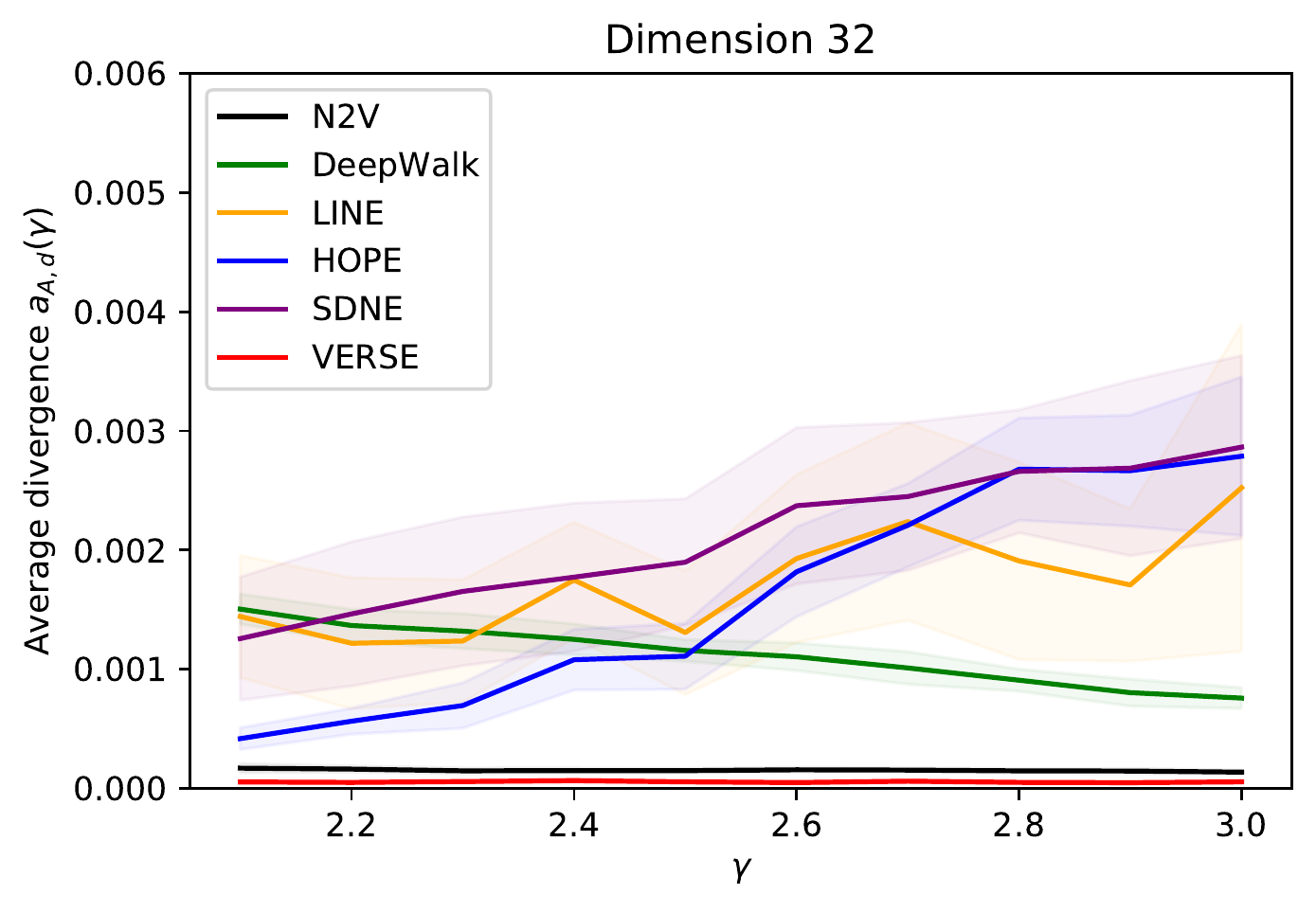}
\includegraphics[width=0.3\textwidth]{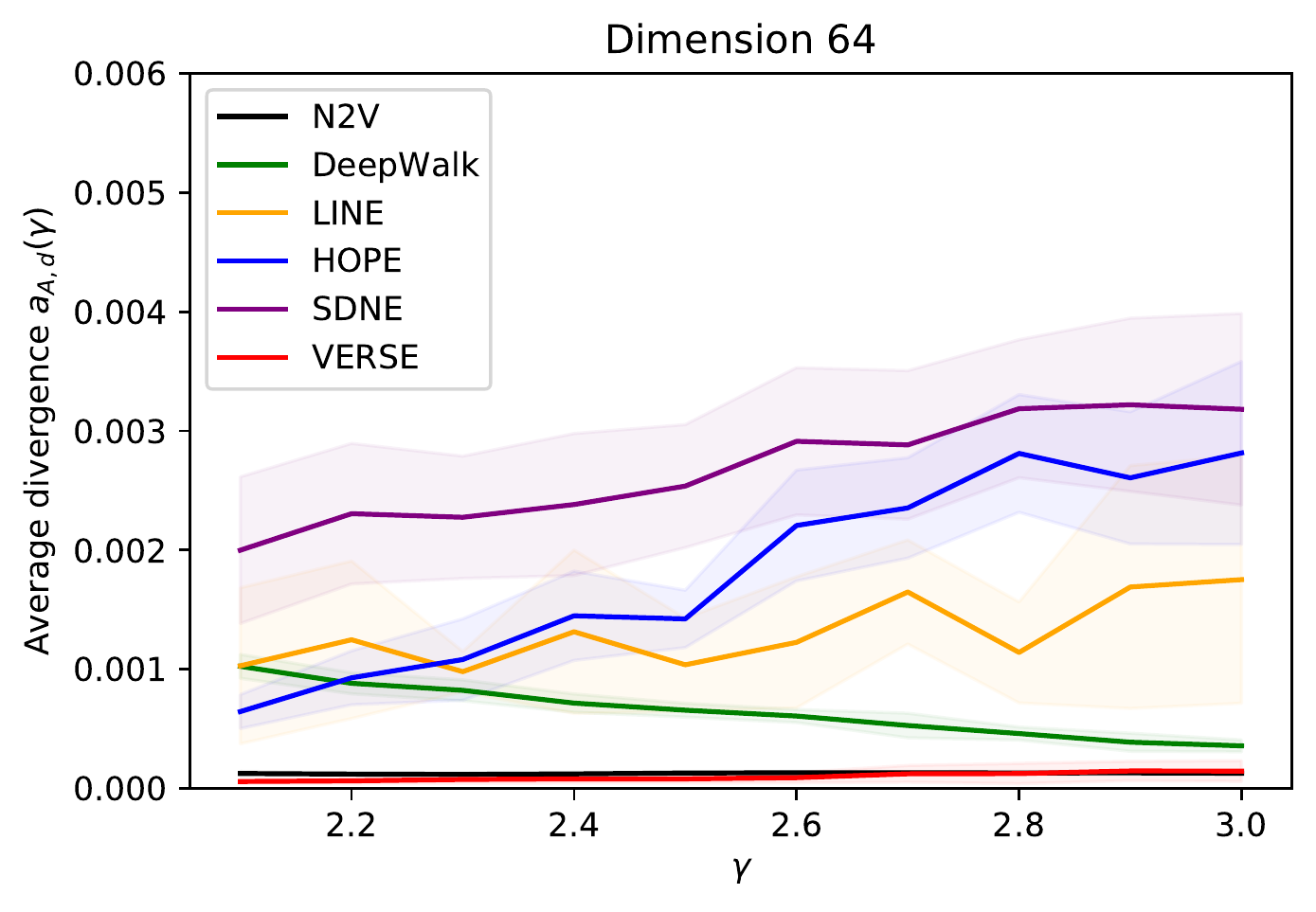}
\includegraphics[width=0.3\textwidth]{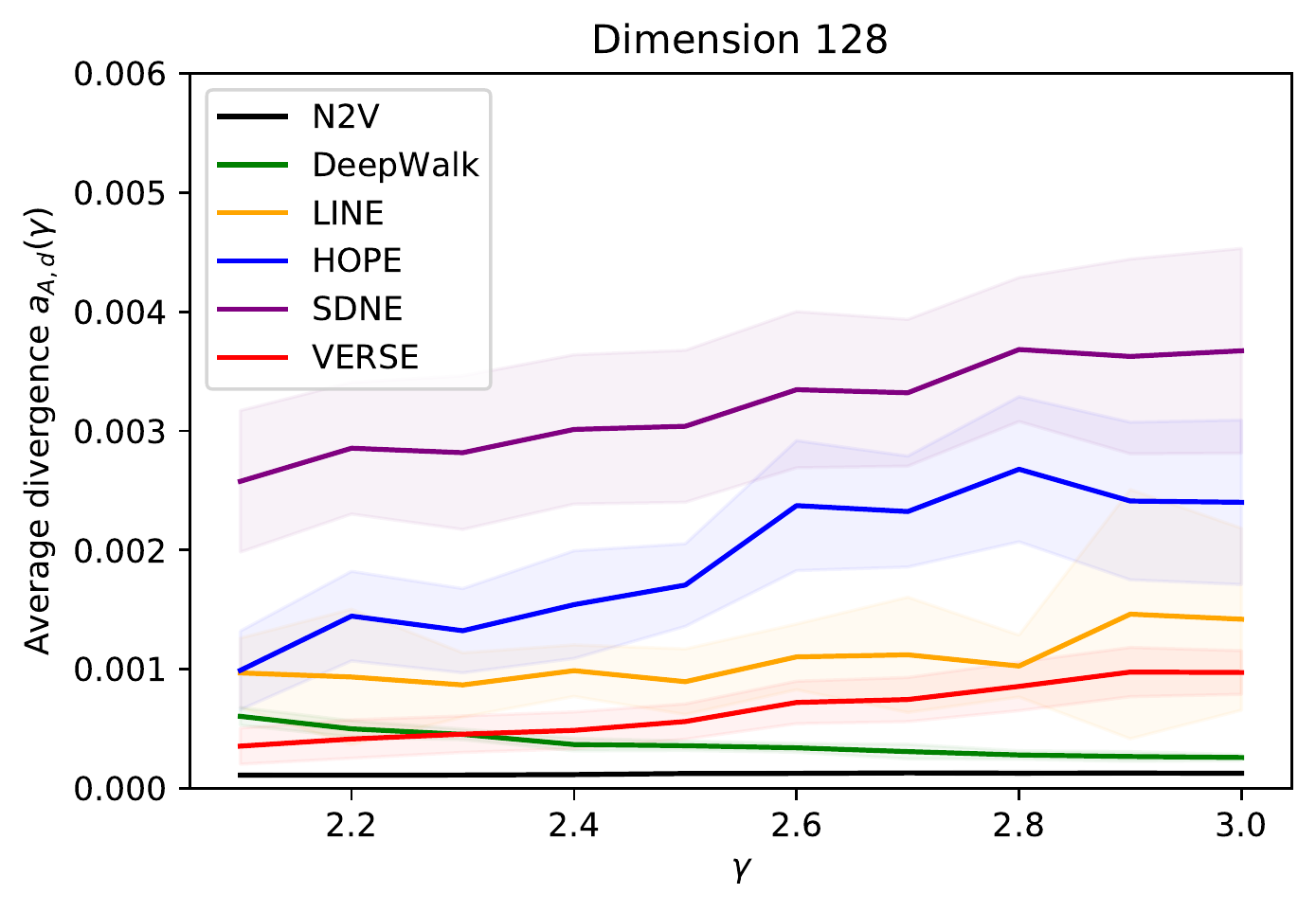} \\
Plots 1 and  2: $a_{A,d}(\gamma) \pm s_{A,d}(\gamma)$ \\

\includegraphics[width=0.2\textwidth]{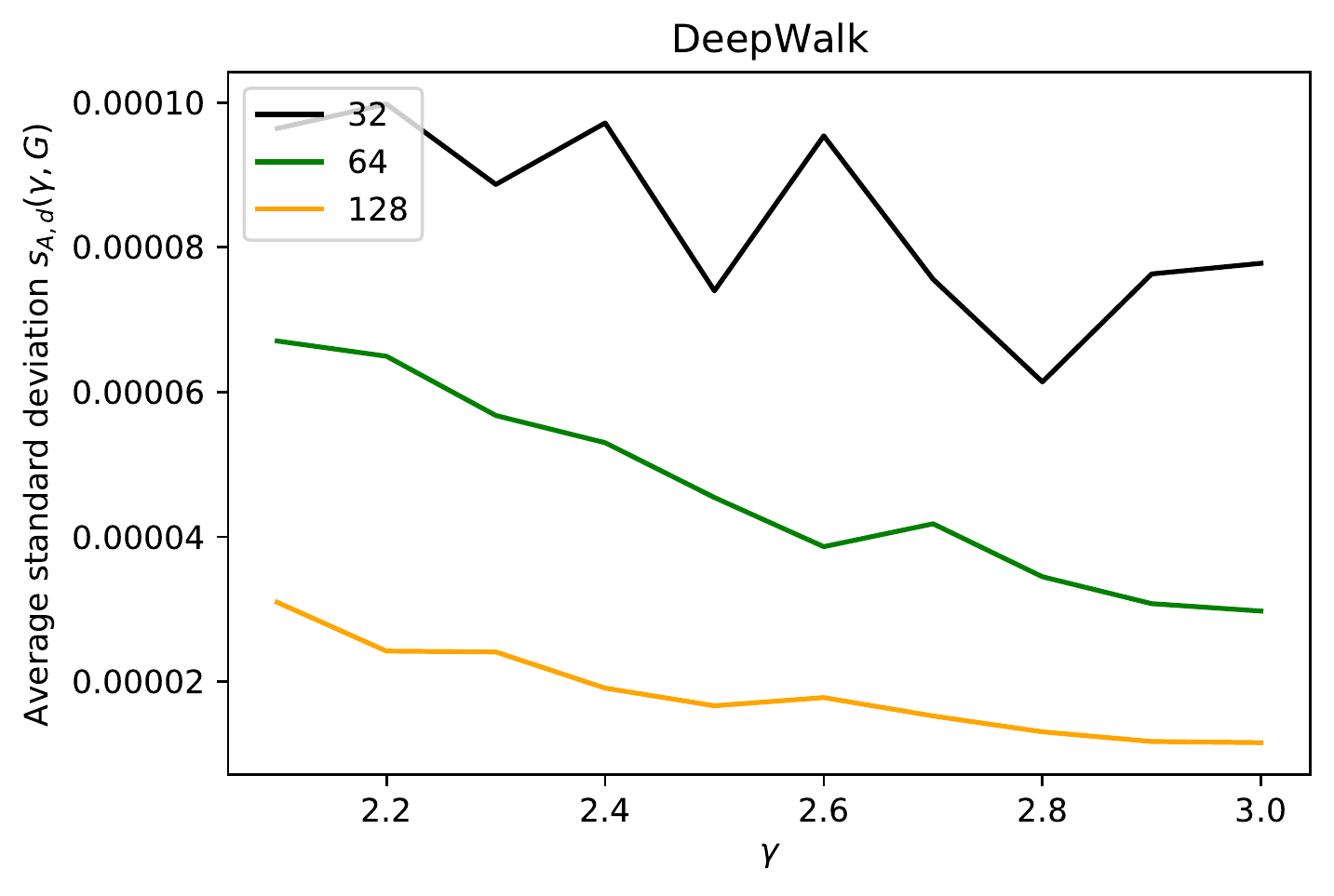}
\includegraphics[width=0.2\textwidth]{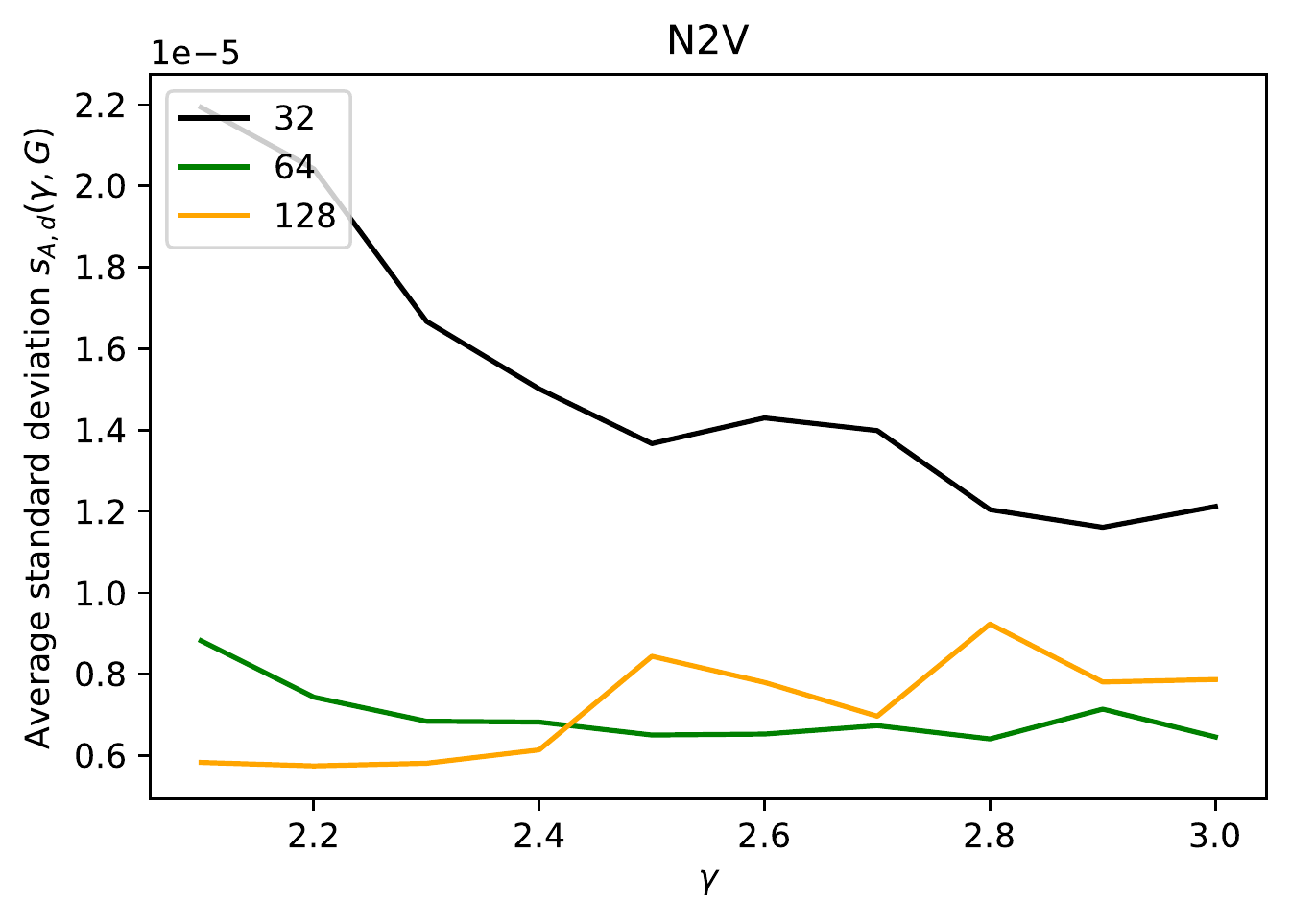}
\includegraphics[width=0.2\textwidth]{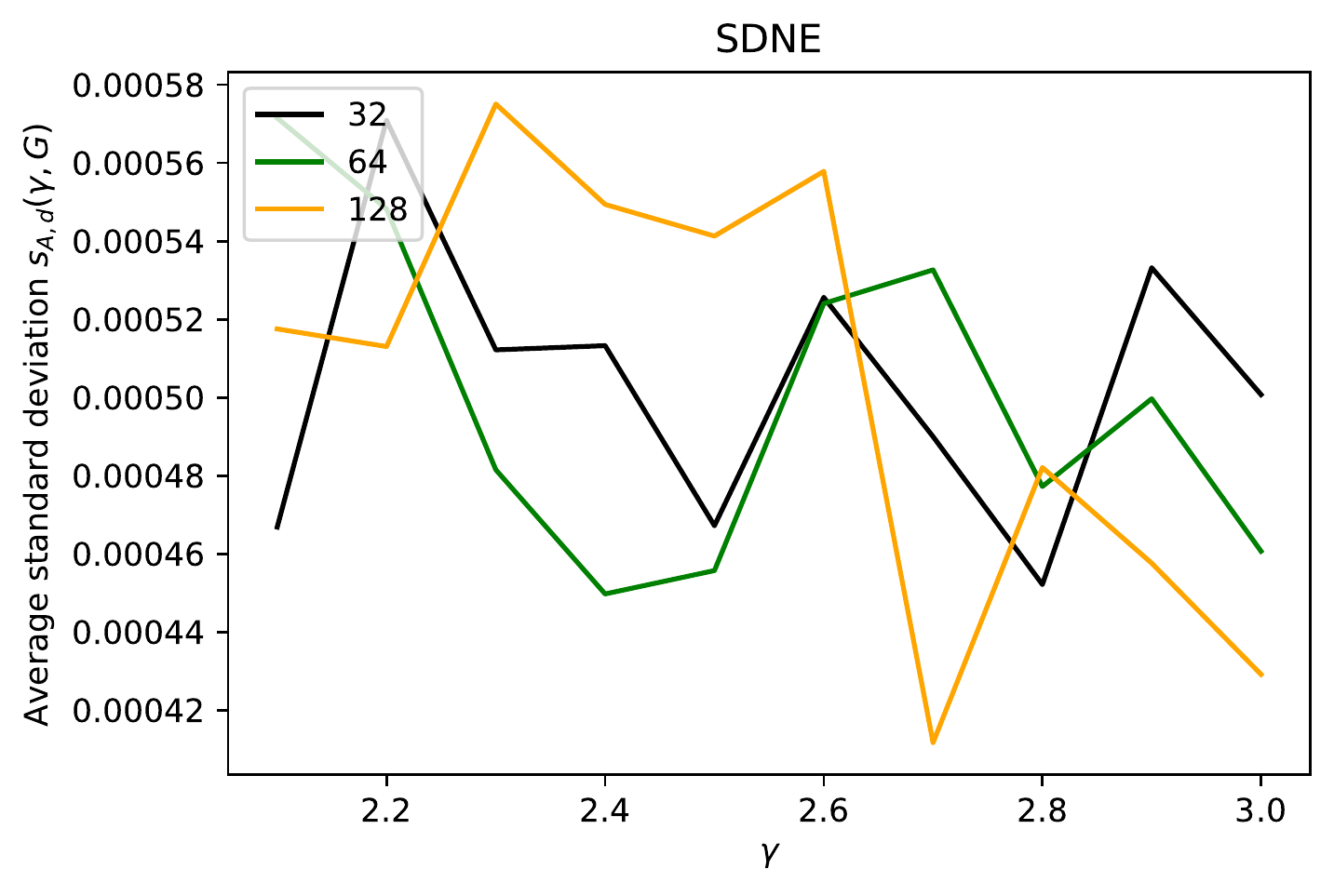} 
\includegraphics[width=0.2\textwidth]{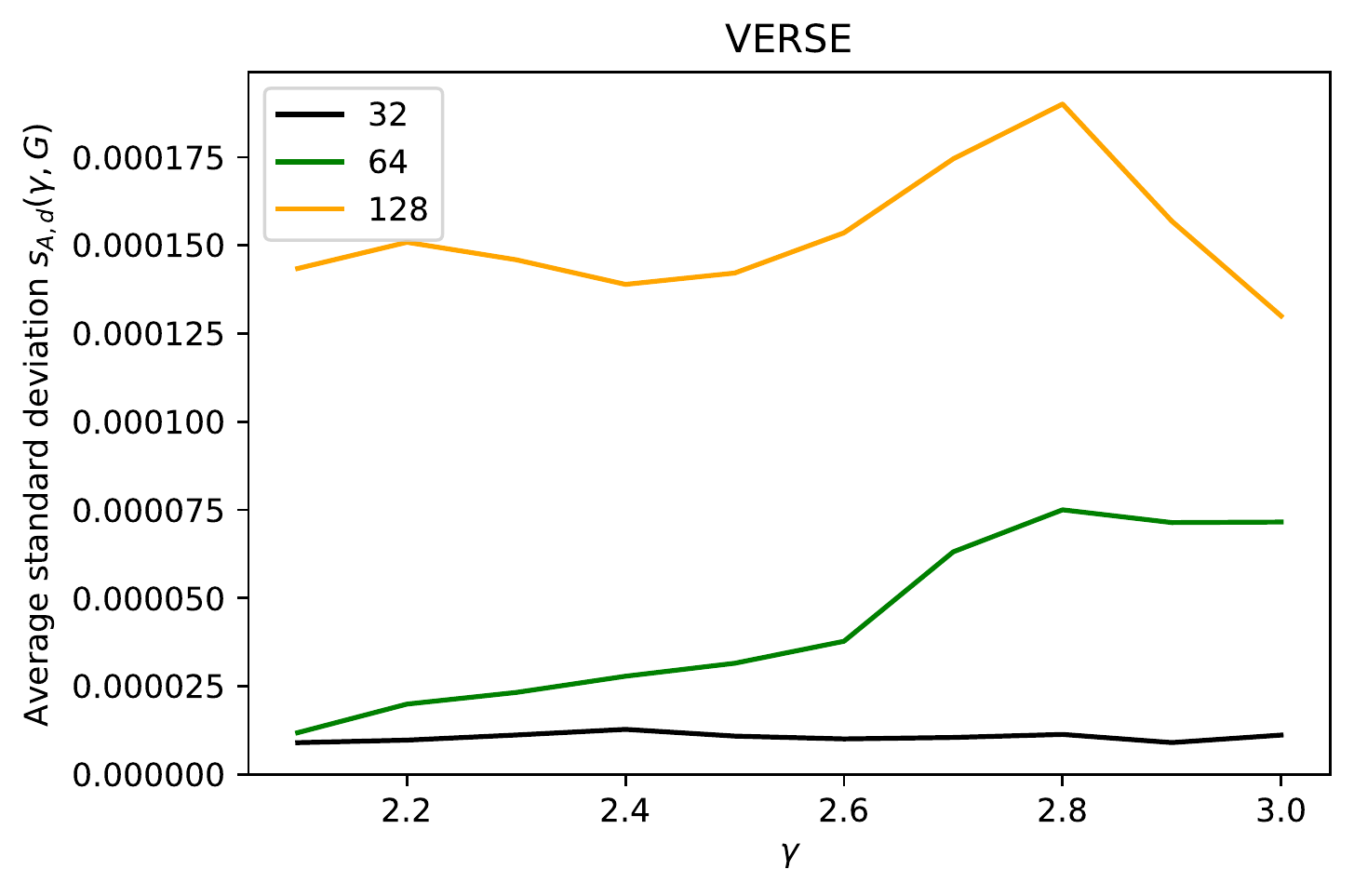} \\
\includegraphics[width=0.3\textwidth]{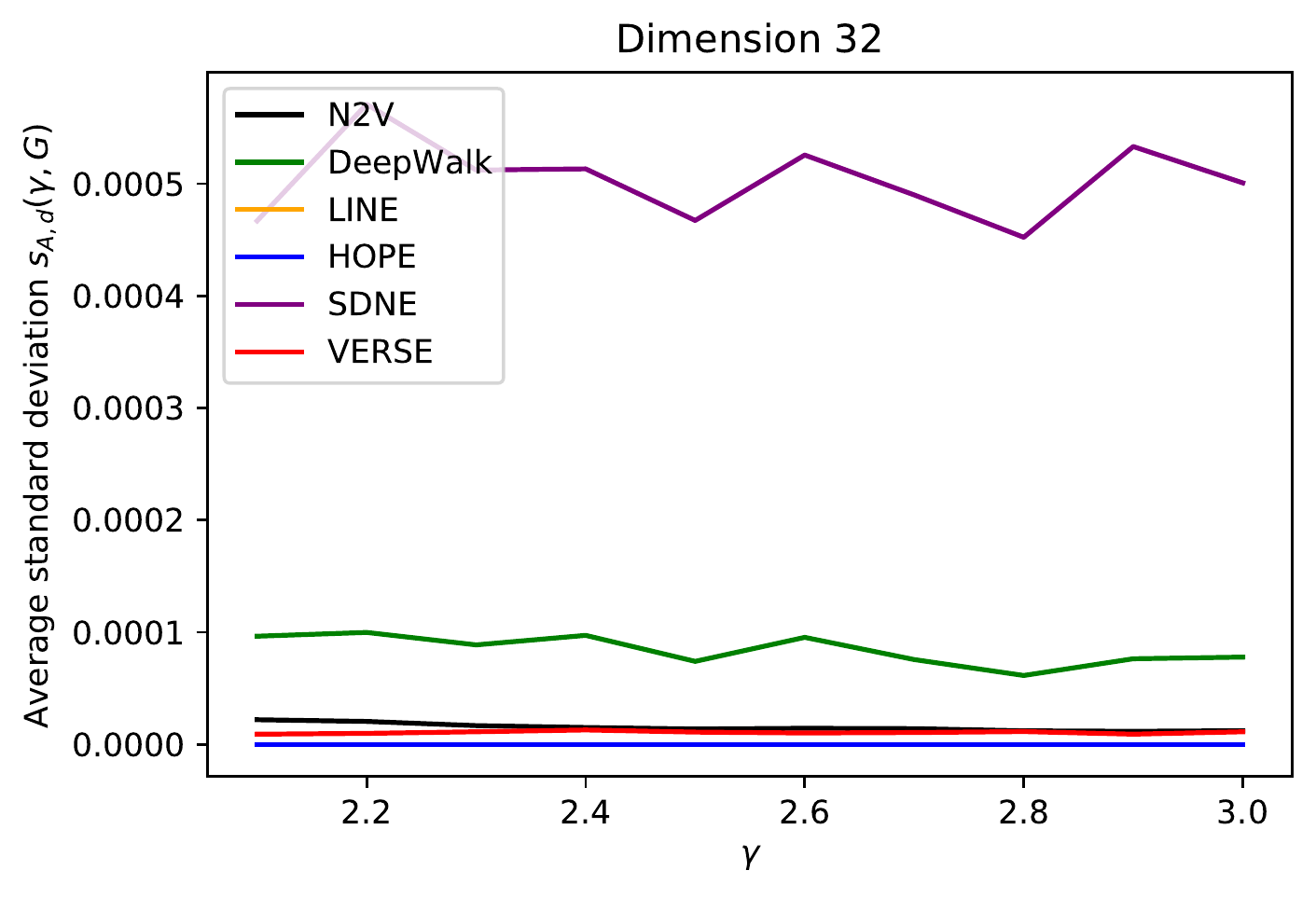}
\includegraphics[width=0.3\textwidth]{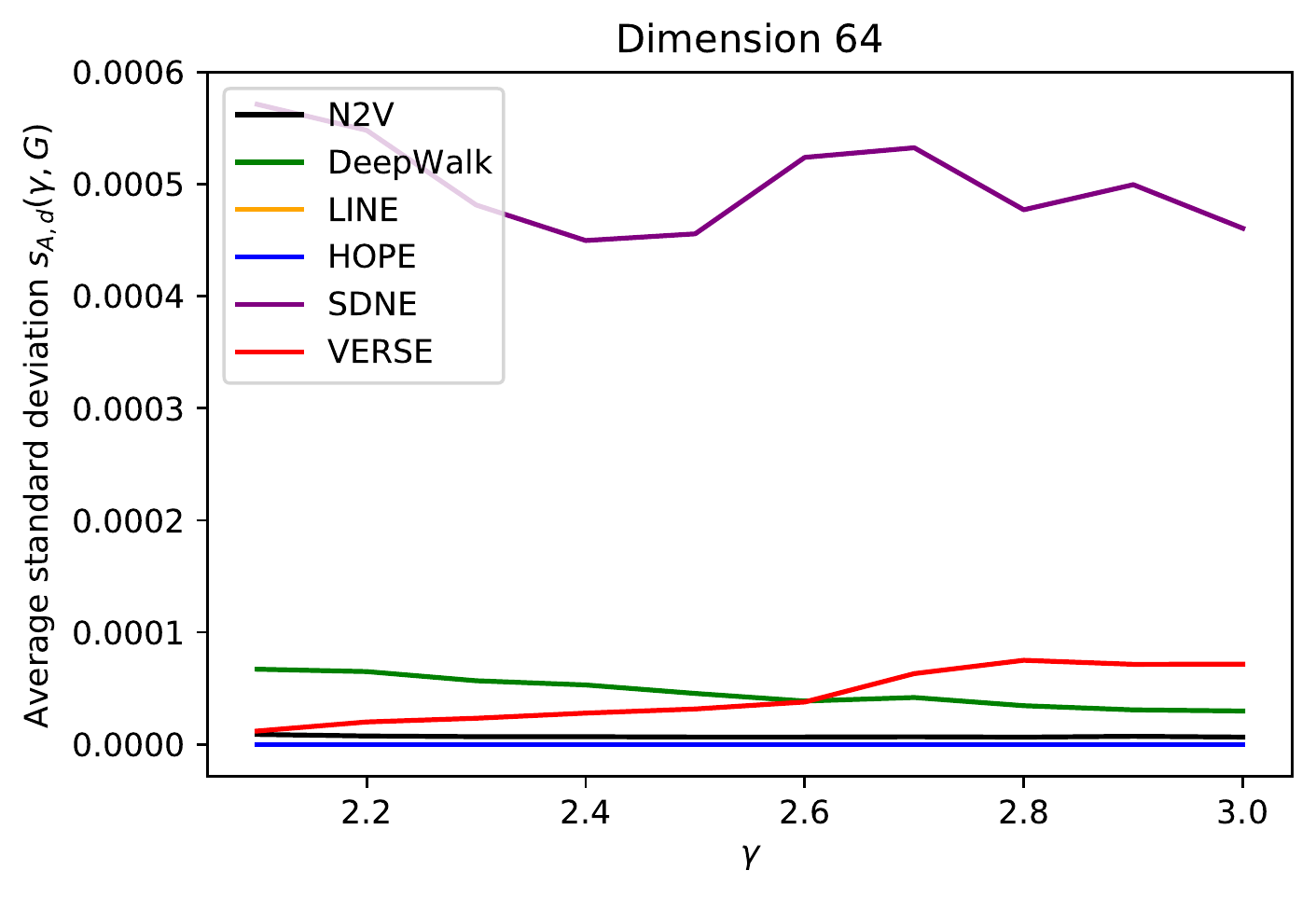}
\includegraphics[width=0.3\textwidth]{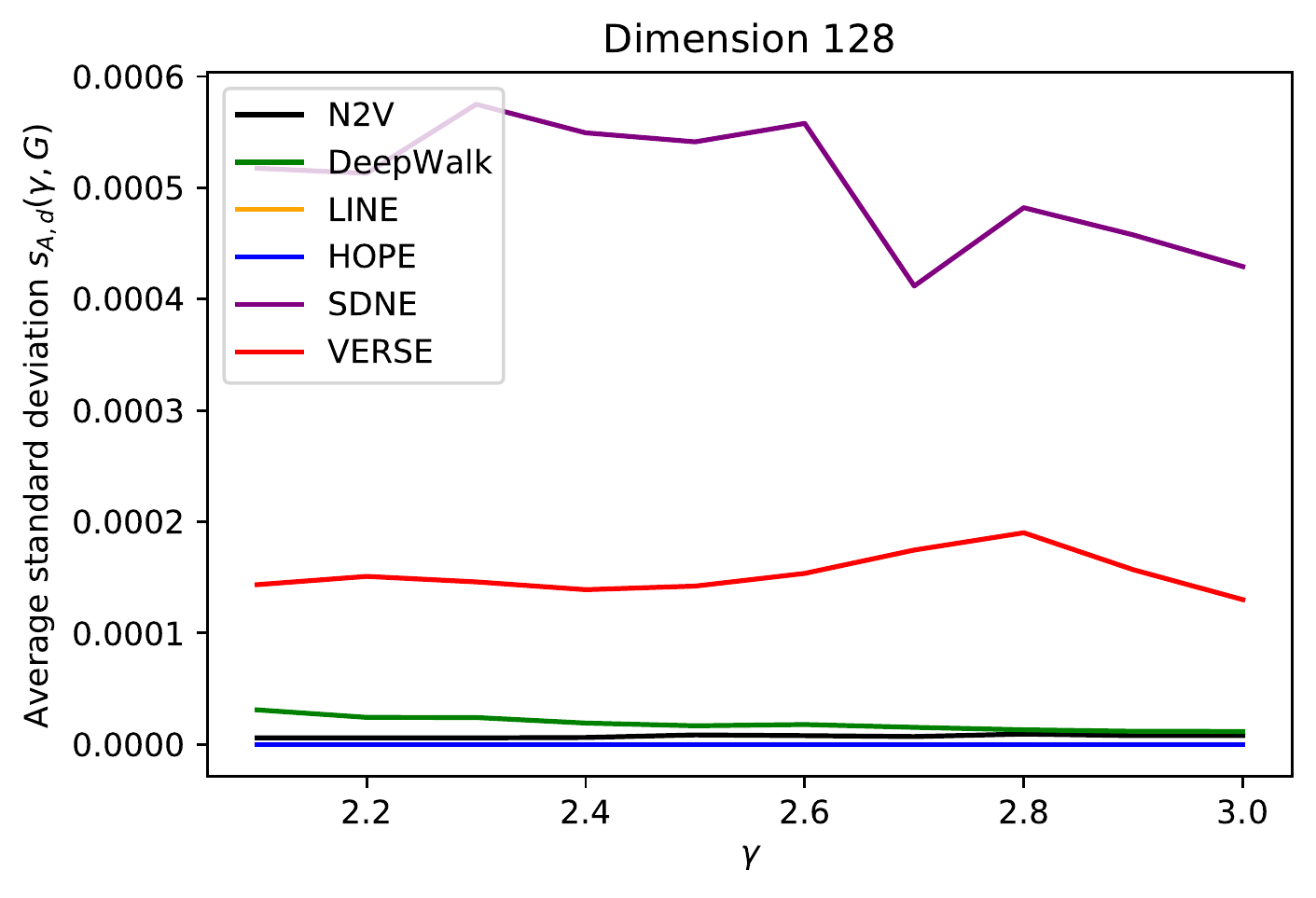}\\
Plots 3 and 4: average $s_{A,d}(\gamma, G)$ (over 10 graphs) 

\includegraphics[width=0.2\textwidth]{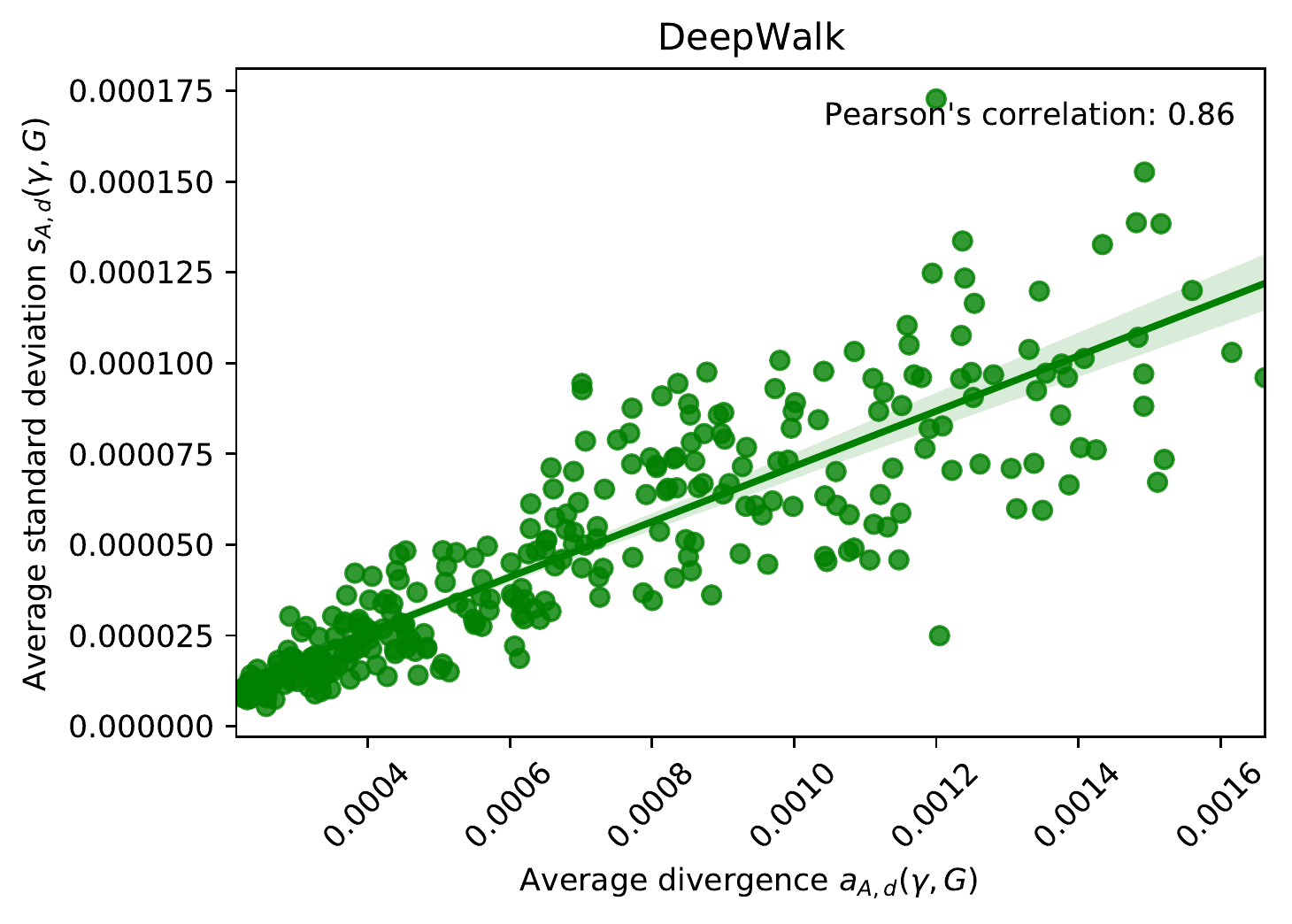}
\includegraphics[width=0.2\textwidth]{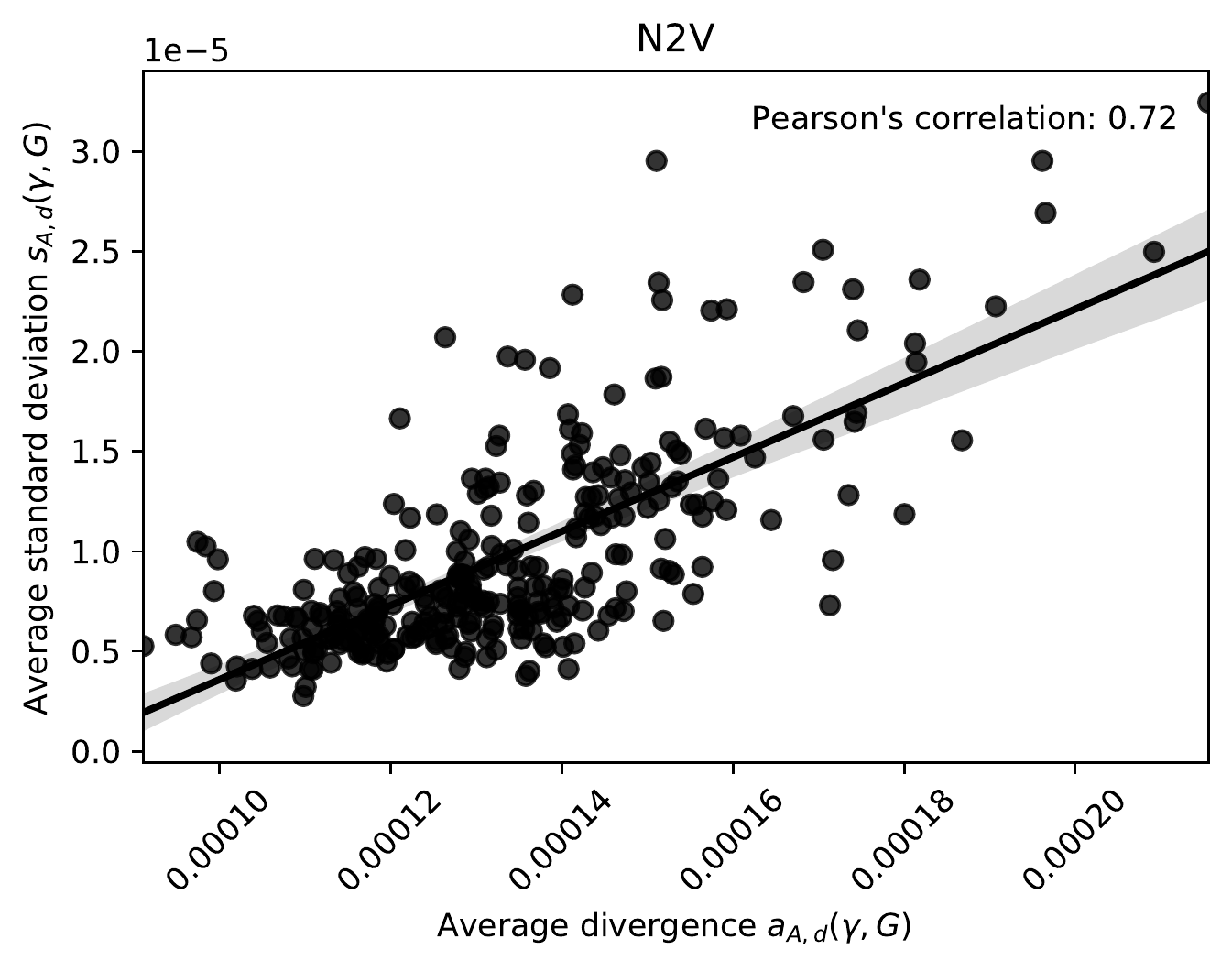}
\includegraphics[width=0.2\textwidth]{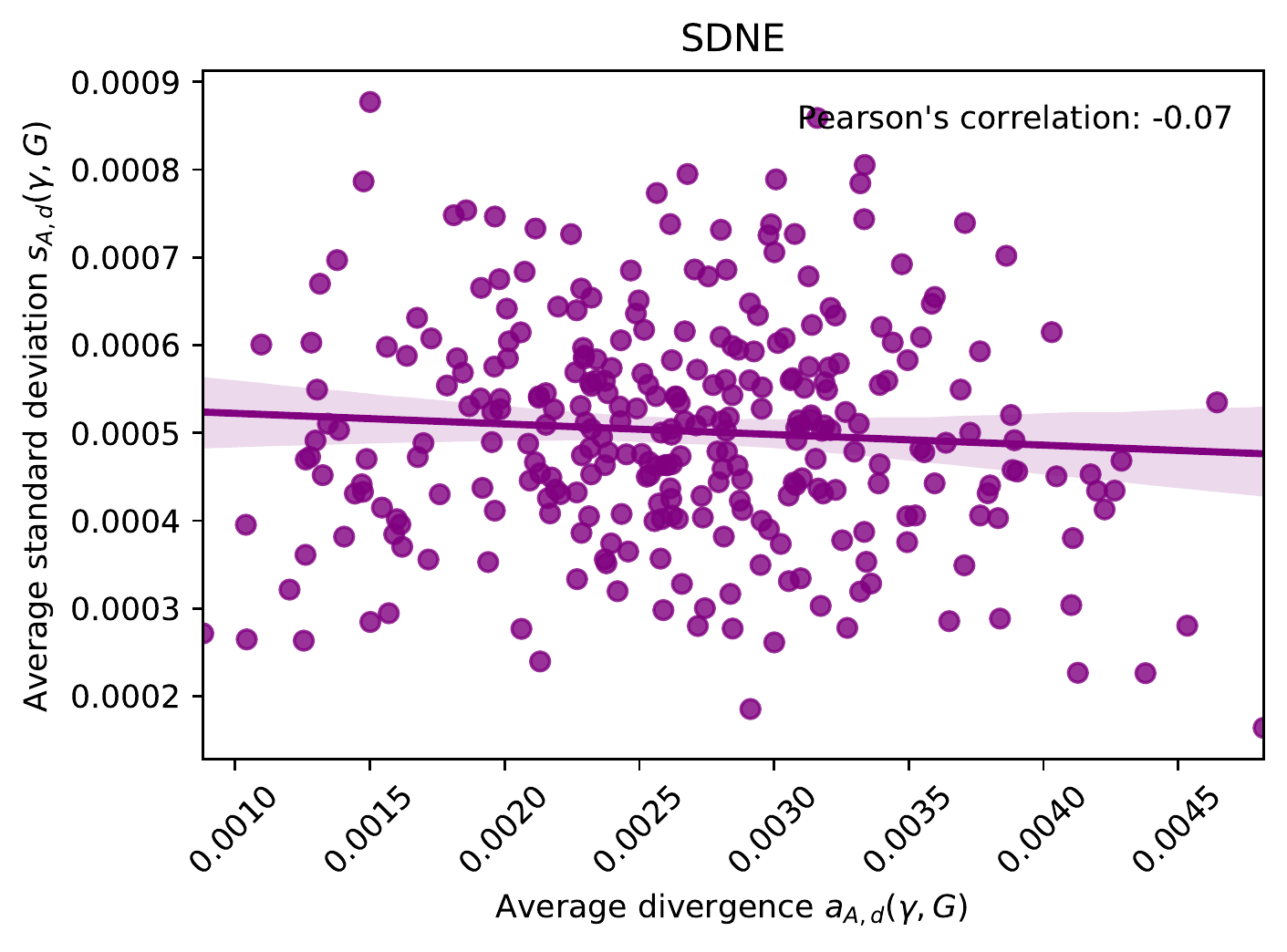}
\includegraphics[width=0.2\textwidth]{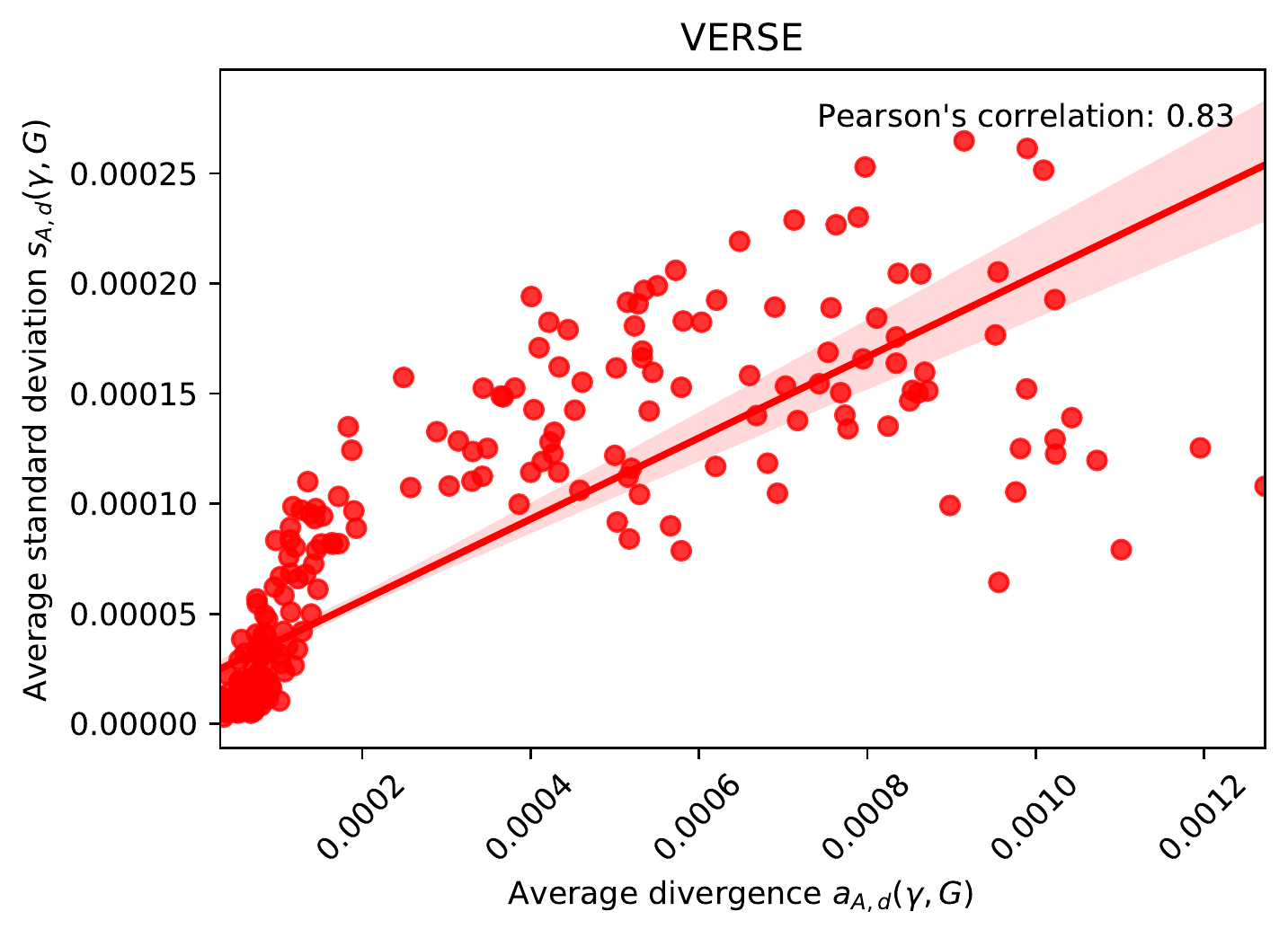} \\
Plot 5: correlation between $a_{A,d}(\gamma, G)$ and $s_{A,d}(\gamma, G)$ 

\caption{Degree Distribution ($\gamma$)}\label{fig:gamma}
\end{center}
\end{figure}

\begin{figure}[htbp!]
\begin{center}
\includegraphics[width=0.3\textwidth]{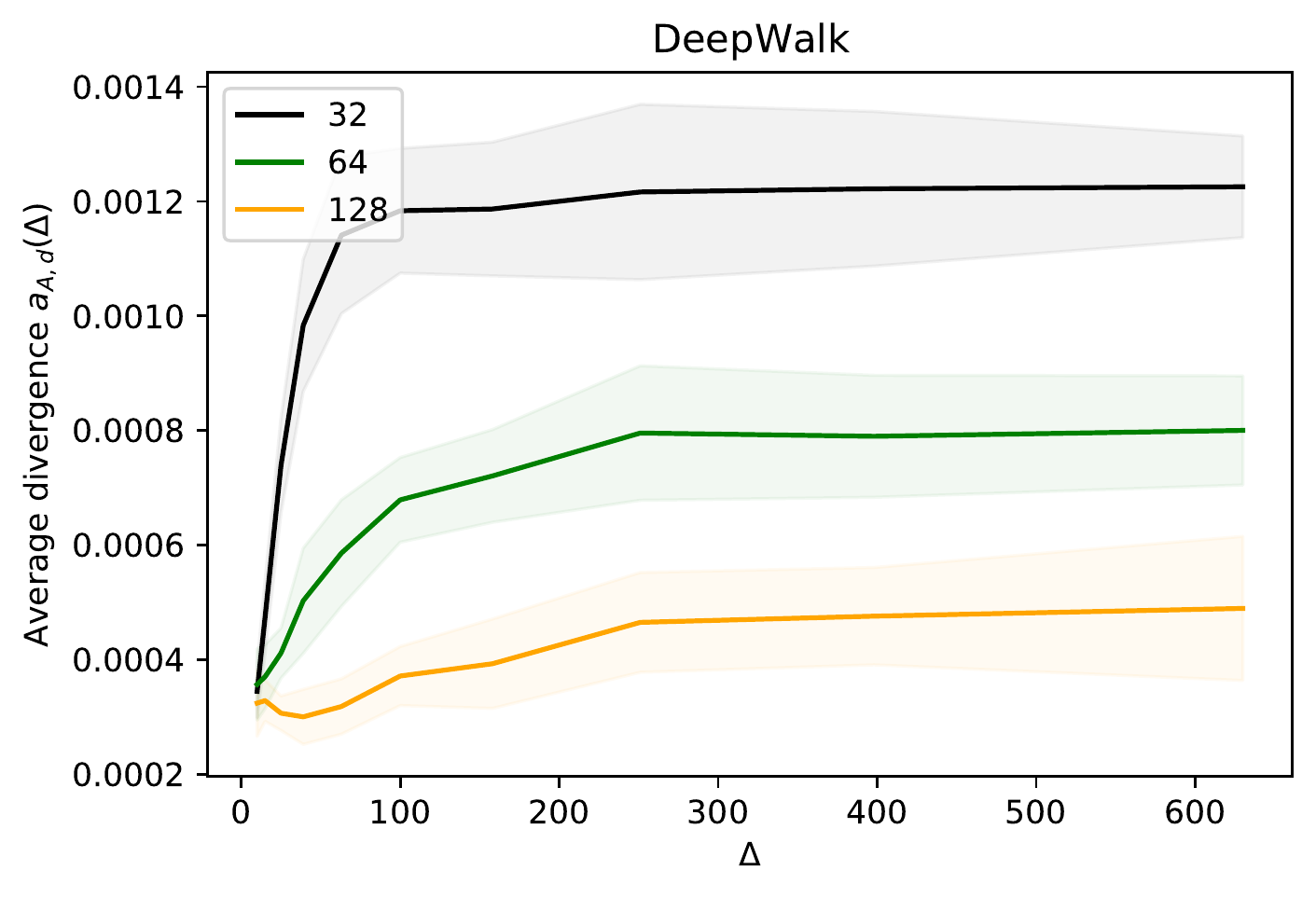}
\includegraphics[width=0.3\textwidth]{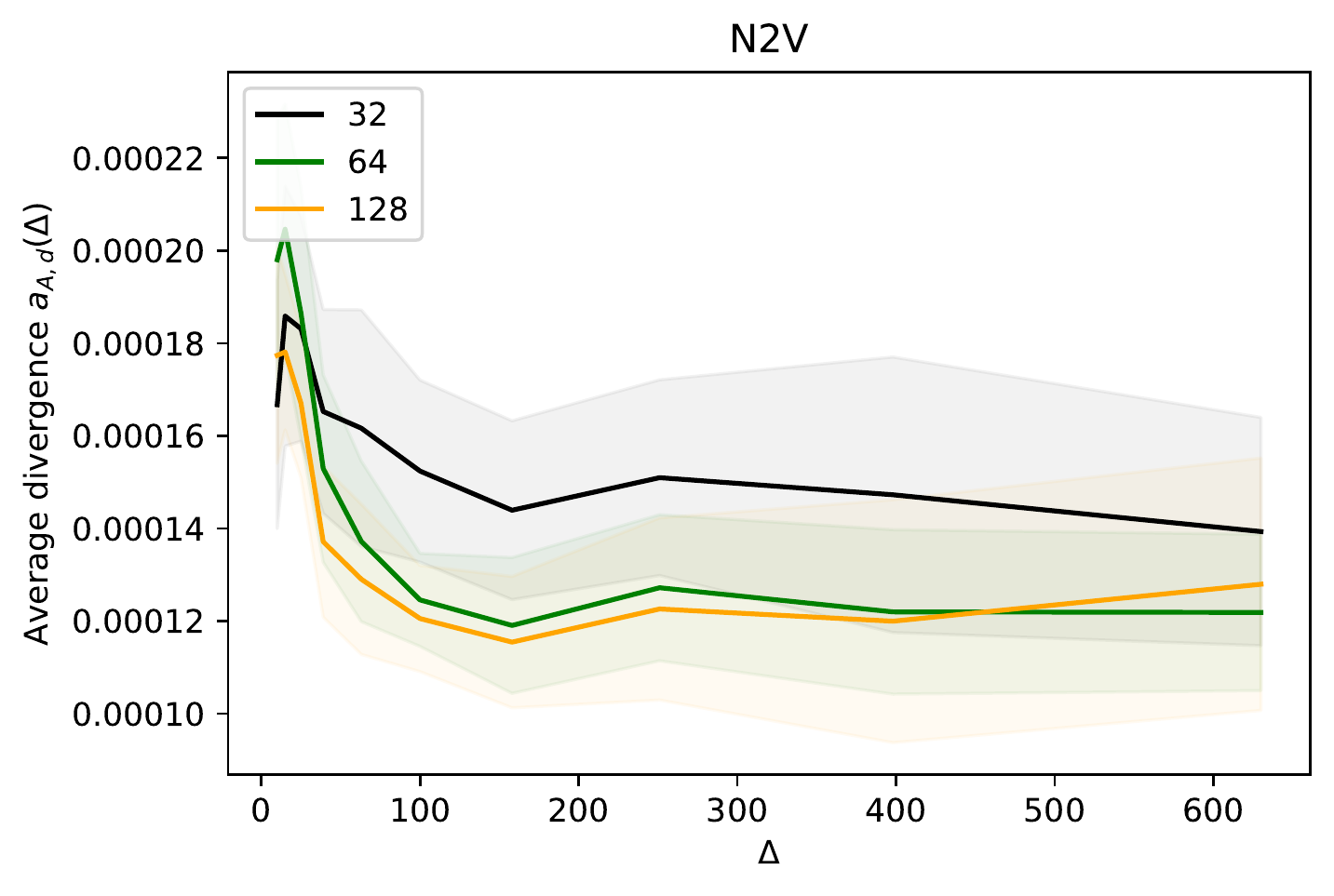}
\includegraphics[width=0.3\textwidth]{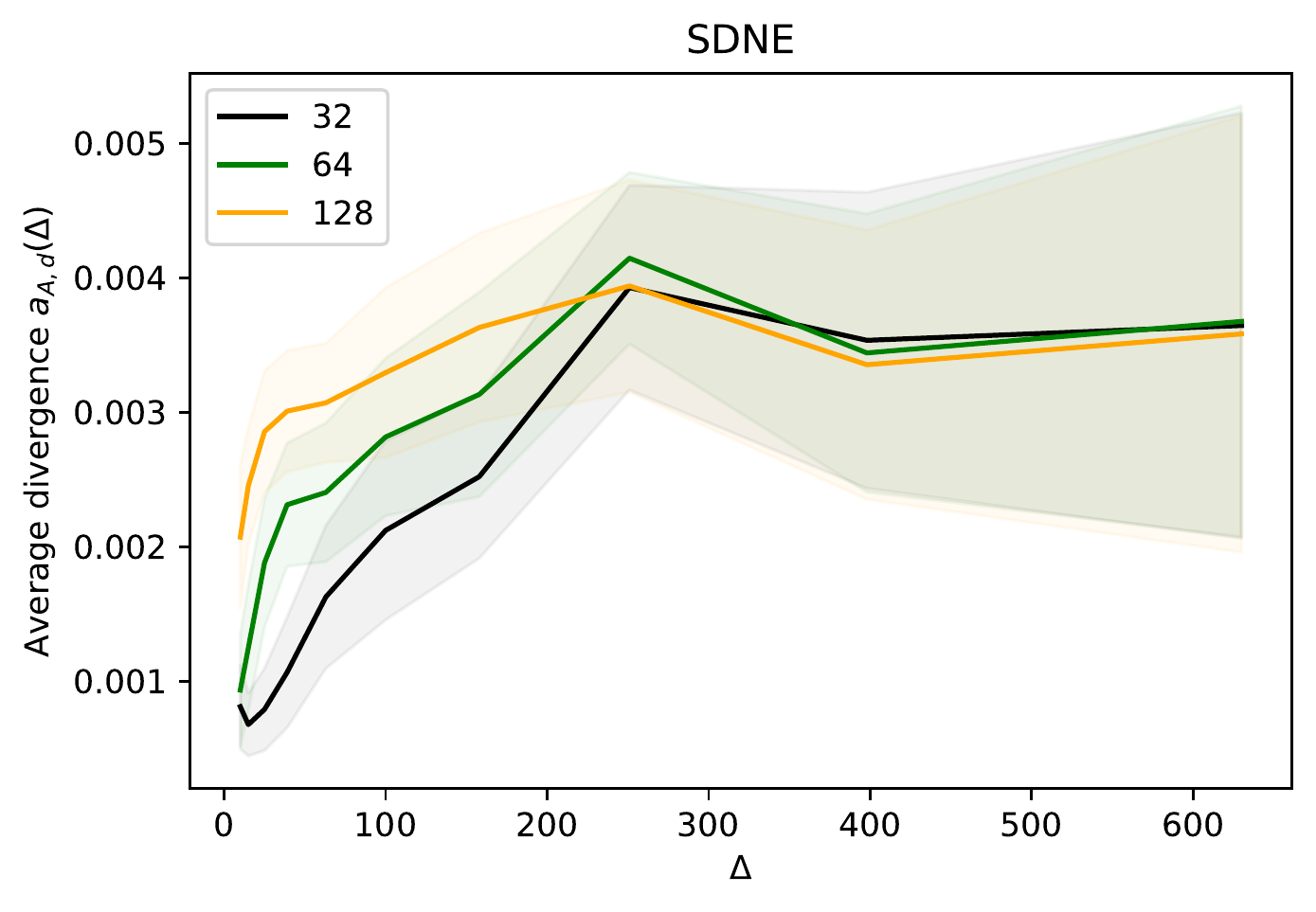} \\
\includegraphics[width=0.3\textwidth]{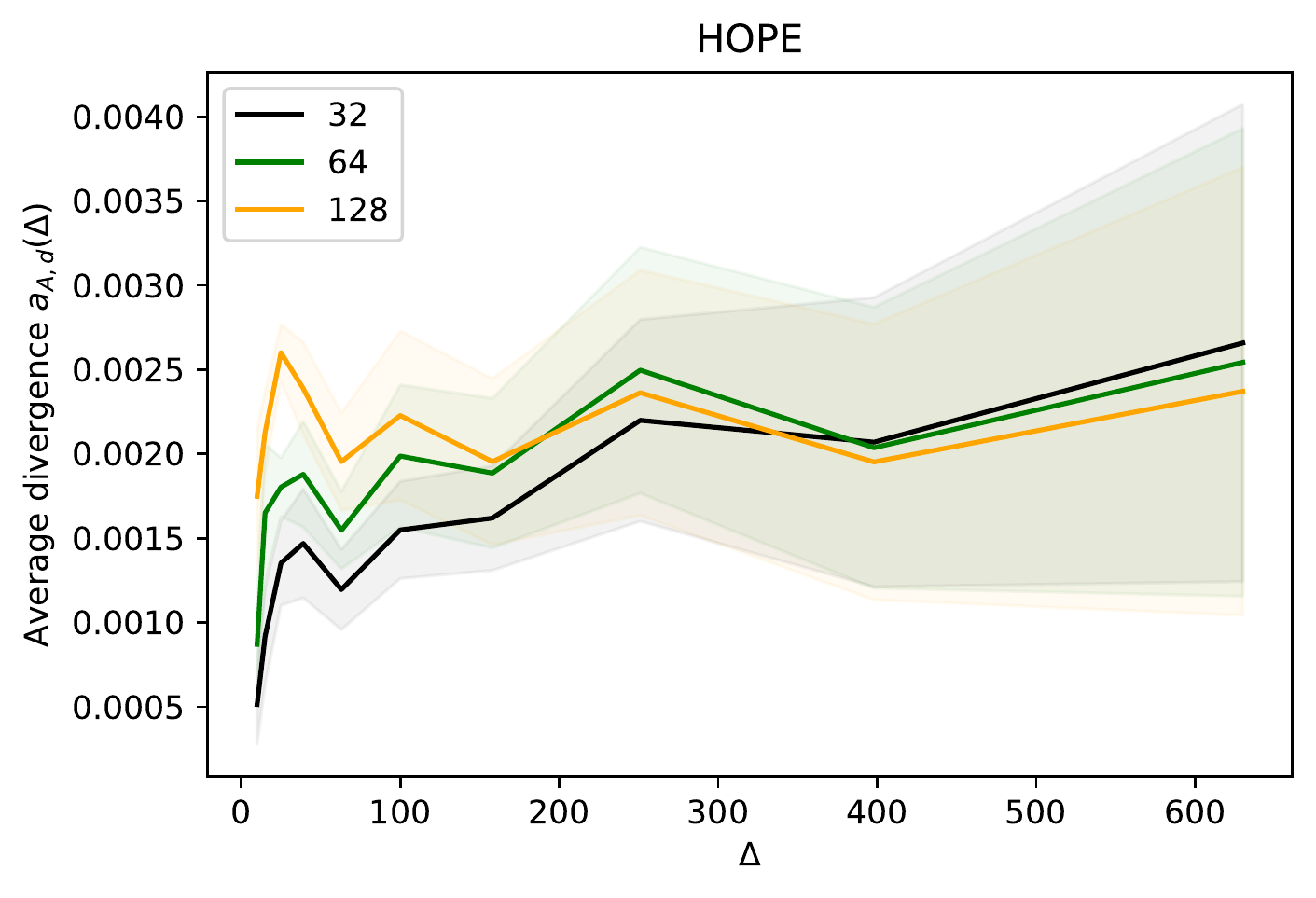}
\includegraphics[width=0.3\textwidth]{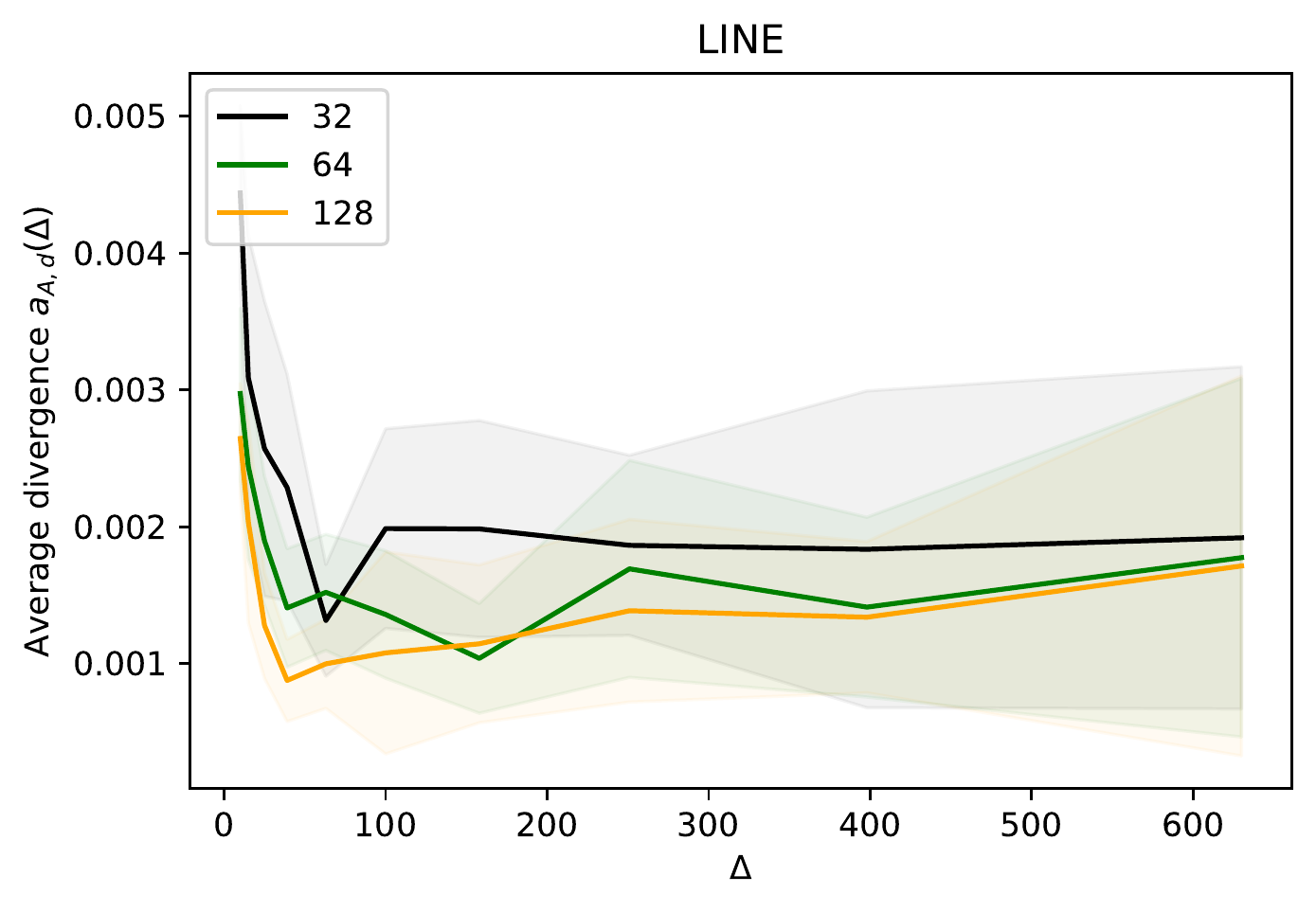}
\includegraphics[width=0.3\textwidth]{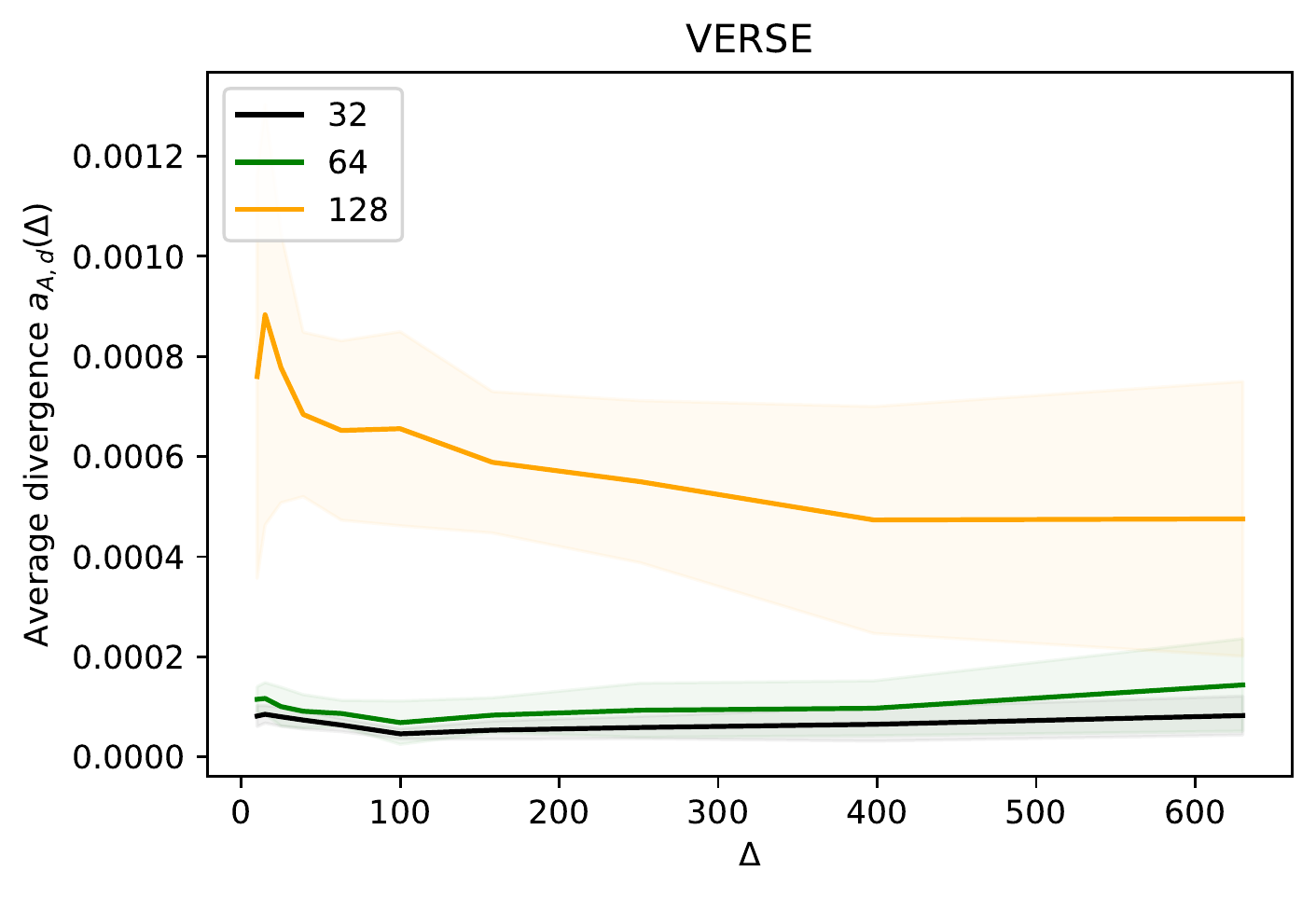} \\
\includegraphics[width=0.3\textwidth]{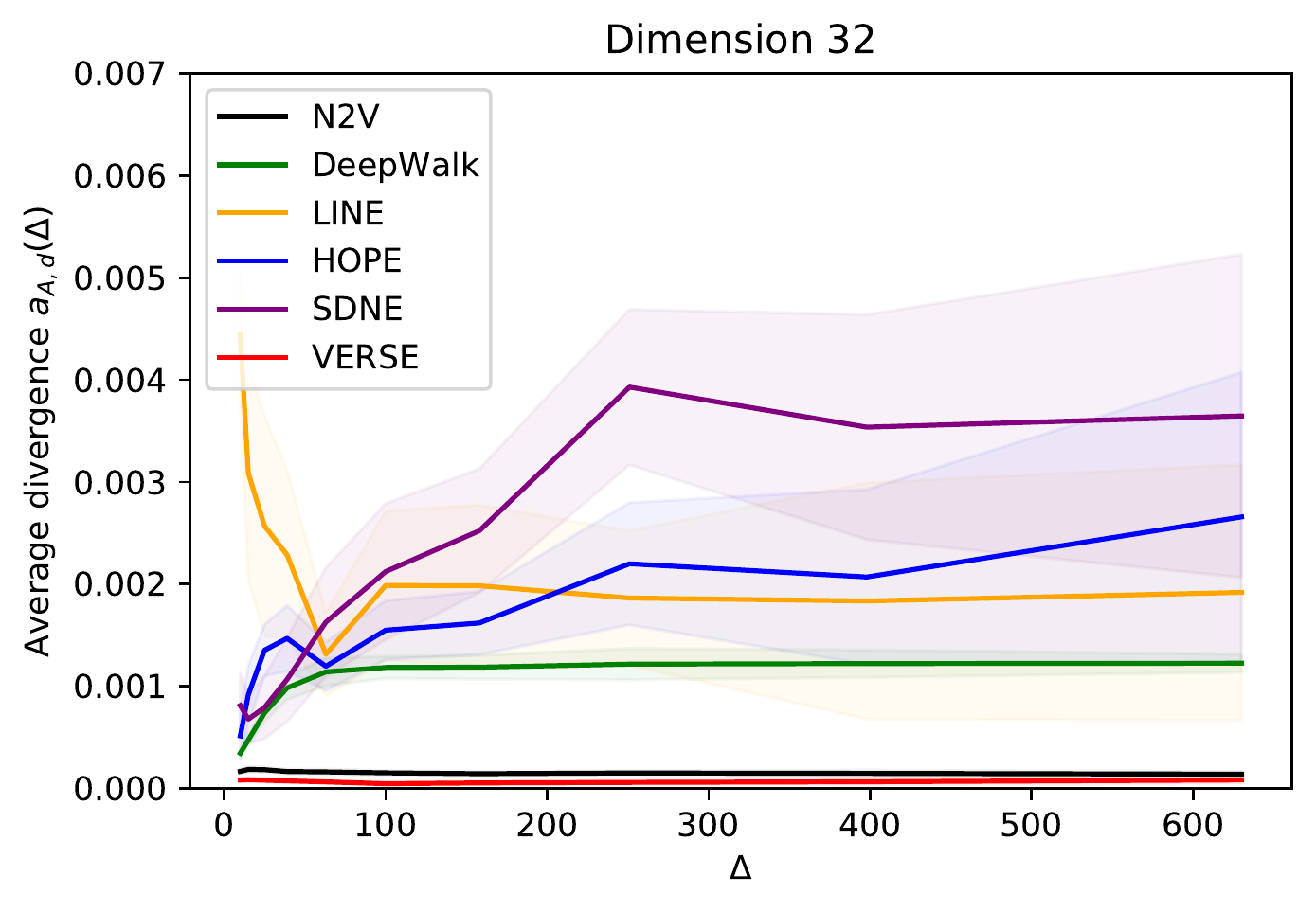}
\includegraphics[width=0.3\textwidth]{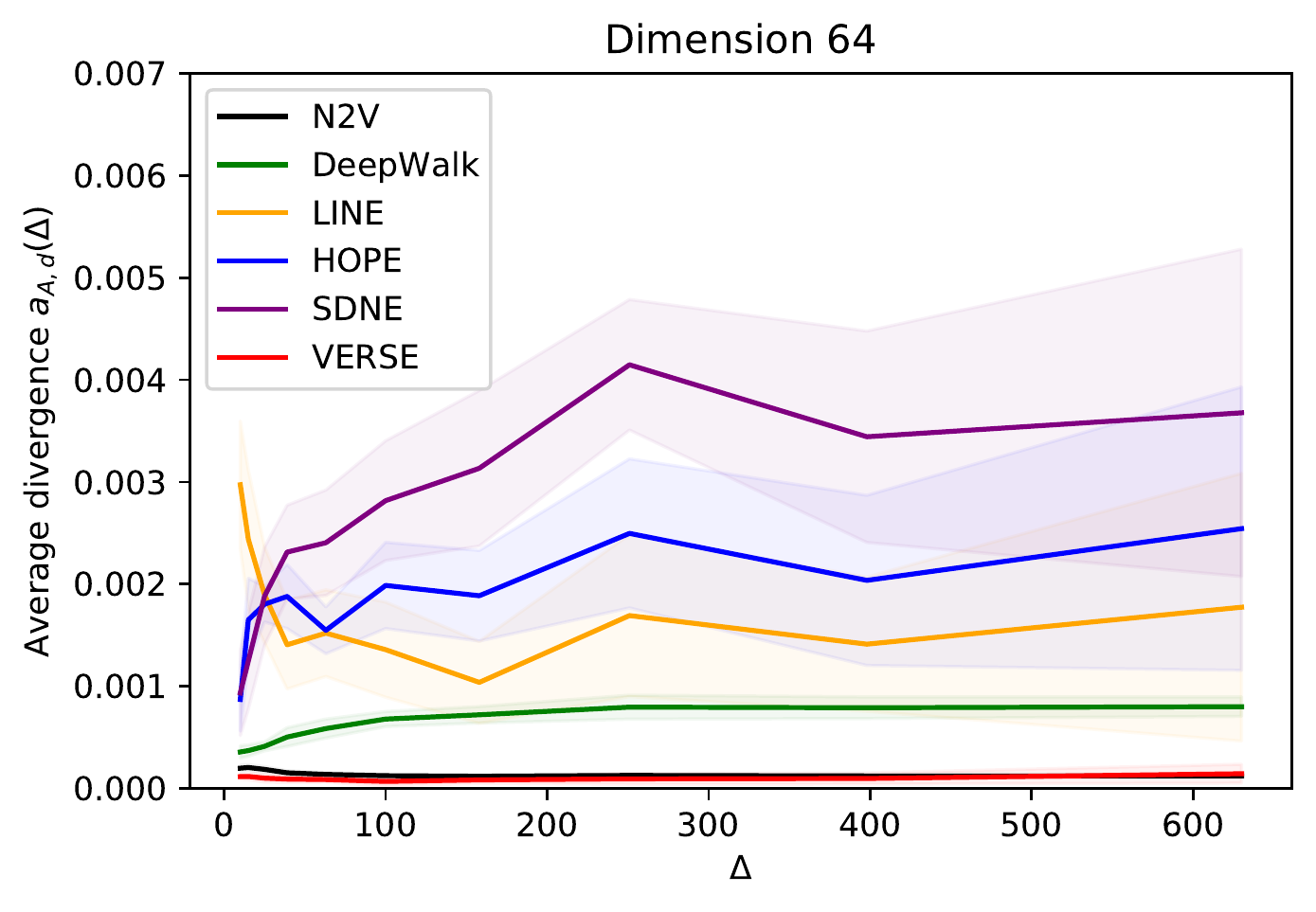}
\includegraphics[width=0.3\textwidth]{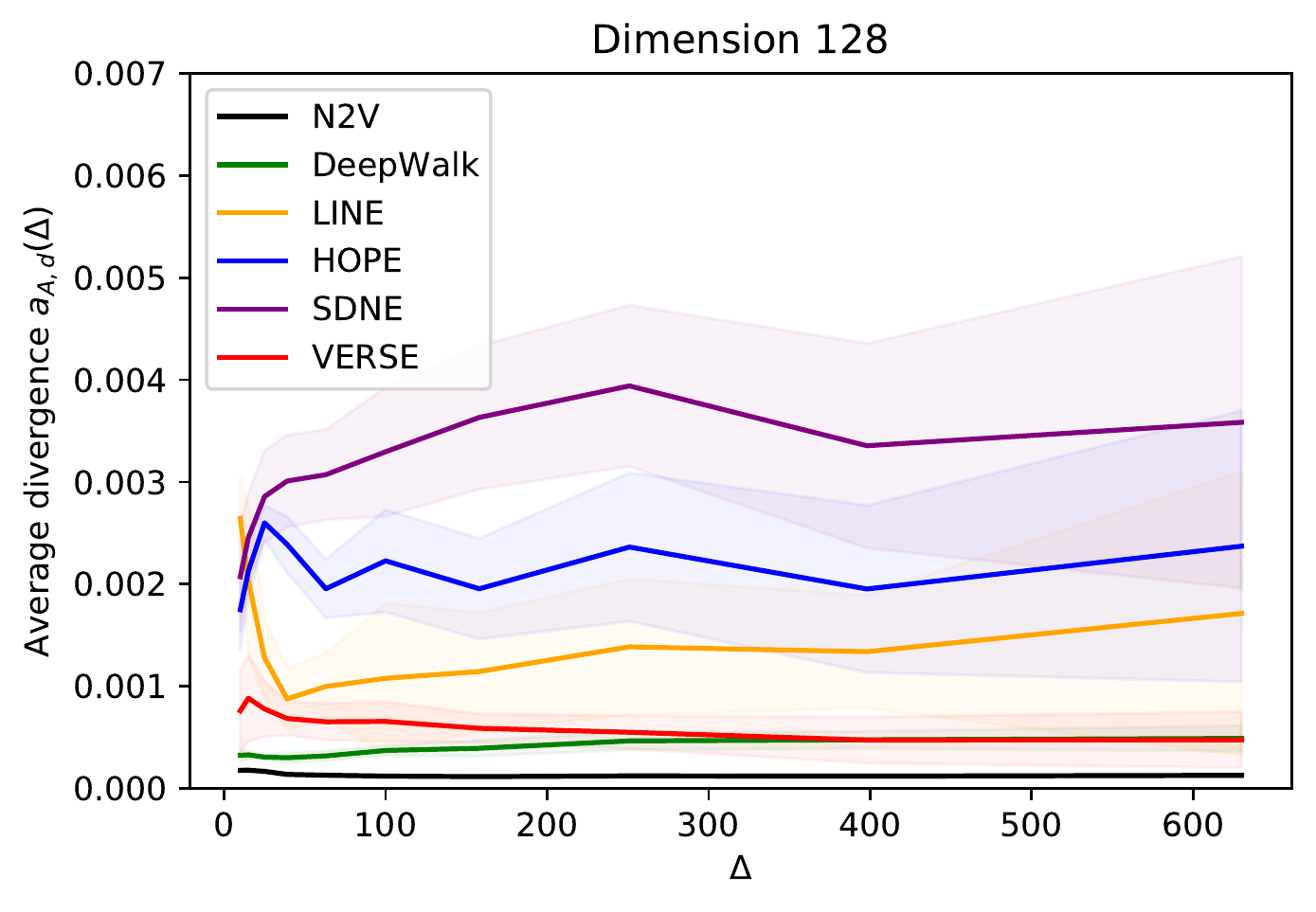}\\
Plots 1 and  2: $a_{A,d}(\Delta) \pm s_{A,d}(\Delta)$ \\

\includegraphics[width=0.2\textwidth]{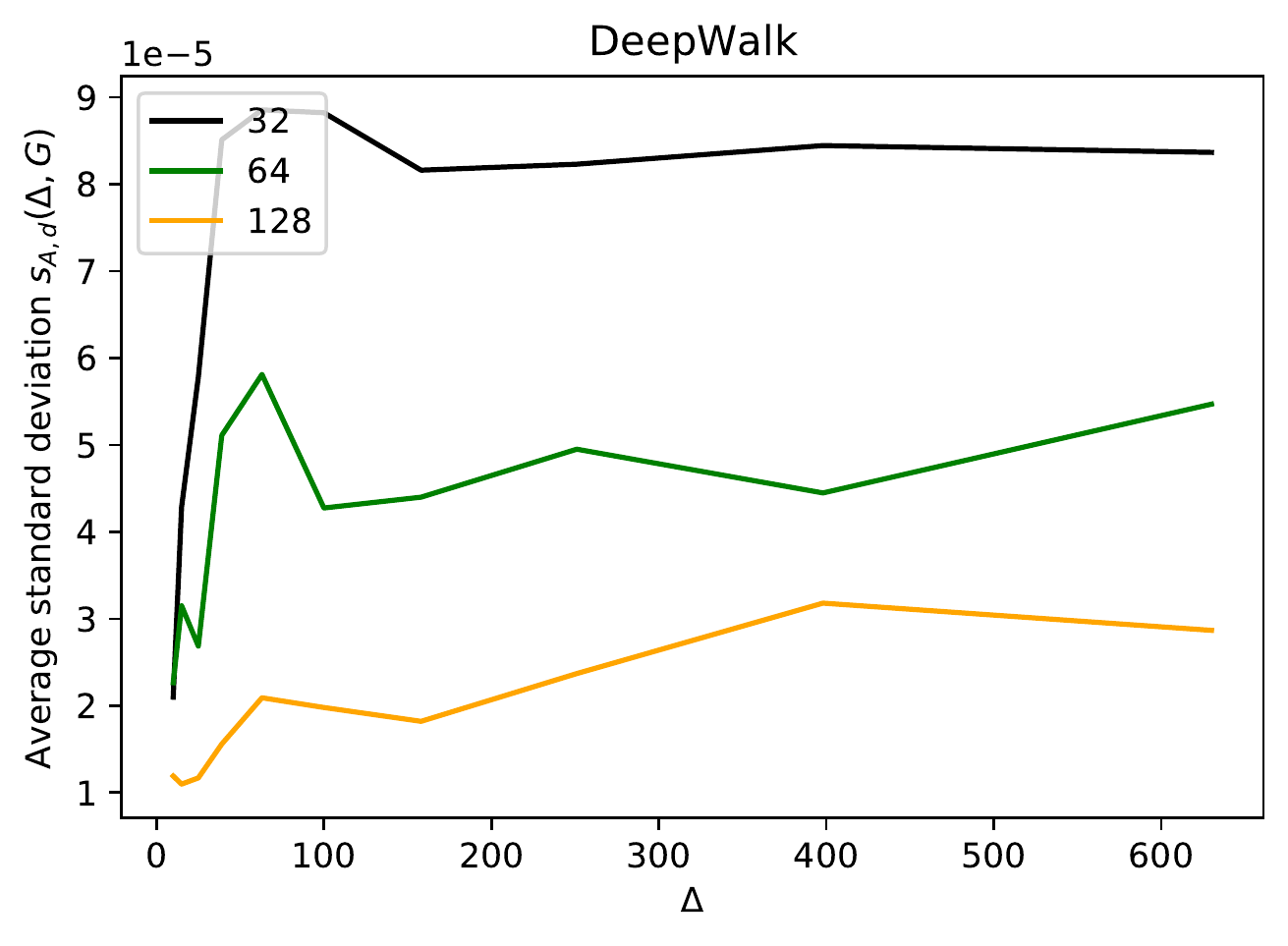}
\includegraphics[width=0.2\textwidth]{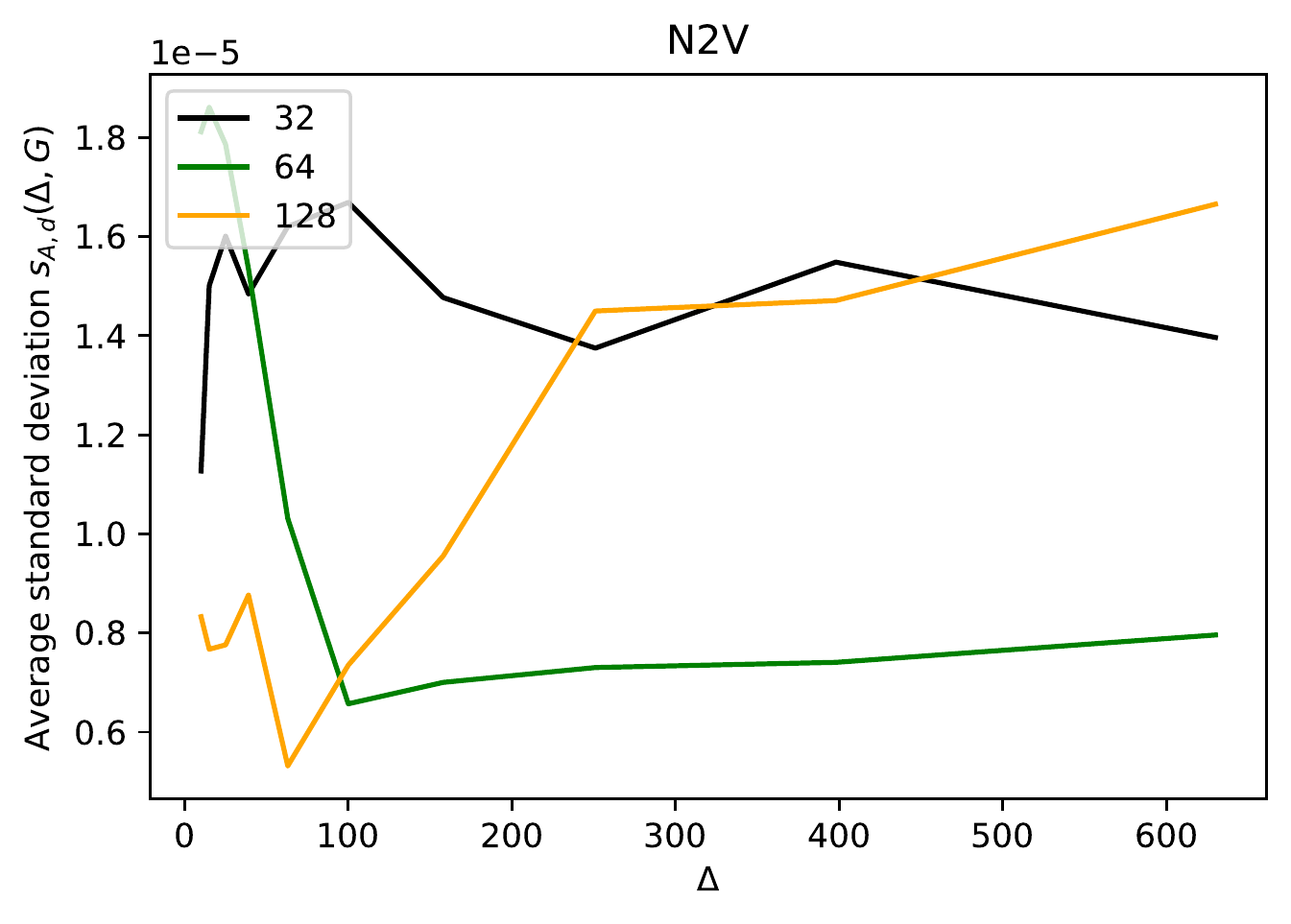}
\includegraphics[width=0.2\textwidth]{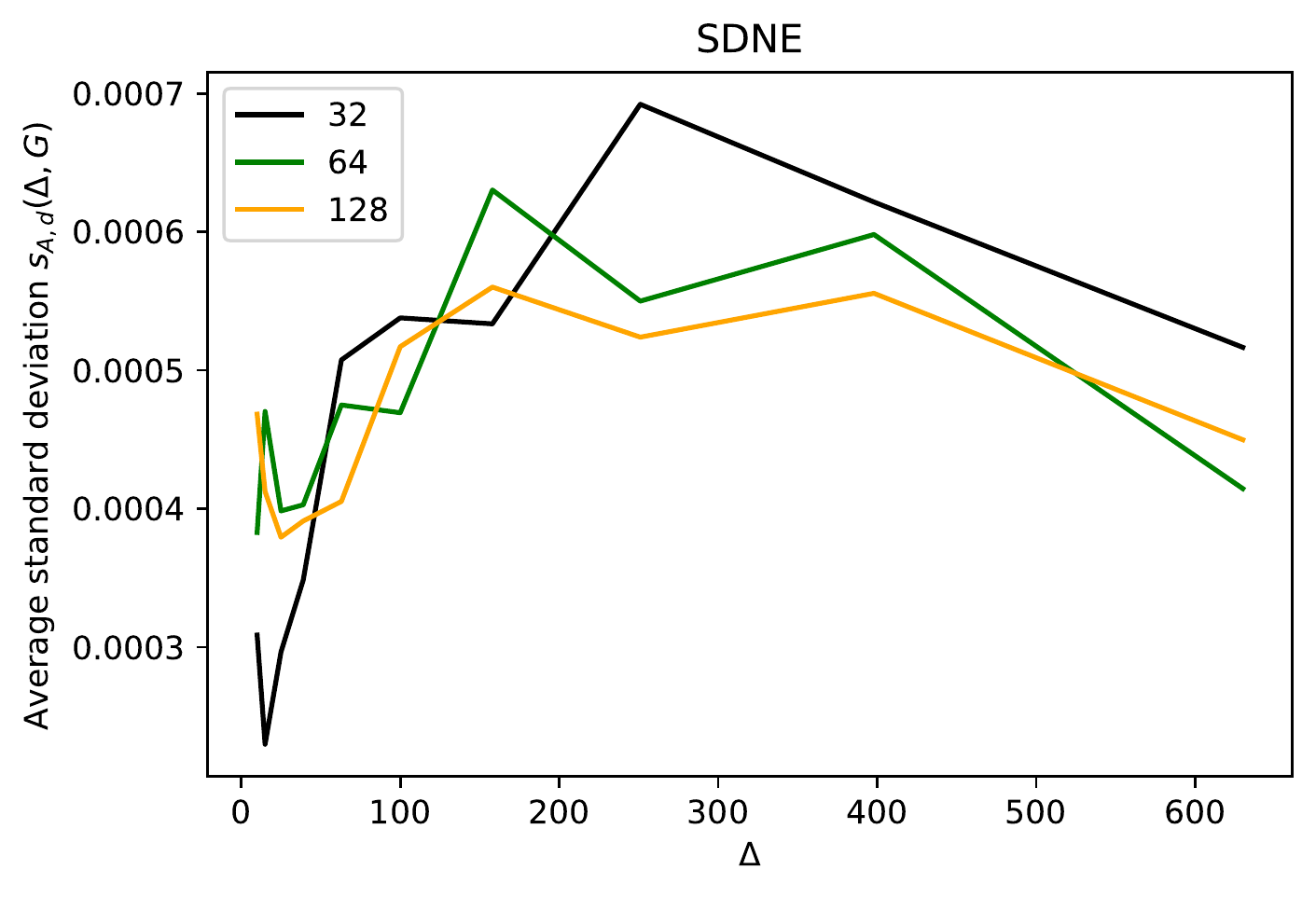}
\includegraphics[width=0.2\textwidth]{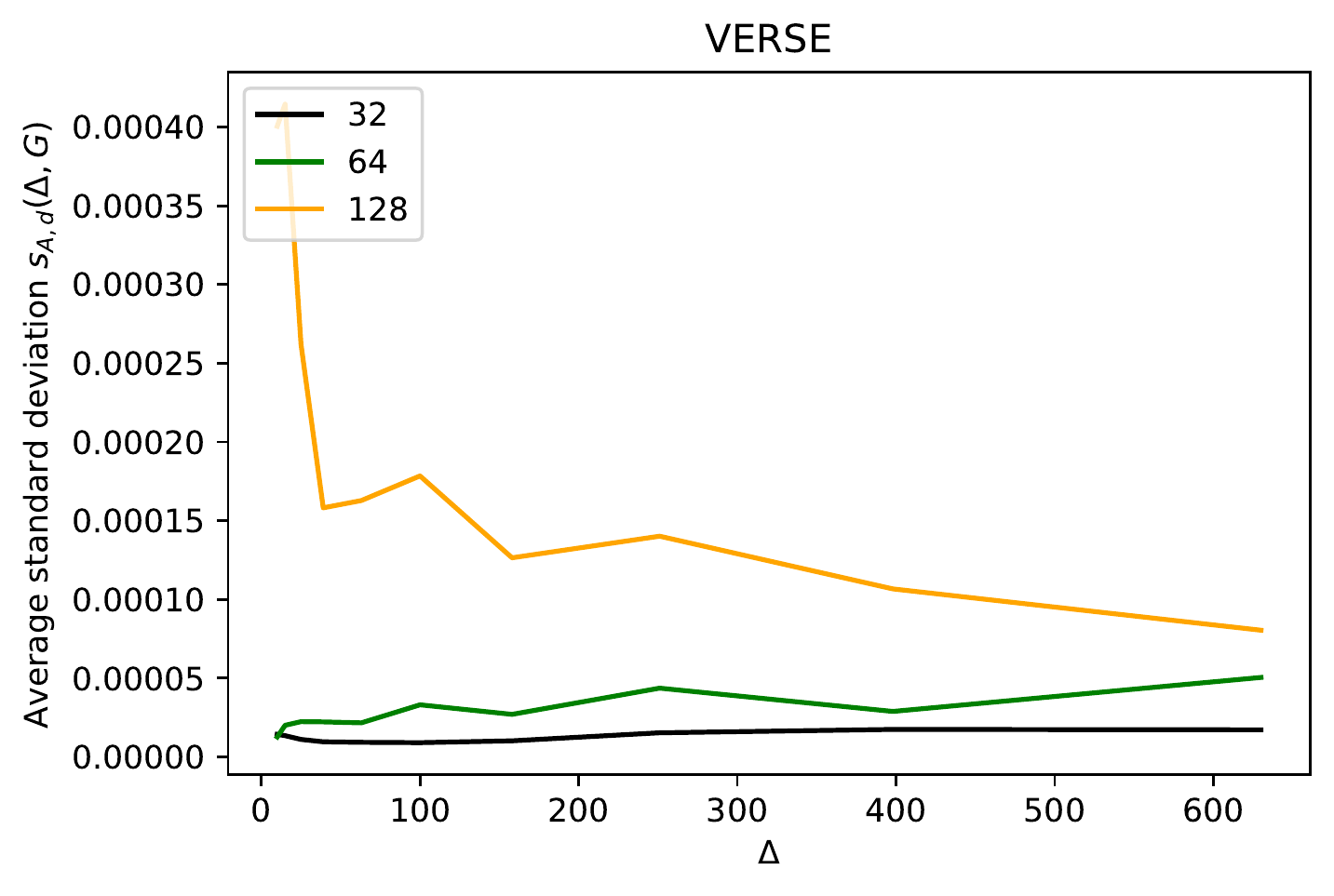} \\
\includegraphics[width=0.3\textwidth]{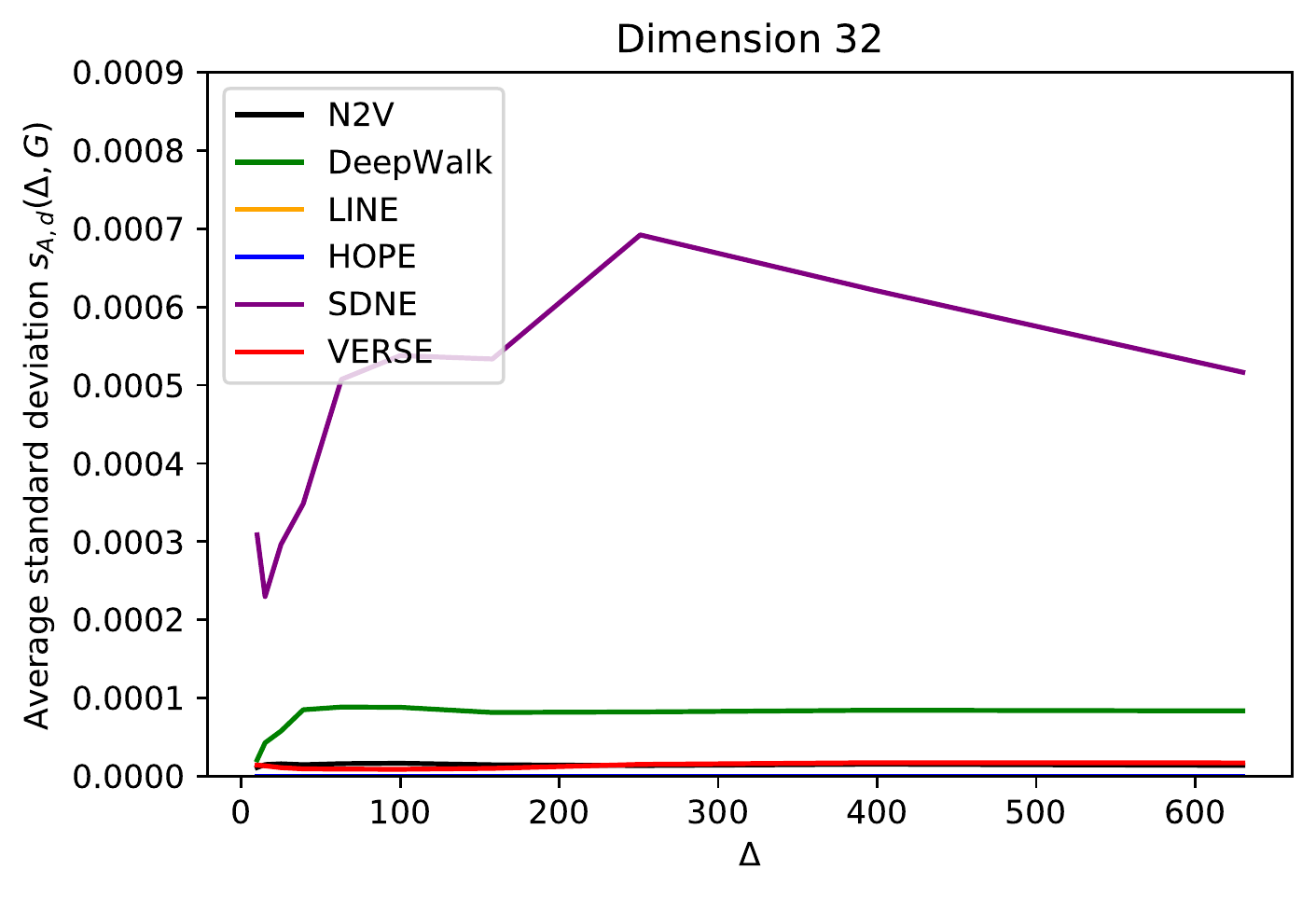}
\includegraphics[width=0.3\textwidth]{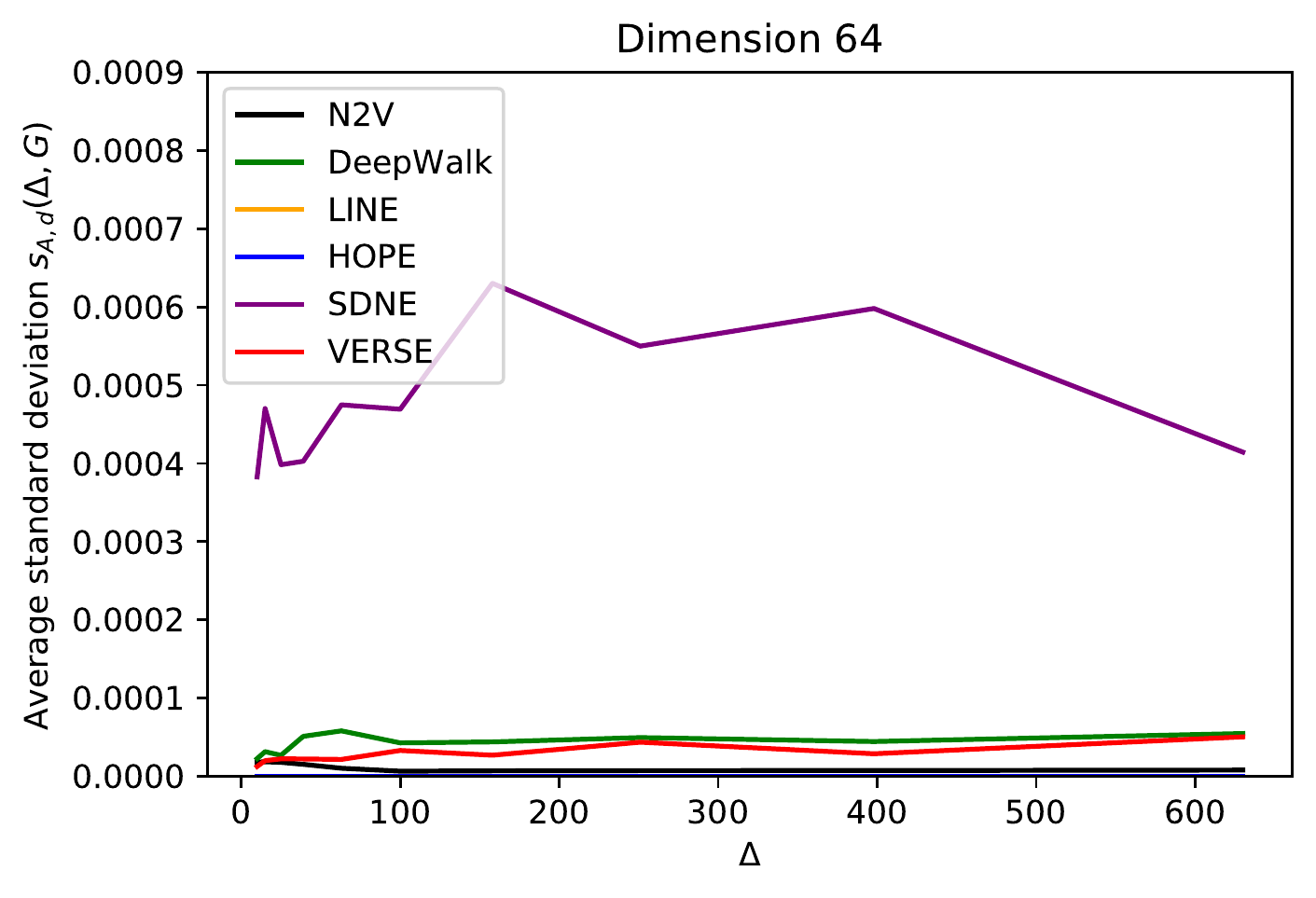}
\includegraphics[width=0.3\textwidth]{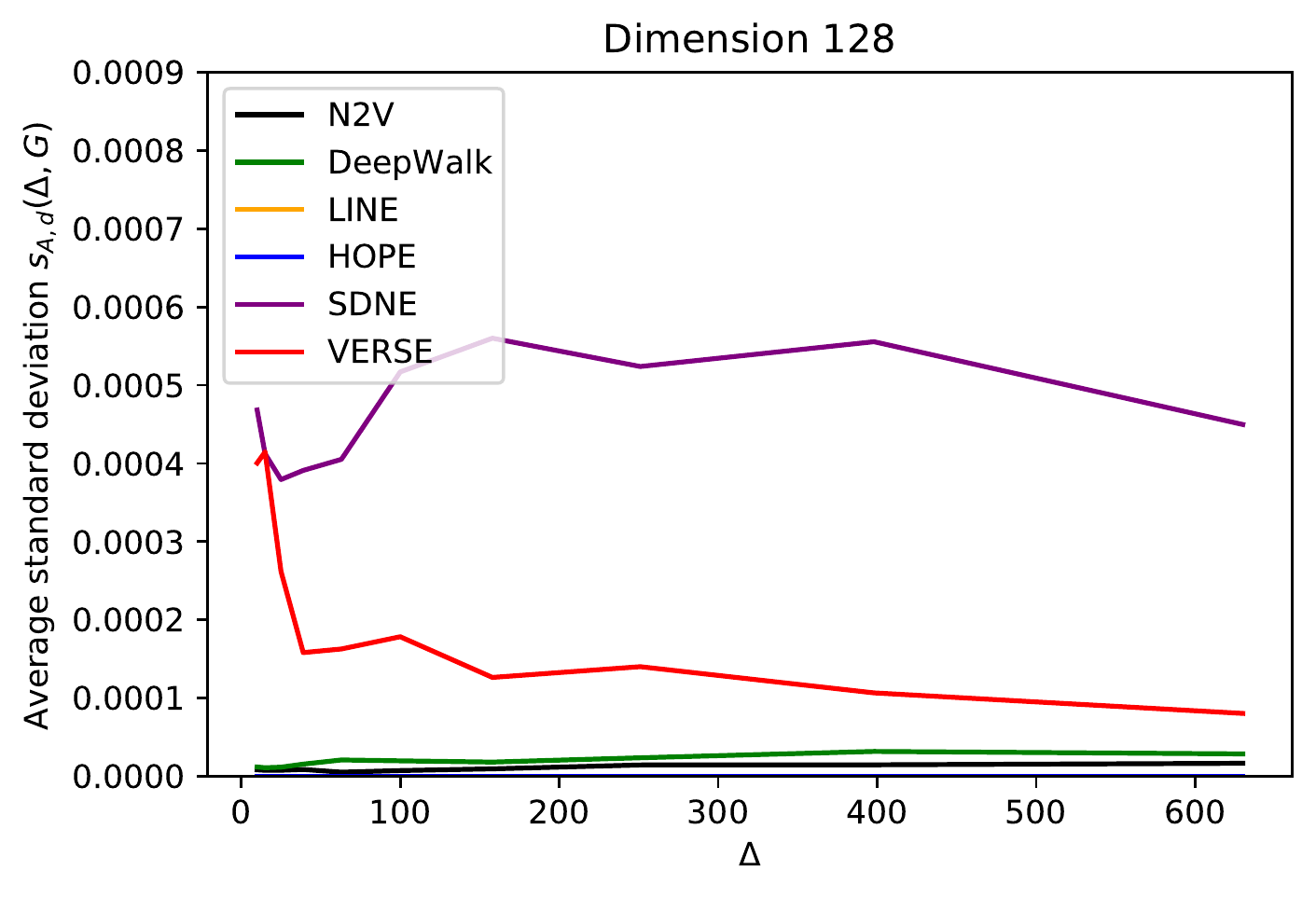} \\
Plots 3 and 4: average $s_{A,d}(\Delta, G)$ (over 10 graphs)  \\

\includegraphics[width=0.2\textwidth]{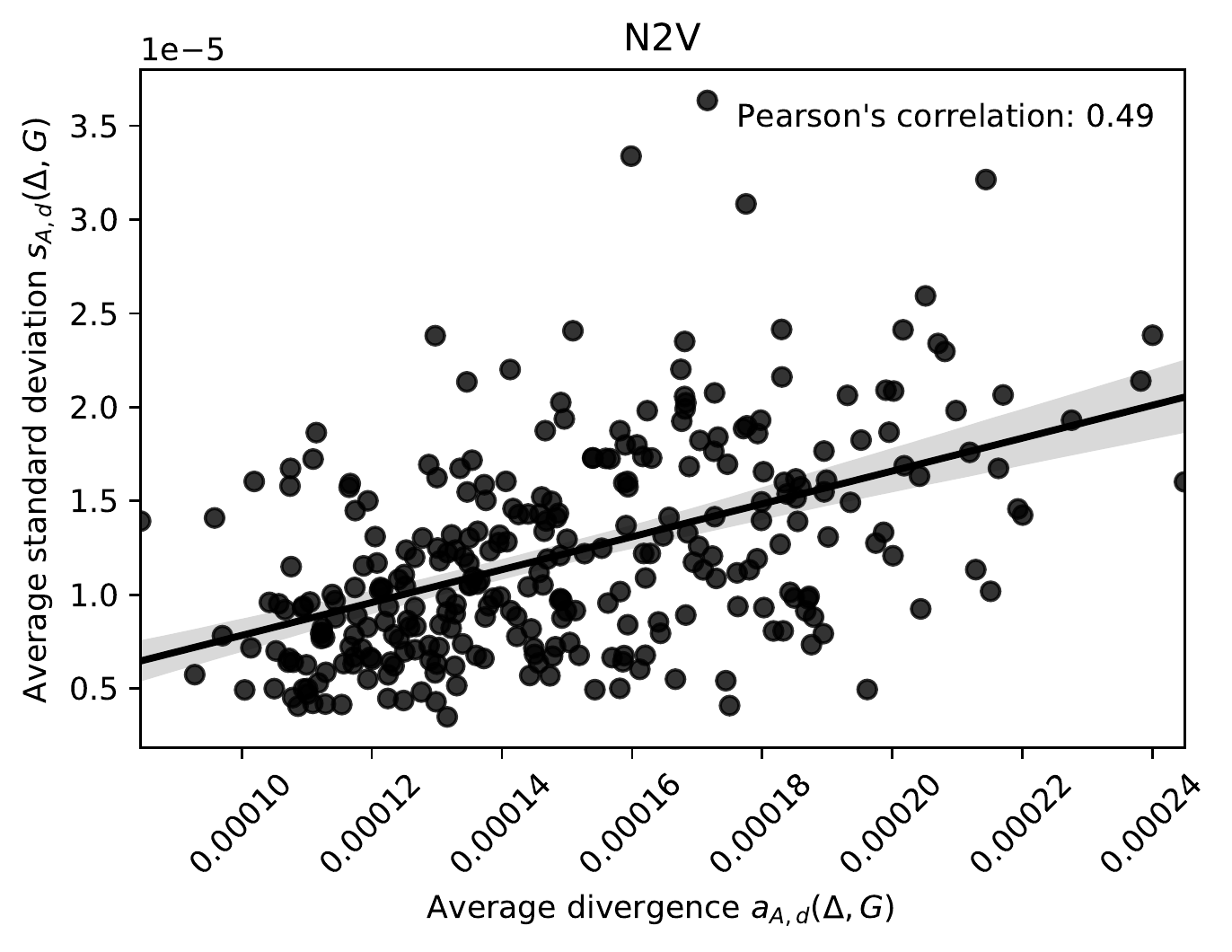}
\includegraphics[width=0.2\textwidth]{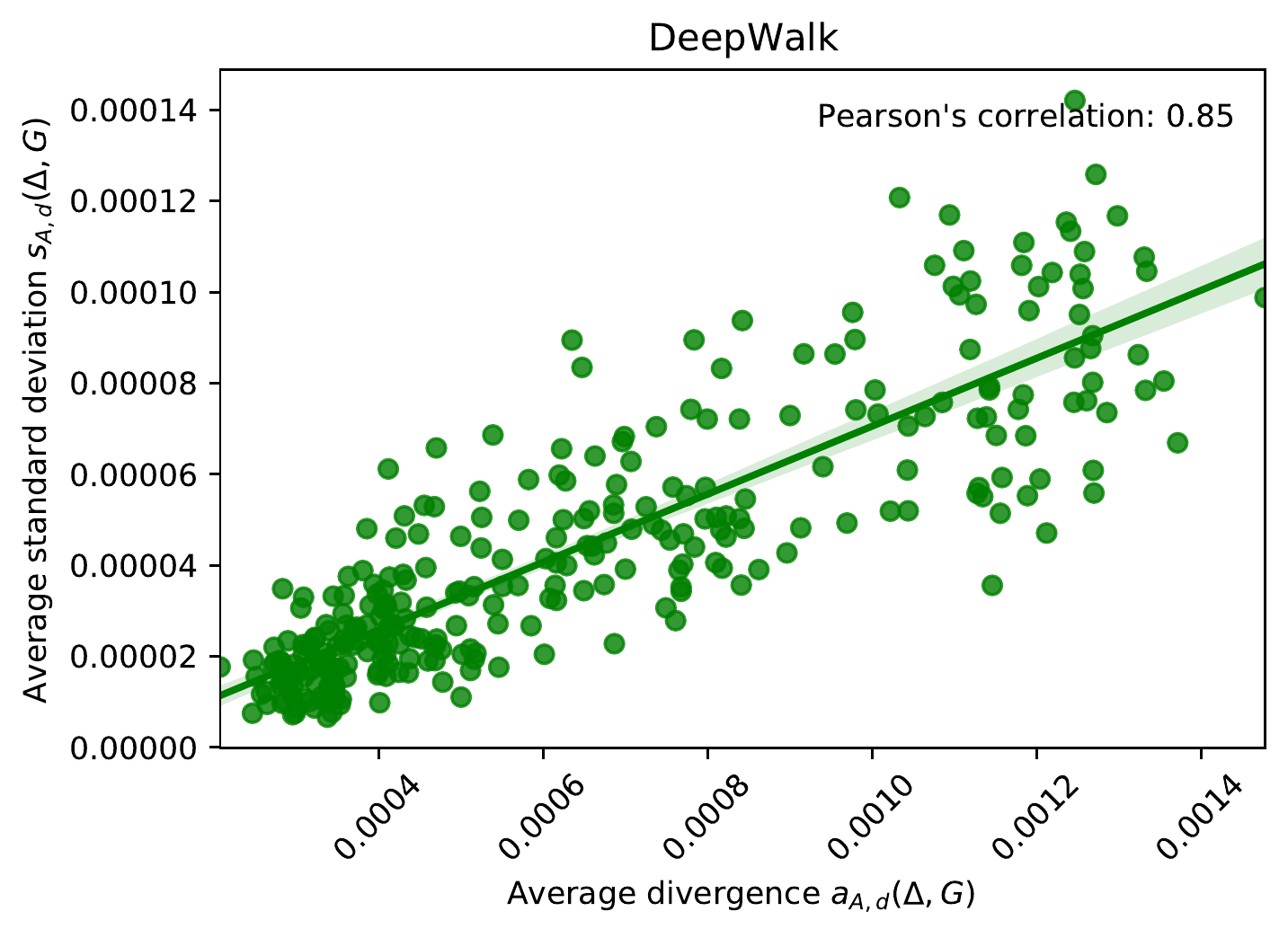}
\includegraphics[width=0.2\textwidth]{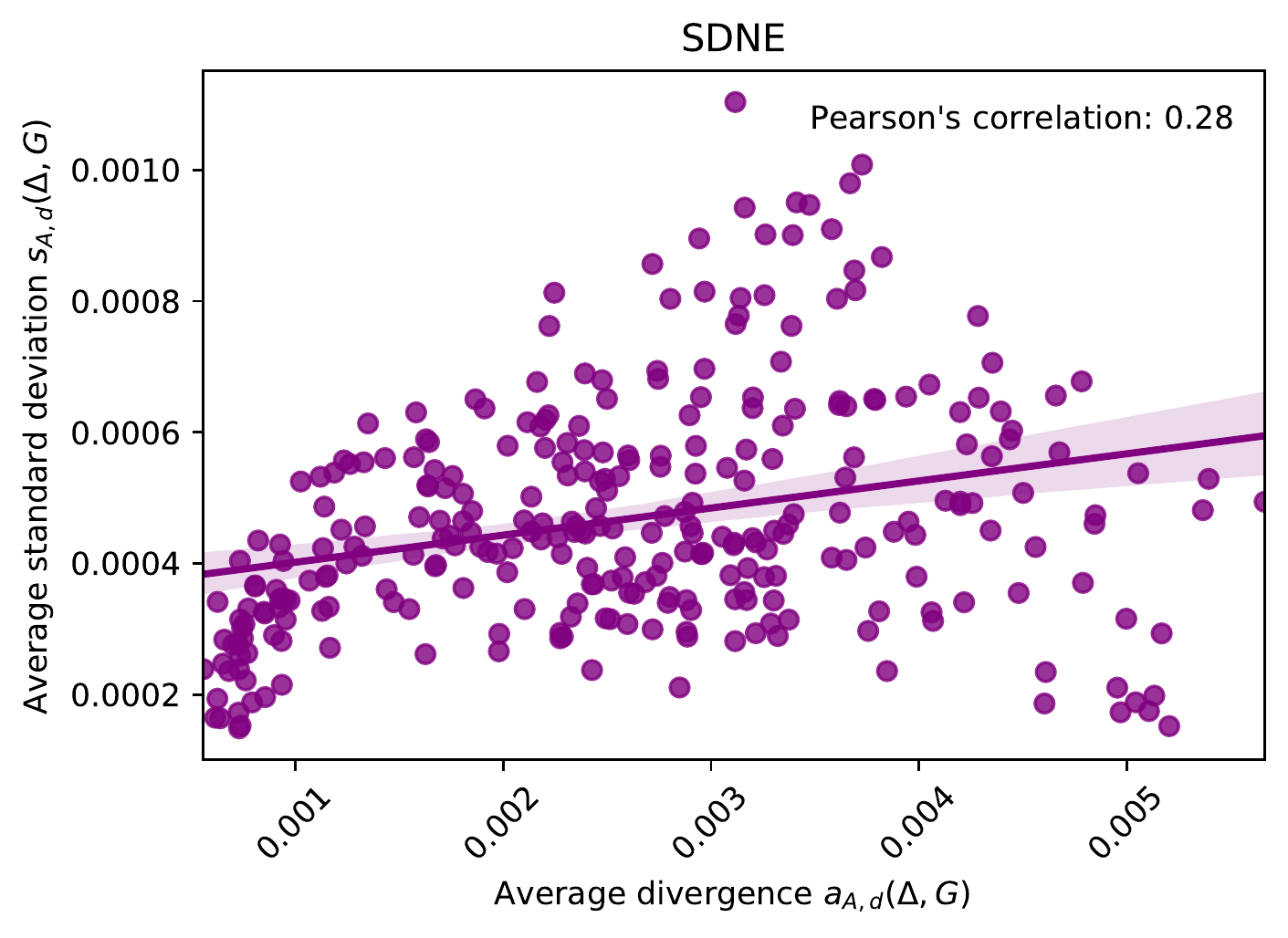}
\includegraphics[width=0.2\textwidth]{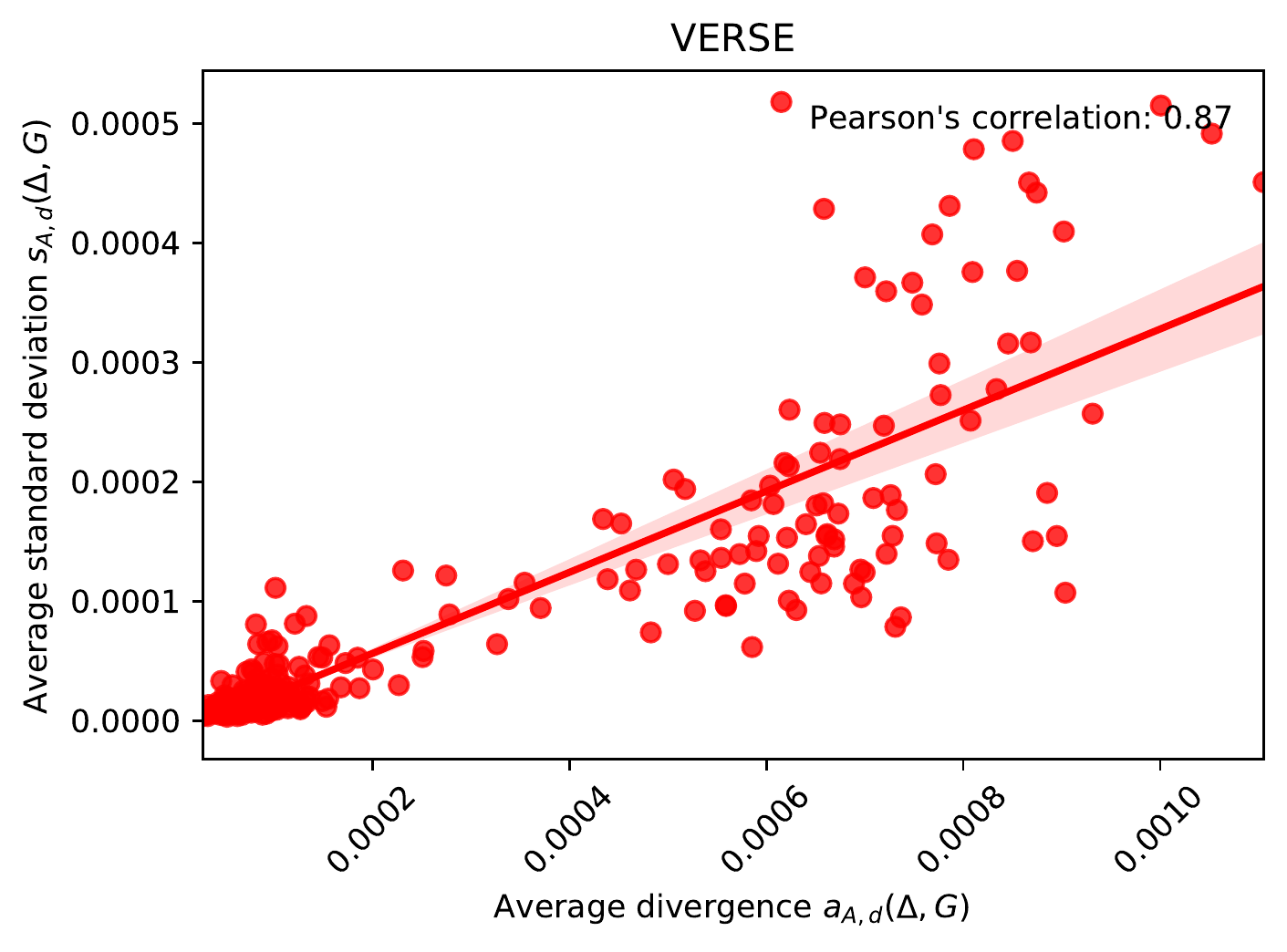} \\
Plot 5: correlation between $a_{A,d}(\Delta, G)$ and $s_{A,d}(\Delta, G)$ 
\caption{Maximum Degree ($\Delta$)}\label{fig:delta}
\end{center}
\end{figure}

\begin{figure}[htbp!]
\begin{center}
\includegraphics[width=0.3\textwidth]{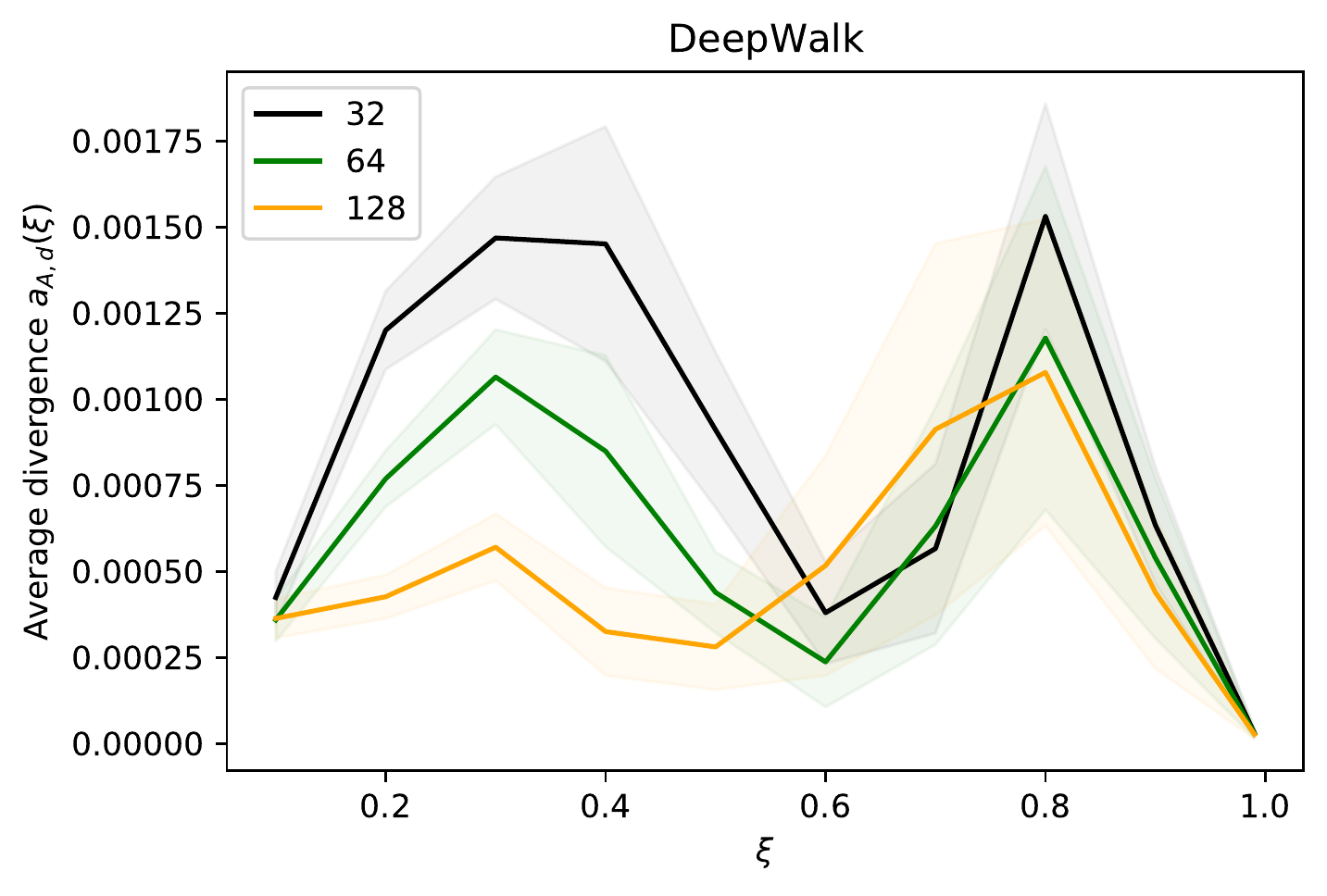} 
\includegraphics[width=0.3\textwidth]{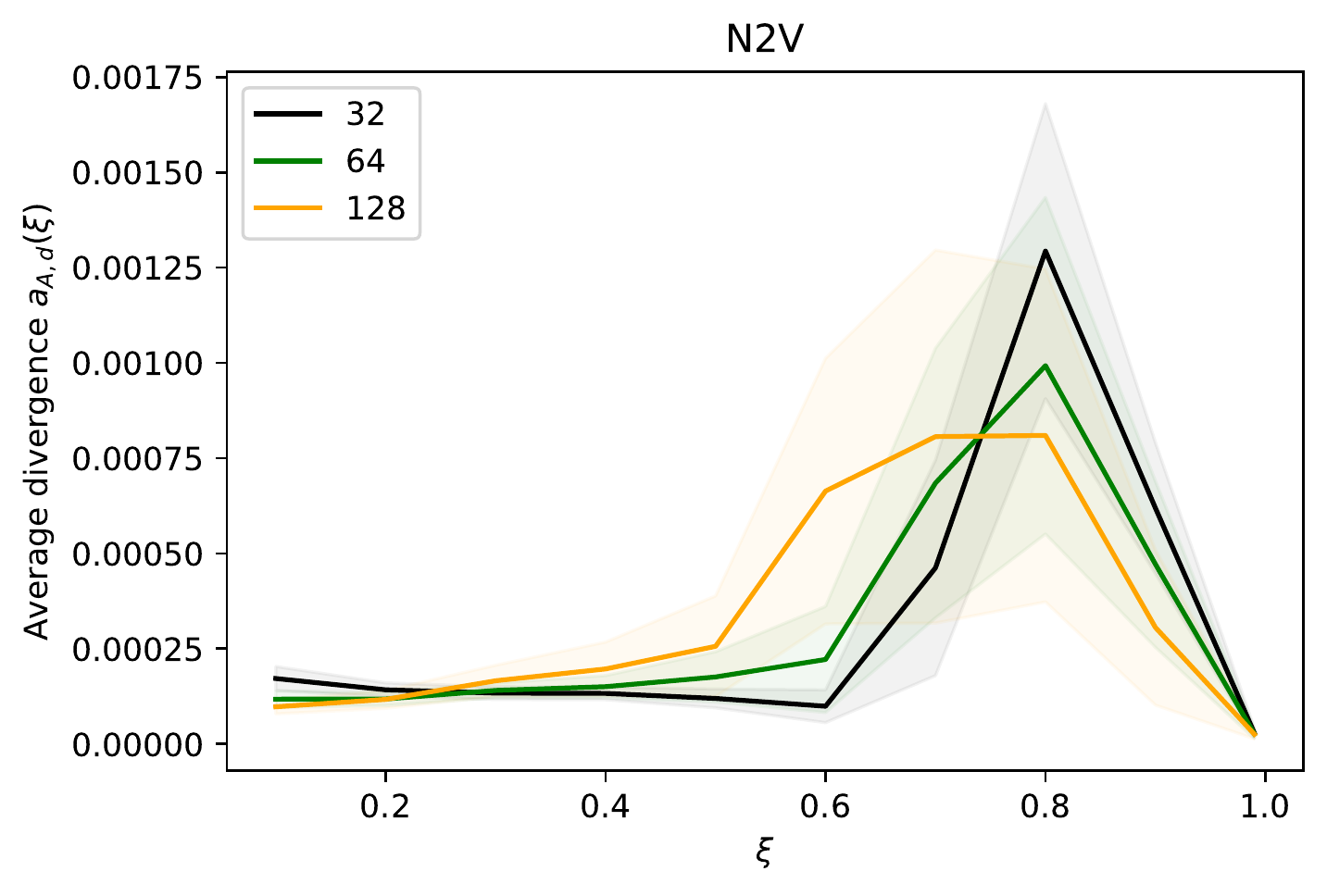}
\includegraphics[width=0.3\textwidth]{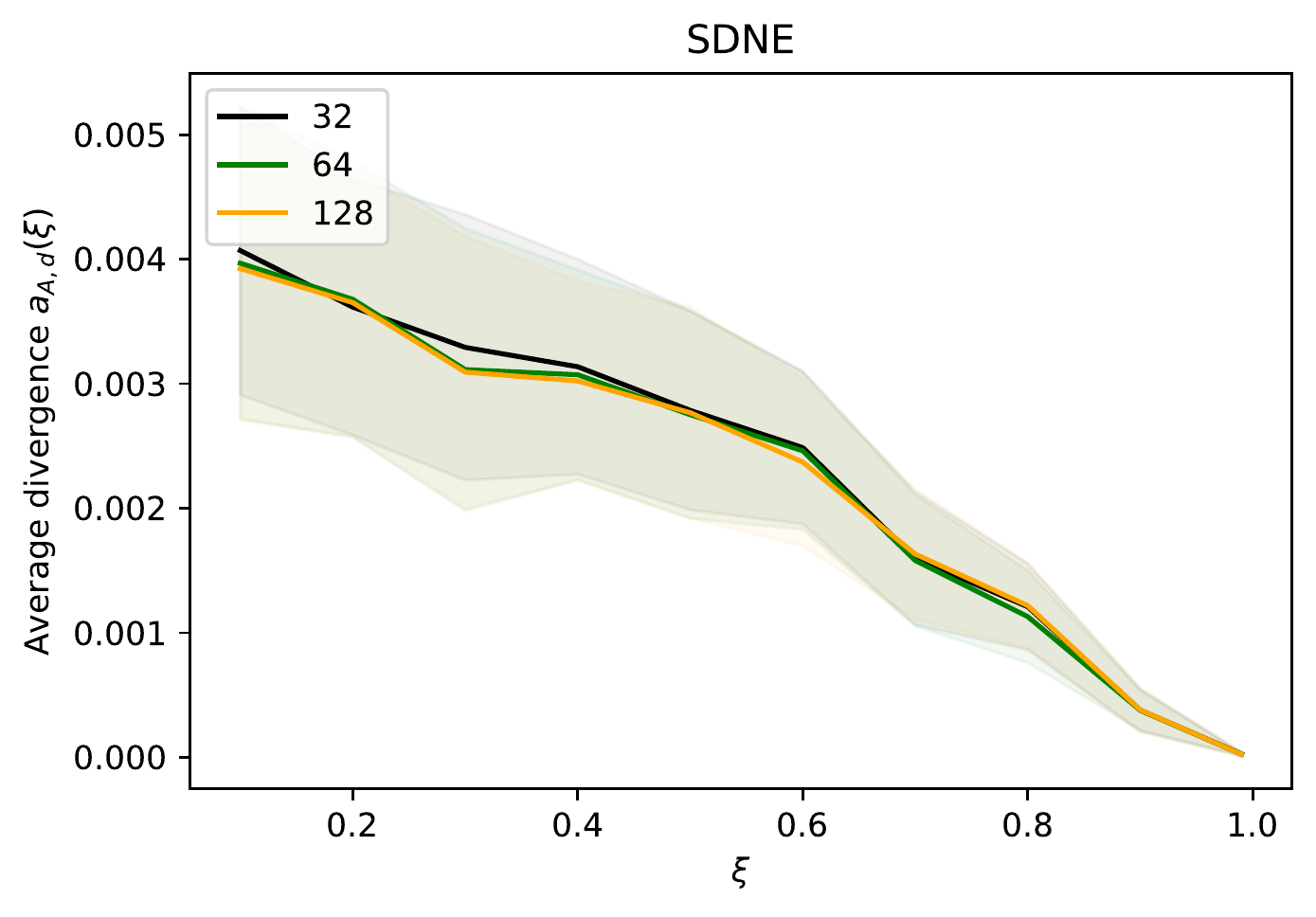} \\
\includegraphics[width=0.3\textwidth]{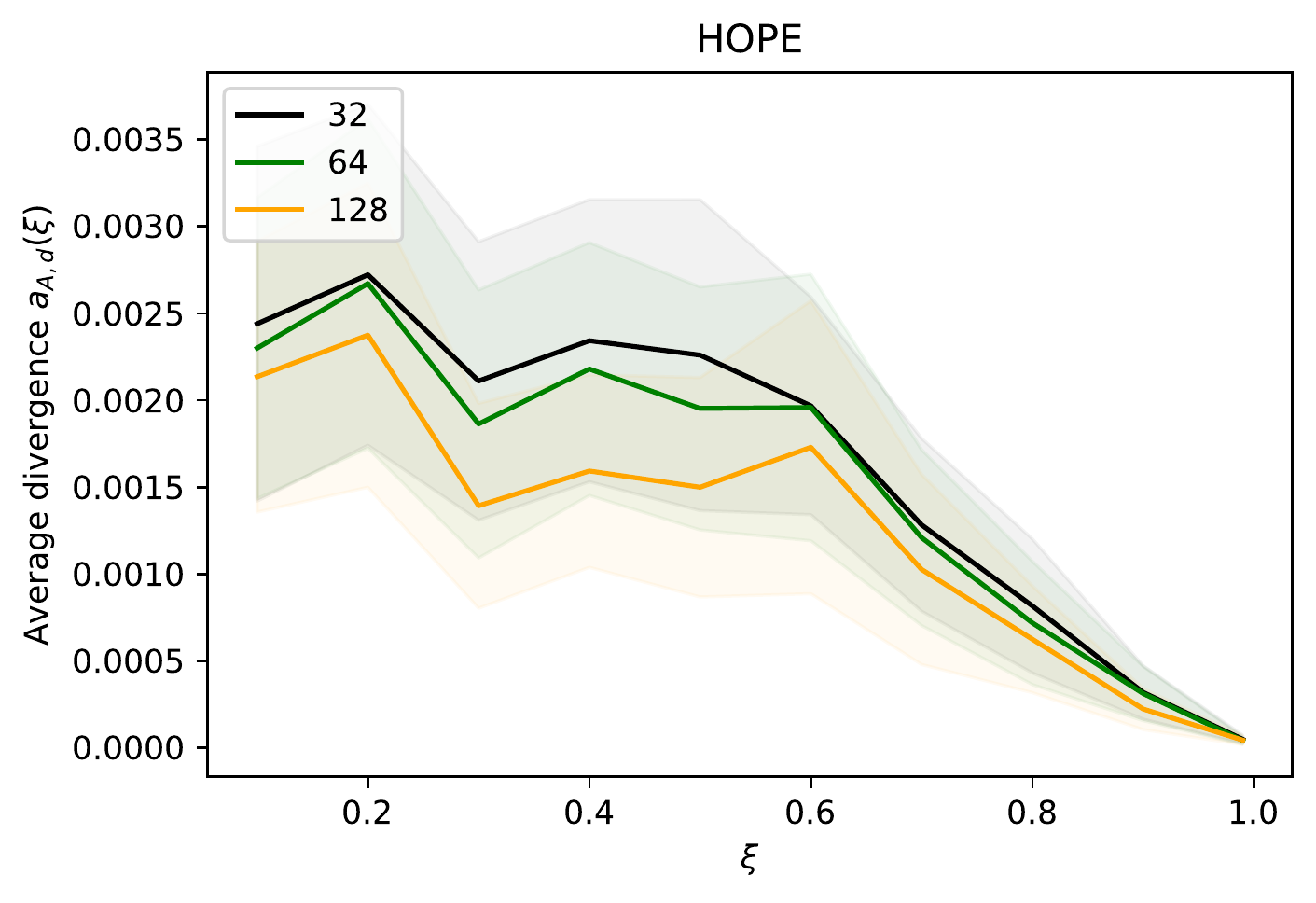}
\includegraphics[width=0.3\textwidth]{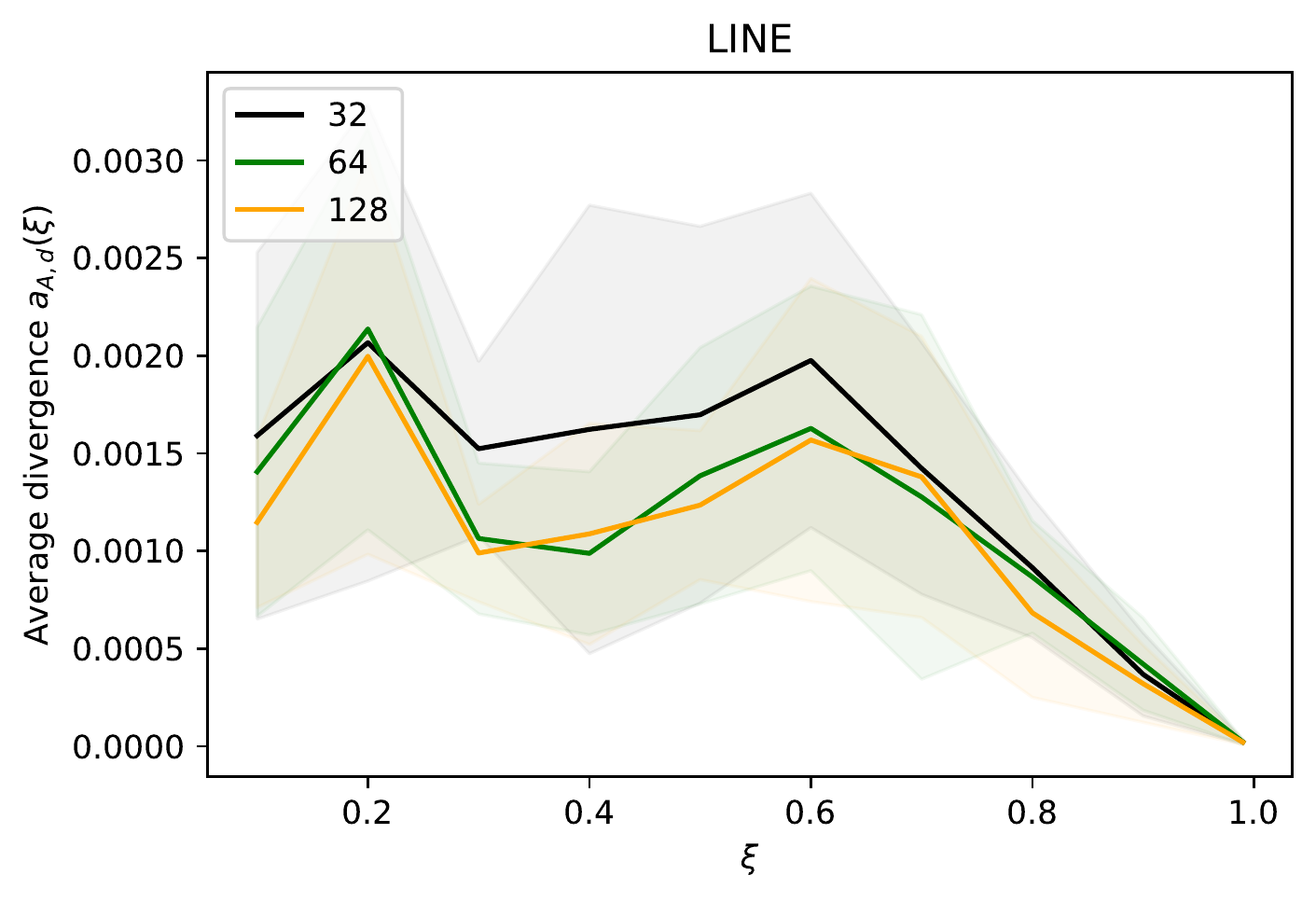} 
\includegraphics[width=0.3\textwidth]{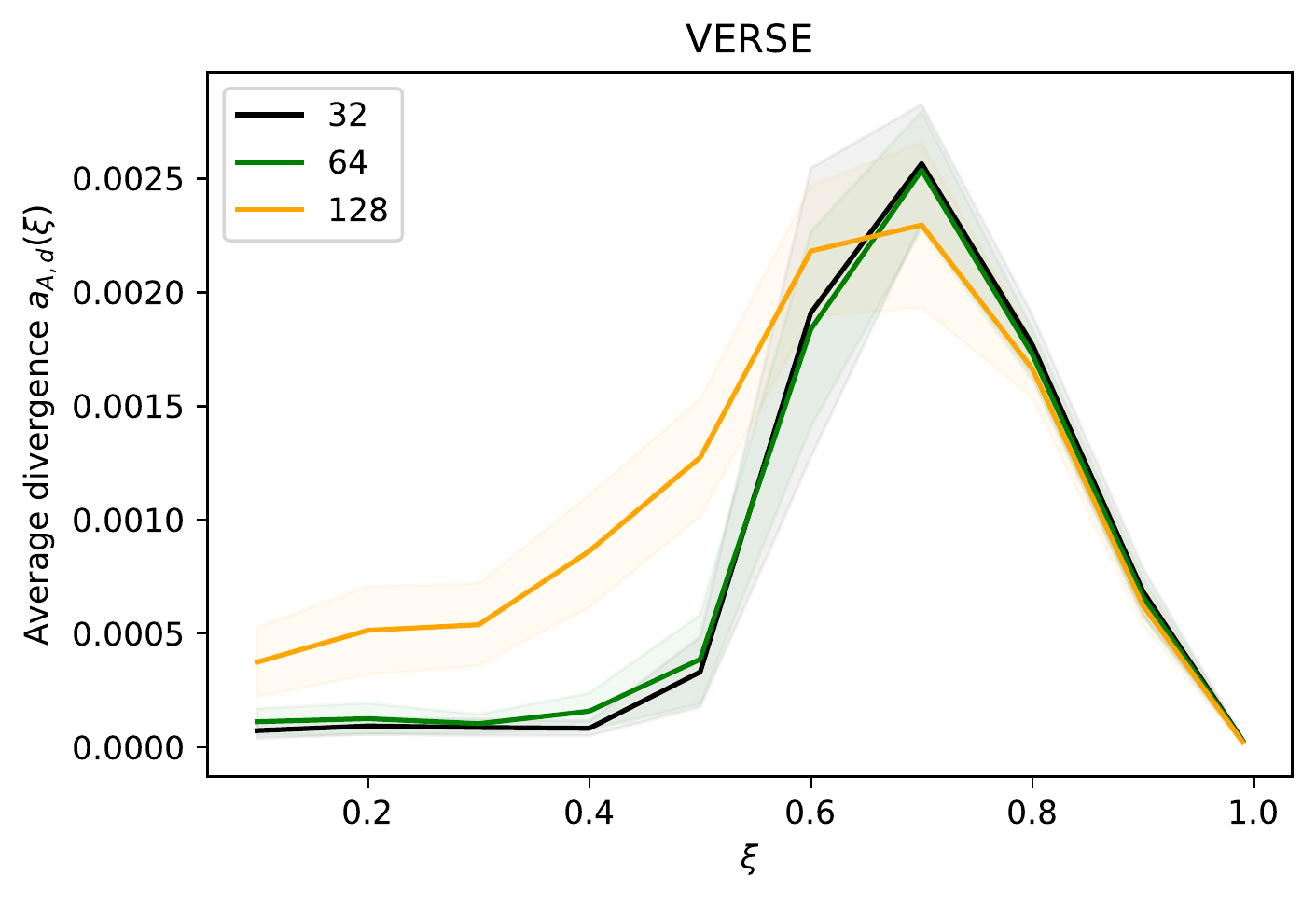} \\
\includegraphics[width=0.3\textwidth]{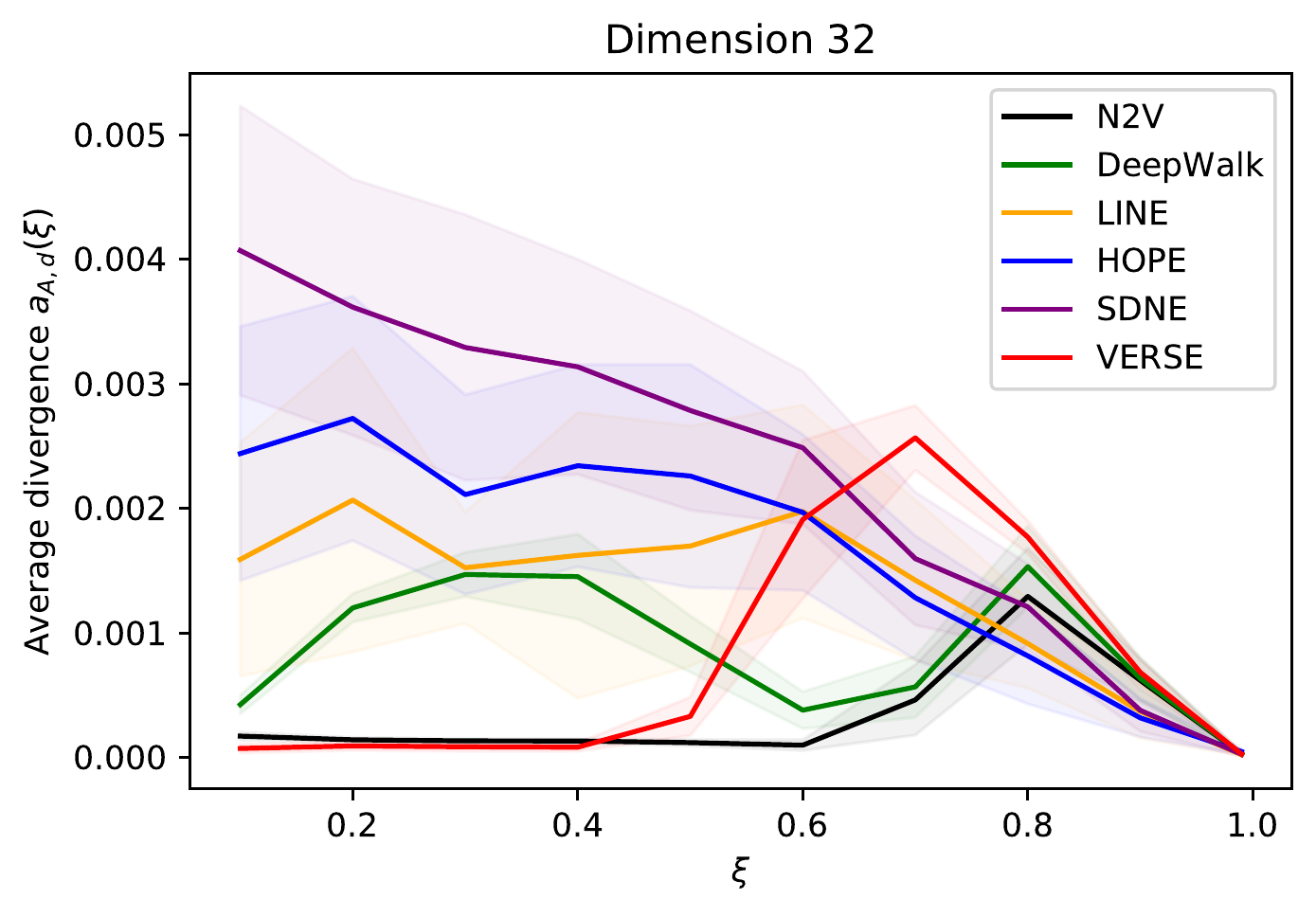}
\includegraphics[width=0.3\textwidth]{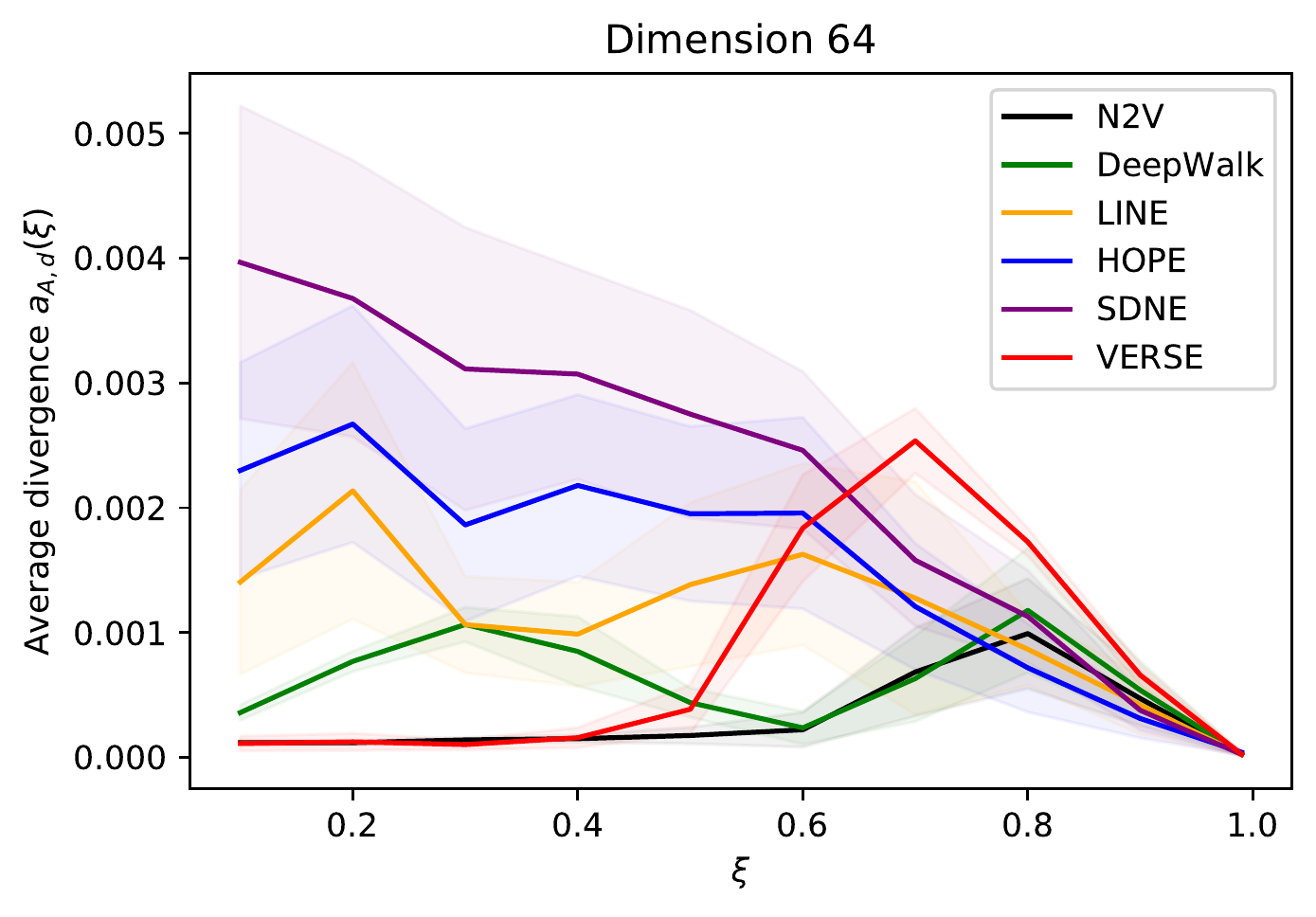}
\includegraphics[width=0.3\textwidth]{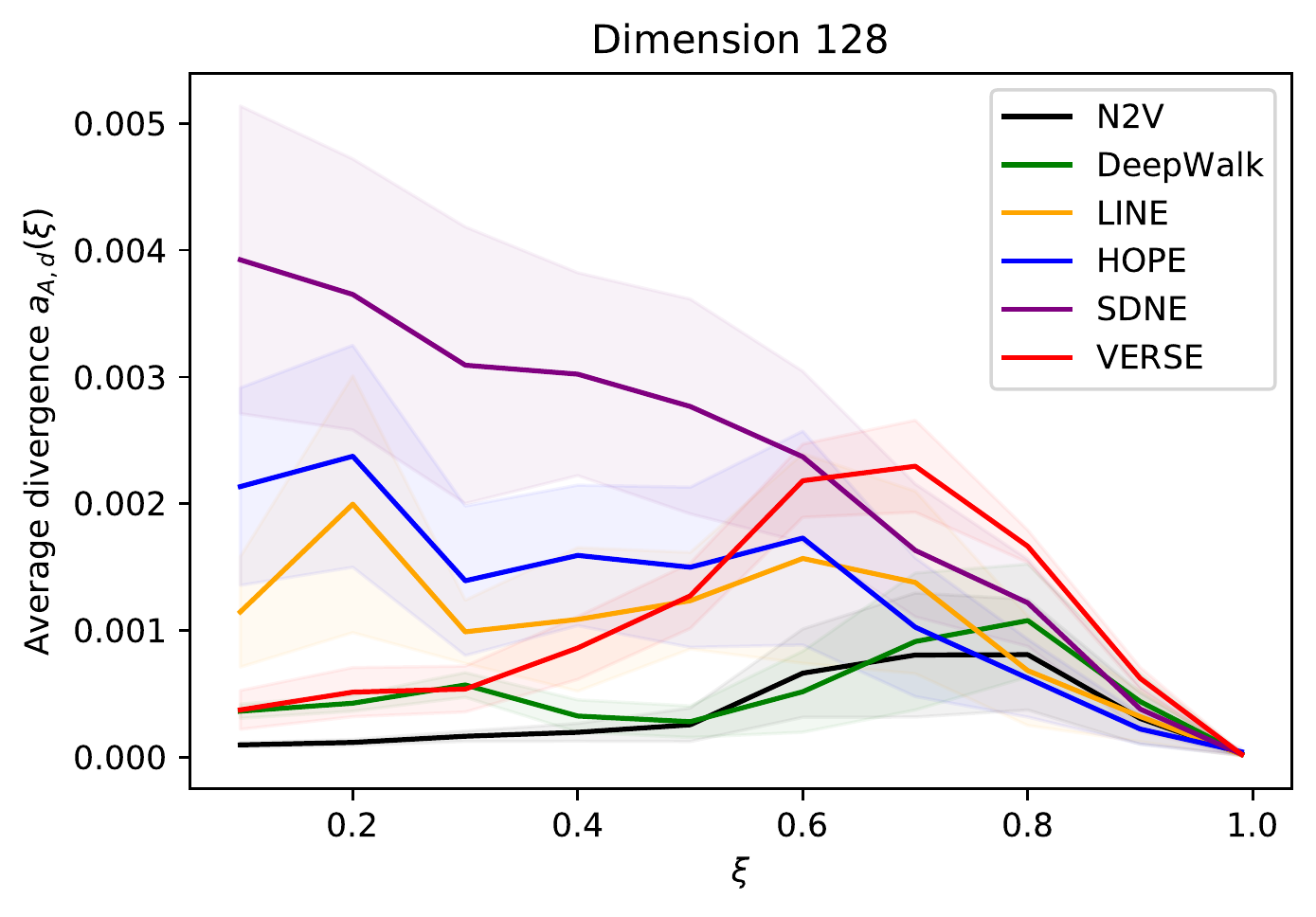}\\
Plots 1 and  2: $a_{A,d}(\xi) \pm s_{A,d}(\xi)$

\includegraphics[width=0.2\textwidth]{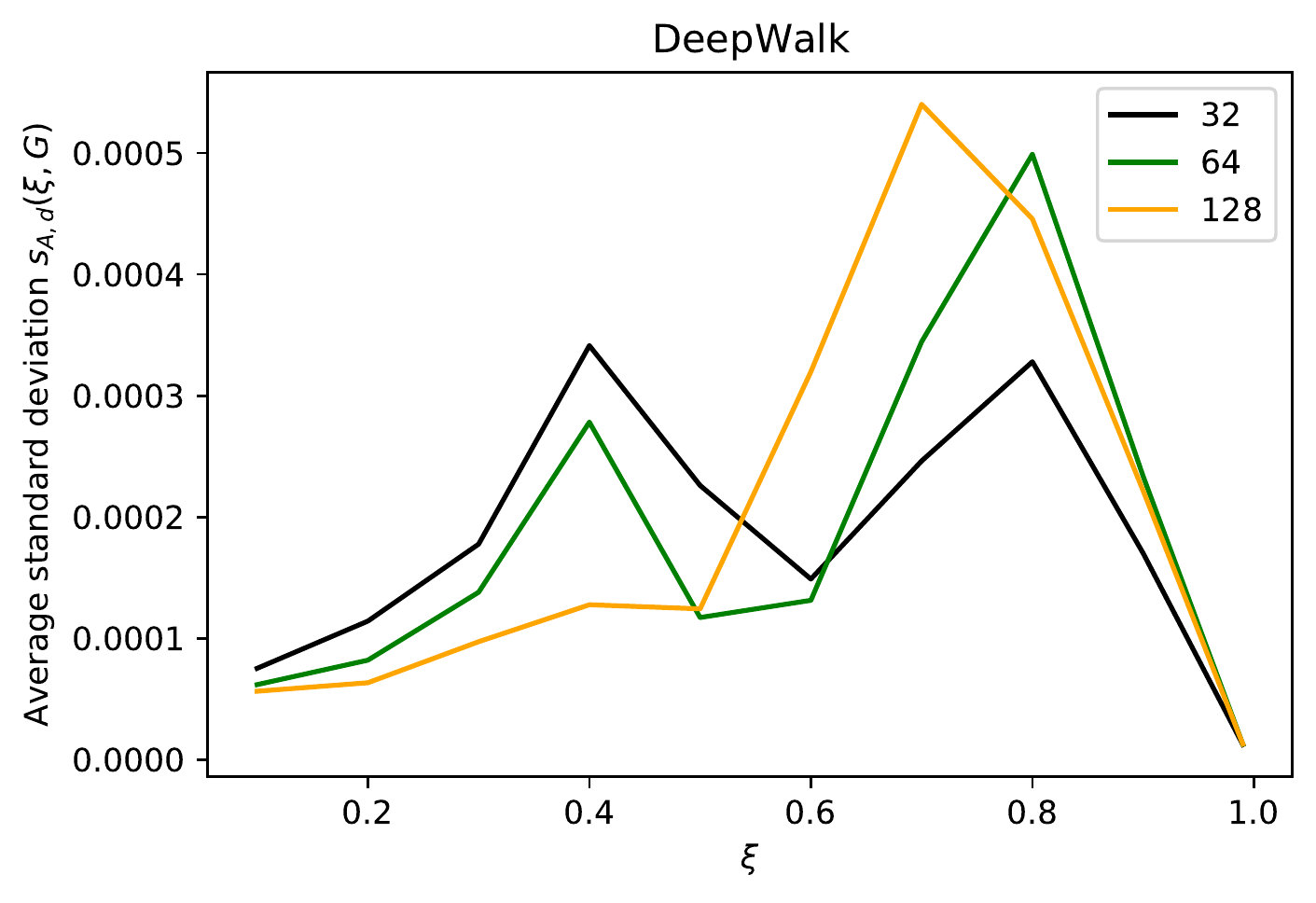} 
\includegraphics[width=0.2\textwidth]{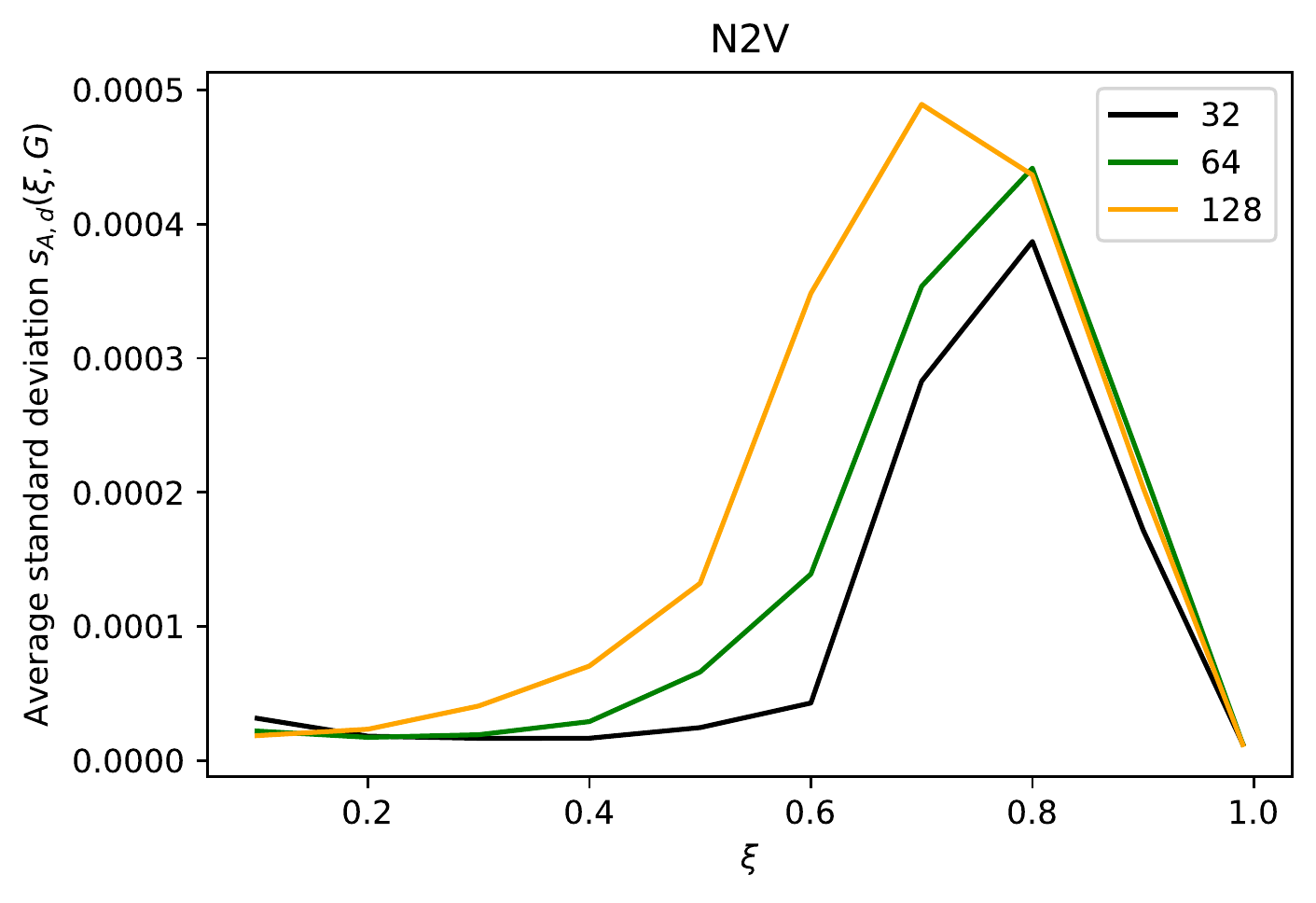}
\includegraphics[width=0.2\textwidth]{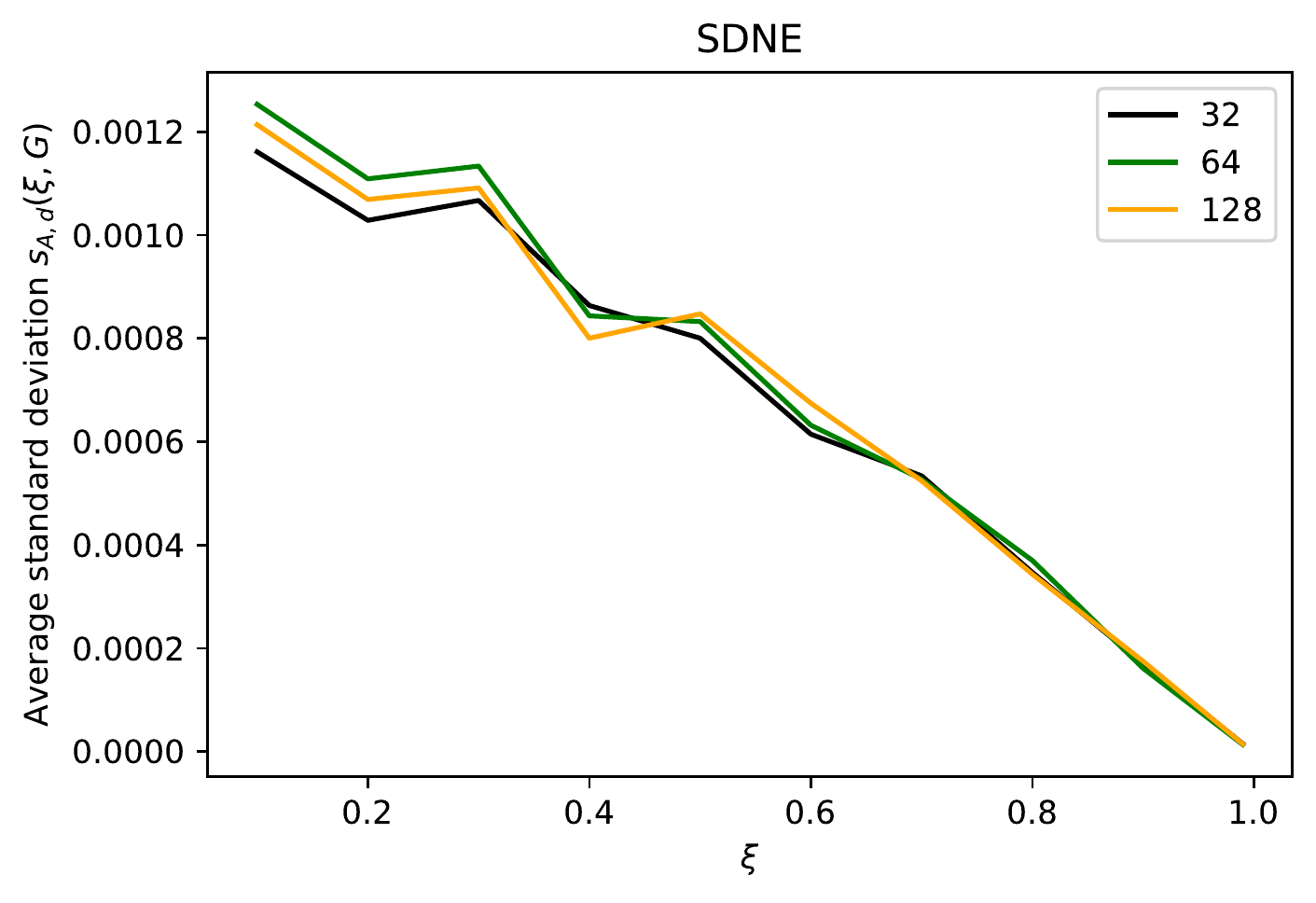}
\includegraphics[width=0.2\textwidth]{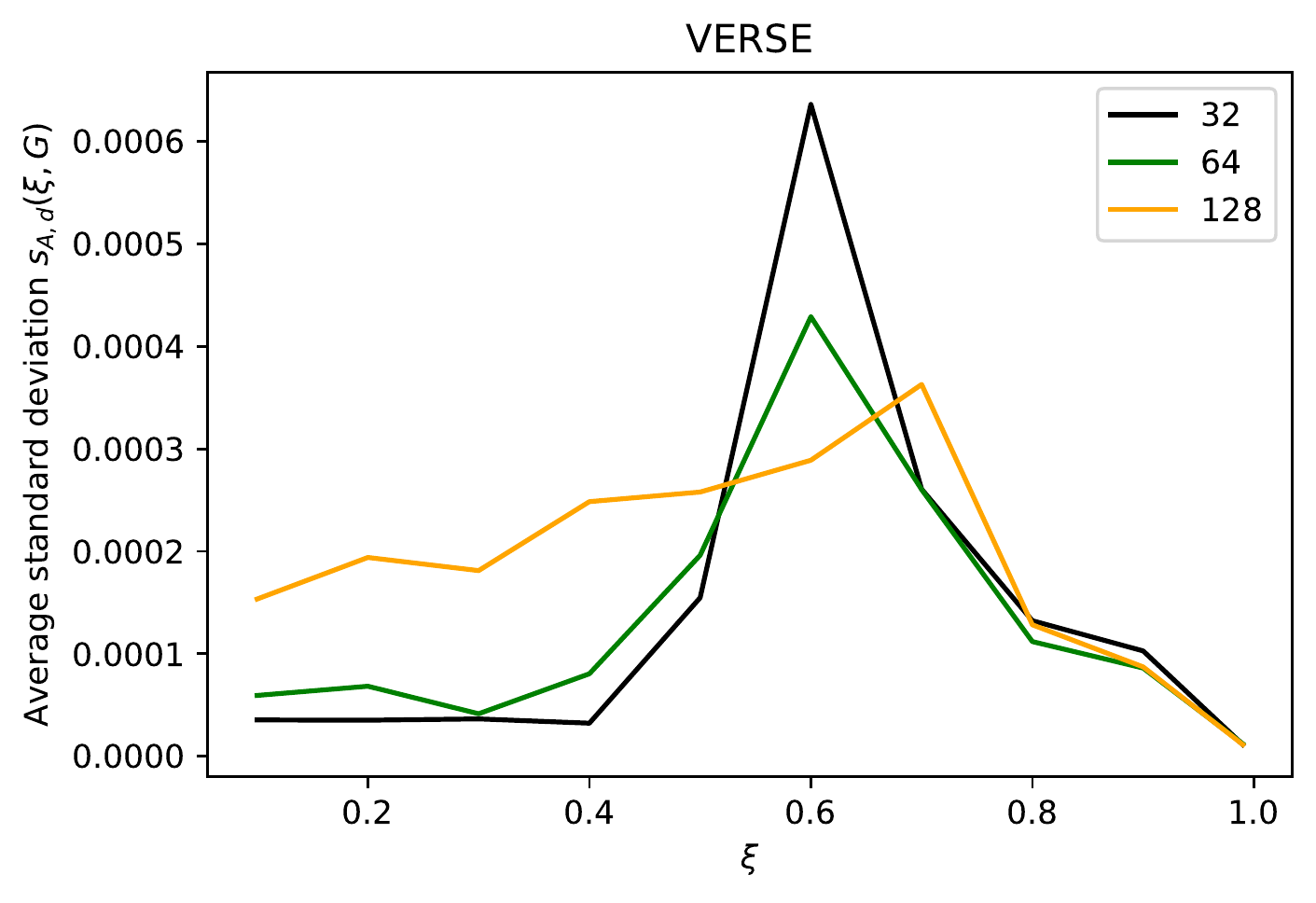} \\
\includegraphics[width=0.3\textwidth]{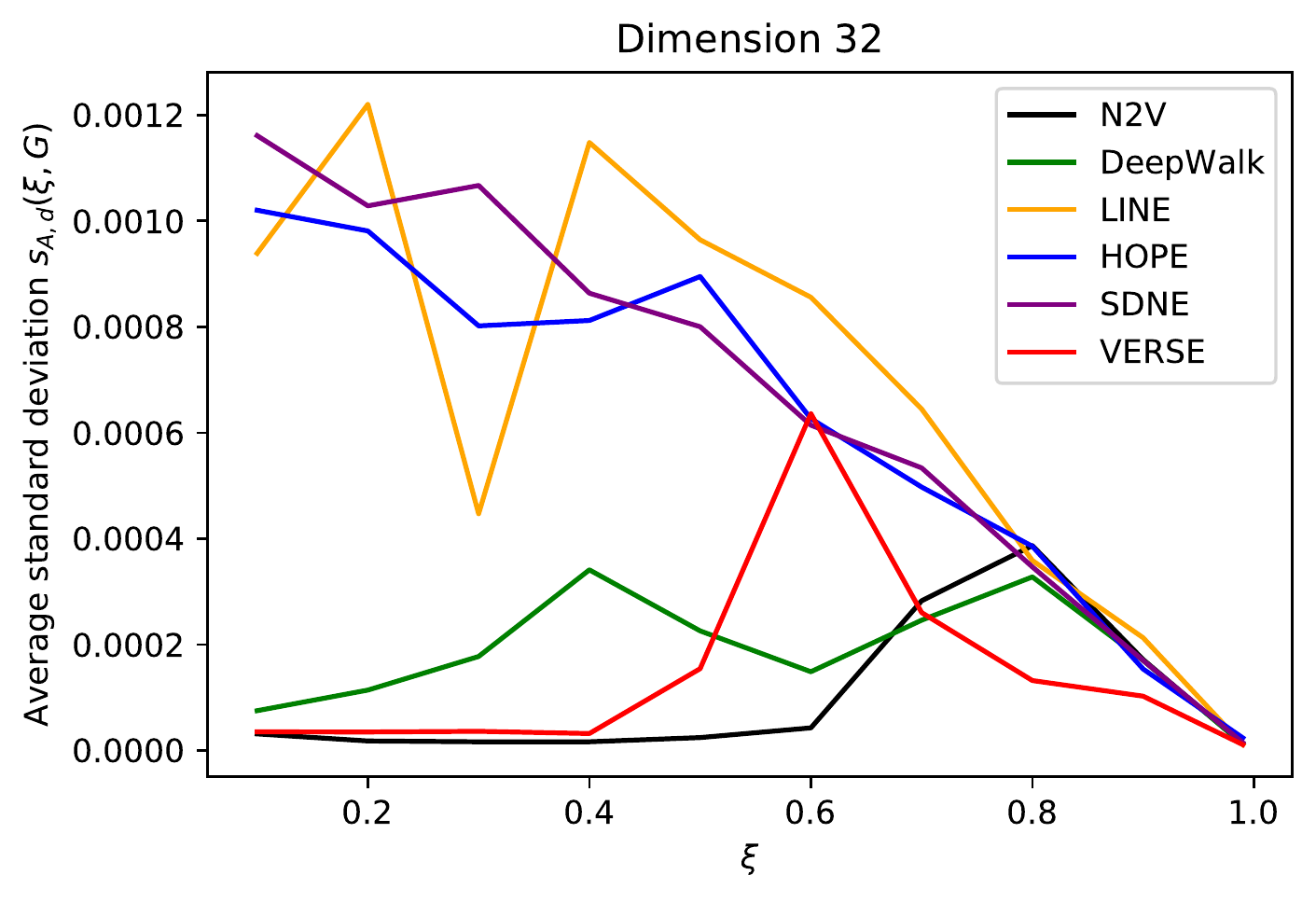}
\includegraphics[width=0.3\textwidth]{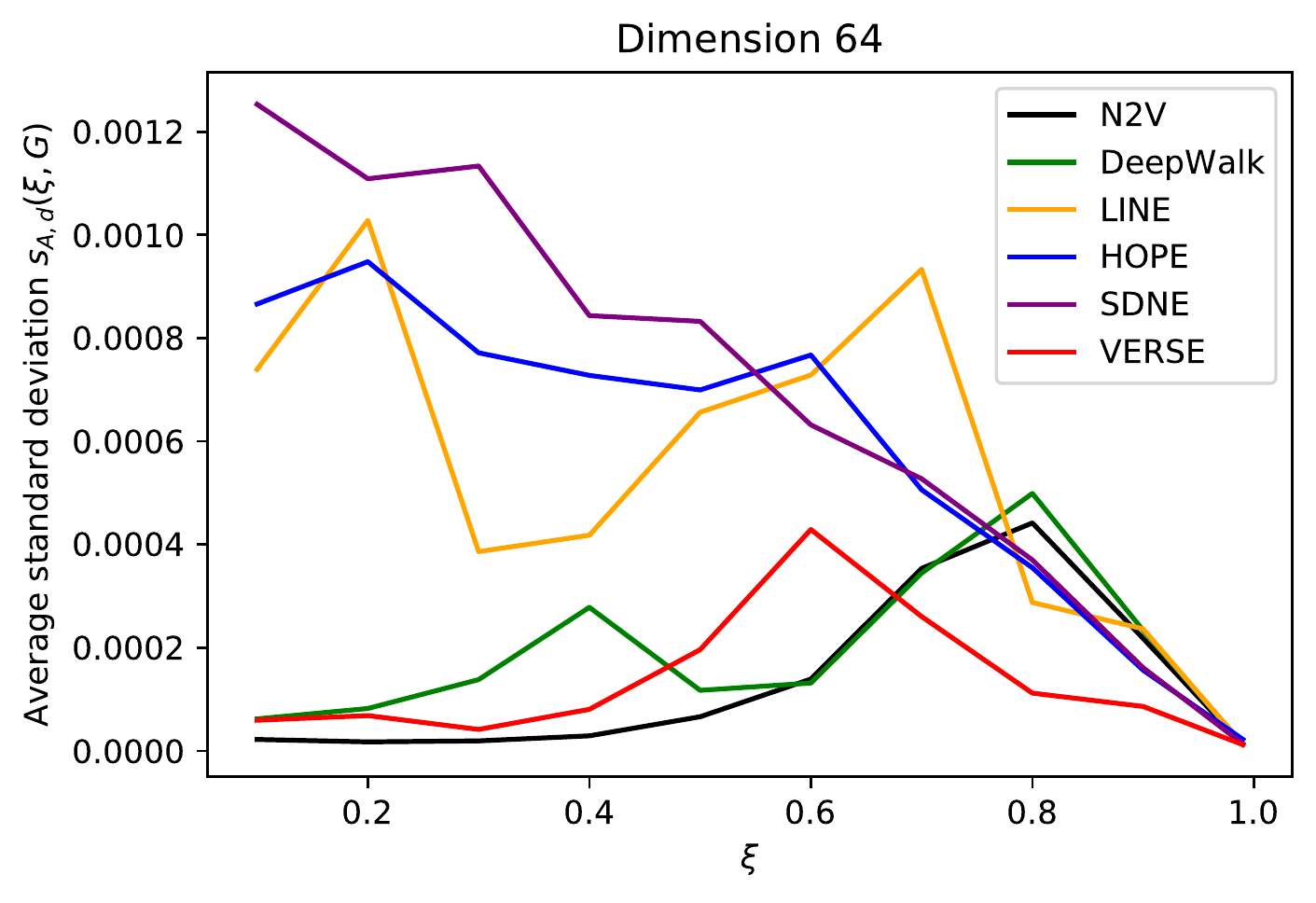}
\includegraphics[width=0.3\textwidth]{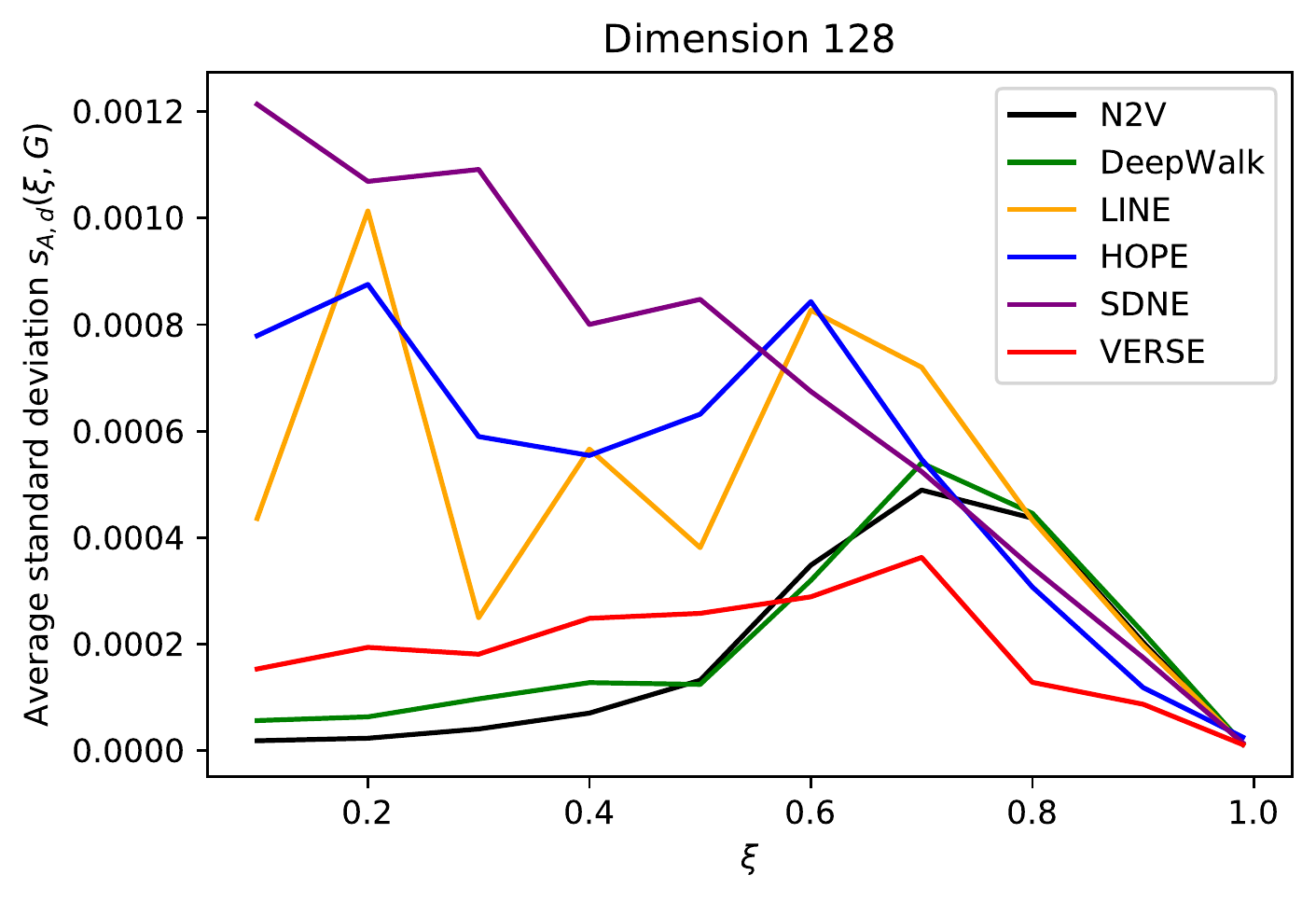} \\
Plots 3 and 4: average $s_{A,d}(\xi, G)$ (over 10 graphs)

\includegraphics[width=0.2\textwidth]{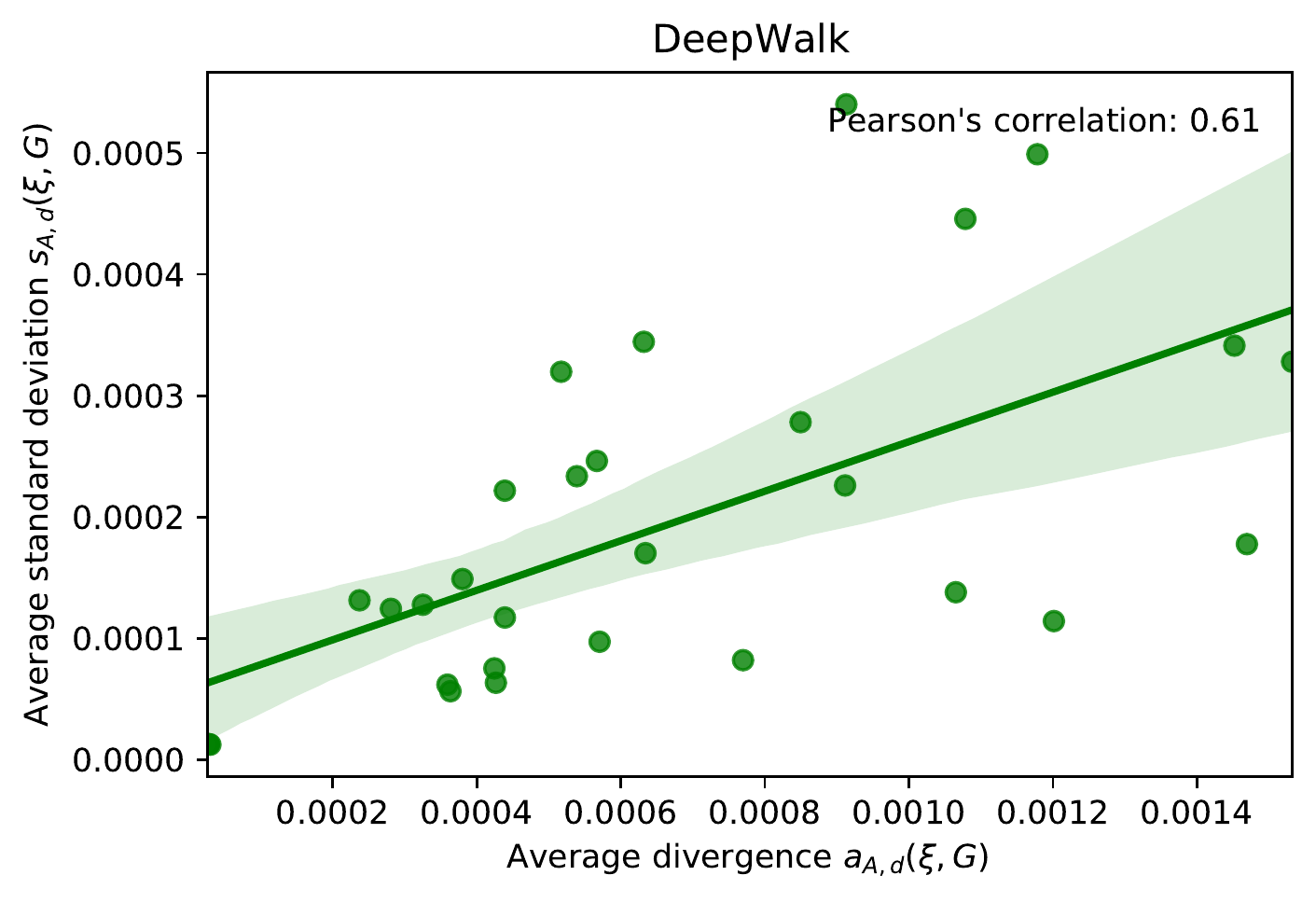}
\includegraphics[width=0.2\textwidth]{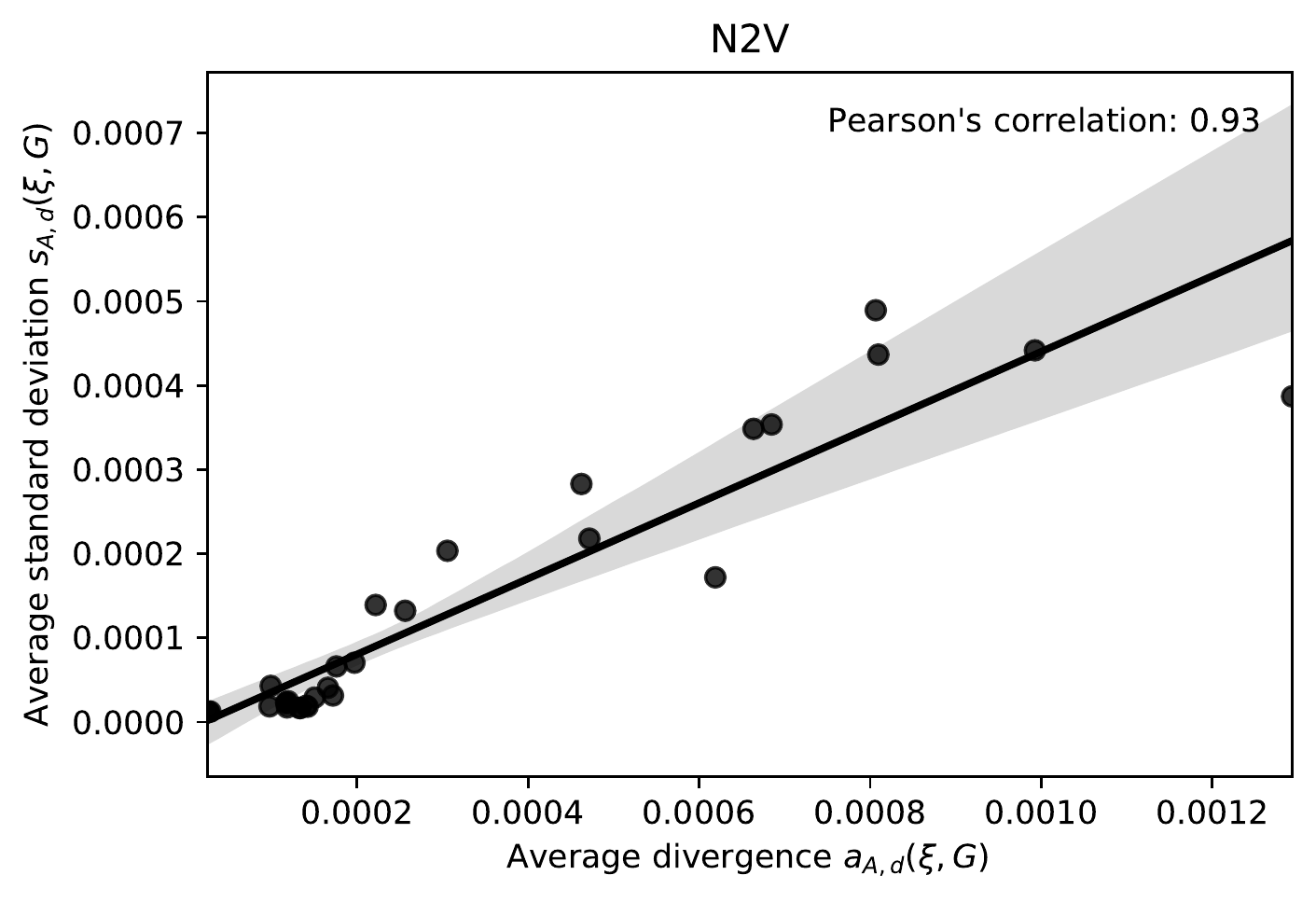}
\includegraphics[width=0.2\textwidth]{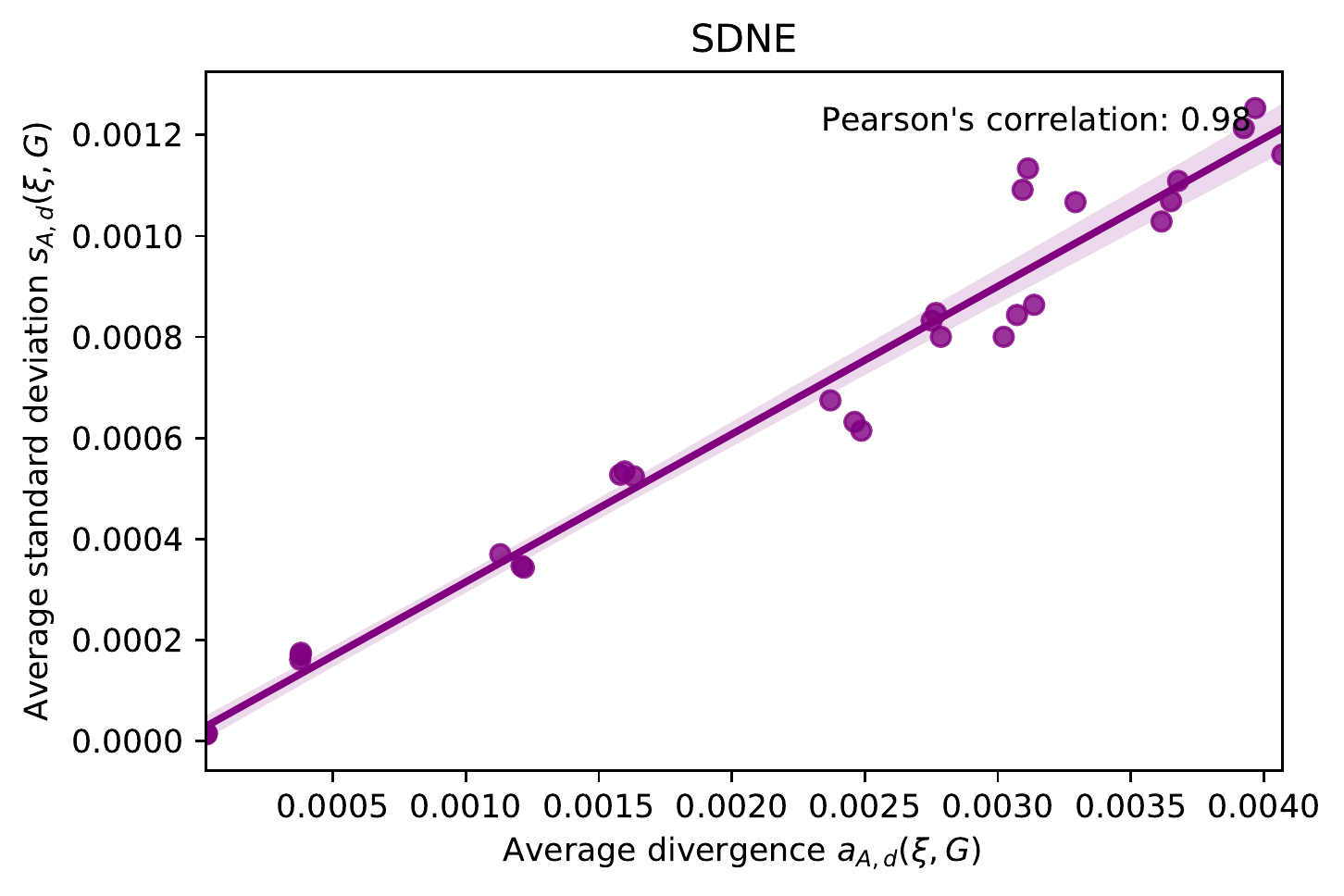}
\includegraphics[width=0.2\textwidth]{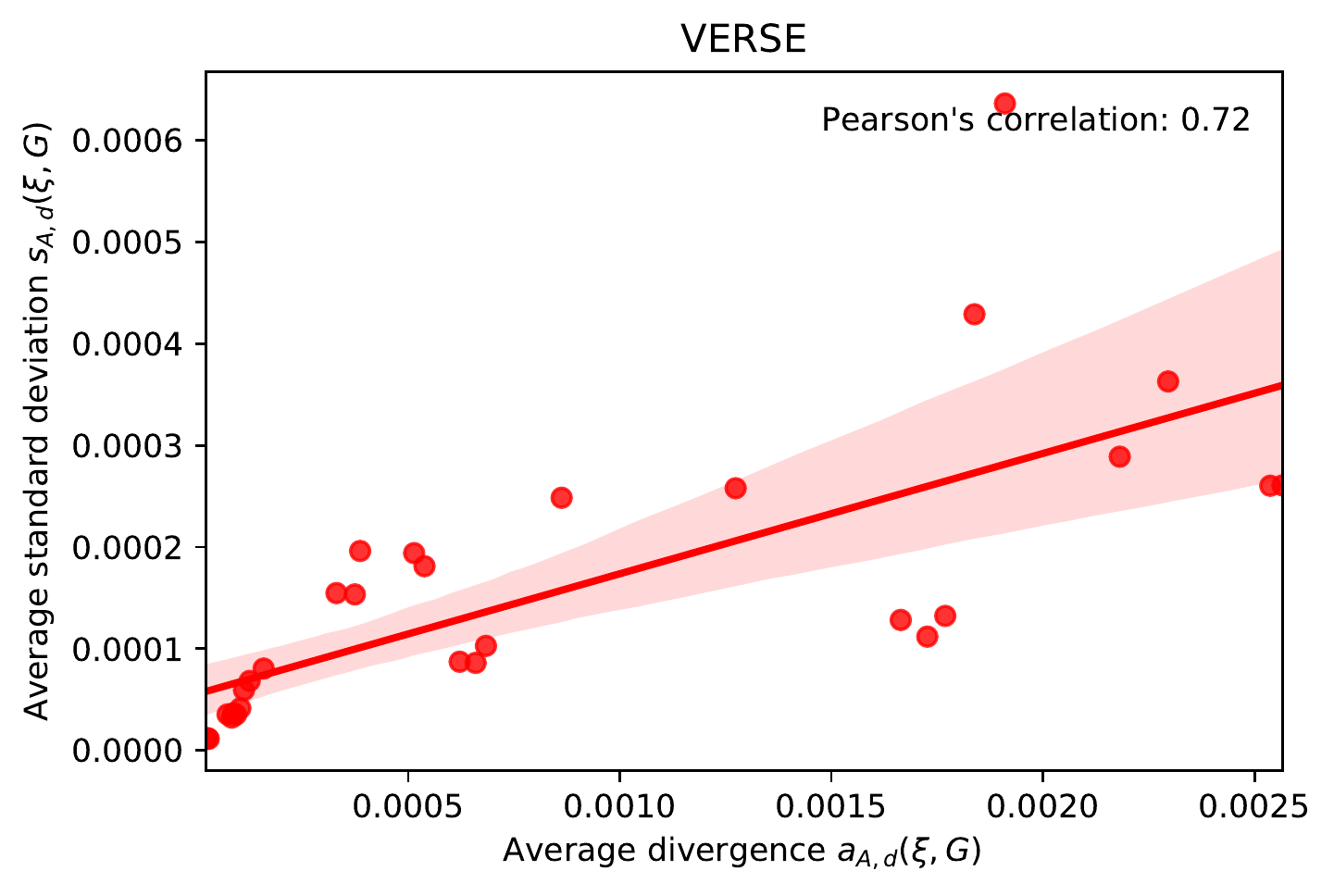} \\
Plot 5: correlation between $a_{A,d}(\xi, G)$ and $s_{A,d}(\xi, G)$
\caption{Level of Noise ($\xi$)}\label{fig:xi}
\end{center}
\end{figure}

\begin{figure}[htbp!]
\begin{center}
\includegraphics[width=0.3\textwidth]{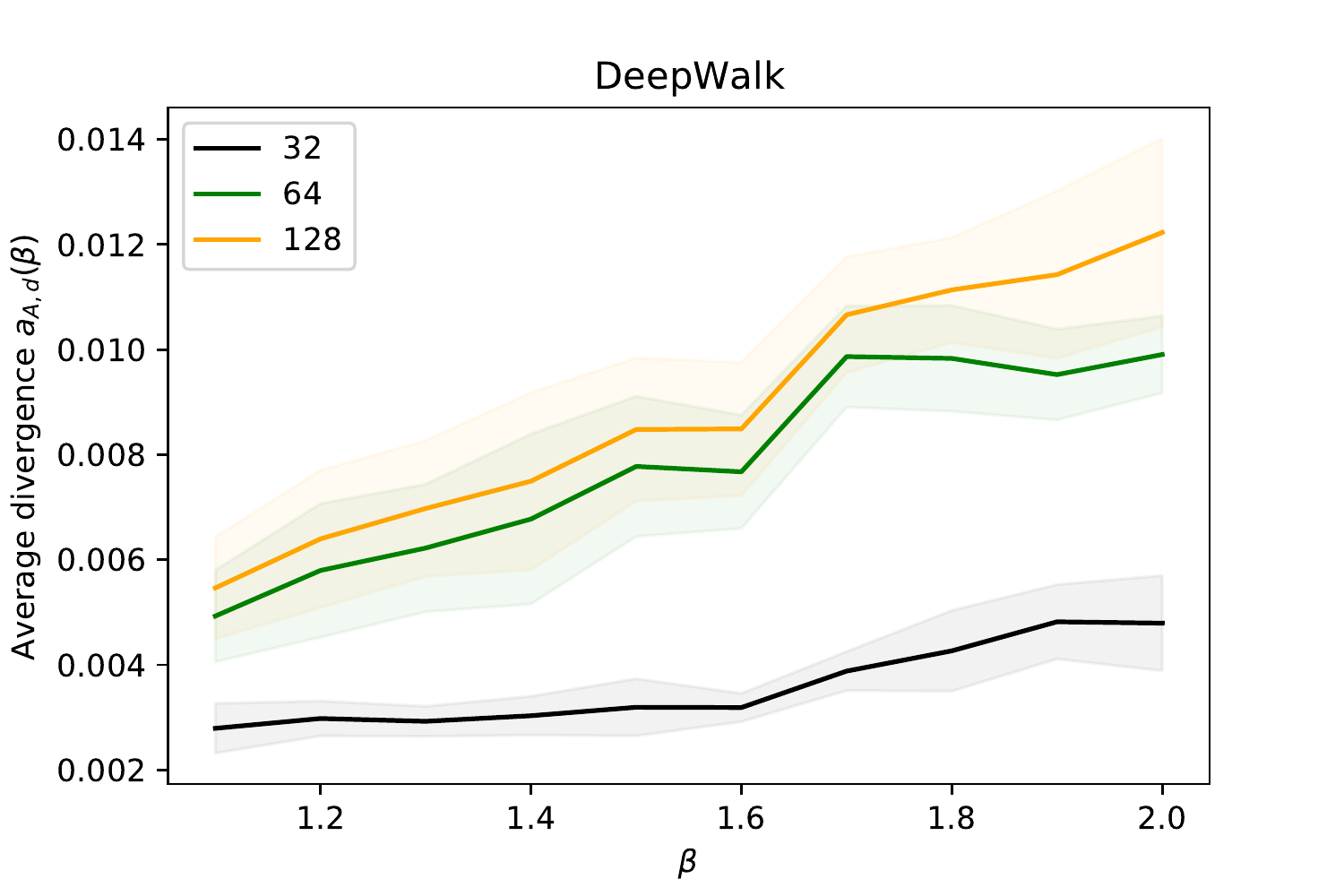}
\includegraphics[width=0.3\textwidth]{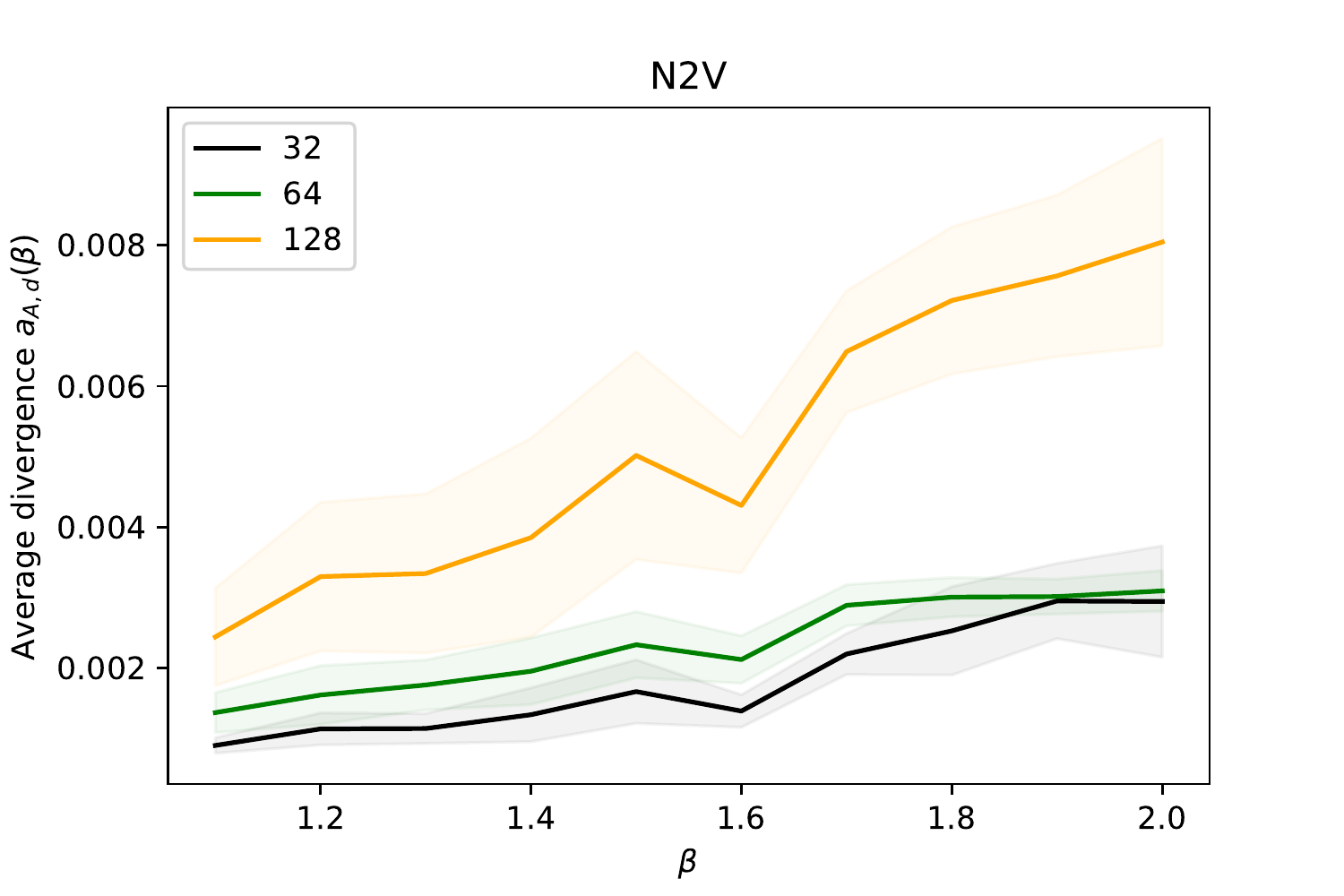}
\includegraphics[width=0.3\textwidth]{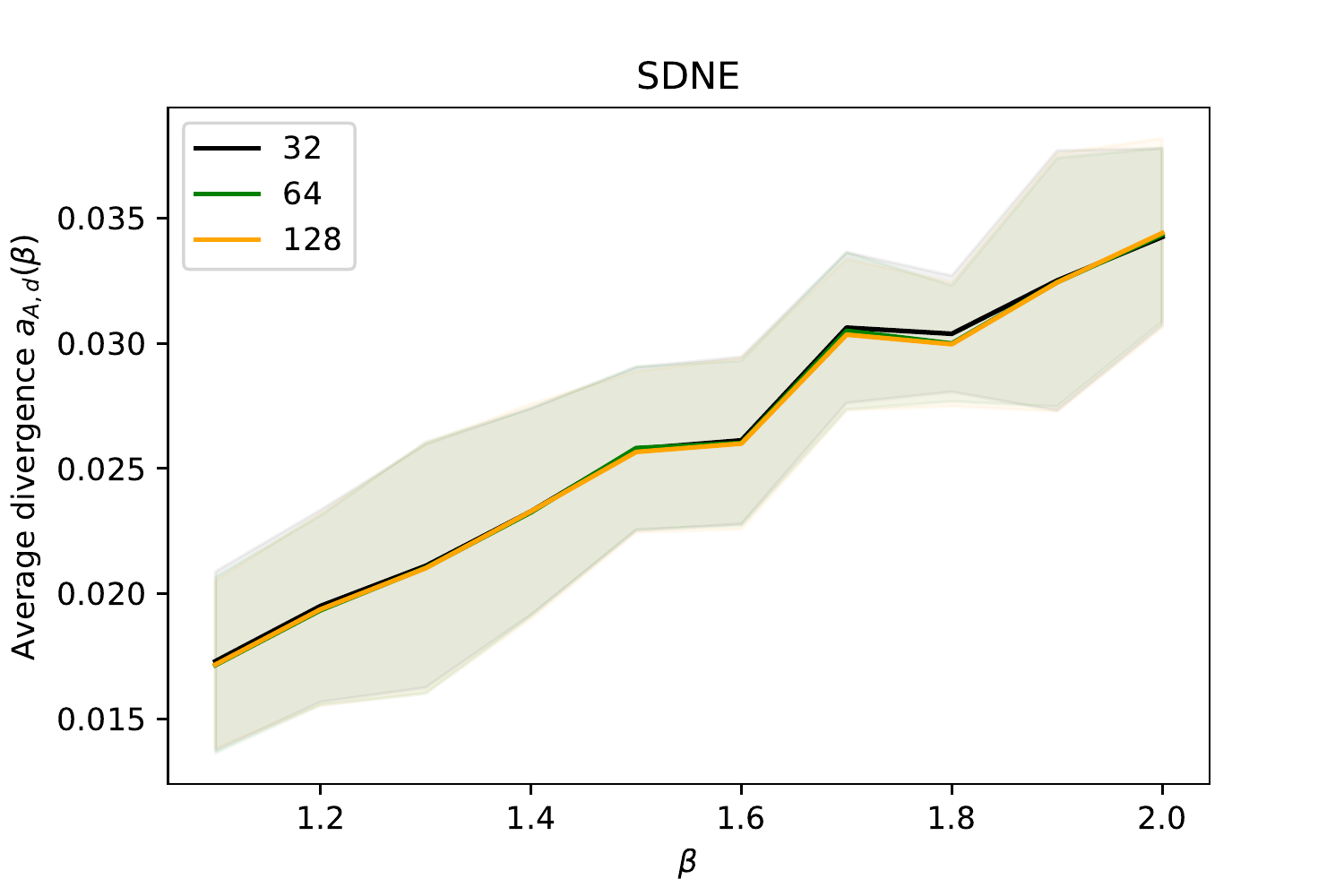} \\
\includegraphics[width=0.3\textwidth]{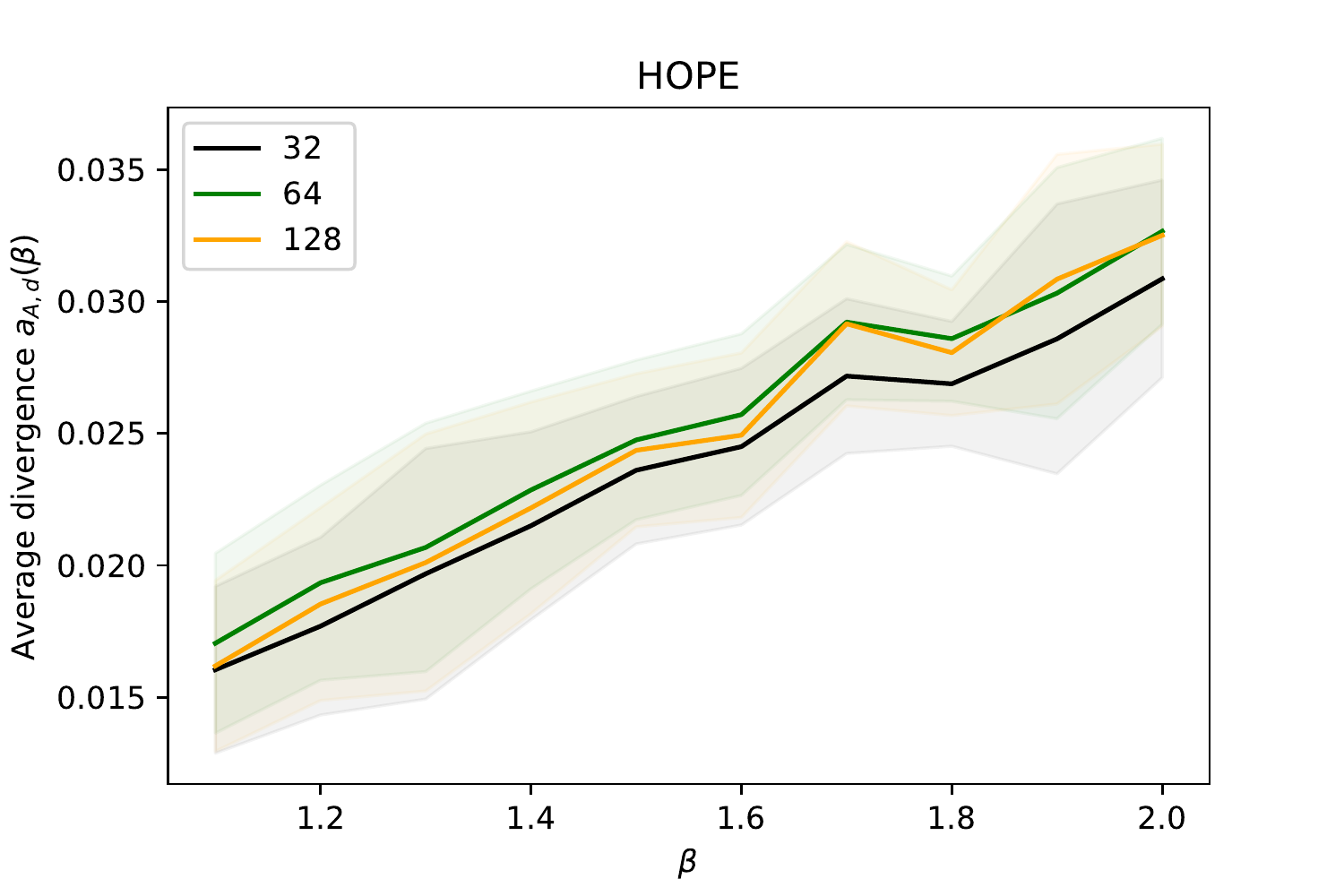}
\includegraphics[width=0.3\textwidth]{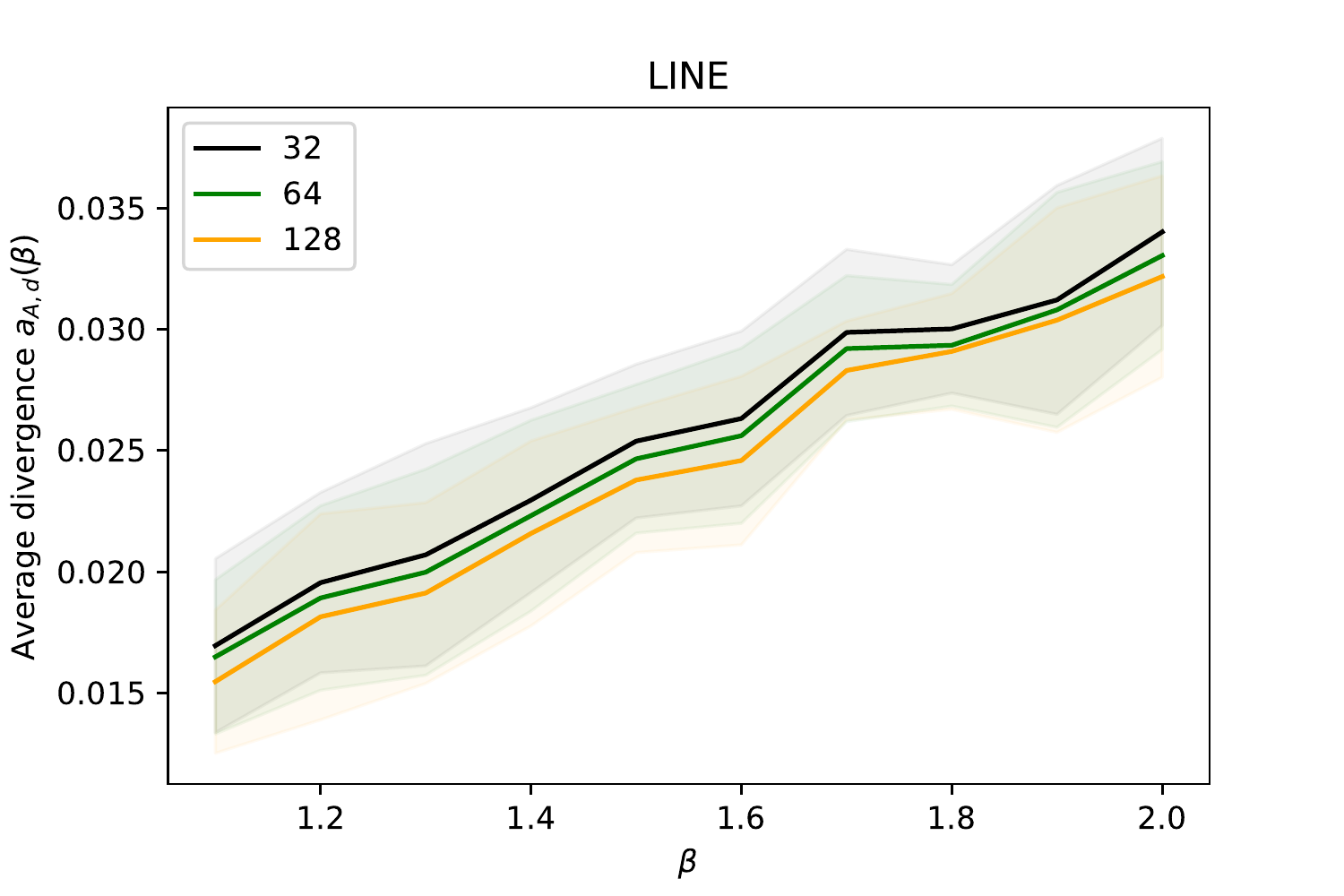} 
\includegraphics[width=0.3\textwidth]{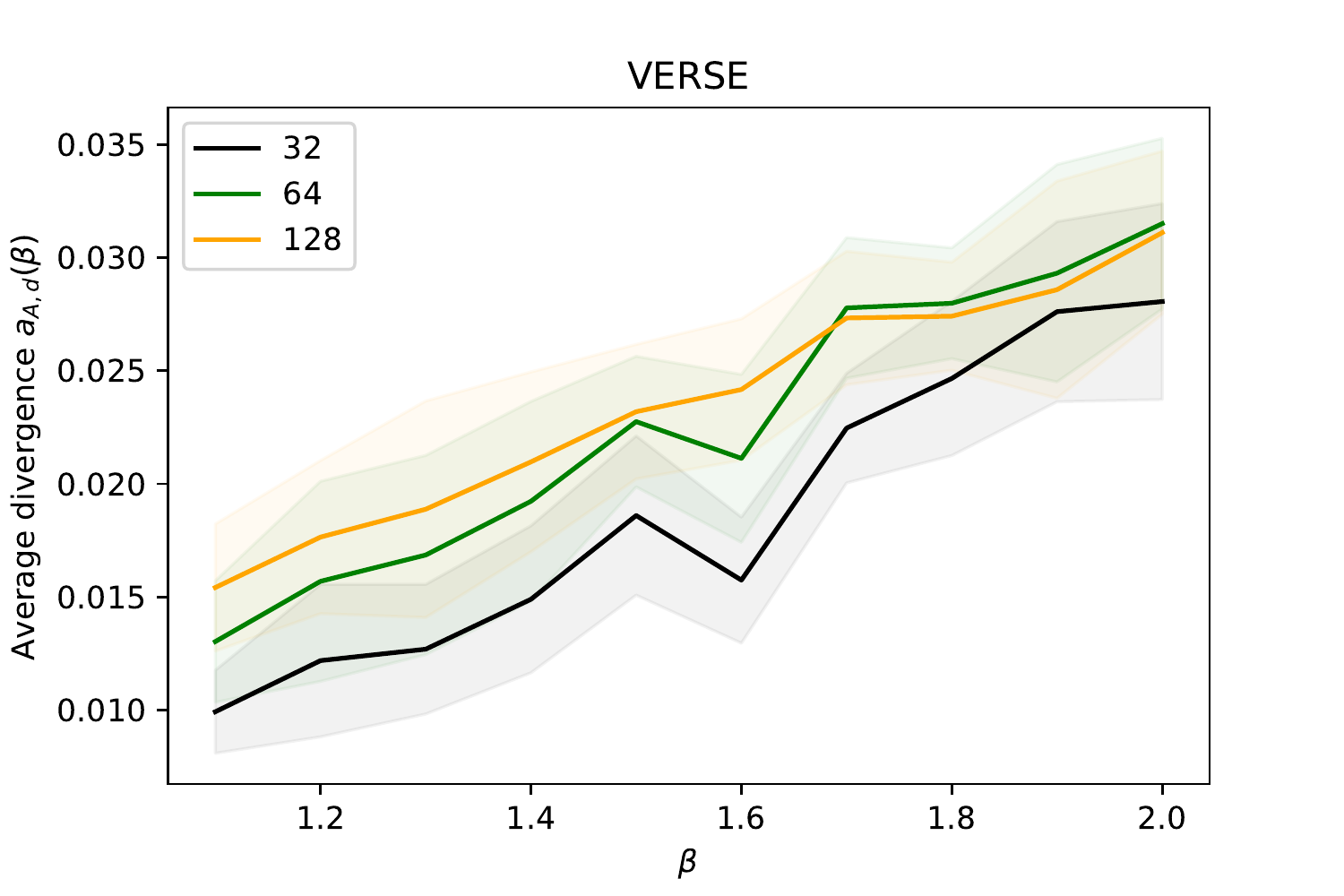} \\
\includegraphics[width=0.3\textwidth]{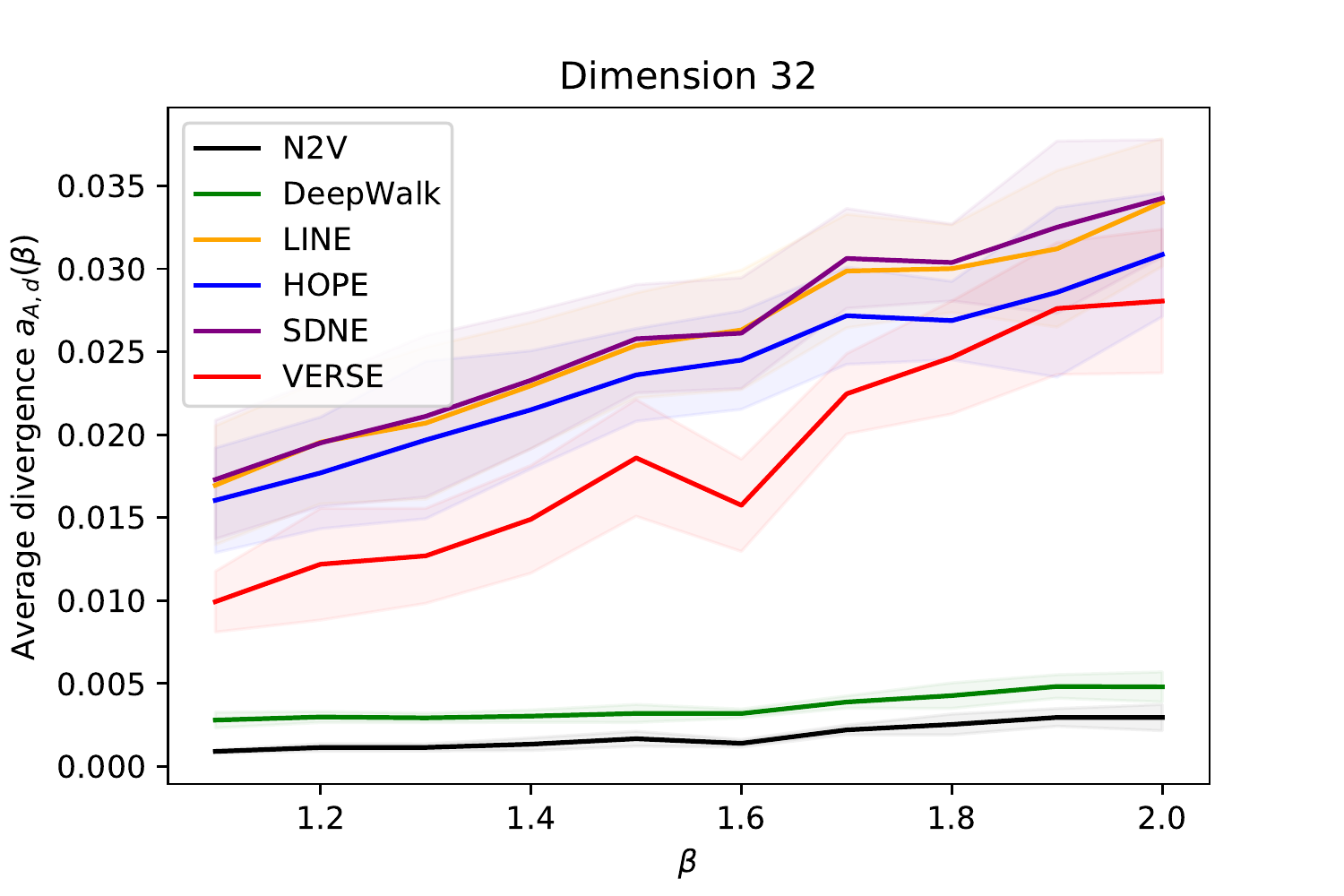}
\includegraphics[width=0.3\textwidth]{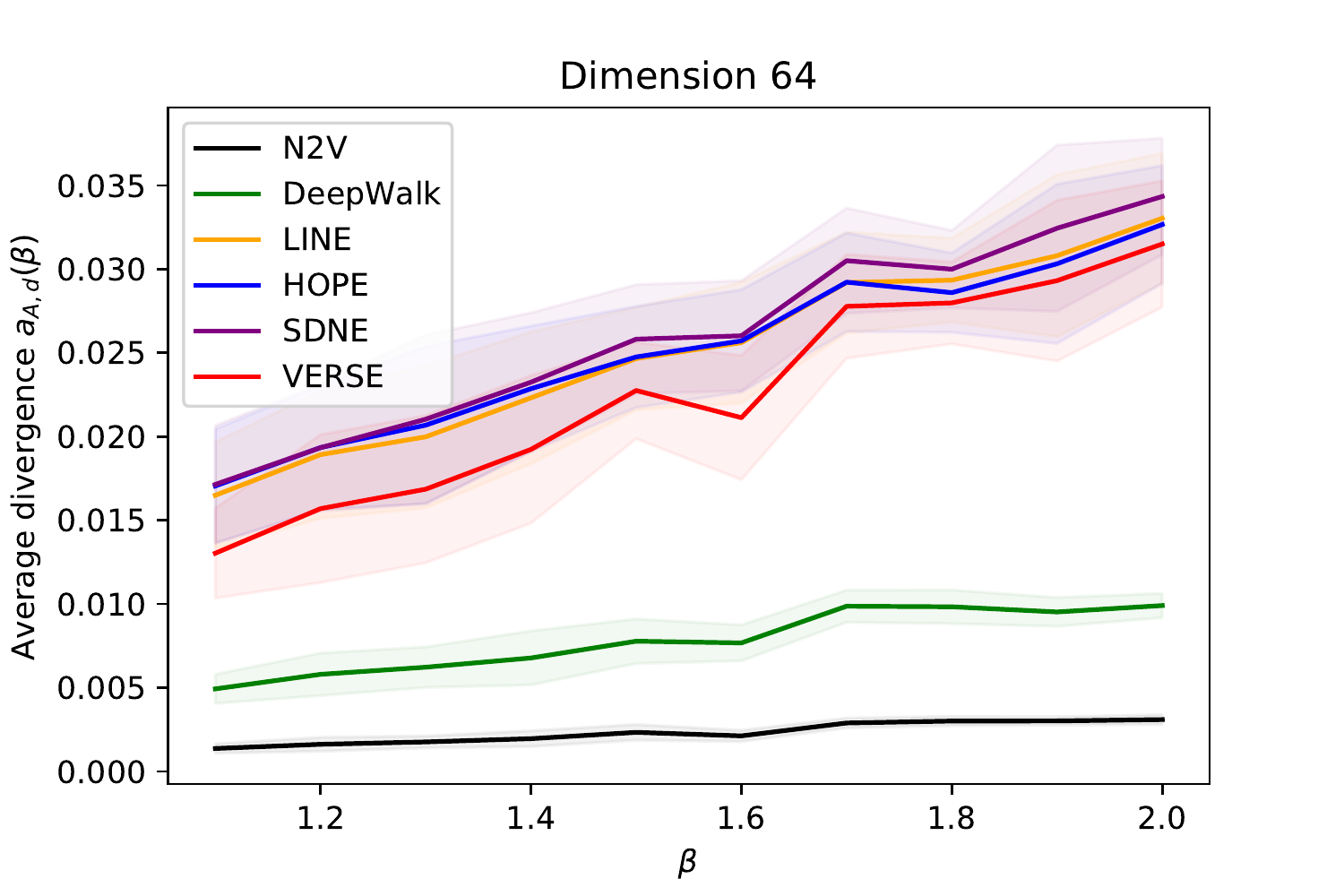}
\includegraphics[width=0.3\textwidth]{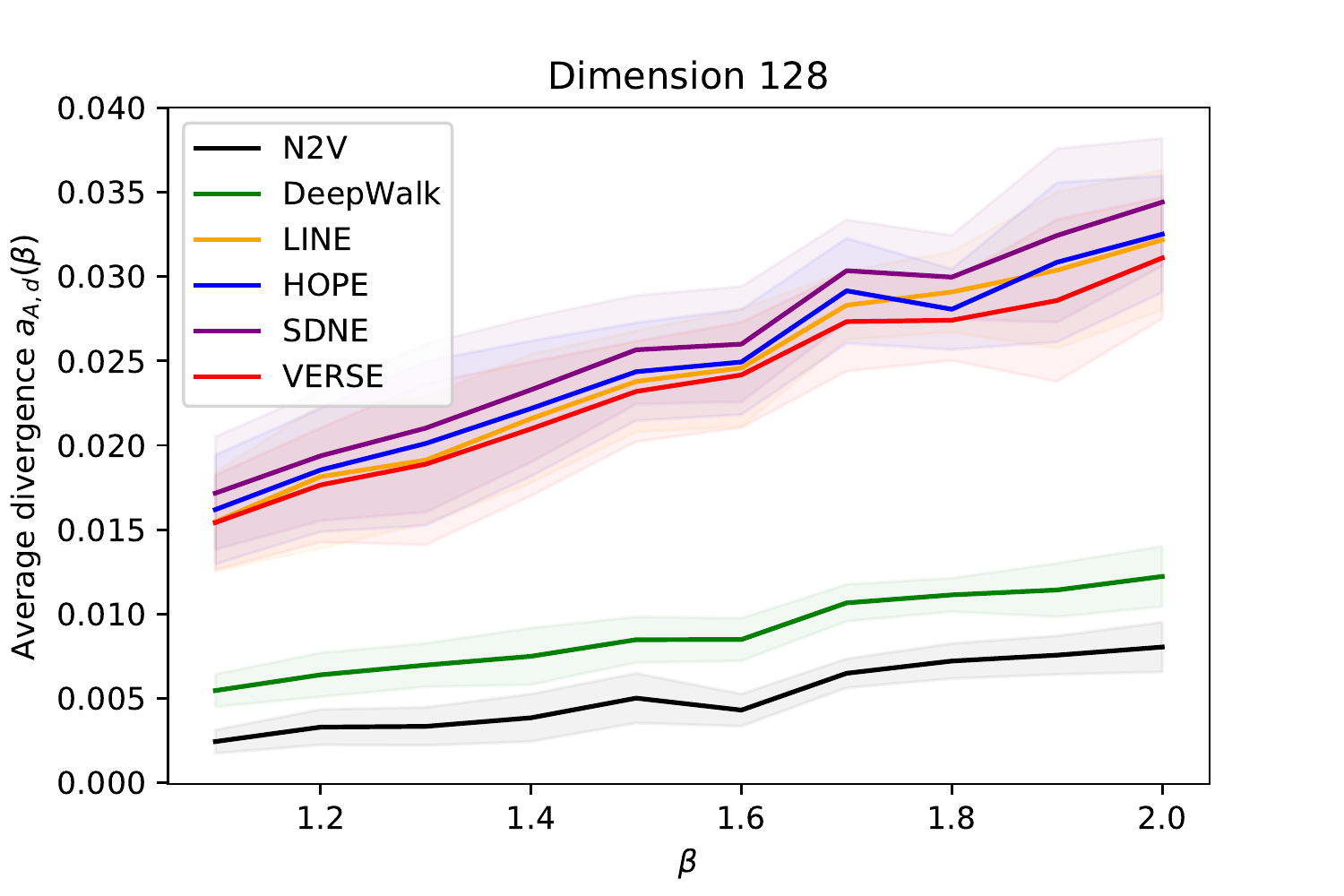} \\
Plots 1 and  2: $a_{A,d}(\beta) \pm s_{A,d}(\beta)$ 

\includegraphics[width=0.2\textwidth]{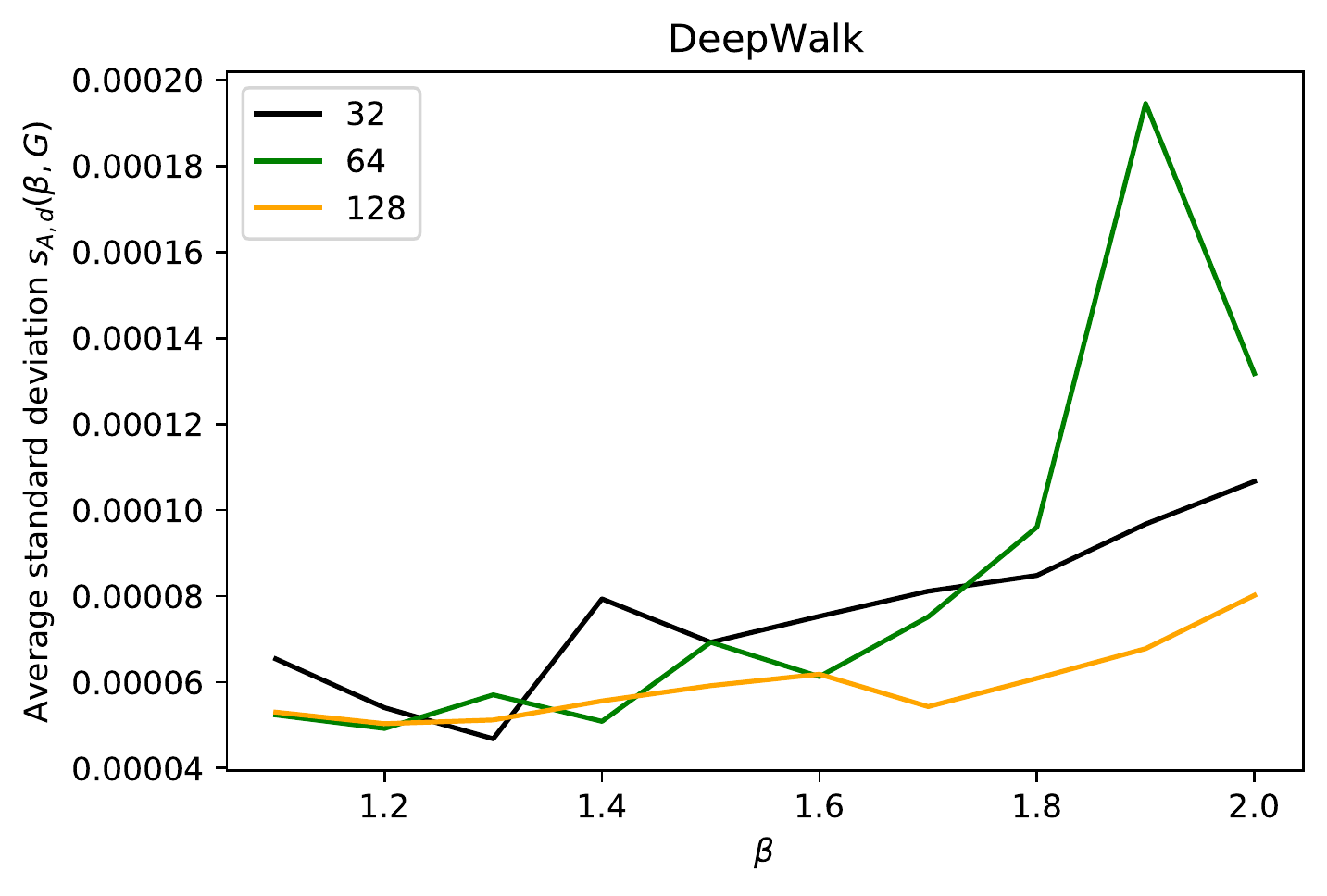}
\includegraphics[width=0.2\textwidth]{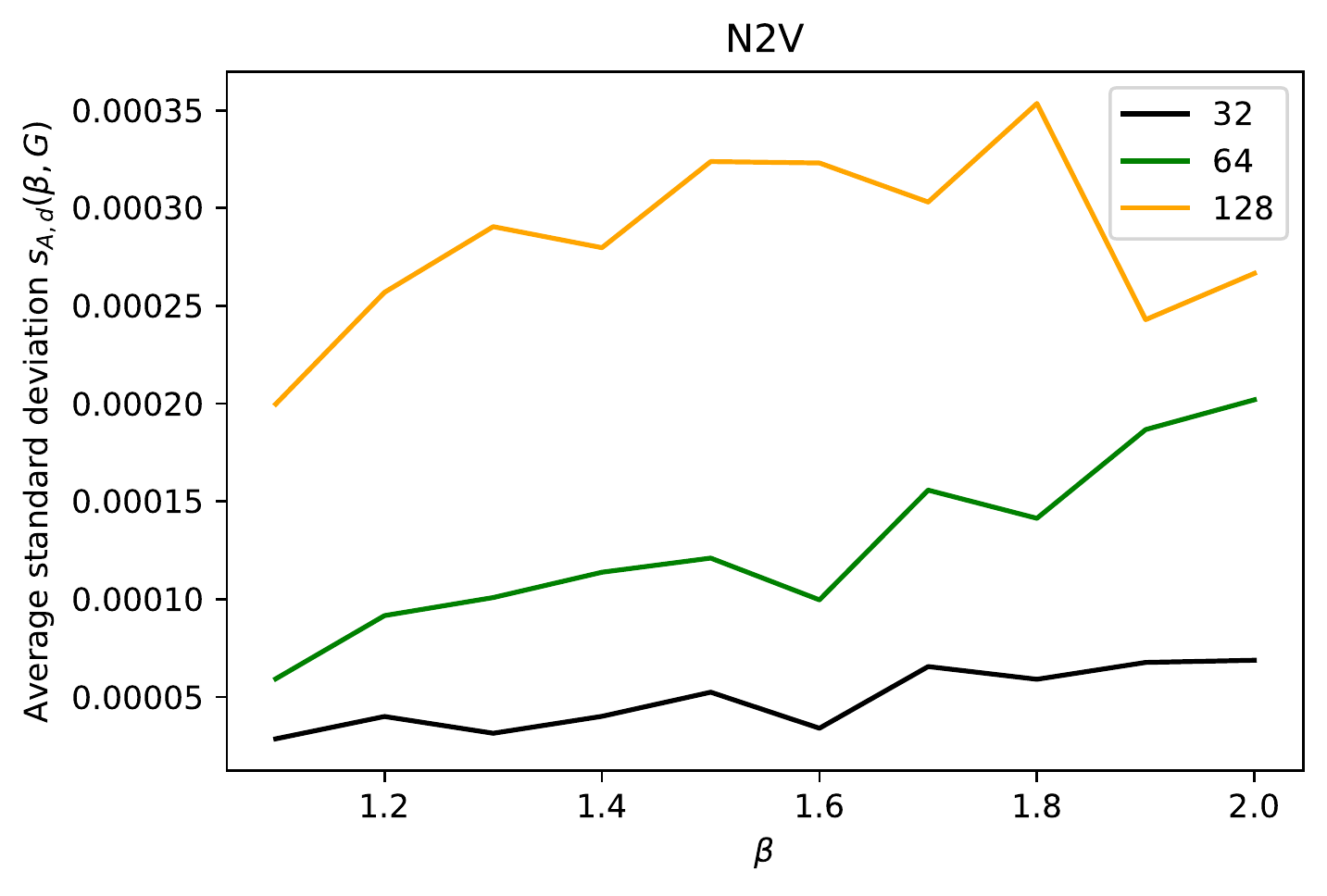}
\includegraphics[width=0.2\textwidth]{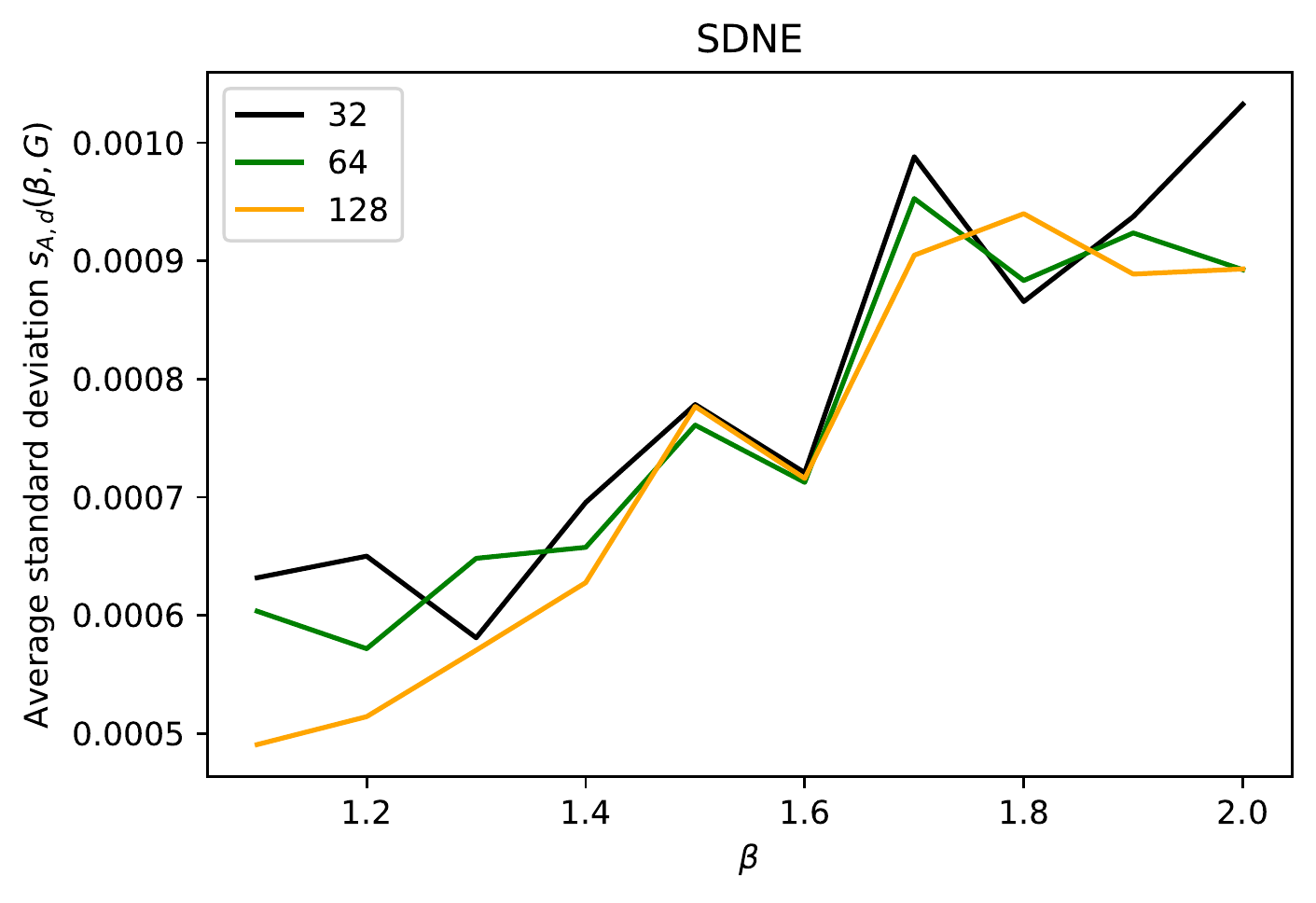}
\includegraphics[width=0.2\textwidth]{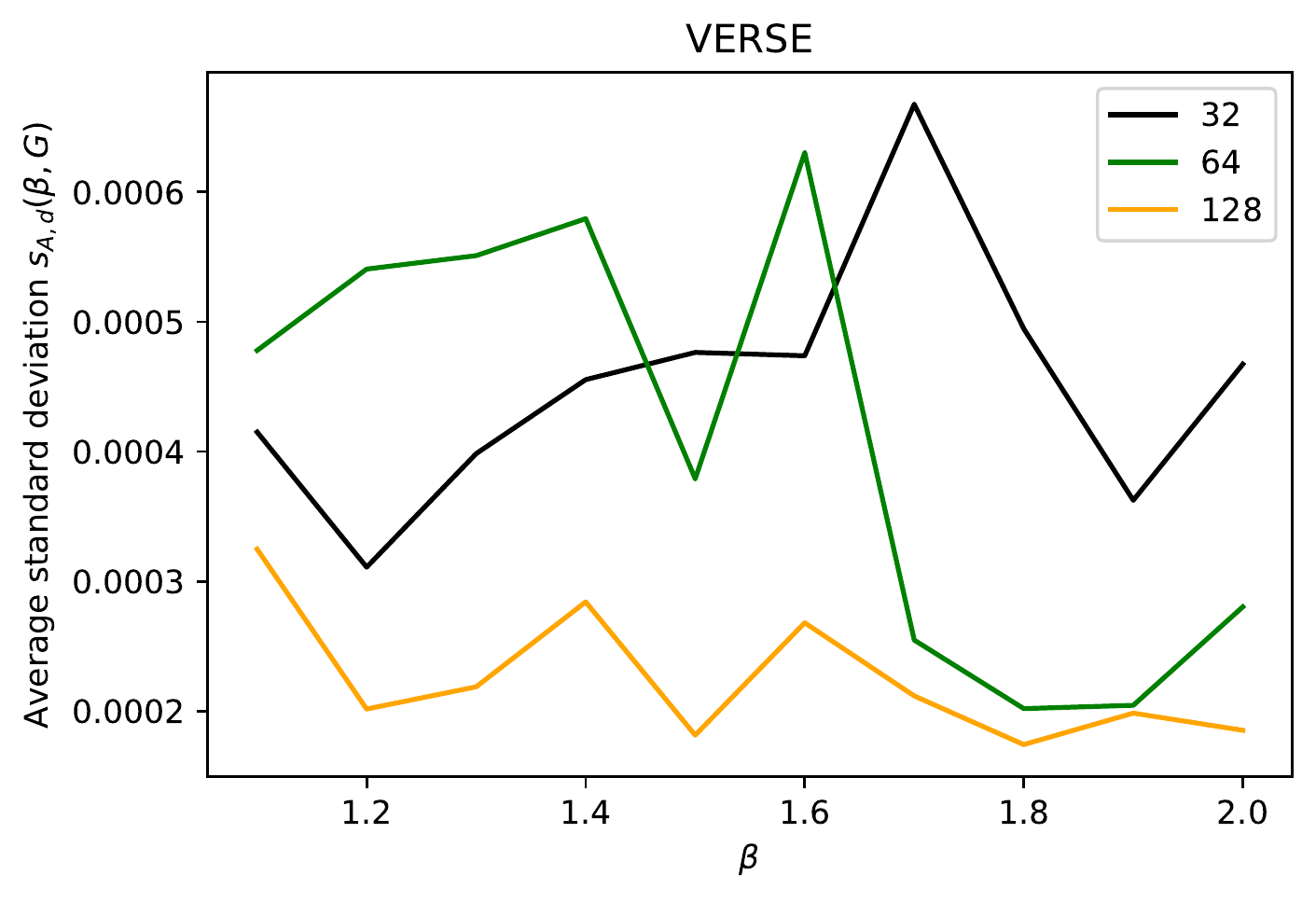} \\
\includegraphics[width=0.3\textwidth]{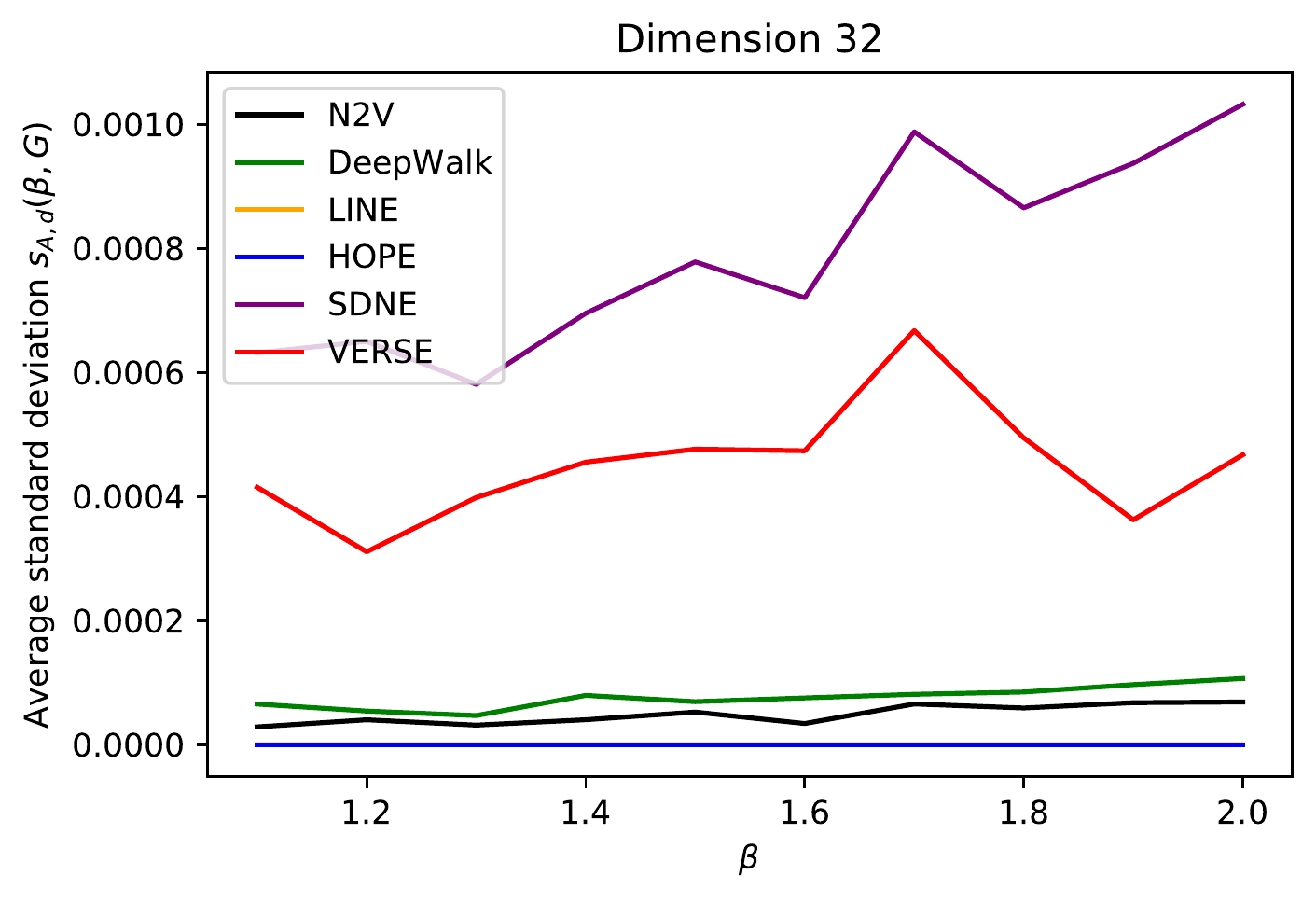}
\includegraphics[width=0.3\textwidth]{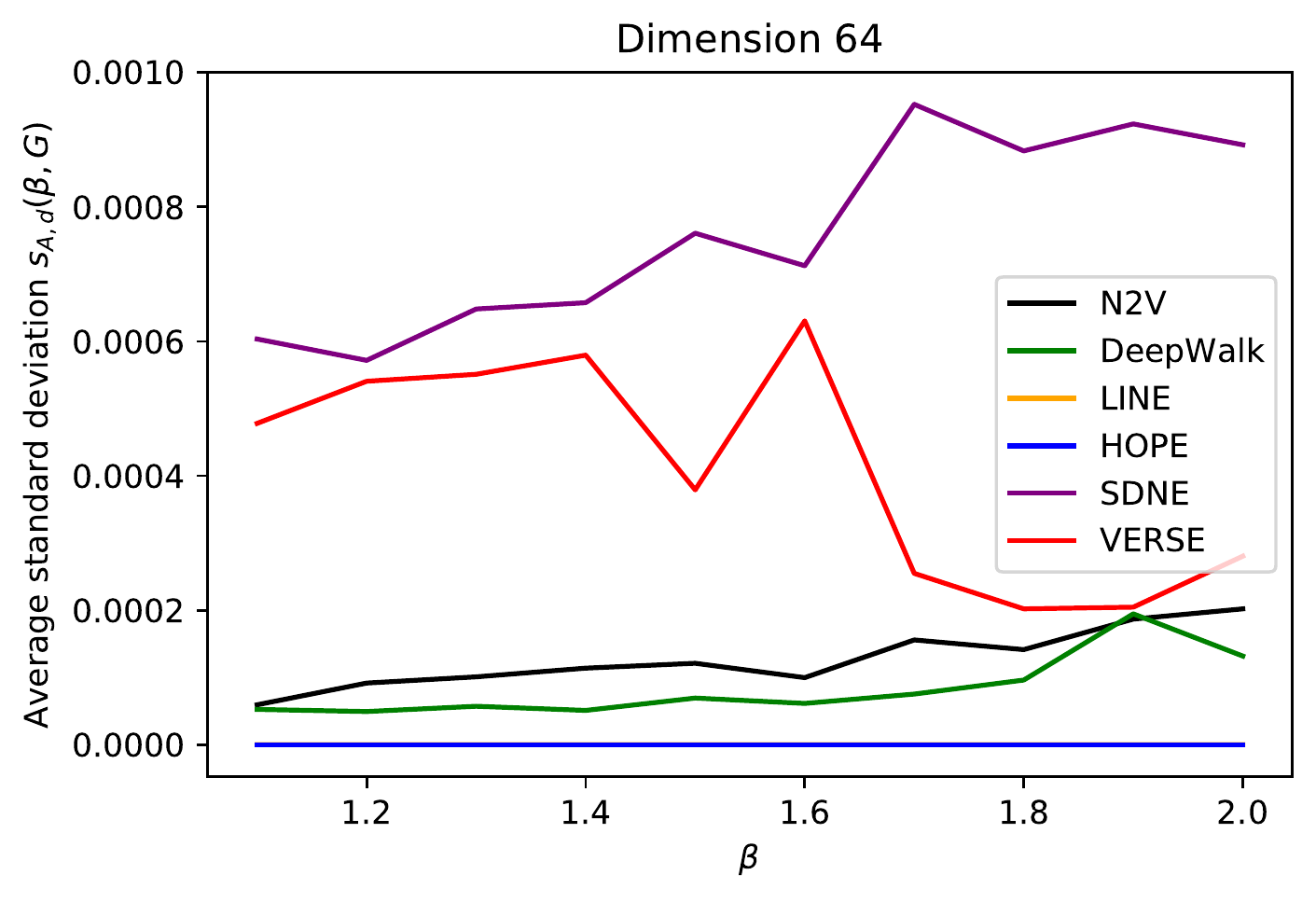}
\includegraphics[width=0.3\textwidth]{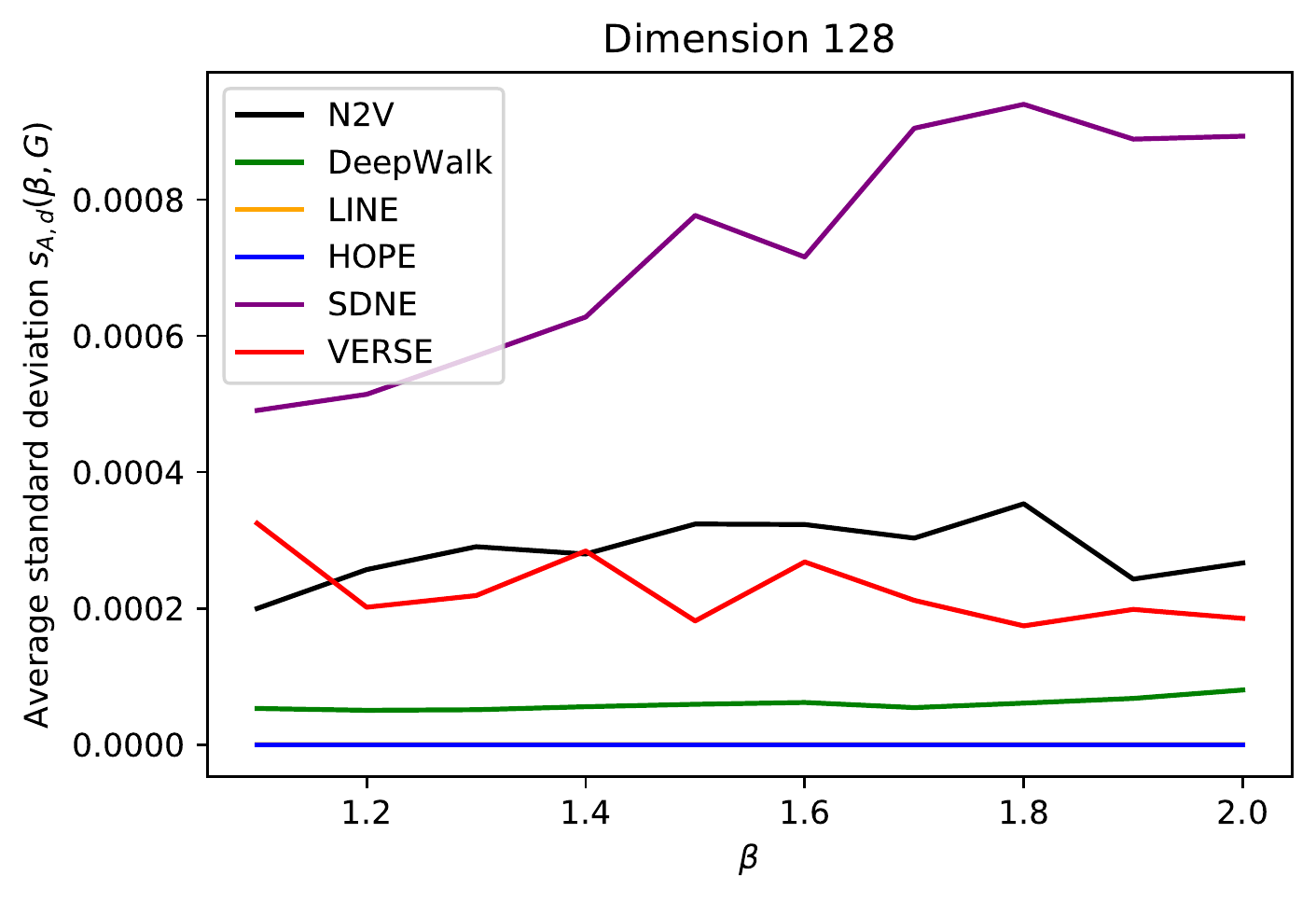} \\
Plots 3 and 4: average $s_{A,d}(\beta, G)$ (over 10 graphs)

\includegraphics[width=0.2\textwidth]{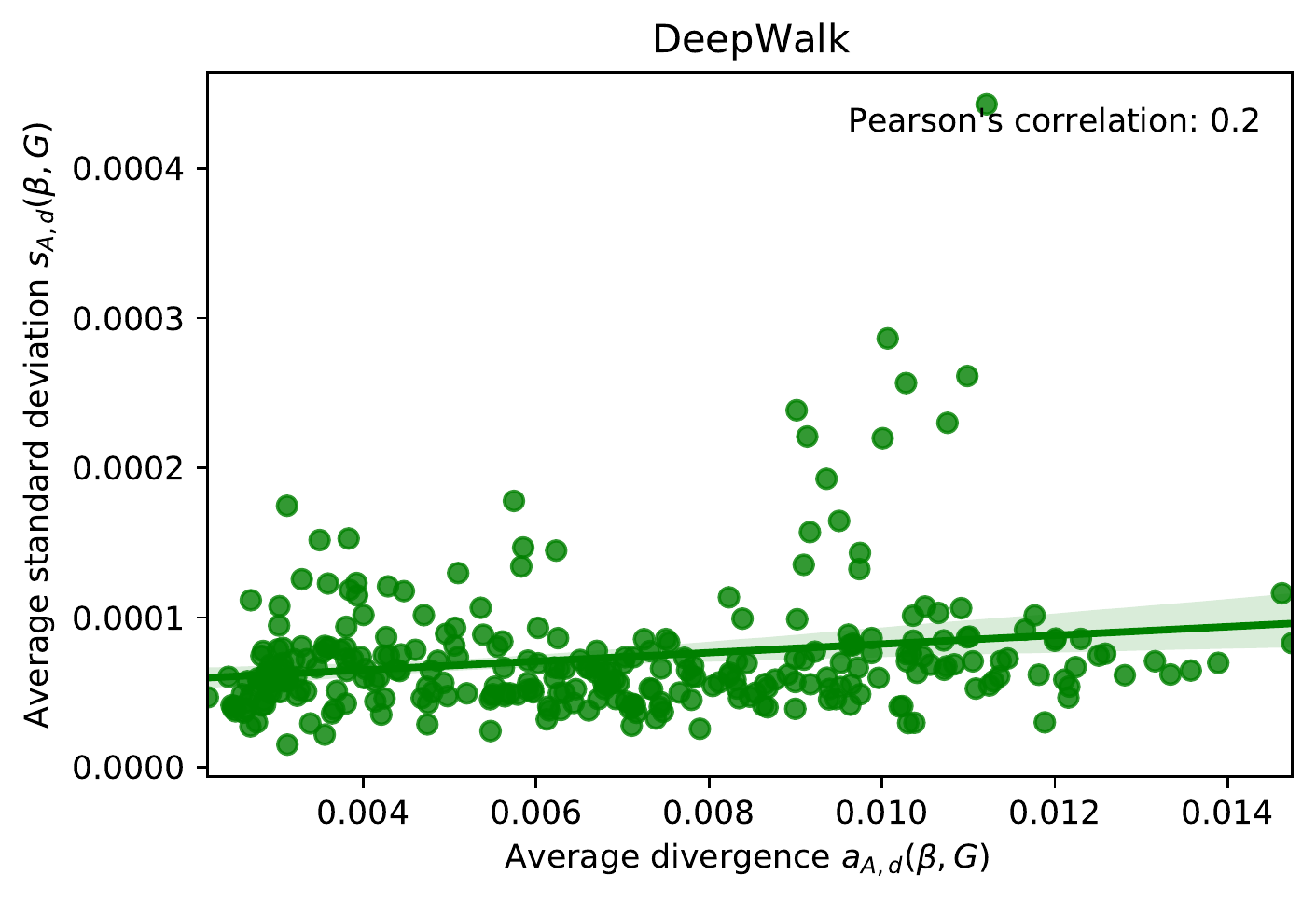}
\includegraphics[width=0.2\textwidth]{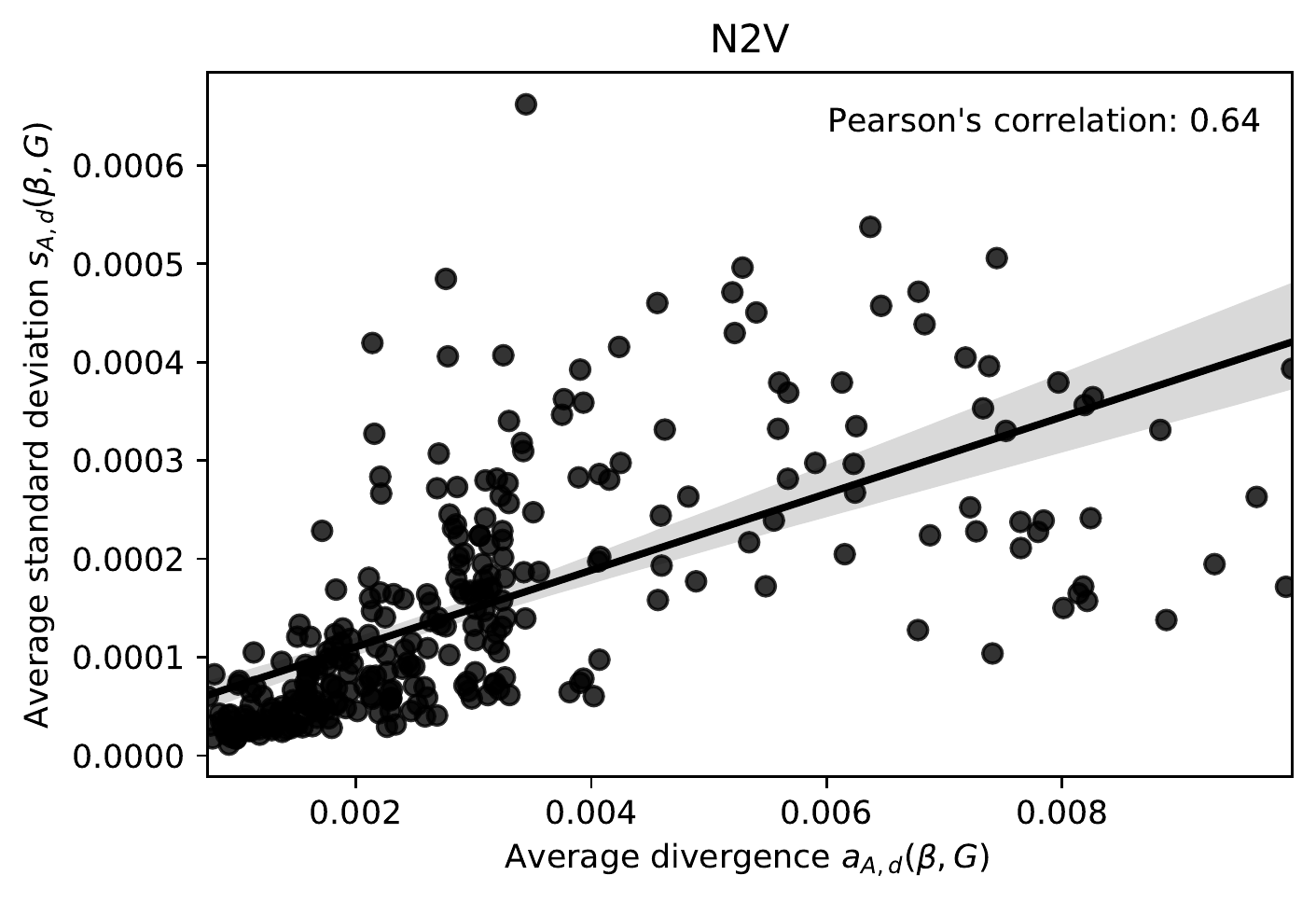}
\includegraphics[width=0.2\textwidth]{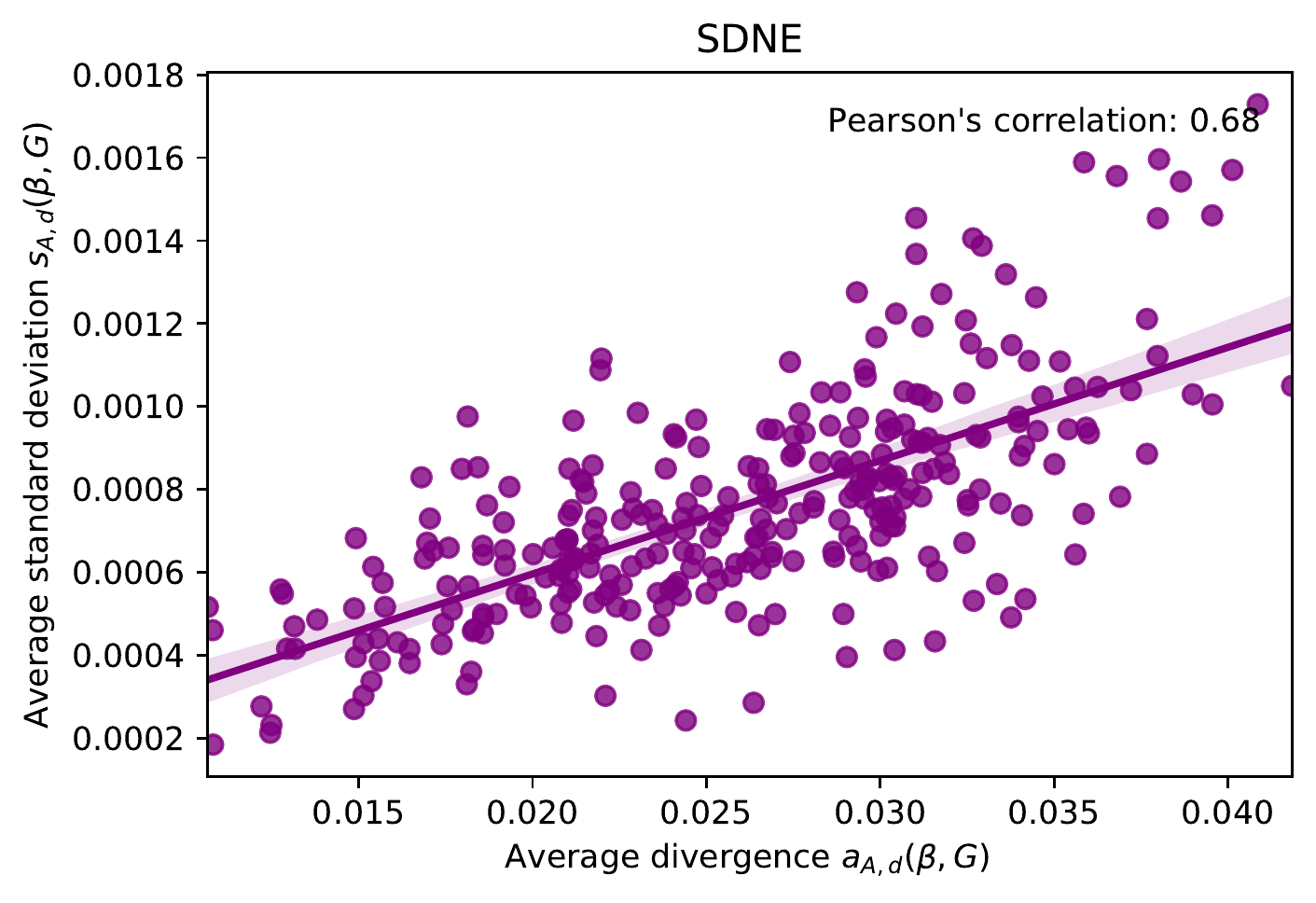}
\includegraphics[width=0.2\textwidth]{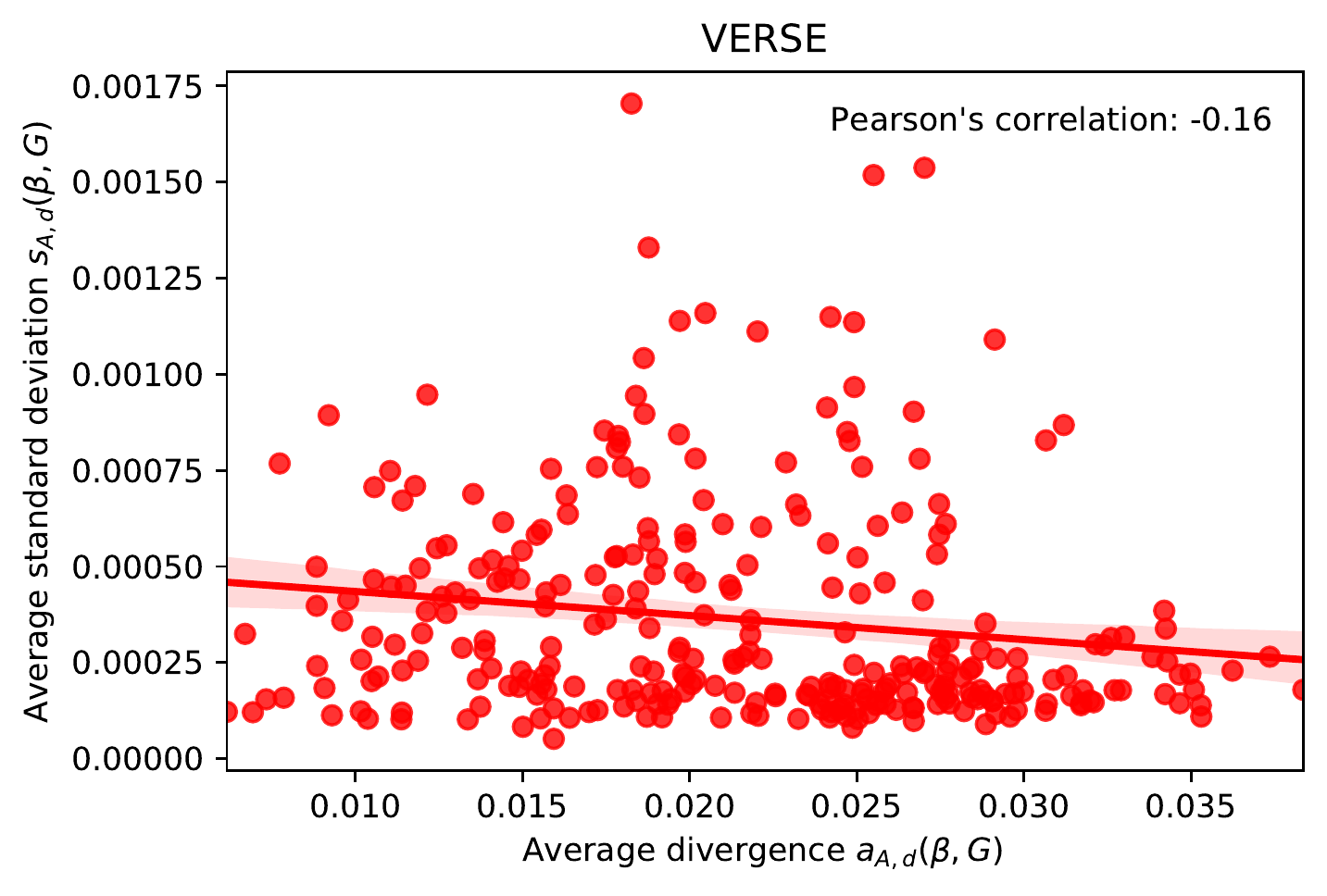} \\
Plot 5: correlation between $a_{A,d}(\beta, G)$ and $s_{A,d}(\beta, G)$
\caption{Community Sizes ($\beta$)}\label{fig:beta}
\end{center}
\end{figure}

In this section we present the following visualizations of experiments:
\begin{enumerate}
    \item Figure~\ref{fig:n} shows a comparison of embedding algorithms as a function of graph size $n$.
    \item Figure~\ref{fig:gamma} shows a comparison of embedding algorithms as a function of degree distribution parameter $\gamma$.
    \item Figure~\ref{fig:delta} shows a comparison of embedding algorithms as a function of the maximum degree $\Delta$.
    \item Figure~\ref{fig:xi} shows a comparison of embedding algorithms as a function of the level of noise $\xi$.
    \item Figure~\ref{fig:beta} shows a comparison of embedding algorithms as a function of community sizes distribution parameter $\beta$.
\end{enumerate}

\end{document}